%% file: main.tex
\documentclass[a4paper,11pt]{article}

\input{preamble.tex}

\title{Higher anomalies, compressing SPTs and cohomology operations}

\author[a]{Shani Nadir Meynet}
\author[b]{and Elias Riedel Gårding}
\affiliation[a]{Department of Physics and Astronomy, University of Pennsylvania,\\
209 South 33rd Street, Philadelphia, PA 19104, USA}
\affiliation[b]{Department of Physics, Massachusetts Institute of Technology, \\
  77 Massachusetts Avenue, Cambridge, MA 02139, USA}
\emailAdd{smeynet@sas.upenn.edu}
\emailAdd{eliasrg@mit.edu}

\abstract{
  We explicitly compute and tabulate the suspension map for cohomology operations, applying it to the dimensional reduction of anomalies and higher group structures of generalised symmetries in quantum field theories and lattice systems.

  In a quantum system with a discrete $p$-form symmetry, we can restrict the symmetry defects to a codimension-$q$ subspace of spacetime, or equivalently confine the gauge field to a slab whose thickness approaches zero. This results in a $(p-q)$-form symmetry on the subspace. We determine the fate of two important properties of symmetries under this process. First, we compute the anomaly of the reduced symmetry, or the higher anomaly of the original symmetry, which acts as an obstruction to higher gauging and onsiteability. Second, for two symmetries forming a higher-group structure, encoded as a Postnikov class, we compute the reduced higher-group structure, which may or may not be trivial. Both cases are captured by the $q$-fold iteration of the cohomology suspension $\Omega$, also known as the loop functor or transgression, in the cohomology of Eilenberg--MacLane spaces. Classical results state that $\Omega$ is an isomorphism in a stable range and annihilates mixed anomalies.

  To calculate $\Omega$ over a base ring which is not a field, we make use of small models for the chain DGAs of Eilenberg--MacLane spaces, pioneered by Cartan and Moore, whose results we review and extend. The critical added step is the improvement of a mod $p$ chain contraction to the $p$-local integers, obtained via a $p$-adic series. We automate the computations in a software package \texttt{emcm}, and provide extensive tables of the results. Finally, to make contact with explicit cochain formulations of cohomology operations, we demonstrate how to calculate cup products and certain cup-$i$ products in this framework.
}

\begin{document}
\maketitle
\flushbottom

\clearpage
\section{Introduction, conclusion and outlook}

\subsection{Cohomology theories and cohomology operations in quantum physics}
Cohomology theories are a fundamental tool of algebraic topology and its applications to theoretical physics. A cohomology theory $h$ assigns homotopy invariants $h^n(X)$ to a topological space $X$. In quantum field theory, the most familiar example is ordinary cohomology $h^n(X) = H^n(X; G)$ where $G$ is an abelian group. In the context of higher-form symmetries \cite{Gaiotto:2014kfa}, the cohomology group $H^{p+1}(X; G)$ corresponds to the set of flat background/gauge fields for an anomaly-free $p$-form symmetry $G$. But $h$ need not be restricted to ordinary cohomology; replacing it with a \emph{generalised cohomology theory} provides a powerful framework for classifying refined topological phenomena across theoretical physics. For $G = \UU(1)$, as relevant for electromagnetism and its higher-form cousins, ordinary cohomology can be replaced with differential cohomology \cite{cheegerDifferentialCharactersGeometric1985,Hopkins:2002rd}\footnote{See \cite{Sharpe:2015mja, McGreevy:2022oyu, Freed:2022iao, Gomes:2023ahz, Schafer-Nameki:2023jdn, Brennan:2023mmt, Bhardwaj:2023kri, Shao:2023gho, Carqueville:2023jhb, Costa:2024wks} for reviews.} to allow for non-flat gauge fields. Other prominent examples include cobordism theories, which classify invertible field theories and symmetry-protected topological (SPT) phases of matter beyond group cohomology \cite{kapustin2014symmetryprotectedtopologicalphases, Kapustin_2014, Chen_2013, Gu_2014, Chen_2021}, while also playing a central role in quantum gravity through the swampland cobordism conjecture \cite{McNamara:2019rup}. Furthermore, cohomotopy theory has emerged as an appropriate generalised refinement for certain applications to M-theory \cite{Sati:2013rxa,Grady:2020yhc,Sati:2025vjw} and abelian anyons \cite{Sati:2024drl,Sati:2025mln}. Similarly, specialised generalised cohomology theories have recently been formulated to characterise the topological structure of Quantum Cellular Automata (QCAs) \cite{Tu:2025bqf,Czajka:2025mme}.

According to Brown's representability theorem \cite{brownCohomologyTheories1962}, cohomology theories can be represented by spectra: The invariant $h^n(X)$ can be expressed as the set $[X, E^n]$ of homotopy classes of (pointed) maps $X \lto E^n$ into some classifying space $E^n$. The spaces $E^n$ form an $\Omega$-spectrum, meaning that $\Omega E^{n+1} \simeq E^n$ where $\Omega$ denotes the loop space. The most basic instance of this fact is that ordinary cohomology is represented by Eilenberg--MacLane spaces: $H^h(X; G) \cong [X, B^h G]$ where $B^h G = K(G, h)$ is a space such that $\pi_h(B^h G) \cong G$ and $\pi_n(B^h G) = 0$ for $n \neq h$ (such a space is unique up to homotopy). Under this isomorphism, a map $A\colon X \lto B^h G$ is identified with the pullback $A^* \iota_h \in H^h(X; G)$ of the fundamental class $\iota_h \in H^h(B^h G; G) \cong \Hom(G, G)$ given by the identity map of $G$. On the cocycle level, gauge transformations $a \lto a + \dd\lambda$ of elements $a \in Z^h(X; G)$ (flat background fields for a $(h-1)$-form $G$ symmetry) correspond abstractly to homotopies of the map $A$. This is made concrete in certain explicit models of $B^hG$, such as the $K(G, h)$ complex of \textcite{eilenbergGroupsHP1953}, where simplicial cocycles $a$ directly correspond to simplicial maps $A$ \cite{maySimplicialObjectsAlgebraic1967}.\footnote{See \cite{Kapustin:2014gua, Kapustin:2014zva} for a review of these concepts in physics literature.}

The viewpoint of classifying spaces provides a way to study \emph{cohomology operations}, which are natural transformations $h_1^m(\blank) \lto h_2^n(\blank)$ between cohomology theories. The Yoneda lemma gives the natural isomorphism
\begin{equation}
  \Nat(h_1^m, h_2^n) \cong \Nat([\blank, E_1^m], [\blank, E_2^n])
  \cong [E_1^m, E_2^n] \cong h_2^n(E_1^m).
\end{equation}
This reduces the classification of cohomology operations $h_1 \to h_2$, which may at first seem like an abstract and intractable problem, to the calculation of the $h_2$ cohomology groups of the particular spaces representing $h_1$. Physically, cohomology operations appear in the classification of SPT phases \cite{Gu_2009, Pollmann_2012, Chen_2011, Senthil_2015} as universal actions: An SPT action for a $p$-form $G$ symmetry in $D$ spacetime dimensions is roughly speaking an (anomaly-free) assignment of a $\UU(1)$-valued Lagrangian $D$-cocycle to each background field configuration $A$, i.e.~a natural transformation $H^{p+1}(\blank; G) \lto H^D(\blank; \UU(1))$, and these are classified by the group $H^D(B^{p+1} G; \UU(1))$. This cohomological classification, however, is not the full story. In general, obtaining a complete classification of $G^{(p)}$-SPTs remains an active area of research, as ordinary group cohomology fails to capture distinct invertible phases of matter that exist even in the absence of symmetry. These phenomena give rise to ``beyond-cohomology'' SPTs \cite{Chen_2013, kapustin2014symmetryprotectedtopologicalphases,Gu_2014, Chen_2021}. To properly incorporate these phases, the ordinary cohomology functor $H^{\bullet}(\blank; \UU(1))$ must be replaced by an appropriate generalised cohomology theory $h^\bullet(\blank)$. In this refined framework, the invertible topological phases without symmetry are classified by the value $h^D(\mathrm{pt})$ at a point, while the full set of SPT phases for a symmetry $G$ is classified by the generalised cohomology group $h^D(B^{p+1} G)$. The modern understanding is that $h$ should be a cobordism theory (or more precisely, its Anderson dual) respecting the underlying tangential structure of the spacetime manifold, such as spin or pin cobordism for fermionic systems \cite{Lu:2012dt,Gu_2014,Freed:2016rqq,Kapustin:2017jrc,Gaiotto:2017zba}.

As is well known, invertible field theories or SPTs are also related to anomalous symmetries via anomaly inflow \cite{Wen_2013, Kapustin_2014, three_lectures_2016}. A $G^{(p)}$ symmetry in a $D$-dimensional QFT is anomalous if the partition function in the presence of background fields fails to be invariant under background gauge transformations, instead changing by a phase: $\cZ[X, a + \dd\lambda] = \e^{\ii \oint_X \varphi(a, \lambda)} \cZ[X,a]$. This anomalous phase can be cancelled by the boundary variation of an SPT system in $D+1$ dimensions, such that the bulk-boundary system is anomaly-free. Then, the possible anomalies are classified by $h^{D+1}(B^{p+1} G)$, or $H^{D+1}(B^{p+1} G; \R/\Z)$ if we disregard beyond-cohomology SPTs.\footnote{In the remainder of the paper, we write $\UU(1)$ as the additive group $\R/\Z$.} The anomaly of a symmetry places severe constraints on its realisation in the ground state of a theory, so it is a crucial step in characterising the theory's infrared behaviour \cite{tHooft:1979rat, harvey2005tasi2003lecturesanomalies, Witten_2016, Jian_2018}.

During the last century, many mathematicians approached the problem of computing the groups $H^D(B^h G; A)$: Firstly, the problem was approached by geometrical means, computing the simplicial cohomology of an explicit realisation of $B^h G$, such as the simplicial model $K(G, h)$ or the bar construction \cite{eilenbergGroupsHP1953}, but the computational complexity grows quickly. Using a more algebraic approach, the first efficient algorithmic calculation was described by Henri Cartan~\cite{doi:10.1073/pnas.40.6.467,doi:10.1073/pnas.40.8.704,cartanAlgebreDEilenbergMacLane1954}, but the results do not seem to be very well-known, at least to physicists, and tables are not readily available.\footnote{This may be because the original seminar \cite{cartanAlgebreDEilenbergMacLane1954} is a lengthy exposition in French with few examples.} \Textcite{Clement2002} performed Cartan's calculation explicitly, but only for $G = \Z_{2^\ell}$. Most other calculations that have appeared are based on the Serre spectral sequence, which is rather laborious and prone to errors. For instance, for $G = \Z$, the calculations by Percy and Jaleel \cite{Percy2010,Jaleel2017} seem to have missed a factor $\Z_4$ in $H^{13}(B^4 \Z; \Z)$\footnote{The first integral cohomology operation whose order is a nontrivial prime power; see \zcref{tab:cohomology-BhZ-Z}.} and a corresponding factor $\Z_2$ in $H^{14}(B^5 \Z; \Z)$. Another preprint \cite{salehiCohomologyGroupKZ42021} presents results for the cohomology of $B^4 \Z$ and $B^5 \Z$ which seem quite implausible throughout. As part of this work, we provide the software \texttt{emcm} \cite{riedelgardingEmcm2026}, written in Lean 4 \cite{demouraLean4Theorem2021}, which computes these cohomology groups reliably and efficiently.\footnote{Lean is known for its theorem-proving capabilities, but used here simply as a practical functional programming language. Concerning efficiency, \texttt{emcm} calculates
\begin{equation}
  H^{1000}(B^3 \Z_{12}; \Z) \cong
  (\Z_2)^{5533159188918052298} \oplus (\Z_3)^{174455060698} \oplus (\Z_4)^{935369} \oplus (\Z_9)^{3}
\end{equation}
in around twenty seconds, despite not being extensively optimised.} We hope that this tool will serve as a useful reference for mathematicians as well as physicists. Tables of its output are given in \zcref{app:tables}.

\subsection{Higher anomalies, compressing SPTs and cohomology suspension}
\label{sec:intro-higher-anomalies}
\Textcite{Roumpedakis:2022aik} introduced the notion of higher gauging of higher-form symmetries. To $q$-gauge a $p$-form symmetry (for $0 \leq q \leq p + 1$) means to gauge it, not in the whole spacetime, but only in a subspace of codimension $q$. Focusing on a discrete symmetry group $G$, we can break the process into two steps: First, we obtain a $(p - q)$-form symmetry in the $(D - q)$-dimensional subspace by viewing its background field as a mesh of symmetry defects of dimension $D - (p + 1) = (D - q) - (p - q + 1)$, and restricting to only those backgrounds for which the symmetry defects lie inside the subspace. Second, we gauge this restricted symmetry as usual. We will focus on the first step. The anomaly $\cA \in h^{D+1}(B^{p+1} G)$ of the original symmetry must somehow determine the anomaly of the restricted symmetry, known as the $q$-anomaly, an element of $h^{D-q+1}(B^{p-q+1} G)$, so there must exist a \emph{higher anomaly map}
\begin{equation}
  \label{eq:higher-anomaly-map}
  \begin{tikzcd}
    h^{D+1}(B^{p+1} G) \ar[r] & h^{D-q+1}(B^{p-q+1}G).
  \end{tikzcd}
\end{equation}
As we will argue carefully in \zcref{sec:compressing-suspension}, this map is the \emph{cohomology suspension} or \emph{loop functor}\footnote{Also sometimes formulated as the ``transgression'' on the cochain level, argued to be the higher anomaly map in ref.~\cite{Feng:2025yge}. Another equivalent formulation was identified with the higher anomaly map in \cite{Bah:2025oxi}.} $\Omega$ iterated $q$ times. We will refer to $\Omega$ as the \emph{cosuspension} for short. The explicit calculation of $\Omega$, for ordinary cohomology $h = H^{\bullet}(\blank; A)$ (with $A = \R/\Z$ the relevant case for anomalies), is the main mathematical problem solved in this work, making use of the extension of Cartan's methods by \textcite{doi:10.1073/pnas.43.5.409}. The calculation is done automatically by \texttt{emcm} \cite{riedelgardingEmcm2026}. Moore's paper is a terse statement of results without explanations, and extracting $\Omega$ from his results requires significant effort. We describe the calculation in detail in \zcref{sec:computing,app:cartan-moore} and include extensive tables in \zcref{app:tables}.

Certain classical properties of $\Omega$ can be derived without a full calculation of the map, and impose severe constraints on the fate of anomalies under higher gauging. Notably, all mixed anomalies—which physically stem from the mutual linking of distinct symmetry defects—are systematically annihilated by $\Omega$. This algebraic vanishing reflects the geometric fact that non-trivial defect linking cannot be sustained when restricted to positive codimension. Furthermore, within the stable range (where the spacetime dimension satisfies $D < 2p$), the cosuspension map is an isomorphism (the stable groups are listed in \zcref{tab:stable-homology}). In this regime, an anomalous $p$-form symmetry is necessarily an anomalous $(p-k)$-form symmetry for all valid $k$; consequently, rather than merely constraining restricted sub-systems, the higher anomaly map faithfully preserves the complete anomaly data of the bulk theory across codimensions.

The most basic example of a higher anomaly, considered in \cite{Roumpedakis:2022aik} (see also \cite[app.~B]{Kaidi:2021xfk}), is a 3d theory with topological lines generating a $\Z_N$ 1-form symmetry. In a 2d surface in spacetime, the lines inside it have codimension $1$ and generate a $\Z_N$ 0-form symmetry. The anomaly in $H^4(B^2 \Z_N; \R/\Z) \cong \Z_{\gcd(2,N)N}$ is mapped via $\Omega$ to the 1-anomaly in $H^3(B\Z_N; \R/\Z) \cong \Z_N$ by the adjoint of multiplication by $N$ (so $\Omega$ is $0$ when $N$ is odd, and has order $2$ when $N$ is even). For $N = 2$ this means that only bosonic lines are free of 0-anomaly, bosonic and fermionic lines are free of 1-anomaly, and semionic or antisemionic lines are 1-anomalous. When the 1-anomaly vanishes, the symmetry can be 1-gauged, producing a two-dimensional condensation defect, or topological mesh, implementing a (possibly non-invertible) 0-form symmetry in the original theory. This procedure has been shown to reproduce the whole group of automorphisms for various 3d theories, providing evidence that in topological theories, the topological operators are either anyonic lines or obtained from them by higher gauging. Returning to the $N = 2$ case, a remarkable consequence of this analysis is that theories with semionic lines do not have a well-defined charge conjugation operator.

As can be seen from this simple example, controlling the cohomology suspension map allows one to constrain the set of allowed operators in a topological theory. This becomes of high relevance when trying to use topological field theories to construct a quantum computer. Recently, it has been shown \cite{Kobayashi:2025cfh, Hsin:2025zgn, Barbar:2025krh} that it is possible to reproduce the algebra of Clifford gates via the automorphism group of certain topological theories. From \cite{Roumpedakis:2022aik} we know that these operators can be constructed via higher gauging and therefore they are constrained by the possibility of performing such gauging.

Analogues of the 3d phenomenon above occur in other dimensions. Consider theories with a $G = \Z_N$ $p$-form symmetry, and let us focus first on spacetime dimension $2p+1$. This case occurs when the charged objects are the holonomies of a $p$-form gauge field in a Chern--Simons type theory. The possible group-cohomology anomalies are classified by $H^{2p+2}(B^{p+1} \Z_N; \R/\Z) \cong H_{2p+2}(B^{p+1} \Z_N)$. This group lies just outside the stable range, but the difference from the stable groups $H^s_{p+1}(\Z_N)$ can be quantified. Using Cartan's formalism, it is not hard to see that $H_{2p+2}(B^{p+1} \Z_N) \cong H^s_{p+1}(\Z_N)$ for even $p$, and that for odd $p$, $H_{2p+2}(B^{p+1} \Z_N)$ is obtained from $H^s_{p+1}(\Z_N)$ by replacing one factor of $G/2G = \Z_{\gcd(2,N)}$ with $\Z_{\gcd(2,N)N}$. This factor is generated by the cup square for odd $N$, and the Pontryagin square for even $N$. All the other possible anomalies are given by stable cohomology operations. While our language expresses these as the dual of certain homology classes (described in detail in \zcref{sec:general-Omega}), it is possible to identify them with expressions in the standard Steenrod algebra \cite[sec.~4.L]{Hatcher2002} (which \texttt{emcm} also has a facility to generate). The facts that the coefficient change $H^n(B^h G; \Z_p) \lto H^n(B^h G; \R/\Z)$ can be identified with the integral Bockstein $\beta_\Z\colon H^n(B^h G; \Z_p) \lto H^{n+1}(B^h G; \Z)$ and that $\beta_\Z \circ \Sq^1 = \beta_\Z \circ \beta = 0$ (where $\beta$ denotes the Bockstein associated to $\Z_p \lto \Z_{p^2} \lto \Z_p$) are enough to determine the expressions for these anomalies and their cosuspensions, listed in \zcref{tab:cs-anomalies} for $0 \leq p \leq 5$. The cosuspension of the Pontryagin square $\mathcal{P}$ is the Postnikov square $\mathfrak{P}(B) = B \cupp \beta(B)$ \cite[thm.~2]{browderAxiomsGeneralizedPontryagin1962}. Exactly as in the three-dimensional case, a non-trivial image under $\Omega$ obstructs higher gauging and implies that the theory lacks a well-defined charge conjugation operator for its membrane defects.

\begin{table}[t]
  \[
  \begin{IEEEeqnarraybox}[][c]{rClrClCs}
  0 &\cong& H^2(B\Z_N; \R/\Z) \\
  \Z_{(2,N)N} &\cong& H^4(B^2\Z_N; \R/\Z) \colon & \frac{1}{N} B \cupp B &\longmapsto& 0 \\
  &&& \frac{1}{2N} \mathcal{P}(B) &\longmapsto& \frac{1}{2} A^3 &\quad& ($2 \mid N$) \\
  \Z_{(2,N)} &\cong& H^6(B^3\Z_N; \R/\Z) \colon & \frac{1}{2} \Sq^2 \Sq^1\! B &\longmapsto& \frac{1}{2} \Sq^2 \Sq^1\! A &\quad& ($2 \mid N$) \\
  \Z_{(2,N)N} \oplus \Z_{(3,N)} &\cong& H^8(B^4\Z_N; \R/\Z) \colon & \frac{1}{N} B \cupp B &\longmapsto& 0 \\
  &&& \frac{1}{2N} \mathcal{P}(B) &\longmapsto& \frac{1}{2} A \cupp \Sq^1\! A &\quad& ($2 \mid N$) \\
  &&& \frac{1}{3} P^1\! B &\longmapsto& \frac{1}{3} P^1\! A &\quad& ($3 \mid N$) \\
  \Z_{(2,N)} \oplus \Z_{(3,N)} &\cong& H^{10}(B^5\Z_N; \R/\Z) \colon & \frac{1}{2} \Sq^4 \Sq^1\! B &\longmapsto& \frac{1}{2} \Sq^4 \Sq^1\! A &\quad& ($2 \mid N$) \\
  &&& \frac{1}{3} P^1 \beta B &\longmapsto& \frac{1}{3} P^1 \beta A &\quad& ($3 \mid N$) \\
  \Z_{(2,N)N} \oplus \Z_{(2,N)} &\cong& H^{12}(B^6\Z_N; \R/\Z) \colon & \frac{1}{N} B \cupp B &\longmapsto& 0 \\
  &&& \frac{1}{2N} \mathcal{P}(B) &\longmapsto& \frac{1}{2} A \cupp \Sq^1\! A &\quad& ($2 \mid N$) \\
  &&& \frac{1}{2} \Sq^4 \Sq^2\! B &\longmapsto& \frac{1}{2} \Sq^4 \Sq^2\! A &\quad& ($2 \mid N$)
  \end{IEEEeqnarraybox}
  \]
  \caption{The cohomology operations relevant as anomalies for $\Z_N$ Chern--Simons theories, and their fate under cosuspension. Operations are written in terms of $\Z_N$ valued cocycles $B^{(p+1)}$ and $A^{(p)}$, and the denominator is the order of the anomaly.}
  \label{tab:cs-anomalies}
\end{table}

Another application of higher anomalies is in the onsiteability of higher-form symmetry operators on the lattice. It has recently been understood that, in certain cases, onsiteability is equivalent to being free of 1-anomaly \cite{Feng:2025yge,Seifnashri:2025vhf,Kawagoe:2025ldx,Shirley:2025yji}. Onsite symmetries are the natural formulation of symmetry operators acting on local degrees of freedom for spin systems. This becomes relevant also in the context of quantum computing and error correction, since transversal single-qubit logic gates can be viewed as the quantum information theory analogue of onsite symmetries \cite{Hsin:2025zgn}.

In addition to anomalies, the cohomology of Eilenberg--MacLane spaces also classifies the statistics of extended excitations \cite{Feng:2025bww,Xue:2024opg,Kobayashi:2024dqj,Feng:2025mdg}. Here, the higher anomaly map $\Omega$ computes the statistics of excitations whose movement is restricted to codimension one \cite{Xue:2024opg,Feng:2025mdg}.

The way we see that the higher anomaly is given by $\Omega$ is from the inflow picture, ``compressing'' the $(D+1)$-dimensional SPT by placing it on a slab $X \times [0,1]$ with Dirichlet boundary conditions and shrinking its width to zero;\footnote{A less dry text than this one would phrase this as ``mashing the mesh'' of symmetry defects.} special cases of this were studied in \cite{Roumpedakis:2022aik,Choi:2022zal}. The process corresponds mathematically to the loop-suspension adjunction (see \zcref{sec:compressing-suspension}). The same argument applies to general topological $\sigma$-models, replacing $B^{p+1} G$ with some target space $M$: Placed on a slab with appropriate boundary conditions, the $(D+1)$-dimensional theory is equivalent to one in $D$ dimensions with target space $\Omega M$. This topological reduction highlights a natural physical application for the cohomology suspension $\Omega\colon H^{D+1}(B^{p+1} G; A) \lto H^D(B^p G; A)$ with coefficient groups $A$ other than $\R/\Z$: characterising the effective symmetry $\Omega B\bbG$ that emerges when a theory with higher-group symmetry $\bbG$ is compressed onto a slab.\footnote{For examples of such theories see \cite{Kapustin:2013uxa, Benini:2018reh, Choi:2022fgx, DelZotto:2024ngj}.} Because the extension structure of a higher group is encoded in its Postnikov invariants, compressing the theory maps these $k$-invariants under the cosuspension functor. In the simplest case of a two-stage higher group classified by a single Postnikov class $k$, the non-trivial extension trivialises on the slab—decoupling into a direct product of independent higher-form symmetries—precisely when $\Omega k = 0$. We will explore the physical consequences of this phenomenon in \zcref{sec:higher-groups}.

As an example, consider a 3-group symmetry where a $\Z_2$ 1-form symmetry is extended by a 2-form symmetry $A$, so that the extension is classified by a Postnikov invariant $k \in H^4(B^2 \Z_2; A)$. The behaviour of this higher group upon slab compression depends drastically on $A$. If $A = \Z_2$, the only nontrivial Postnikov class is given by the Steenrod square $k = \Sq^2(\iota_2)$. As seen in \zcref{tab:Omega-B2Z2-Z2}, its cosuspension is identically zero ($\Omega k = \Sq^2 \iota_1 = 0$; see also \zcref{app:Zp}), meaning that the higher-group structure trivialises on a slab and reduces to a split product $\Z_2^{(2)} \times \Z_2^{(1)}$. On the other hand, if the 2-form symmetry is extended to $A = \Z_4$, the Postnikov invariant can be chosen as the Pontryagin square $k = \mathcal{P}(\iota_2)$ which generates $H^4(B^2 \Z_2; \Z_4) \cong \Z_4$. Then, according to \zcref{tab:Omega-B2Z2-Z4}, $\Omega k$ does not vanish, but is the generator of $H^3(B\Z_2; \Z_4) \cong \Z_2$. Consequently, the higher-group structure does not trivialise on the slab, but rather descends to a non-trivial 2-group.

Calculating $\Omega$ with other coefficients turns out to require a more detailed understanding of Moore's work than needed for the $\R/\Z$ case; nevertheless, we will explain this and have included the capability in \texttt{emcm}.

The previous paragraph applies to dynamical gauge fields (or general $\sigma$-model fields) as well as to background fields. Our calculation of $\Omega$ therefore not only determines the compression (slab reduction) of SPTs, but also dynamical discrete gauge theories (Dijkgraaf--Witten theories). \Textcite{Dijkgraaf:1989pz} discussed the cohomology suspension $H^4(BG; \Z) \lto H^3(G; \Z)$ for a Lie group $G$, interpreting it as a map from Chern--Simons actions (DW twists) to Wess--Zumino--Witten actions. In contrast, we deal only with discrete $G$, and interpret $\Omega$ as mapping the DW twist before compression to that after compression.

\subsection{Structure of the paper}

In \zcref{sec:mathematics} we provide a review of the mathematics behind anomalies for higher gauging and their relation to the cosuspension map. We describe how backgrounds for symmetry operators of the bulk theory reduce on a codimension $q$ surface upon setting suitable boundary conditions. This operation turns out to be nothing but the suspension isomorphisms applied to the symmetry background.

In \zcref{sec:cup-products} we discuss cup products and cup-$i$ products. The Cartan--Moore constructions, despite being powerful, are somewhat abstract. To make a connection between our algebra and the more geometrical cochain constructions usually used in physics, we indicate how to compute the cup-$i$ product structure, often used to write Lagrangian actions, in our constructions. We give a complete algorithm to compute cup products, and demonstrate some examples. Cup-$i$ products turn out to be more difficult; we compute them only in $BG$; doing so in $B^2 G$ appears to require a machinery of homotopy coherencies which we do not delve into.

The main takeaway of this work is twofold: First, we review and extend the mathematical works of Cartan and Moore, applying the results to provide a complete description of higher gauging anomalies for any finitely generated abelian group, as well as a characterisation of higher group structures viewed in positive codimension. Second, we provide a software package, \texttt{emcm} \cite{riedelgardingEmcm2026}, capable of efficiently computing the (co)homology groups and (co)suspension maps in the Eilenberg--MacLane spaces associated to any finitely generated abelian group.

\subsection{Further directions}
We conclude this introduction by mentioning some further directions that this work can open up. From the mathematical point of view, the Cartan--Moore approach we will describe and use is elegant and computationally tractable, but it does not immediately translate into the formulations of cohomology operations, based on simplicial cochains (the $K(G, h)$ complex of \cite{eilenbergGroupsHP1953}) or---imperfectly---differential forms, that are popular in physics. In particular, the approach is ``homology-first'', based on the Pontryagin product instead of intrinsically cohomological operations like cup products. It is possible to calculate the cohomology cup product (with e.g.~$\Z$ coefficients) in our language, but the procedure is more complicated; we have outlined it in \zcref{sec:cup-products} and done the simplest case as an example, but we have not yet automated it in \texttt{emcm}. Using the same machinery \cite[exposé~4, thm.~5]{cartanAlgebreDEilenbergMacLane1954}, one can in principle construct a map translating simplicial cochains into our language, but we defer an implementation of this until the need arises.

Having implemented an efficient computational approach to the cohomology of Eilenberg--MacLane spaces, or the anomalies of ordinary higher-form symmetries, it is natural to wonder whether we can extend it to deal with more general spaces, classifying the anomalies of higher groups and computing their higher anomalies. Despite some recent results \cite{Xue:2026yqc,Gu:2026trc}, a systematic approach to the study of the cohomology of higher groups is still lacking. If such calculations could be done efficiently, they would be of great interest for constraining the low-energy physics of quantum systems with higher-group symmetry. The first step would be to algorithmically compute the cohomology of a two-stage Postnikov tower, classified by a single Postnikov class $k\colon B^n G \lto B^{m+1} A$ as in \eqref{eq:simple-higher-group}. In our language, it is straightforward to construct $k$ as a chain map given its representation as a class in $H^{m+1}(B^n G; A)$. It is less clear whether this information is enough to easily obtain a chain complex for the total space, or if not, whether it can be done with some small amount of extra data (e.g.~cup products), short of a full representation of $k$ as a map of simplicial sets. With coefficients in a field, it may be possible to use the Serre or Eilenberg--Moore spectral sequences (see e.g.~\cite[thm.~7.1]{mcclearyUsersGuideSpectral2001}), but with more general coefficients, one would run into the usual extension problem of spectral sequences (see \cite[app.~I]{Xue:2026yqc} for an example of overcoming this).

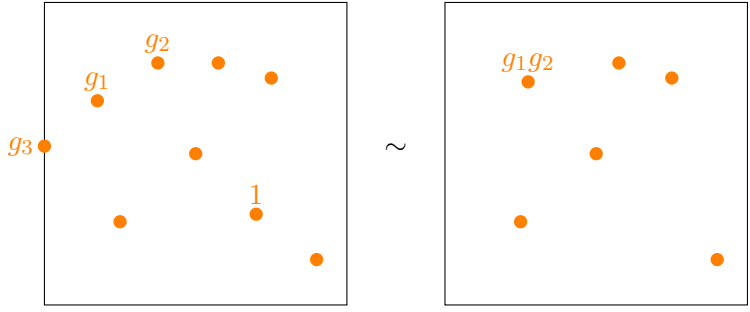
\begin{figure}[h]
  \[
    \newcommand{\points}{{(0,0)},{(1,1)},{(-1,-.9)},{(.3,1.2)},{(1.6,-1.4)}}
    \begin{tikzpicture}[baseline=0]
      \draw (-2, -2) rectangle (2, 2);
      \foreach \point in \points {
        \node[dotnode,orange] at \point {};
      }
      \node[dotnode,orange] (3) at (.8,-.8) {};
      \node[dotnode,orange] (g1) at (-1.3,.7) {};
      \node[dotnode,orange] (g2) at (-.5,1.2) {};
      \node[dotnode,orange] (g3) at (-2,.1) {};
      \node[above,orange] at (3) {$1$};
      \node[above,orange] at (g1) {$g_1$};
      \node[above,orange] at (g2) {$g_2$};
      \node[left,orange] at (g3) {$g_3$};
    \end{tikzpicture}
    \quad\sim\quad
    \begin{tikzpicture}[baseline=0]
      \draw (-2, -2) rectangle (2, 2);
      \foreach \point in \points {
        \node[dotnode,orange] at \point {};
      }
      \coordinate (g1) at (-1.3,.7);
      \coordinate (g2) at (-.5,1.2);
      \node[dotnode,orange] (g1g2) at ($(g1)!.5!(g2)$) {};
      \node[above,orange] at (g1g2) {$g_1 g_2$};
    \end{tikzpicture}
  \]

  \caption{The configuration space $\cC(G, d)$ with $d = 2$.}
  \label{fig:conf-space}
\end{figure}

\begin{figure}[h]
  \[
    \newcommand{\rad}{1.4}
    \begin{tikzpicture}[baseline=0]
      \draw (-2, -2) rectangle (2, 2);
      \node[dotnode,orange] (g1) at (-\rad,0) {};
      \node[dotnode,orange] (g2) at (\rad,0) {};
      \node[left,orange] at (g1) {$g_1$};
      \node[right,orange] at (g2) {$g_2$};
      \draw[->] (10:\rad) arc[radius=\rad,start angle=10,end angle=170];
      \draw[->] (190:\rad) arc[radius=\rad,start angle=190,end angle=350];
      \foreach \ang in {30,60,...,360} {
        \draw[gray,very thin,-{Latex[length=2pt]}] (\ang:1.25) -- (\ang:.15);
      }
    \end{tikzpicture}
    \quad\rightsquigarrow\quad
    \begin{tikzpicture}[baseline=0]
      \draw (-2, -2) rectangle (2, 2);
      \node[dotnode,orange] (g1g2) at (0,0) {};
      \node[above,orange] at (g1g2) {$g_1 g_2$};
    \end{tikzpicture}
  \]

  \caption{Particle exchange is null-homotopic, so the particles are bosons.}
  \label{fig:conf-space-contract}
\end{figure}
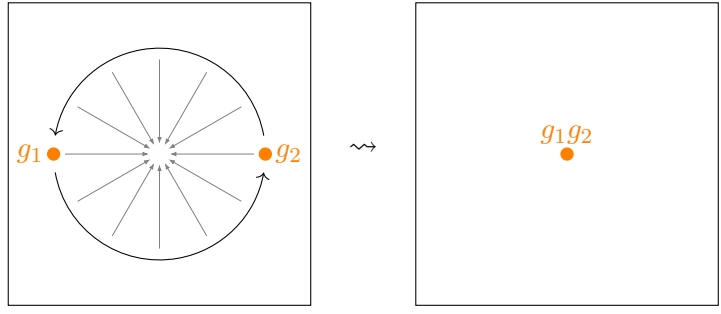

In most physical contexts where the groups $H^n(B^p G; A)$ appear, $n$ is the dimension of some spacetime. Since physically relevant spacetimes are low-dimensional (even in string theory, $n \lesssim 11$ or $26$), it is reasonable to ask whether knowledge of the cases with larger $n$ has any physical applications. Apart from their role as classifying spaces for gauge fields, one setup where the Eilenberg--MacLane spaces appear is as configuration spaces of bosons with a group fusion law. Specifically, let $\cC(G, d)$ be the configuration space of points in the $d$-cube $[0,1]^d$, each labelled by a group element $g \in G$, such that two coincident points labelled by $g_1$ and $g_2$ are identified with a single point labelled by $g_1 g_2$,\footnote{Consistency of this fusion law requires $G$ to be abelian if $d \geq 2$.} and such that points labelled by $1 \in G$ and points at the boundary of the cube are identified with the absence of a point (\zcref{fig:conf-space}). This is a model for particles in $d$ spatial dimensions charged under the Pontryagin dual group $\widehat{G}$, surrounded on all sides by a condensate that breaks charge conservation; they are bosons because the loop that exchanges two particles is contractible to a constant loop through fusing the particles (\zcref{fig:conf-space-contract}). Now, it is a striking fact that $\cC(G, d)$ is an Eilenberg--MacLane space $B^d G$ (an element $g \in \pi_d(\cC(G, d)) \cong G$ corresponds to a particle of charge $g$ sweeping the $d$-cube along the identity map $S^d \overset{\sim}{\lto} [0,1]^d / \partial[0,1]^d$), and in fact the standard bar construction is a CW-complex structure on this space \cite{milgramBarConstructionAbelian1967,mccordClassifyingSpacesInfinite1969,sinhaGeometryEilenbergMacLaneSpaces2024,chaiserUsingParticleModel2021}. This means that $H^n(B^d G; A)$ has an interpretation in terms of $n$-parameter families of particle configurations in $d$ dimensions, in contrast to the more usual one of topological actions in $n$ dimensions. In particular, large $n$ corresponds to a large number of particles rather than spatial dimensions. Although we are currently unaware of a direct application of the (co)homology groups to many-body physics, we point out the connection for future investigations.

\section{Mathematics of restricted symmetries and their anomalies}\label{sec:mathematics}
In this section we will give a formal definition of the symmetry restriction procedure, and we will carefully show why the cohomology suspension map captures the higher anomaly. In doing so, we will also show how the same map can be used to describe the boundary higher group structure inherited from a bulk higher group symmetry.

\subsection{Higher anomalies and compressing $G^{(p)}$ SPTs as cohomology suspension}
\label{sec:compressing-suspension}

We will derive the higher anomaly in the inflow picture, which we now review briefly. By a \emph{background} for a $G^{(p)}$-symmetry on a $D$-dimensional spacetime $X_D$, we mean a cocycle $A \in Z^{p+1}(X; G)$, \emph{not} its equivalence class under gauge transformations $A \lto A + \dd\lambda$. If there is an anomaly, the partition function is not invariant under these transformations, but rather $\cZ[X, A + \dd{\lambda}] = \e^{\ii\oint_X \varphi(A, \lambda)} \cZ[X, A]$. We take a $(D+1)$-dimensional spacetime $Y_{D+1}$ with $\partial Y = X$ and extend the background fields to $Y$.\footnote{This works only when $(X_D, A)$ is null-bordant in $\Omega^D(B^{p+1} G)$, i.e. $X$ is one boundary component of $Y$.} We find an inflow action $\int_{Y_{D+1}} \alpha(A)$ such that $\alpha(A + \dd\lambda) - \alpha(A) = -\dd{\varphi(A, \lambda)}$; then the bulk-boundary system $\cZ\qty[X_D, \eval{A}_X] \e^{\ii \int_Y \alpha(A)}$ is anomaly-free. Then, we can safely represent the background as its cohomology class $[A] \in H^{p+1}(Y; G)$. Alternatively, by the Poincaré--Lefschetz duality $H^{p+1}(Y;G) \cong H_{D-p}(Y,X;G)$, such backgrounds are symmetry defects in $Y$ allowed to end freely on $X$.

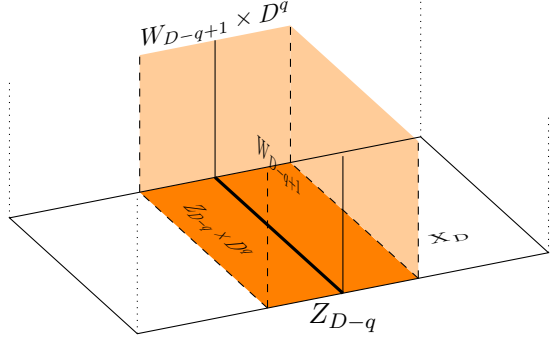
\begin{figure}[t]
  \centering
  \tdplotsetmaincoords{65}{-25}
  \begin{tikzpicture}[tdplot_main_coords]
      \newcommand{\slabW}{1.1}
      \colorlet{lightorange}{orange!40!white}
      \fill[orange,canvas is xy plane at z=0]
      (-\slabW,-2) rectangle (\slabW,2);
      \fill[canvas is yz plane at x=\slabW,lightorange]
      (-2,0) rectangle +(4,2);
      \fill[canvas is xz plane at y=2,lightorange]
      (-\slabW,0) rectangle (\slabW,2);

      \draw[canvas is xy plane at z=0] (-3,-2) rectangle (3,2);
      \draw[dotted] (-3,-2,0) -- +(0,0,2);
      \draw[dotted] (3,-2,0) -- +(0,0,2);
      \draw[dotted] (-3,2,0) -- +(0,0,2);
      \draw[dotted] (3,2,0) -- +(0,0,2);

      \draw[very thick] (0,-2,0) -- +(0,4,0);
      \draw[dashed] (-\slabW,-2,0) -- +(0,4,0);
      \draw[dashed] (\slabW,-2,0) -- +(0,4,0);
      \draw (0,-2,0) -- +(0,0,2);
      \draw (0,2,0) -- +(0,0,2);
      \draw[dashed,thin] (-\slabW,-2,0) -- +(0,0,2);
      \draw[dashed,thin] (-\slabW,2,0) -- +(0,0,2);
      \draw[dashed,thin] (\slabW,-2,0) -- +(0,0,2);
      \draw[dashed,thin] (\slabW,2,0) -- +(0,0,2);

      \node[canvas is xy plane at z=0,transform shape]
      at (2,-1) {$X_D$};
      \node[below] at (0,-2,0) {$Z_{D-q}$};
      \coordinate (Y) at (0,0,1);
      \node[canvas is zy plane at x=0,rotate=-90,transform shape]
      at (Y) {$W_{D-q+1}$};

      \node[canvas is xy plane at z=0,rotate=-90,transform shape]
      at (-0.7,0.5) {\small $Z_{D-q} \times D^q$};

      \coordinate (YD) at (0,2,2.2);
      \node[canvas is xz plane at y=0,transform shape]
      at (YD) {\small $W_{D-q+1} \times D^q$};
    \end{tikzpicture}
  \caption{A submanifold $Z_{D-q} \subset X_D$ and its tubular neighbourhood in
    the inflow picture ($D = 2$, $q = 1$ is shown).}
  \label{fig:inflow-tubular}
\end{figure}

Consider now a submanifold $Z_{D-q} \subset X_D$, so that it lies on the boundary of $W_{D-q+1} \subset Y_{D+1}$. The bulk--boundary subsystem can be embedded in the original one by considering a tubular neighbourhood $(W_{D-q+1}\times D^q, Z_{D-q}\times D^q)$ of $(W_{D-q+1},Z_{D-q})$, as in \zcref{fig:inflow-tubular}.\footnote{We assume that $Z_{D-q}$ has a trivial normal bundle in $X_D$, so that the tubular neighbourhood is as well a trivial disk bundle. When one drops this assumption, a possible interplay between gravitation and symmetry anomaly may arise; see \cite{Debray_2026, Hason:2020yqf}.}
To restrict the $G^{(p)}$-symmetry to $(W,Z)$, we first consider backgrounds supported only inside the tubular neighbourhood $(W \times D^q, Z \times D^q)$, with the condition $\eval{A}_{W \times S^{q-1}} = 0$; in other words classes in $H^{p+1}(W \times D^q, W \times S^{q-1}; G)$. In the limit where the radius of the tube shrinks to zero, we obtain the backgrounds for the $G^{(p-q)}$-symmetry on $(W, Z)$. In terms of symmetry defects, this statement takes the form of the geometrically obvious isomorphism
\begin{equation}
  \label{eq:susp-iso-homology}
  H_{D-p}(W \times D^q, Z \times D^q; G) \cong H_{D-p}(W, Z; G),
\end{equation}
which says that the nontrivial defects in the slice $W$ are the same as those in its thickening. Applying  a generalised form of Lefschetz duality \cite[thm.~3.43]{Hatcher2002}\footnote{Given an $n$-dimensional orientable manifold $M$ with $\partial M = A \cup B$ and $\partial A = \partial B = A \cap B$, then $H^{p}(M,A;G) \cong H_{n-p}(M,B;G)$. In our case $\partial (W \times D^q) = Z \times D^q \cup W \times S^{q-1}$, with common boundary $Z \times S^{q-1}$.} to both sides of \eqref{eq:susp-iso-homology}, we obtain the corresponding statement for background fields:
\begin{equation}
  \label{eq:susp-iso-cohomology}
   H^{p+1}(W \times D^q, W \times S^{q-1}; G) \cong H^{p-q+1}(W; G).
\end{equation}

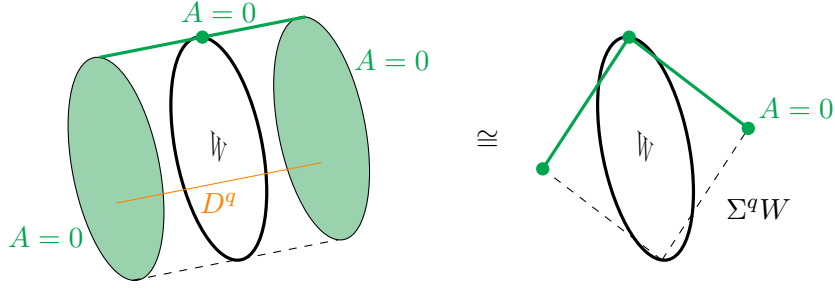
\begin{figure}[t]
  \[
    \tdplotsetmaincoords{65}{-25}
    \newcommand{\slabW}{1.5}
    \def\myAngle{70}
    \begin{tikzpicture}[tdplot_main_coords,baseline=0]
      \colorlet{basepoint}{Green}
      \colorlet{basepointl}{basepoint!40!white}

      \draw[very thick,canvas is yz plane at x=0] (0,0) circle[radius=1.5];
      \draw[fill=basepointl,canvas is yz plane at x=-\slabW] (0,0) circle[radius=1.5];
      \draw[fill=basepointl,canvas is yz plane at x=\slabW] (0,0) circle[radius=1.5];

      \coordinate (top1) at (-\slabW,{1.5*cos(\myAngle)},{1.5*sin(\myAngle)});
      \coordinate (bot1) at (-\slabW,{-1.5*cos(\myAngle)},{-1.5*sin(\myAngle)});
      \draw[dashed,thin] (bot1) -- +(2*\slabW,0,0);
      \draw[very thick,basepoint] (top1) -- +(2*\slabW,0,0);
      \node[dotnode,basepoint] at ($(top1)+(\slabW,0,0)$) {};

      \node[above right,basepoint] at (\slabW,-0.7,1) {$A = 0$};
      \node[above,basepoint] at (0,0,1.7) {$A = 0$};
      \node[below left,basepoint] at (-\slabW,0.7,-1) {$A = 0$};

      \coordinate (Y) at (0,0,0);
      \node[canvas is zy plane at x=0,rotate=-90,transform shape]
      at (Y) {\Large$W$};

      \draw[orange] (-\slabW,0,-0.5) -- +(2*\slabW,0,0);
      \node[orange,below] at (0,0,-0.5) {$D^q$};
    \end{tikzpicture}
    \quad\cong\quad
    \begin{tikzpicture}[tdplot_main_coords,baseline=0]
      \colorlet{basepoint}{Green}
      \colorlet{basepointl}{basepoint!40!white}

      \draw[very thick,canvas is yz plane at x=0] (0,0) circle[radius=1.5];

      \coordinate (top) at (0,{1.5*cos(\myAngle)},{1.5*sin(\myAngle)});
      \coordinate (bot) at (0,{-1.5*cos(\myAngle)},{-1.5*sin(\myAngle)});
      \draw[very thick,basepoint] (top) -- (-\slabW,0,0);
      \draw[very thick,basepoint] (top) -- (\slabW,0,0);
      \draw[dashed,thin] (bot) -- (-\slabW,0,0);
      \draw[dashed,thin] (bot) -- (\slabW,0,0);
      \node[dotnode,basepoint] at ($(top1)+(\slabW,0,0)$) {};

      \node[dotnode,basepoint] at (-\slabW,0,0) {};
      \node[dotnode,basepoint] at (\slabW,0,0) {};

      \node[above right,basepoint] at (\slabW,0,0) {$A = 0$};

      \coordinate (Y) at (0,0,0);
      \node[canvas is zy plane at x=0,rotate=-90,transform shape]
      at (Y) {\Large$W$};

      \node at (1.1*\slabW,0,-1.2) {$\Sigma^q W$};
    \end{tikzpicture}
  \]
  \caption{Confining the gauge field to a tubular neighbourhood of $W$ is
    equivalent to confining it to the $q$-fold suspension of $W$.}
  \label{fig:shrink-suspension}
\end{figure}

The isomorphism \eqref{eq:susp-iso-cohomology} is the \emph{suspension isomorphism} in cohomology, and is most cleanly expressed in the homotopical formulation of background fields as maps between the base space and the classifying space of the $G^{(p)}$-symmetry. We work with pointed spaces,  taking all spaces to be equipped with basepoints $*$ and considering only basepoint-preserving maps. In this formalism, a background is given as a pointed map $A\colon W \times D^q \to B^{p+1} G$ with the Dirichlet boundary condition $\eval{A}_{W \times S^{q-1}} = *$. Because of this boundary condition, the background is equivalently a pointed map $A \colon \Sigma^q W \to B^{p+1} G$, where $\Sigma^q W = W \wedge S^q$ is the $q$-fold reduced suspension of $W$ and $\wedge$ is the smash product (see \zcref{fig:shrink-suspension}). By the currying adjunction
\begin{equation}
  \label{eq:currying-adjunction}
  \Maps_*(M \wedge N, H) \cong \Maps_*(M, \Maps_*(N, H)),
\end{equation}
or its special case with $N = S^1$, the loop--suspension adjunction,
\begin{equation}
  \label{eq:loop-suspension-adjunction}
  \Maps_*(\Sigma M, H) \cong \Maps_*(M, \Omega H),
\end{equation}
this equivalently means that
\begin{equation}
  A \in [\Sigma^q W, B^{p+1} G] \cong [W, \Omega^q B^{p+1} G]
  \cong [W, B^{p-q+1} G],
\end{equation}
which is the suspension isomorphism $H^{p+1}(\Sigma^q W; G) \cong H^{p-q+1}(W; G)$.

Having understood the fate of backgrounds under compression, we now turn to the inflow theory by replacing $H^\bullet(\blank; G)$ with some generalised cohomology theory $h$.\footnote{While the following arguments work for any $W$ regardless of whether it has a boundary, it may be most straightforward to imagine that $\partial W = \emptyset$. We can arrange this through a ``mirror charge'' trick, replacing the original QFT on $Z = \partial W$ with its inflow theory on $\overline{W}$ (the orientation-reversal of $W$), which has the same anomaly. We are then working with the inflow theory on the closed manifold $\overline{W} \sqcup_Z W$.} The value of the original inflow theory on $\Sigma^q W$ is an element of $h^{D+1}(\Sigma^q W)$. By Brown's representability theorem, any generalised cohomology theory $h$ can be represented by an $\Omega$-spectrum, a collection of spaces $\{E^n\}$ with $\Omega E^{n+1} \simeq E^n$, such that $h^n(X) = [X, E^n]$. Therefore this group can be equivalently expressed as
\begin{align}
\begin{gathered}
  h^{D+1}(\Sigma^q W) = [\Sigma^q W, E^{D+1}] \cong [W, \Omega^q E^{D+1}] \\
  \cong [W, E^{D-q+1}] = h^{D-q+1}(W);
\end{gathered}
\end{align}
this is, once again, the suspension isomorphism $h^{D+1}(\Sigma^q W) \cong h^{D-q+1}(W)$.

Thus, given the original anomaly $\alpha \in h^{D+1}(B^{p+1} G) \cong \Nat(H^{p+1}(\blank; G), h^{D+1}(\blank))$ and a background $A \in H^{p-q+1}(W; G)$, we associate the $q$-anomaly $\alpha_q(W)(A) \in h^{D-q+1}(W)$ using the suspension isomorphisms as follows:
\begin{equation}
  \label{eq:anomaly-map-2}
  H^{p-q+1}(W; G) \cong H^{p+1}(\Sigma^q W; G)
  \xlongrightarrow{\alpha(\Sigma^q W)} h^{D+1}(\Sigma^q W)
  \cong h^{D-q+1}(W).
\end{equation}
We can reformulate this to make it independent of $W$. Using the loop--suspension adjunction \eqref{eq:loop-suspension-adjunction} and viewing $\alpha$ as an element of $[B^{p+1} G, E^{D+1}]$, \eqref{eq:anomaly-map-2} becomes
\begin{equation}
  [W, B^{p-q+1} G]
  \simeq [W, \Omega^q B^{p+1} G]
  \xlongrightarrow{\Omega^q \alpha \circ \blank}
  [W, \Omega^q E^{D+1}]
  \simeq [W, E^{D-q+1}],
\end{equation}
where now the $q$-anomaly $\alpha_q \in h^{D-q+1}(B^{p-q+1} G) \cong [B^{p-q+1} G, E^{D-q+1}]$ is simply $\Omega^q \alpha$. Here $\Omega^q = \Maps_*(S^q, \blank)$ is the $q$-fold loop functor, sending a map $f\colon M \to N$ to $\Omega^q f = f \circ \blank \colon \Omega^q M \to \Omega^q N$, i.e.~given an element $g \in \Omega^qM$, we have $(\Omega^qf)g = f \circ g \in \Omega^qN$.

In summary, we have found that the higher anomaly map \eqref{eq:higher-anomaly-map} is the $q$-fold loop functor
\begin{equation}
  h^{D+1}(B^{p+1} G) = [B^{p+1} G, E^{D+1}]
  \xlongrightarrow{\Omega^q}
  [B^{p-q+1} G, E^{D-q+1}] = h^{D-q+1}(B^{p-q+1} G),
\end{equation}
which is also known as the (iterated) cohomology suspension.

In the remainder of this paper, we will take $h$ to be ordinary cohomology with coefficients in some abelian group $A$, i.e.~$H^{\bullet}(\blank; A)$. The case $A = \R/\Z$ is relevant for the classification of anomalies and SPT phases. While an accurate classification is known to require more advanced cohomology theories, we view ordinary cohomology as a first step to understanding the loop functor in those theories.

\subsection{Compressing background fields for higher group symmetries}
\label{sec:higher-groups}
Apart from its interpretation as the higher anomaly map, the cohomology suspension defined above appears naturally when considering higher group symmetries, as pointed out in \cite{Bah:2025oxi}. For a discrete higher group $\bbG$, background fields for an anomaly-free $\bbG$ symmetry are understood as homotopy classes of maps $A\colon X_D \lto B\bbG$, where $B\bbG$ is its classifying space. We restrict to the case when $\pi_1(B\bbG)$ acts trivially on $\pi_{>1}(B\bbG)$; then, $B\bbG$ has a Postnikov tower of principal fibrations:
\begin{equation}
  \label{eq:postnikov-general}
  \begin{tikzcd}
    & \vdots \ar[d] \\
    B^3 \pi_3 \ar[r,"i_2"]
    & (B\bbG)_{\leq 3} \ar[d,"f_2"] \ar[r,"k_3"]
    & B^5 \pi_4 \\
    B^2 \pi_2 \ar[r,"i_1"]
    & (B\bbG)_{\leq 2} \ar[d,"f_1"] \ar[r,"k_2"]
    & B^4 \pi_3 \\
    & B\pi_1 \ar[r,"k_1"]
    & B^3 \pi_2.
  \end{tikzcd}
\end{equation}
Here each $(i_i, f_i, k_i)$ is a fibration sequence, so that the fibration $f_i$ has fibre $B^{i+1} \pi_{i+1}$ and is classified by $k_i \in [(B\bbG)_{\leq i}, B^{i+2} \pi_{i+1}] \cong H^{i+2}((B\bbG)_{\leq i}; \pi_{i+1})$.

Now, if the spacetime $X_D$ is a slab $X_{D-q} \times D^q$, with boundary
conditions $A = 0$ on $X_{D-q} \times \partial D^q$, we can
identify $X_D$ with $\Sigma^q X_{D-q}$ as before. The background field $A$ is then an element of $[\Sigma^q X_{D-q}, B\bbG] \cong [X_{D-q}, \Omega^q B\bbG]$. Applying $\Omega^q$ to \eqref{eq:postnikov-general} gives the Postnikov tower of $\Omega^q B\bbG$:\footnote{This is \cite[exercise~4.3.20]{Hatcher2002}.}
\begin{equation}
  \label{eq:postnikov-general-loop}
  \begin{tikzcd}
    & \vdots \ar[d] \\
    B^{3-q} \pi_3 \ar[r,"\Omega^q i_2"]
    & (\Omega^q B\bbG)_{\leq 3-q} \ar[d,"\Omega^q f_2"] \ar[r,"\Omega^q k_3"]
    & B^{5-q} \pi_4 \\
    B^{2-q} \pi_2 \ar[r,"\Omega^q i_1"]
    & (\Omega^q B\bbG)_{\leq 2-q} \ar[d,"\Omega^q f_1"] \ar[r,"\Omega^q k_2"]
    & B^{4-q} \pi_3 \\
    & B^{1-q} \pi_1 \ar[r,"\Omega^q k_1"]
    & B^{3-q} \pi_2.
  \end{tikzcd}
\end{equation}
In particular, the new Postnikov invariants are obtained from the original ones by the iterated cohomology suspension.

In general, computing $H^{i+2}((B\bbG)_{\leq i}; \pi_{i+1})$ is difficult.
However, when there are only two non-trivial homotopy groups $\pi_n(B\bbG) \cong G$ and $\pi_m(B\bbG) \cong A$ ($m > n$), the Postnikov tower of $B\bbG$ takes the simpler form
\begin{equation}
  \label{eq:simple-higher-group}
  \begin{tikzcd}
    B^m A \ar[r,"i"] & B\bbG \ar[d,"f"] \\
    & B^n G \ar[r,"k"] & B^{m+1} A,
  \end{tikzcd}
\end{equation}
and the one of $\Omega^q B\bbG$ is correspondingly
\begin{equation}
  \label{eq:simple-higher-group-loop}
  \begin{tikzcd}
    B^{m-q} A \ar[r,"\Omega^q i"] & \Omega^q B\bbG \ar[d,"\Omega^q f"] \\
    & B^{n-q} G \ar[r,"\Omega^q k"] & B^{m-q+1} A.
  \end{tikzcd}
\end{equation}

The new Postnikov invariant is determined by the map
\begin{equation}
  \Omega^q\colon H^{m+1}(B^n G; A) \lto H^{m-q+1}(B^{n-q} G; A),
\end{equation}
and we will see below that we can compute this map explicitly when $G$ and $A$ are finitely generated abelian. For example, when this map has a kernel, it is possible that $k \neq 0$ but $\Omega^q k = 0$. In such a case, although $B\bbG$ is a nontrivial higher group, $\Omega^q B\bbG$ is just a product $B^{n-q} G \times B^{m-q} A$. This means that the higher group structure cannot be detected on a slab $\Sigma^q X_{D-q}$, but only on the bulk spacetimes $X_D$. We may call it an \emph{intrinsically top-dimensional} higher group.

For example, using Steenrod operations one can easily see that e.g.~$\Omega\colon H^4(B^2 \Z_2; \Z_2) \lto H^3(B \Z_2; \Z_2)$ is zero (see \eqref{eq:Z2-coeff-suspension}). Therefore, for any anomaly-free higher group symmetry $\bbG$ which is a twisted product of a $\Z_2$ 1-form symmetry and a $\Z_2$ 2-form symmetry, i.e.~which has a Postnikov tower of the form
\begin{equation}
  \begin{tikzcd}
    B^3 \Z_2 \ar[r] & B\bbG \ar[d] \\
    & B^2 \Z_2 \ar[r,"k"] & B^4 \Z_2,
  \end{tikzcd}
\end{equation}
we have $\Omega B\bbG \simeq B \Z_2 \times B^2 \Z_2$. This means that when spacetime is a slab $\Sigma X_{D-1}$, the higher group structure is not detectable and $\bbG$ reduces to a product $\Z_2^{(0)} \times \Z_2^{(1)}$; $\bbG$ is intrinsically top-dimensional. On the other hand, as we mentioned in \zcref{sec:intro-higher-anomalies}, there are 2-groups mixing $\Z_2^{(1)}$ with $\Z_4^{(2)}$ with a nontrivial higher group structure in codimension 1, because $\Omega\colon H^4(B^2 \Z_2; \Z_4) \lto H^3(B \Z_2; \Z_4)$ is nonzero.

\subsection{Basic features of the cosuspension}
\label{sec:basic-features}
Equivalently to its formulation as the loop functor above, the cohomology suspension can be defined as a pullback in terms of the evaluation map\footnote{This map sends a pair $(\ell, t) \in \Sigma\Omega X = \Maps_*(S^1, X) \wedge S^1$ to the point $\ell(t) \in X$. It is the adjoint of $\id\colon \Omega X \lto \Omega X$ under \eqref{eq:loop-suspension-adjunction}.} $\evMap\colon \Sigma\Omega X \lto X$ together with the suspension isomorphism:
\begin{equation}
  \label{eq:Omega-as-pullback}
  \Omega\colon H^{D+1}(X; A) \xrightarrow{\evMap^*}
  H^{D+1}(\Sigma\Omega X; A) \cong H^D(\Omega X; A).
\end{equation}
There is also a dual notion of \emph{homology suspension}:
\begin{equation}
  \label{eq:sigma-as-pushforward}
  \sigma\colon H_D(\Omega X; A) \cong H_{D+1}(\Sigma\Omega X; A)
  \xrightarrow{\evMap_*} H_{D+1}(X; A).
\end{equation}

It is a classical result \cite[VII, cor.~6.5]{whiteheadElementsHomotopyTheory1978} that when $X$ is a $p$-connected space (such as $B^{p+1} G$, or the classifying space $B\bbG$ of any higher-group symmetry $\bbG$ where the lowest-degree symmetry is a $p$-form symmetry), these maps are \emph{isomorphisms} in the ``stable range'' $D < 2p$. Furthermore, for $D = 2p$, $\Omega$ is a monomorphism while $\sigma$ is an epimorphism.\footnote{This result can be stated equivalently as $\evMap$ being a $(2p+1)$-connected map \cite[ex.~4.2.13]{munsonCubicalHomotopyTheory2015a}.}  This follows from inspecting the Leray--Serre spectral sequence in (co)homology for the pathspace fibration $\Omega X \lto PX \lto X$: The transgressions $\tau^{D+1} = \partial\colon E^{D+1}_{D+1,0} \lto E^{D+1}_{0, D}$ and $\tau_{D+1} = \dd\colon E_{D+1}^{D+1,0} \lto E_{D+1}^{0, D}$ are the isomorphisms
\begin{equation}
    \tau^{D+1}\colon \im\sigma \overset{\cong}{\lto} \coim\sigma
    \qq{and}
    \tau_{D+1}\colon \im\Omega \overset{\cong}{\lto} \coim\Omega
\end{equation}
respectively \cite[thm.~6.6]{mcclearyUsersGuideSpectral2001}. Using that $E^{D+1}_{i,j} = E_{D+1}^{i,j} = 0$ for $i \leq p$ and $j < p$ (except $(i,j) = (0,0)$) by the Hurewicz theorem, one finds that these are isomorphisms $\sigma^{-1}$ and $\Omega^{-1}$ in the ranges in question, and similarly for the mono(epi)morphism statement.\footnote{\label{foot:serre}The result follows from the simpler Serre exact sequence \cite[VII, cor.~6.3]{whiteheadElementsHomotopyTheory1978}, but the spectral sequence is more general, and can sometimes be used to compute $\sigma$ and $\Omega$ outside the stable range. This is useful as an independent check on the results of this paper.} The fact that $\Omega$ is a monomorphism for all $D \leq 2p$ has immediate interesting consequences for anomalies and higher-group structures:
\begin{quote}
  An anomalous $p$-form symmetry in $D \leq 2p$ spacetime dimensions is also $1$-anomalous, and indeed $q$-anomalous for all $q \leq 2p - D + 1$. The same is true for an anomalous higher group symmetry involving only $p'$-form symmetries for $p' \geq p$.

  A nontrivial higher group structure mixing a $p$-form symmetry with a single $p'$-form symmetry such that $p < p' < 2p$ is detectable in codimension 1, and indeed in all codimensions $q \leq 2p - p'$.
\end{quote}
A heuristic interpretation of the first fact in particular is that the anomalies that trivialise when one dimension is removed are those that come from the linking of two or more symmetry defects. For such a linking to occur, the dimension must be at least large enough to accommodate the linking of two defects, that is, $D \geq 2p + 1$.

Beware that the anomaly statement holds for higher group cohomology anomalies, but not necessarily for anomalies beyond group cohomology, when ordinary cohomology is replaced by another cohomology theory accounting for invertible phases without symmetry. This is because the argument above uses the Hurewicz theorem, which is only valid for ordinary cohomology.

Another classical result \cite[thm.~20.2]{eilenbergGroupsHPII1954} states that $\Omega$ and $\sigma$ annihilate ``cross-effects''; that is, when $G = G_1 \oplus G_2 \oplus \cdots$, they send to zero any (co)homology class mixing the different summands. Furthermore, the image of $\Omega$ and $\sigma$ does not contain cross-effects; we will see this clearly in equation~\eqref{eq:sbar-tensor}. The physical consequence of this fact is:
\begin{quote}
  Mixed anomalies do not give rise to higher anomalies, nor do they appear as higher anomalies.
\end{quote}
In other words, while a mixed anomaly may be an obstruction to ordinary gauging, it is never an obstruction to higher gauging. Intuitively, mixed anomalies come strictly from the linking of symmetry defects of different species, and linking cannot occur in positive codimension.

\section{Computing the cohomology suspension}
\label{sec:computing}

In this section, we give a complete calculation of the cohomology suspension
\begin{eqnarray}
  \label{eq:ordinary-omega}
  \Omega\colon H^{D+1}(B^{h+1} G; A) \lto H^D(B^h G; A)
\end{eqnarray}
for general $D, h \geq 1$ and finitely generated abelian groups $G$ and $A$, as well as $A = \R/\Z$.

Before explaining our methods, let us motivate why they are necessary. The difficulty of the problem actually depends heavily on the choice of coefficient group $A$. In the case of $A = \Z_p$ (by which we mean $\Z/p\Z$) for a prime $p$, it is simple to calculate \eqref{eq:ordinary-omega} using a well-known algorithm based on Steenrod operations, summarised in \zcref{app:Zp}. For more general $A$, calculating the cosuspension \eqref{eq:ordinary-omega} is more difficult, due to the presence of unstable cohomology operations which do not appear over $\Z_p$. However, it can be done algorithmically and efficiently using Henri Cartan's notion of \emph{construction} \cite{doi:10.1073/pnas.40.8.704}, explained in \zcref{sec:cartan-constructions}.

Cartan described an explicit construction for $A = \Z_p$, in which $\Omega$ is easy to read off. He also used this result to compute the groups $H_D(B^h G; \Z)$ by an indirect argument. This determines the groups $H^D(B^h G; A)$ for any $A$ by the Universal Coefficient Theorem, but does not immediately determine $\Omega$. At least in some cases, $\Omega$ can be deduced from the transgressions in the Leray--Serre spectral sequence associated to the path space fibration $B^h G \lto P(B^{h+1} G) \lto B^{h+1} G$ (see \zcref{foot:serre}), but this technique is laborious and ad-hoc, and at most determines $\Omega$ up to isomorphism. We view it as a consistency check on our results, but not as a practical general method of calculation.

Instead, we will make use of Moore's extension \cite{doi:10.1073/pnas.43.5.409} of Cartan's work, which allows us to compute small chain complexes for $B^h G$ with coefficients in the ring $\Lambda = \Z_{(p)}$ of integers localised at $p$, that is, the rational numbers with denominators not divisible by $p$. We can also obtain complexes with coefficients in $\Z_{p^\ell} \cong \Z_{(p)} / p^\ell \Z_{(p)}$ by simply reducing these mod $p^\ell$.\footnote{It is important that we do this \emph{before} passing to (co)homology.} The (co)homology groups with such coefficient rings $\Lambda$ are easy to read off, and we can also calculate the (co)homology suspension maps because these are \emph{special constructions}---they are equipped with a chain contraction. In \zcref{sec:coefficients} we explain why this is enough to obtain these groups and maps with all coefficient groups of interest.

\subsection{Cartan's notion of construction}
\label{sec:cartan-constructions}
Recall that $B^h G$ is an $H$-space, i.e. it is a space equipped with a multiplication $B^h\mu\colon B^h G \times B^h G \to B^h G$, induced from the group multiplication $\mu$ in $G$. This equips the chains of $B^h G$ with an algebra structure. The $H$-space multiplication induces a product in homology, the \emph{Pontryagin product}~\cite[sec.~3.C]{Hatcher2002}
\begin{equation}
  \label{eq:pontryagin-product}
  H_D(B^h G) \times H_{D'}(B^h G) \xrightarrow{\times}
  H_{D+D'}(B^h G \times B^h G) \xrightarrow{B^h \mu_*} H_{D+D'}(B^h G).
\end{equation}
It is associative and, for abelian $G$, graded commutative. Cartan's constructions provide a model for the chain algebra of $B^h G$ where the multiplication is strictly associative and graded commutative even on the chain level.

Cartan's work \cite{cartanAlgebreDEilenbergMacLane1954} relies on some preliminary notions, which we now review. Given a base ring $\Lambda$, we want to consider \emph{commutative differential graded augmented $\Lambda$-algebras} (CDGA-algebras)\footnote{Following Cartan, the ``A'' in ``DGA'' stands for ``augmented'', not ``algebra''.} $\cA$. Such an algebra has a differential $\partial\colon \cA \lto \cA$ of degree -1 satisfying $\partial^2 = 0$ and the graded Leibniz rule, as well as an \emph{augmentation}: a ring morphism $\eps\colon \cA \to \Lambda$, zero on elements of positive degree, such that $\eps\partial = 0$.

A CDGA-algebra is \emph{acyclic} if the sequence
\begin{equation}
  0 \longleftarrow \Lambda \xlongleftarrow{\eps} \cA_0
  \xlongleftarrow{\partial} \cA_1 \xlongleftarrow{\partial} \cA_2
  \xlongleftarrow{\partial} \cdots
\end{equation}
is exact.

A \emph{multiplicative construction} on a CDGA-algebra $\cA$ is a choice of graded augmented $\Lambda$-algebra $\cN$, together with an extension of the differential of $\cA$ to $\cM \defeq \cA \otimes_\Lambda \cN$, making $\cM$ into a CDGA-algebra with augmentation $\eps_\cM \defeq \eps_\cA \otimes \eps_\cN$. This defines a differential $\partialbar$ on $\cN$ by quotienting $\cM$ by the augmentation ideal $I=\langle \ker( \eps_\cA \otimes \text{Id}_{\cN}) \rangle$,
\begin{equation}
  \cM \lto \cM/I \cM \cong A/\ker(\eps_{\cA}) \otimes_{\Lambda} \cN \cong \cN
\end{equation}
(which we denote as $m \mapsto \overline{m}$), making $\cN$ into a new CDGA-algebra. We call $\cA$ the \emph{initial algebra} and $\cN$ the \emph{final algebra} of the construction or the \emph{delooping} of $\cA$; we may also call $\cM$ the \emph{total algebra}. Furthermore we say that the construction is \emph{acyclic} if $\cM$ is.

A construction $\cA \longhookrightarrow \cM \lto \cN$ is an algebraic analogue of a fibration. In particular, an acyclic construction corresponds to a fibration where the total space is contractible, for example the path space fibration $\Omega X \lto PX \lto X$.

A multiplicative construction is \emph{special} if it is equipped with a chain contraction: a $\Lambda$-linear (but not necessarily multiplicative in any sense) map $s\colon \cM \lto \cM$ that raises degree by 1, such that
\begin{equation}
  \label{eq:contraction}
  \partial s + s \partial = \id - \eps, \qquad
  s^2 = 0, \qquad
  s(\Lambda) = 0,
\end{equation}
and such that the sub-$\Lambda$-module $\tildeN \defeq \Lambda + \im s \subset \cM$ is a subalgebra (i.e.~it is closed under multiplication) containing $\cN$. A special construction is necessarily acyclic; in fact, any closed element has a canonical antiderivative, obtained by applying $s$. Precisely speaking, for every $m \in \cM$ with $\partial m = \eps m = 0$, $sm$ is its unique antiderivative in $\im s$ (that is, its unique antiderivative in $\tildeN$ up to scalars).\footnote{If $\partial m = \eps m = 0$, then $\partial (s m) = (1 - \eps - s \partial)m = m$. For any other $m'$ with $\partial (sm') = m$, we have $sm' = (\partial s + s \partial)sm' = s\partial s m' =
sm$.}

Given two special constructions $(\cA_1, \cN_1, \eps_1, \partial_1, s_1)$ and $(\cA_2, \cN_2, \eps_2, \partial_2, s_2)$, their tensor product gives a
special construction $(\cA_1 \otimes \cA_2, \cN_1 \otimes \cN_2, \eps, \partial, s)$ with the following augmentation, differential and contraction:
\begin{equation}
  \label{eq:tensor-construction}
  \eps = \eps_1 \otimes \eps_2, \qquad
  \partial = \partial_1 \otimes 1 + 1 \otimes \partial_2, \qquad
  s = s_1 \otimes 1 + \eps_1 \otimes s_2
\end{equation}
(note the asymmetric definition of $s$).\footnote{Here we implicitly use the
  isomorphism $\cM = (\cA_1 \otimes \cA_2) \otimes (\cN_1 \otimes \cN_2)
  \cong (\cA_1 \otimes \cN_1) \otimes (\cA_2 \otimes \cN_2) = \cM_1 \otimes
  \cM_2$, which introduces signs from anticommutation.}

Composing $s$ with the inclusion $\cA \longhookrightarrow \cM$ and the quotient map $\cM \lto \cN$, we get a map $\sbar\colon \cA \lto \cN$ sending $a$ to $\overline{s(a \otimes 1)}$. Knowing the map $\sbar$ is the easiest way to calculate the homology and cohomology suspensions $\sigma$ and $\Omega$ in practice, as we will discuss in later sections.

\subsection{Constructions for Eilenberg--MacLane spaces}

The starting point of Cartan's calculation is the group algebra $\Lambda[G]$, viewed as a CDGA $\Lambda$-algebra. Its elements are finite formal sums $\sum_g \lambda_g g$, all of degree 0, equipped with zero differential and the augmentation $\eps\qty\big(\sum_g \lambda_g g) = \sum_g \lambda_g$. Given a decomposition $G \cong \Z^r \oplus \bigoplus_i \Z_{p_i^{f_i}}$, the group algebra correspondingly decomposes as
\begin{equation}
  \label{eq:G-decomposition}
  \Lambda[G] \cong \Lambda[\Z]^{\otimes r}
  \otimes \bigotimes_i \Lambda[\Z_{p_i^{f_i}}].
\end{equation}
We can view $\Lambda[\Z]$ as the algebra $\Lambda[a, a^{-1}]/(a a^{-1} - 1)$, and $\Lambda[\Z_N]$ as $\Lambda[u]/(u^N - 1)$. To construct our model for $C_{\bullet}(B^h G; \Lambda)$, we simply need to find an $h$-fold delooping of $\Lambda[G]$.

Every CDGA-algebra $\cA$ over any ring $\Lambda$ admits at least one special construction with $\cA$ as the initial algebra, the \emph{bar construction} \cite{eilenbergGroupsHP1953,cartanAlgebreDEilenbergMacLane1954}, which furthermore has $\tildeN = \cN$. In principle, one can start with the group algebra $\Lambda[G]$ and iterate the bar construction $h$ times to obtain a model of the chain algebra $C_{\bullet}(B^h G; \Lambda)$, but the resulting algebra is too large and unwieldy for many practical calculations. However, the point of Cartan's theory is that homological properties do not depend on the choice of construction, and in fact it is possible to replace these large algebras with very simple ones, yielding a model of $C_{\bullet}(B^h G; \Lambda)$ small enough to be manipulated by hand, or at least by computer. These will be our main tool.

The small constructions needed for delooping the group algebras are built from two basic types of CDGAs: One is the exterior algebra $E(x, n)$, with a single generator of odd degree $n$, subject to $x^2 = 0$. The other is the \emph{divided polynomial algebra} $\Gamma(x, n)$, given by a generator of even degree $n$ and its \emph{divided powers} $\gp_k(x)$ of degree $kn$. When $k!$ is invertible in $\Lambda$, $\gp_k(x)$ is equal to $x^k/k!$, otherwise it is a new adjoined element satisfying similar equations. In particular we have $\gp_0(x) = 1$, $\gp_1(x) = x$ and
\begin{align}
  \label{eq:gp-sum}
  \gp_k(x + y) &= \sum_{i + j = k} \gp_i(x) \gp_j(y), \\
  \label{eq:gp-product}
  \gp_k(x) \gp_{k'}(x) &= {k + k' \choose k} \gp_{k + k'}(x), \\
  \label{eq:gp-d}
  \partial \gp_{k+1}(x) &= \gp_k(x) \partial x.
\end{align}

Using $E(x, n)$ and $\Gamma(x, n)$ as basic ingredients, we define the following four types of CDGA-algebras, Cartan's \emph{elementary complexes}:
\begin{IEEEeqnarray}{lClClCl}
  \AI(x, n) &\defeq& E(x, n)
  &\qquad&\partial x = 0 &\qquad& \text{($n$ odd)} \\
  \BI(x, n) &\defeq& \Gamma(x, n)
  &&\partial x = 0 && \text{($n$ even)} \\
  \label{eq:AII-definition}
  \AII(x, y, n, \eta) &\defeq& E(x, n) \otimes \Gamma(y, n + 1),
  &&\partial x = 0,\ \partial y = \eta x && \text{($n$ odd)} \\
  \label{eq:BII-definition}
  \BII(x, y, n, \eta) &\defeq& \Gamma(x, n) \otimes E(y, n + 1),
  &&\partial x = 0,\ \partial y = \eta x && \text{($n$ even)}.
\end{IEEEeqnarray}
These elementary complexes turn out to be the only algebras that we need to handle when computing the (co)homology of Eilenberg--MacLane spaces, as follows: We first decompose $\Lambda[G]$ as in \eqref{eq:G-decomposition}. We will see in \zcref{sec:simple-constructions} that each factor has a delooping which is an elementary complex (of type $\AI$ or $\AII$). By virtue of the tensor product construction \eqref{eq:tensor-construction}, we get a delooping of $\Lambda[G]$ as a tensor product of elementary complexes. Moore's result \cite{doi:10.1073/pnas.43.5.409} states that over $\Z_{(p)}$, each elementary complex remarkably has a delooping which is again a tensor product of elementary complexes (Cartan established this over $\Z_p$). We can therefore iterate these deloopings $h$ times together with the tensor product construction to obtain the desired small model of $C_{\bullet}(B^h G; \Lambda)$.

\subsection{Some simple constructions for $B\Z$, $B\Z_N$ and $B^2 \Z$}
\label{sec:simple-constructions}

To warm up, we begin with some classical examples of simple constructions that work over any base ring $\Lambda$, and show how they reproduce the homology of $B\Z \simeq S^1$, the infinite lens space $B\Z_N$ and $B^2\Z \simeq \mathbb{CP}^\infty$. These constructions will serve as the base cases of the general calculation.

We start by considering the fibration $G \lto P(BG) \lto BG$, with $G = \mathbb{Z}$ or $G = \mathbb{Z}_N$. Algebraically, this is modelled by the construction where the initial algebra is the group algebra $\Lambda[G]$ and the final algebra is determined according to $G$ being finite or infinite.
\begin{itemize}
\item For $G = \Z$, the group algebra $\cA = \Lambda[\Z] = \Lambda[a,a^{-1}]$ is the algebra of Laurent polynomials $P(a)$, with augmentation $\eps(P(a)) = P(1)$. Let the final algebra be $\cN = \AI(x, 1) = E(x, 1)$ and define the differential on $\cM = \cA \otimes \cN$ by $\partial x = a - 1$, which makes $\cM$ an acyclic CDGA-algebra. To make this construction special we need a contraction $s$, satisfying $\partial s (P(a)) = P(a) - P(1)$ and $s \partial (x P(a)) = x P(a)$. To this end, we can define
\begin{equation}
    s(P(a)) = \frac{P(a) - P(1)}{a - 1}\, x, \qquad s(P(a)\, x) = 0 \, ,
\end{equation}
and since the submodule $\tildeN = \Lambda + \cA \otimes x$ is closed under multiplication and contains $\cN$, this defines a special construction. The quotient map $\cM \lto \cN$ sets $a$ to $1$, so the induced differential on $\cN$ is $\partialbar x = 0$, and the contraction becomes $\sbar(P(a)) = P'(1)\, x$; in particular $\sbar(a^n) = nx$.

As an aside, this algebraic construction mirrors the topological path space fibration $\Z \lto \R \lto S^1 \simeq B\Z$: We begin with $\Z$, viewed as a multiplicative discrete topological group with a generator $a$. The group element $a$ corresponds to a loop in the base of the fibration -- the loop space. The base loop $a$ can be lifted in the total space to a path $x$ from the basepoint $1$ to $a$, satisfying $\partial x = a - 1$. The path $a^n \to a^{n+1}$ is obtained by transporting $x$ along the loop labelled by $a$; we encode this transport as multiplication by $a$ and write these paths as products $a^n x$, which agrees with the Leibniz rule $\partial(a^n x) = a^{n+1} - a^n$. A contraction $s$ assigns a path from the basepoint to any given point, meaning that $\partial s(a^n) = a^n - 1$. Taking $s(a^n) = x + ax + a^2x + \cdots + a^{n-1}x = \frac{a^n - 1}{a - 1} x$  fulfils the requirement. Finally, the quotient by the fibre collapses every group element to the basepoint $1$, so the base space has a single loop $x$---it is a circle.

The final algebra $\cN$ and its differential $\partialbar$ now agree with the standard chain complex for $B\Z \simeq S^1$:
\begin{equation}
    \label{eq:S1-chain-complex}
    1
    \xlongleftarrow{0} x
    \longleftarrow 0
    \longleftarrow 0
    \longleftarrow \cdots.
\end{equation}
In particular, its homology $H_{\bullet}(B\mathbb{Z}; \Lambda) \cong \{\Lambda, \Lambda, 0, 0, \dots\}$ agrees with $H_{\bullet}(S^1; \Lambda)$. Notice that \eqref{eq:S1-chain-complex} is much smaller than the bar construction, where the group of $D$-chains is $\Lambda[\Z^D]$.

\item For $G = \Z_N$, the group algebra is $\cA = \Lambda[\Z_N] = \Lambda[u] / (u^N - 1)$. We write its elements as polynomials $P(u)$ of degree $< N$. Let the final algebra be $\cN = E(x, 1) \otimes \Gamma(y, 2)$, and by defining the differential on $\cM$ by
\begin{equation}
  \label{eq:LambdaZN-partial}
  \partial x = u - 1, \qquad \partial y = \frac{u^N - 1}{u - 1}\, x,
\end{equation}
compatible with divided powers, we make $\cM$ acyclic. A contraction is given by
\begin{equation}
  \label{eq:LambdaZN-s}
  \eqbox{c}{rCl}{
    s\qty\big(P(u) \gp_k(y)) &=& \frac{P(u) - P(1)}{u - 1}\, x \gp_k(y), \\
    s\qty\big(u^n x \gp_k(y)) &=&
    \begin{cases}
        0 & 0 \leq n < N - 1 \\
        \gp_{k + 1}(y) & n = N - 1,
    \end{cases}}
\end{equation}
and one checks that this defines a special construction. The induced differential on $\cN$ is
\begin{equation}
  \label{eq:BZN-d}
  \partialbar x = 0, \qquad \partialbar y = Nx
\end{equation}
so that $\cN = \AII(x, y, 1, N)$, and the induced contraction is $\sbar(P(u)) = P'(1)\, x$. In particular $\sbar(u^n) = nx$ for $0 \leq n < N$.

Topologically, the differential $\partialbar$ is the boundary map in the standard CW-complex for the lens space $B\Z_N$:
\begin{equation}
    \label{eq:BZN-CW}
    1
    \xlongleftarrow{0} x
    \xlongleftarrow{N} y
    \xlongleftarrow{0} xy
    \xlongleftarrow{N} \gp_2(y)
    \xlongleftarrow{0} x\gp_2(y)
    \xlongleftarrow{N} \gp_3(y)
    \xlongleftarrow{0} \cdots \,
\end{equation}
from which we recover the known result $H_{\bullet}(B\mathbb{Z}_N; \Lambda) \cong \left\{\Lambda,\Lambda/N \Lambda,{}_N\Lambda,\dots \right\}$, where $_N\Lambda = \Tor_{\Z}(\Lambda, \Z_N)$.
\end{itemize}

In addition to the group algebras above, it is also easy to write down a construction on one of the four elementary complexes, namely the exterior algebra $\AI(x, n)$:
\begin{itemize}
\item For $\cA = \AI(x, n) = E(x, n)$, let $\cN = \Gamma(y, n + 1)$ with $\partial y = x$. A contraction is given by
  \begin{equation}
    \label{eq:AI-s}
    s(x\gp_k(y)) = \gp_{k+1}(y), \qquad s(\gp_k(y)) = 0.
  \end{equation}
  The induced differential is $\partialbar y = 0$, and $\sbar x = y$.\footnote{Like the bar construction, this construction has $\tildeN = \cN$. In fact, the two are isomorphic.}

  For $n = 1$, the final algebra
  \begin{equation}
    1 \longleftarrow 0 \longleftarrow
    y \longleftarrow 0 \longleftarrow
    \gp_2(y) \longleftarrow 0 \longleftarrow
    \gp_3(y) \longleftarrow 0 \longleftarrow \cdots
  \end{equation}
  agrees with the standard chain complex for $\C P^\infty \simeq B^2 \Z$. In particular, the homology groups $H_{\bullet}(\Gamma(y, 2); \Lambda) \cong \{\Lambda, 0, \Lambda, 0, \dots\} \cong H_{\bullet}(\mathbb{CP}^\infty; \Lambda)$ agree.
\end{itemize}

\subsection{The cohomology suspension in a construction}
\label{sec:suspension-construction}

Before continuing with the computation, let us see how to obtain the cohomology suspension in an acyclic construction.

For simplicity, let us start with the dual case of the homology suspension $\sigma\colon H_D(\Omega X; \Lambda) \lto H_{D+1}(X; \Lambda)$. We can define this on cycles by considering the path space fibration $\Omega X \xlongrightarrow{i} PX \xlongrightarrow{\pi} X$, as follows: Given a cycle $a \in Z_D(\Omega X; \Lambda)$, we obtain a cycle $i_* a \in Z_D(PX; \Lambda)$. Since $PX$ is contractible, we can find a chain $m \in C_{D+1}(PX; \Lambda)$ such that $\partial m = a - \eps a$, where $\eps$ is the standard augmentation on $C_{\bullet}(\blank; \Lambda)$. Then
\begin{equation}
  \label{eq:sigma-on-cocycles}
  \sigma[a] = [\pi_* m].
\end{equation}
It is simple to check that this is a well-defined map in homology (see \cite[exposé~6, sec.~1]{cartanAlgebreDEilenbergMacLane1954}).

Dually, the cohomology suspension $\Omega\colon H^{D+1}(X; \Lambda) \lto
H^D(\Omega X; \Lambda)$ can be defined on cocycles: Given $\nu \in Z^{D+1}(X; \Lambda)$, we obtain $\pi^* \nu \in Z^{D+1}(PX; \Lambda)$, and $\pi^*\nu = \dd{\mu}$ for some cochain $\mu \in C^D(PX; \Lambda)$; for $D = 0$ we additionally require $\mu(1) = 0$ where $1 \in C_0(PX; \Lambda)$ is the basepoint.\footnote{This condition, and the appearance of $\eps$ in the definition of $\sigma$, are artefacts of the fact that these maps are more naturally defined in reduced (co)homology.} The pullback $i^* \mu \in C^D(\Omega X; \Lambda)$ is closed: $\dd(i^* \mu) = (\pi \circ i)^* \nu = 0$. We then have
\begin{equation}
  \label{eq:omega-on-cocycles}
  \Omega[\nu] = [i^* \mu].
\end{equation}

In any acyclic construction, the homology suspension $\sigma\colon H_D(\cA) \lto H_{D+1}(\cN)$ and the cohomology suspension $\Omega\colon H^{D+1}(\cN) \lto H^D(\cA)$ are defined as for acyclic fibrations in \eqref{eq:sigma-on-cocycles} and \eqref{eq:omega-on-cocycles} (with $\dd \defeq \partial^*$). That is,
\begin{IEEEeqnarray}{lClCsl}
    \label{eq:sigma-def-construction}
    \sigma[a] &=& [\overline{m}] &\qquad& where & \partial m = a - \eps a, \\
    \label{eq:Omega-def-construction}
    \Omega[\nu] &=& [i^* \mu] && where & \dd{\mu} = \overline{\nu}
    \qq{and} \mu(1) = 0
\end{IEEEeqnarray}
(writing $\overline{\nu}(m) = \nu(\overline{m})$). However, we can also obtain more direct expressions using the contraction $s$, as the induced maps of $\sbar\colon \cA \lto \cN$ in (co)homology:
\begin{equation}
  \sigma = \sbar_* \qq{and} \Omega = \sbar^* \, .
\end{equation}
It is not difficult to see that these definitions agree with \eqref{eq:sigma-def-construction} and \eqref{eq:Omega-def-construction}: Given $a \in A$ with $\partial a = 0$, we need to pick an arbitrary antiderivative $m$ of $a \otimes 1 - \eps a$. But $s$ provides a canonical antiderivative: $\partial(s(a \otimes 1)) = a \otimes 1 - \eps a$, so we can pick $m = s(a\otimes 1)$. Then $\sigma[a] = [\overline{m}] = \qty\Big[\overline{s(a \otimes 1)}] = [\sbar a]$. Similarly, since \eqref{eq:contraction} implies $\dd s^* + s^* \dd = \id - \eps^*$ on $C^{\bullet}(\cM) = \Hom(\cM, \Lambda)$, if $\nu \in C^{D+1}(\cN)$ is closed, we have $\nu = \dd(s^*\nu)$ and $s^*\nu(1) = \nu(s(1)) = 0$, so $\Omega[\nu] = \qty[\overline{s^*\nu}] = [\sbar^* \nu]$.

For example, consider the simple constructions of \zcref{sec:simple-constructions}. For $\sigma\colon \Lambda[\Z] \lto H_1(B\Z; \Lambda)$, we obtain $\sigma[P(a)] = [P'(1) x]$. Dually for $\Omega\colon H^1(B\Z; \Lambda) \lto \Hom(\Lambda[\Z], \Lambda)$, we have $\Omega[\nu] = \qty\big[P(a) \mapsto \nu\qty\big(P'(1) x)]$. The same formulae hold for $B\Z_N$ (with $a$ replaced by $u$). For the construction over $\AI(x, n)$, we get $\sigma[x] = [y]$, $\Omega[y^*] = [x^*]$ and $\Omega[\gp_k(y)^*] = 0$ for $k \geq 2$.

We now list a few properties of the map $\sbar$. First, the (co)homology suspension annihilates \emph{decomposable elements}: If $a, a' \in \cA$ are both closed and have zero augmentation, then
\begin{equation}
  s(aa') = s\qty((\partial s a) a') = s\partial\qty((s a) a')
  = (s a) a' - \partial s\qty((s a) a'),
\end{equation}
taking the quotient into $\cN$ we have $\sbar(aa') = -\partialbar \overline{s\qty\big((s a) a')}$, which is exact, and thus $\sigma\qty\big([a][a']) = 0$.\footnote{This actually holds in all acyclic constructions, not only special ones.}
Similarly, if $\dd\nu = \dd\nu' = \eps^*\nu = \eps^*\nu' = 0$, then $\Omega\qty\big([\nu] \cupp [\nu']) = 0$.\footnote{A cup product is extra structure not included in the data of a construction (see \zcref{sec:more-structures}), but the proof works for any product that satisfies the graded Leibniz rule.}

Second, the map $\sbar$ annihilates tensor products: In a tensor product construction \eqref{eq:tensor-construction}, we have
\begin{equation}
  \label{eq:sbar-tensor}
  \sbar = \sbar_1 \otimes \eps_2 + \eps_1 \otimes \sbar_2,
\end{equation}
(which is symmetric, in contrast to \zcref{eq:tensor-construction}) so that $\sbar(a \otimes b) = 0$ whenever both $a$ and $b$ have zero augmentation, e.g.~if both $a$ and $b$ have positive degree. This means that $\sigma\colon H_D(\cA_1 \otimes \cA_2) \lto H_{D+1}(\cN_1 \otimes \cN_2)$ also decomposes as $\sigma_1 \otimes \eps_2 + \eps_1 \otimes \sigma_2$. For example, in the Künneth formula when $\Lambda$ is a PID,
\begin{equation}
  \label{eq:kunneth}
  0 \lto \bigoplus_{i + j = D} H_i(\cA_1) \otimes_\Lambda H_j(\cA_2)
  \lto H_D(\cA_1 \otimes \cA_2)
  \lto \bigoplus_{i + j + 1 = D} \Tor_\Lambda(H_i(\cA_1), H_j(\cA_2)) \lto 0,
\end{equation}
any homology class coming from a term with $i \neq D \neq j$ has zero suspension.

Finally, to obtain the contractions in the constructions over $\Z_{(p)}$ that we will encounter, we will start from a given contraction mod $p$, and improve it to one over $\Z_{(p)}$, as follows. Begin with some $s_1$ such that $\id - \eps - (\partial s_1 + s_1 \partial) = 0 \bmod{p}$ and $s_1^2 = s_1 \eps = \eps s_1 = 0$ exactly. Then define
\begin{equation}
  e = \partial s_1 + s_1 \partial + \eps,
\end{equation}
so that $e = \id \bmod{p}$. In each degree, $e$ is a square matrix\footnote{Finite-dimensional as our constructions are degreewise finitely generated.} whose determinant $\det e = 1 \bmod{p}$ is a unit in $\Z_{(p)}$; therefore $e$ is invertible over $\Z_{(p)}$. Now we obtain a contraction
\begin{equation}
    s = e^{-1} s_1.
\end{equation}
To check that $s$ is a contraction, first note that $\partial$ and $s_1$ both commute with $e$ (by virtue of $\partial^2 = s_1^2 = 0$), and therefore also with $e^{-1}$. Similarly, $\eps e = \eps = e \eps$ implies $e^{-1} \eps = \eps = \eps e^{-1}$. Then
\begin{equation}
    \partial s + s \partial = e^{-1} (\partial s_1 + s_1 \partial)
    = e^{-1} (e - \eps) = \id - \eps,
\end{equation}
and it is also easy to see that $s^2 = s\eps = \eps s = 0$. To compute $s$ in practice, we will use the geometric series expansion
\begin{equation}
  \label{eq:p-local-improvement}
  s = \sum_{\ell = 0}^\infty (\id - e)^\ell s_1
  = \sum_{\ell = 0}^\infty (\id - s_1 \partial)^\ell s_1
\end{equation}
where the second equality is due to $s_1^2 = 0$. Since $\id - e = 0 \bmod{p}$, this gives an iterative $p$-adic improvement of $s_1$, in the sense that the $\ell$th partial sum $s_\ell$ is a good contraction over $\Z_{p^\ell}$. Over other rings, if \eqref{eq:p-local-improvement} converges when applied to some $x$, it converges to $s(x)$. By definition, it always converges over the $p$-adic integers $\hat{\Z}_p$. Over $\Z_{(p)}$, we conjecture that, in the specific constructions of Moore that we use, $\id - e$ is pointwise nilpotent, so that \eqref{eq:p-local-improvement} in fact terminates and allows us to compute $s(x)$ as a finite sum for all $x$. It seems likely that this can be proven using a filtration argument. However, as the sum is finite for all the $x$ we apply it to in \zcref{app:moore-constructions}, which is all that is needed for our purposes, we do not pursue the general proof further.

We are not aware of an explicit mention of the iteration \eqref{eq:p-local-improvement} in previous literature, although the idea is natural, and $p$-adic approximations in general (as in Hensel's lemma) are ubiquitous. The method is closely related to, but not the same as, the \emph{homological perturbation lemma}
\cite{shihHomologieEspacesFibres1962,brownTwistedEilenbergZilberTheorem,gugenheimChaincomplexFibration1972,lambeApplicationsPerturbationTheory1987,barnesFixedPointApproach1991,realHomologicalPerturbationTheory2000}. In a specialised form, this states that if $(\cM, \eps, \partial)$ is a DGA-algebra with contraction $s$, and we perturb the differential into another differential $\partial' = \partial + \delta$, a new contraction is given by
\begin{eqnarray}
  \label{eq:pert-lemma}
   s' = (\id + s \delta)^{-1} s = \sum_{\ell = 0}^\infty (-s\delta)^\ell s,
\end{eqnarray}
so long as $s\delta$ is pointwise nilpotent. In fact, this method could be used to derive contractions for the constructions over the algebras $\AII(x, y, n, \eta)$ and $\BII(x, y, n, \eta)$ of \eqref{eq:AII-definition} and \eqref{eq:BII-definition}. These agree as algebras with the tensor products $\AI(x, n) \otimes \BI(y, n + 1)$ and $\BI(x, n) \otimes \AI(y, n + 1)$ respectively, but the differentials are ``perturbed'' or ``twisted'' by the parameter $\eta$, with respect to those of the tensor products. Accordingly, we can adjust the tensor product contraction \eqref{eq:tensor-construction} to obtain a contraction for the perturbed construction, using the perturbation lemma with $\delta = \partial - \eval{\partial}_{\eta = 0}$. However, since the perturbation lemma does not directly allow us to lift contractions from $\Z_p$ to $\Z_{(p)}$, we forgo it in favour of \eqref{eq:p-local-improvement}.

\subsection{Coefficient groups}
\label{sec:coefficients}

In the coming sections, we will compute the (co)suspensions $\sigma_A$ and $\Omega_A$ where $A = \Z_{p^\ell}$ or the ring of $p$-local integers $\Z_{(p)}$. Let us first see that this information is sufficient to recover the maps for any finitely generated abelian group $A$, as well as $\Omega_{\R/\Z}$.

First, we reconstruct $\sigma_{\Z}$. To access homology with $\Z$ coefficients, note that the suspension fits into the (split) short exact sequence provided by the Universal Coefficient Theorem as follows:
\begin{equation}
  \label{eq:UCT-sigma}
  \small
  \begin{tikzcd}
    0 \ar[r]
    & H_D(B^h G) \otimes A \ar[r] \ar[d,"\sigma_\Z \otimes \id_A"]
    & H_D(B^h G; A) \ar[r]
      \ar[d,"\sigma_A"]
    & \Tor(H_{D-1}(B^h G), A) \ar[r] \ar[d]
    & 0 \\[2em]
    0 \ar[r]
    & H_{D+1}(B^{h+1} G) \otimes A \ar[r]
    & H_{D+1}(B^{h+1} G; A) \ar[r]
    & \Tor(H_D(B^{h+1} G), A) \ar[r]
    & 0
  \end{tikzcd}
\end{equation}
This diagram is natural in $A$ (as well as in $G$) and when $A$ is a torsion-free group (like $\Z$ or $\Z_{(p)}$), the $\Tor$ term vanishes, simplifying \eqref{eq:UCT-sigma} to
\begin{equation}
  \label{eq:UCT-sigma-torsion-free}
  \small
  \begin{tikzcd}
    0 \ar[r]
    & H_D(B^h G) \otimes A \ar[r,"\cong"] \ar[d,"\sigma_\Z \otimes \id_A"]
    & H_D(B^h G; A) \ar[r]
      \ar[d,"\sigma_A"]
    & 0 \\[2em]
    0 \ar[r]
    & H_{D+1}(B^{h+1} G) \otimes A \ar[r,"\cong"]
    & H_{D+1}(B^{h+1} G; A) \ar[r]
    & 0
  \end{tikzcd}
\end{equation}
and we can identify $\sigma_A$ with $\sigma_\Z \otimes \id_A$. Since $G$ is
finitely generated, so is $H_D(B^h G)$ by \cite[Lemma~5.10]{hatcherSpectralSequences2004}. Therefore, knowledge of $\sigma_\Z \otimes \id_{\Z_{(p)}}$ for all $p$ is enough to reconstruct $\sigma_\Z$.

In particular, suppose that $G$ is $p$-primary (the order of each element is a power of $p$). Then $H_D(B^h G; \Z)$ are also $p$-primary for $D \geq 1$ \cite[Lemma~5.10]{hatcherSpectralSequences2004}. Since $H \otimes \Z_{(p)} \cong H$ for a $p$-primary group, this means that
\begin{equation}
  \label{eq:primary-local-Z}
   H_D(B^h G; \Z) \cong H_D(B^h G; \Z_{(p)}) \qq{and} \sigma_\Z = \sigma_{\Z_{(p)}}
   \qquad (D \geq 1).
\end{equation}

Next, we consider cohomology. The cosuspension enters in the UCT as
follows:
\begin{equation}
  \label{eq:UCT-Omega}
  \small
  \begin{tikzcd}
    0 \ar[r]
    & \Ext(H_D(B^{h+1} G), A) \ar[r] \ar[d]
    & H^{D+1}(B^{h+1} G; A) \ar[r]
      \ar[d,"\Omega_A"]
    & \Hom(H_{D+1}(B^{h+1} G), A) \ar[r] \ar[d,"(\sigma_\Z)^*"]
    & 0 \\[2em]
    0 \ar[r]
    & \Ext(H_{D-1}(B^h G), A) \ar[r]
    & H^D(B^h G; A) \ar[r]
    & \Hom(H_D(B^h G), A) \ar[r]
    & 0
  \end{tikzcd}
\end{equation}
For cohomology with $A = \R/\Z$ (or any other injective $\Z$-module),
$\Ext(\blank, \R/\Z) = 0$ and \eqref{eq:UCT-Omega} simplifies to
\begin{equation}
  \label{eq:UCT-Omega-U1}
  \small
  \begin{tikzcd}
    0 \ar[r]
    & H^{D+1}(B^{h+1} G; \R/\Z) \ar[r,"\cong"]
      \ar[d,"\Omega_{\R/\Z}"]
    & \widehat{H_{D+1}(B^{h+1} G)} \ar[r] \ar[d,"(\sigma_\Z)^*"]
    & 0 \\[2em]
    0 \ar[r]
    & H^D(B^h G; \R/\Z) \ar[r,"\cong"]
    & \widehat{H_D(B^h G)} \ar[r]
    & 0
  \end{tikzcd}
\end{equation}
where $\widehat{\blank} = \Hom(\blank, \R/\Z)$ is the Pontryagin dual, and we can identify $\Omega_{\R/\Z} = (\sigma_\Z)^*$. Thus, computing the higher anomaly map $\Omega_{\R/\Z}$ amounts to computing the integral suspension $\sigma_\Z$.

For cohomology with other coefficients (as relevant for the application to
higher groups), Moore's constructions will directly give us $\Omega_{\Z_{(p)}}$
and $\Omega_{\Z_{p^k}}$. We can compute $\Omega_\Z$ from the knowledge of all
$\Omega_{\Z_{(p)}}$ by making use of the natural isomorphism
\begin{equation}
  \label{eq:p-local-cohomology-isomorphism}
  H^{\bullet}(X; \Z) \otimes \Z_{(p)} \cong H^{\bullet}(X; \Z_{(p)})
\end{equation}
which holds when $X$ has finitely generated homology (see
\zcref{app:p-local-cohomology}). Naturality means that the top square of
\begin{equation}
  \begin{tikzcd}
    H^{D+1}(B^{h+1} G; \Z) \otimes \Z_{(p)} \ar[r,"\cong"]
    \ar[d,"\evMap^* \otimes \id_{\Z_{(p)}}"]
    \ar[dd,bend right=75,"\Omega_\Z \otimes \id_{\Z_{(p)}}"']
    & H^{D+1}(B^{h+1} G; \Z_{(p)})
    \ar[d,"\evMap^*"]
    \ar[dd,bend left=75,"\Omega_{\Z_{(p)}}"]
    \\[2em]
    H^{D+1}(\Sigma B^h G; \Z) \otimes \Z_{(p)} \ar[r,"\cong"] \ar[d,"\cong"]
    & H^{D+1}(\Sigma B^h G; \Z_{(p)}) \ar[d,"\cong"]
    \\[2em]
    H^D(B^h G; \Z) \otimes \Z_{(p)} \ar[r,"\cong"]
    & H^D(B^h G; \Z_{(p)})
  \end{tikzcd}
\end{equation}
commutes, and the bottom square commutes since \eqref{eq:p-local-cohomology-isomorphism} is a natural transformation of cohomology theories. The composites are the cohomology suspensions (see \eqref{eq:Omega-as-pullback}), so this means that we can identify $\Omega_\Z \otimes \id_{\Z_{(p)}}$ with $\Omega_{\Z_{(p)}}$. As for $\sigma_\Z$, knowledge of these maps for all $p$ is sufficient to reconstruct $\Omega_\Z$.

This determines $\sigma$ and $\Omega$ with coefficients in an arbitrary finitely generated abelian group: We simply have $H_n(X; A \oplus B) \cong H_n(X; A) \oplus H_n(X; B)$ and $\sigma_{A \oplus B} = \sigma_A \oplus \sigma_B$, and similarly $H^n(X; A \oplus B) \cong H^n(X; A) \oplus H^n(X; B)$ and $\Omega_{A \oplus B} = \Omega_A \oplus \Omega_B$.

In conclusion, knowing the (co)suspensions $\sigma$ and $\Omega$ for coefficients in $\Z_{(p)}$ and $\Z_{p^\ell}$ is enough to calculate them for any finitely generated coefficient group, as well as the cosuspension with $\R/\Z$ coefficients.

When $G$ is finite, since $H^*(B^h G; \R)=0$, the connecting morphism of the long exact sequence in cohomology associated to the coefficients sequence $\Z \lto \R \lto \R/\Z$ gives a natural isomorphism
\begin{equation}
  \label{eq:connecting-isomorphism}
  H^D(B^h G; \R/\Z) \cong H^{D+1}(B^h G; \Z).
\end{equation}
This isomorphism commutes with $\Omega$ by naturality, identifying $\Omega_{\R/\Z}$ and $\Omega_\Z$. In \zcref{sec:general-Omega}, we compute $\Omega_{\R/\Z}$ and $\Omega_\Z$ by independent means (using $\Omega_{\R/\Z} = (\sigma_\Z)^*$ \eqref{eq:UCT-Omega-U1} rather than its connection to $\Omega_\Z$), and they do agree in the resulting output of \texttt{emcm} \cite{riedelgardingEmcm2026}. This is a nontrivial consistency check on our methods. For infinite $G$, e.g.~$G = \Z$, the connecting morphism is no longer an isomorphism, since now $H^*(B^h G; \R) \neq 0$ in general. The discrepancies between $\Omega_{\R/\Z}$ and $\Omega_\Z$ for $G = \Z$ can be seen from the differences between \zcref{tab:Omega-BZ-R/Z} and \zcref{tab:Omega-BZ-Z}.

\subsection{A first example of the cosuspension}
In this section, we will walk the reader through the calculation of the cosuspension
\begin{equation}
  \Omega\colon H^{D+1}(B^2 \Z_{p^f}; \R/\Z) \lto H^D(B\Z_{p^f}; \R/\Z),
\end{equation}
introducing only the results necessary for this task. We leave the remainder of the general calculation of \eqref{eq:ordinary-omega} to the next section.

The first step is to determine the algebra $C_{\bullet}(B^2 \Z_{p^f}; \Z_{(p)})$ by delooping the group algebra $\Z_{(p)}[\Z_{p^f}]$ twice. We have already determined the first delooping in \zcref{sec:simple-constructions}; it is $\AII(x, y, 1, p^f)$ with $\sbar(u^n) = n\, x$ (where $u$ is a generator of $\Z_{p^f}$). \Textcite{doi:10.1073/pnas.43.5.409} found a further delooping of this algebra, explained in detail in \zcref{app:moore-constructions}. Here we do not enter into the derivation but simply quote the result \eqref{eq:AII-delooping-as-tensor}; the delooping is\footnote{For sake of simplicity, we stray a bit from the notation of the next section, where the generators are written as formal words in certain letters $\sigma$, $\gp_{p^i}$, $\tp_p$, $\psi_N$ and $\eps_{p^i}$; here we employ a simpler but less generalisable naming scheme.}
\begin{equation}
  \label{eq:B2Zpf-algebra}
    \BII(z, a, 2, -p^f) \otimes
    \bigotimes_{i \geq 0} \AII(c_{i+1}, b_i, 2 p^{i+1} + 1, -p \zeta_i).
\end{equation}
The constants $\zeta_i$ are defined precisely below; here we only need to know that they are some integers not divisible by $p$.

The (co)homology groups of $B^2 \Z_{p^f}$ are those of \eqref{eq:B2Zpf-algebra}, and are easy to compute. First, the homology of the individual elementary complexes is easy to calculate. In $\AII(x, y, n, -p\zeta_i)$, the differential reads $\partial(\gp_{k+1}(y)) = -p\zeta_i\, x \gp_k(y)$, so we have the homology classes
\begin{equation}
  \qty\Big[x \gp_k(y)] \qquad\text{of order } p
\end{equation}
(since $\zeta_i$ is invertible in $\Z_{(p)}$ it does not affect the order). In $\BII(z, a, 2, -p^f)$, it reads $\partial(\gp_{k-1}(z)\, a) = -p^f \gp_{k-1}(z)\, z = -p^f k \gp_k(z)$, so we have the homology classes
\begin{equation}
  \qty\Big[\gp_k(z)] \qquad\text{of order } \gcd(p^f k, p^\infty),
\end{equation}
where the notation for the order means the highest power of $p$ in the prime decomposition of $p^f k$.

With the homology groups of the individual complexes in hand, we can compute those of the tensor product \eqref{eq:B2Zpf-algebra} using the Künneth formula; the split short exact sequence \eqref{eq:kunneth}. For $a \in H_i(A)$ of order $\alpha$ and $b \in H_j(B)$ of order $\beta$, we obtain generators $a \otimes b \in H_{i+j}(A \otimes B)$ and $\tau(a, b) \in H_{i+j+1}(A \otimes B)$, both of order $\gcd(\alpha, \beta)$.\footnote{In general, the $\tau$ generator is not naturally determined by the classes $a$ and $b$ as the split in \eqref{eq:kunneth} is not natural, but in our setting there is a preferred choice, given later in \eqref{eq:tau}.}

By this process, we obtain $H_{\bullet}(B^2 G; \Z_{(p)})$, which agrees with $H_{\bullet}(B^2 G; \Z)$ in positive degree by \eqref{eq:primary-local-Z}.

For instance, let us compute $H_{\bullet}(B^2 \Z_2; \Z_{(2)})$ up to degree $10$. The first elementary complex $\BII(z, a, 2, -2)$ has the following homology generators, where the number in angle brackets is the order:
\begin{equation}
  \begin{tikzcd}[row sep=1ex, column sep=1em]
    H_0 & H_2 & H_4 & H_6 & H_8 & H_{10} & \cdots \\
    1 & z\langle2\rangle & \gp_2(z)\langle4\rangle & \gp_3(z)\langle2\rangle & \gp_4(z)\langle8\rangle & \gp_5(z)\langle2\rangle & \cdots
  \end{tikzcd}
\end{equation}
The second complex $\AII(c_1, b_0, 5, -2\zeta_0)$ has an order $2$ class $c_1$ in degree $5$; the next lowest class is $c_1 b_0$ in degree $11$. The third complex $\AII(c_2, b_1, 9, -2\zeta_1)$ has an order $2$ class $c_2$ in degree $9$; the next is $c_2 b_1$ in degree $19$. The remaining complexes are all trivial below degree $10$. Combining the above homologies using the Künneth formula gives the homology groups and generators of \zcref{tab:Z2-homology-manual}. They agree with $H_{\bullet}(B^2 \Z_2; \Z)$ in positive degree.
\begin{table}[h]
  \centering
  \footnotesize
\begin{tikzcd}[row sep=1ex, column sep=3pt]
    H_0 & H_1 & H_2 & H_3 & H_4 & H_5 & H_6 & H_7 & H_8 & H_9 & H_{10} \\
    \Z_{(2)} & 0 & \Z_2 & 0 & \Z_4 & \Z_2 & \Z_2 & \Z_2 &
    \Z_8 \oplus \Z_2 & (\Z_2)^2 & (\Z_2)^2 \\
1
&
& z\langle 2 \rangle
&
& \gamma_{2}(z)\langle 4 \rangle
& c_1\langle 2 \rangle
& \gamma_{3}(z)\langle 2 \rangle
& z \otimes c_1\langle 2 \rangle
& \gamma_{4}(z)\langle 8 \rangle
& c_2\langle 2 \rangle
& \gamma_{5}(z)\langle 2 \rangle \\
&&&&&&&
& \tau(z, c_1)\langle 2 \rangle
& \gamma_{2}(z) \otimes c_1\langle 2 \rangle
& \tau(\gamma_{2}(z), c_1)\langle 2 \rangle
\end{tikzcd}
  \caption{The homology groups $H_{\bullet}(B^2 \Z_2; \Z_{(2)})$ and their generators.}
  \label{tab:Z2-homology-manual}
\end{table}

We now move on to calculate the homology suspension. In \zcref{app:moore-constructions}, we show how to obtain the contraction $s$ using the iteration described in \eqref{eq:p-local-improvement}. The map is quite complicated, but the quotient $\sbar$ is considerably simpler (though neither map was made explicit by Moore). To compute its induced map $\sigma$ in homology, it is enough to know the value of $\sbar$ on closed chains of $C_{\bullet}(B \Z_{p^f}; \Z_{(p)}) = \AII(x, y, 1, p^f)$. The result \eqref{eq:AII-sbar-appendix} happens to be particularly simple:
\begin{equation}
    \sbar\qty\big(x \gp_k(y)) = (-p^f)^k \gp_{k+1}(z).
\end{equation}
For $k = 0$, we simply find $\sigma[x] = [z]$. For larger $k$, recall that the class $[\gp_{k+1}(z)]$ has order $\gcd(p^f (k+1), p^\infty)$. A simple counting of powers of $p$ then shows that $(-p^f)^k [\gp_{k+1}(z)] = 0$ whenever $k \geq 2$ or $p \geq 3$ (the argument is spelled out in the next section). The only remaining nonzero suspension is when $k = 1$ and $p = 2$; then $\sigma[xy] = 2^f[\gamma_2(z)]$. In summary, the homology suspension $\sigma\colon H_D(B\Z_{p^f}; \Z_{(p)}) \lto H_{D+1}(B^2\Z_{p^f}; \Z_{(p)})$ is
\begin{equation}
  \label{eq:sigma-B2Zpf}
  \sigma[x\gp_k(y)] =
  \begin{cases}
    [z] & k = 0 \\
    2^f [\gp_2(z)] = 2^{f-1} [z]^2 & k = 1 \text{ and } p = 2 \\
    0 & \text{otherwise}
  \end{cases}
\end{equation}
(where we have omitted a sign in the middle line since the image has order $2$), and this agrees with the suspension with $\Z$ coefficients by \eqref{eq:primary-local-Z}.

Finally, we compute the cosuspension with $\R/\Z$ coefficients; by \eqref{eq:UCT-Omega-U1} this is simply the adjoint of \eqref{eq:sigma-B2Zpf}. We write cochains in $C^{\bullet}(\blank; \R/\Z) = \Hom(C_{\bullet}(\blank), \R/\Z)$ in terms of the dual basis: For a finite abelian group $G$ with a generating set $\{g\}$, we have a dual basis $\{\frac{1}{\ord(g)} g^*\}$ for $\Hom(G, \R/\Z)$, where for generators $g$ and $g'$,
$g^*(g') = 1$ if $g = g'$ and $0$ otherwise (the map $g \mapsto \frac{1}{\ord(g)} g^*$ is an isomorphism $G \cong \Hom(G, \R/\Z)$). For a homomorphism $f\colon G \lto H$, the adjoint (pullback) $f^*\colon \Hom(H, \R/\Z) \lto \Hom(G, \R/\Z)$ is defined by $f^*(h^*)(g) = h^*(f(g))$; this expands to
\begin{equation}
  f^*(h^*) = \sum_g h^*(f(g)) g^*
\end{equation}
where the sum is over the generating set. In particular, if $f(g) = nh$, where $g$ and $h$ are both generators and $f(g')$ does not contain $h$ for any other generator $g'$, this simplifies to $f^*(h^*) = n g^*$.

In the present case, we have cohomology generators $\frac{1}{p^f k} [\gp_k(z)]^*$, and the nonzero values of $\Omega$ are $\Omega [z]^* = [x]^*$ and, for $p = 2$, $\Omega [\gp_2(z)]^* = 2^f [xy]^*$, so the cosuspension is given by
\begin{equation}
  \label{eq:B2Zpf-Omega}
  \Omega\qty(\frac{1}{p^f k} [\gp_k(z)]^*) = \begin{cases}
    \frac{1}{p^f} [x]^* & k = 1 \\
    \frac{1}{2} [xy]^* & p = k = 2 \\
    0 & \text{otherwise},
  \end{cases}
\end{equation}
and $\Omega = 0$ for all other generators. The middle case is precisely the $\Omega\colon H^4(B^2 \Z_N; \R/\Z) \lto H^3(B\Z_N; \R/\Z)$ mentioned in \zcref{sec:intro-higher-anomalies}, showing the special behaviour for the prime $p = 2$. This shows the algebraic counterpart to the physical computation of \cite{Roumpedakis:2022aik}.

\subsection{General calculation of the cosuspension}
\label{sec:general-Omega}

We now turn to the calculation of the cosuspension \eqref{eq:ordinary-omega} in the general case. The method is the one outlined in the previous sections: starting with the group algebra $\Lambda[G]$, we proceed with its iterated delooping as found by \textcite{doi:10.1073/pnas.43.5.409}.\footnote{Our conventions differ slightly from Moore's original ones, in order to allow us to work with integers rather than general elements of $\Z_{(p)}$ in as many places as possible.} This produces a model of $C_{\bullet}(B^h G; \Lambda)$ from which it is straightforward to compute the (co)homology. Moreover (and this is the laborious part), we equip them with the structure of special constructions, equipped with a chain contraction $s$. Then we obtain the map $\sbar$ from $s$, allowing us to read off $\Omega$ as its induced map in cohomology.

In this section, we aim to be as general as possible, providing the constructions that allow the computation of $H^{\bullet}(B^h G; A)$ for $A = \R/\Z$ (which controls anomalies for higher gauging) and for $A = \Z$ or $\Z_N$ (which controls how higher-group structures behave under topological suspension). To this end, we now quote all the algebra deloopings (denoted $\cA \lto \cN$) needed to compute any (co)chain complex $C_{\bullet}(B^h G; A)$. The details on how they are derived can be found in \zcref{app:cartan-moore}.

Two observations are in order before proceeding. First, we will now transition to a more easily generalisable nomenclature for the elements in the various algebras. While we used a simplified notation in the previous section, it becomes too cumbersome to maintain when dealing with the (co)homology of classifying spaces in arbitrary dimensions. We write the algebra generators as formal words in certain letters $\sigma$, $\gp_{p^i}$, $\tp_p$ and $\psi_N$ (following Cartan's original naming scheme), as well as a new letter $\eps_{p^i}$ that we introduce to accommodate Moore's generalisation. Second, the following deloopings hold for both $\Lambda=\Z_{(p)}$ and $\Lambda=\Z_{p^\ell}$. The relevant difference between the two coefficient rings only enters later, when computing the (co)homology of the final algebra. We will explicitly point out when a result is common to both choices of coefficients and when it is specific to one.

We have the deloopings derived in \zcref{sec:simple-constructions}, valid over any $\Lambda$:
\begin{IEEEeqnarray}{lClctL}
  \label{eq:LambdaZ-delooping}
  \Lambda[\Z] &\lto& \AI(\sigma a, 1)
  &\qquad& with & \sbar(a^n) = n\, \sigma a, \\
  \label{eq:LambdaZN-delooping}
  \Lambda[\Z_N] &\lto& \AII(\sigma u, \tpsi_N\! u, 1, N)
  && with & \sbar(u^n) = n\, \sigma u, \\
  \AI(x, n) &\lto& \BI(\sigma x, n + 1)
  && with & \sbar(x) = \sigma x,
\end{IEEEeqnarray}
and the following ones, valid for $\Lambda = \Z_{(p)}$ or $\Z_{p^\ell}$\footnote{Or any ring such that all integers coprime to $p$ are invertible.} with $\eta \in \Lambda$ such that $\eta = 0$ mod $p$:
\begin{IEEEeqnarray}{lCl}
  \BI(x, n) &\lto&
  \AI(\sigma x, n + 1) \IEEEnonumber \\
  && {} \otimes \bigotimes_{i \geq 0}
  \AII(\sigma\!\gp_{p^{i+1}}\! x,\ \tp_p\!\gp_{p^i}\! x,\
  p^{i+1} n + 1,\ -p\zeta_i) \\
  & \IEEEeqnarraymulticol{2}{l}{\text{with }
    \sbar(\gp_k(x)) = \begin{cases}
      \sigma\!\gp_{p^i}\! x & k = p^i \\
      0 & \text{$k$ not a power of $p$}.
    \end{cases}} \\
  \label{eq:BII-delooping}
  \BII(x, y, n, \eta) &\lto&
  \AII(\sigma x, \sigma y, n + 1, -\eta) \IEEEnonumber \\
  && {} \otimes \bigotimes_{i \geq 0}
  \AII(\sigma\!\gp_{p^{i+1}}\! x,\ \tp_p\!\gp_{p^i}\! x,\
  p^{i+1} n + 1,\ -p\zeta_i) \\
  \label{eq:BII-sbar}
  & \IEEEeqnarraymulticol{2}{l}{\text{with } \eqbox{t}{lCl}{
      \sbar(\gp_k(x)) &=& \begin{cases}
        \sigma\!\gp_{p^i}\! x & k = p^i \\
        0 & \text{$k$ not a power of $p$},
      \end{cases} \\
      \sbar(y \gp_k(x)) &=& \begin{cases}
        \sigma y & k = 0 \\
        \frac{1}{\rho_i} \eta p^i \tp_p\!\gp_{p^i}\! x & k = p^{i+1} - 1 \\
        0 & \text{otherwise}
      \end{cases}
    }} \\
  \label{eq:AII-delooping}
  \AII(x, y, n, \eta) &\lto&
  \BII(\sigma x, \sigma y, n + 1, -\eta) \IEEEnonumber \\
  && {} \otimes \bigotimes_{i \geq 0}
  \AII(\eps_{p^{i+1}} y,\ \tp_p\!\gp_{p^i}\! y,\ p^{i+1} (n + 1) + 1,\ -p\zeta_i) \\
  \label{eq:AII-sbar}
  & \IEEEeqnarraymulticol{2}{l}{\text{with } \eqbox{t}{lCl}{
      \sbar(x \gp_k(y)) &=& (-\eta)^k \gp_{k+1}(\sigma x), \\
      \sbar(\gamma_k(y)) &=&
      {k \choose {p^i}}^{-1} \gp_{k - p^i}(-\eta \sigma x) \sum_{j = 0}^i
      \qty(\prod_{l = 1}^j \theta_{i-l} \gp_{p^{i-l}}(\sigma x)^{p-1})
      \eps_{p^{i-j}} y \\
      && \text{where } i = \nu_p(k).
    }} \qquad
\end{IEEEeqnarray}
By convention, we set $\eps_1 = \sigma$. The constants $\zeta_i$, $\rho_i$ and $\theta_i$ ($i \geq 0$) are defined as
\begin{equation}
  \zeta_i = \frac{(p^{i+1})!}{p((p^i)!)^p} \in \Z, \qquad
  \rho_i = \frac{(p^{i+1} - 1)!}{\qty\big(p! (p^2)! \cdots (p^i)!)^{p-1}} \in \Z
  \qq{and}
  \theta_i = \frac{(-\eta)^{p^i(p - 1)}}{p} \zeta_i^{-1} \in \Z_{(p)}.
\end{equation}
The integers $\zeta_i$ and $\rho_i$ are not divisible by $p$, and therefore invertible in $\Z_{(p)}$ and largely unimportant for cohomology calculations. On the other hand, $\theta_i$ contains at least $p^i(p - 1) - 1$ factors of $p$, and will often also disappear from (co)homology calculations, as we shall see.

The deloopings of the elementary complexes are nothing but (in)finite tensor products of the elementary complexes themselves. To iterate the delooping procedure, we can deloop each such elementary complex individually, and then assemble these deloopings using the tensor product construction \eqref{eq:tensor-construction}. After $h$ iterations, we have a list of complexes, usually infinitely many, but always a finite number that are nontrivial below a fixed degree. The tensor product of all these complexes constitutes the model of $C_{\bullet}(B^h G; \Z_{(p)})$. To get $C_{\bullet}(B^h G; \Z_{p^\ell})$ instead, we can simply reduce mod $p^\ell \Z_{(p)}$.

The homology of elementary complexes is easy to calculate. As $\AI$ and $\BI$ have zero differential, they are their own homology. For the $\AII$ and $\BII$ complexes, we have to distinguish between $\Lambda = \Z_{(p)}$ and $\Lambda = \Z_{p^\ell}$. Since the homology classes share common expressions, we will treat the two cases simultaneously, where the former is denoted by formally taking $\ell = \infty$. The classes that exist for both $\ell < \infty$ and $\ell = \infty$ are highlighted in \both{blue}, while the ones that exist only  for $\ell < \infty$ are highlighted in \only{red}.

In $\AII(x, y, n, \eta)$, the differential reads $\partial(\gp_{k+1}(y)) = \eta x \gp_k(y)$, so we have homology classes
\begin{equation}
  \label{eq:AII-homology-classes}
  \both{\overset{(\ell \leq \infty)}{\qty\Big[x \gp_k(y)]}_{\AII}} \quad\qq{and}\quad
  \only{\overset{(\ell < \infty)}{\qty[\frac{p^\ell}{\gcd(\eta, p^\ell)} \gp_{k+1}(y)]}_{\AII}}
  \qquad\text{of order } \gcd(\eta, p^\ell).
\end{equation}
The differential in $\BII(x, y, n, \eta)$ reads
$\partial(\gp_{k-1}(x)\, y) = \eta \gp_{k-1}(x)\, x = \eta k \gp_k(x)$, and the homology classes are
\begin{equation}
  \label{eq:BII-homology-classes}
  \both{\overset{(\ell \leq \infty)}{\qty\Big[\gp_k(x)]}_{\BII}}
  \quad\qq{and}\quad
  \only{\overset{(\ell < \infty)}{\qty[\frac{p^\ell}{\gcd(\eta k, p^\ell)} y \gp_{k-1}(x)]}_{\BII}}
  \qquad\text{of order } \gcd(\eta k, p^\ell).
\end{equation}
When $\ell = \infty$, the notation $\gcd(n, p^\infty)$ means the $p$-primary component of $n$.

In a tensor product of elementary complexes, we obtain the homology using the Künneth formula \eqref{eq:kunneth}. Any homology class mixing different complexes has zero $\sbar$ by \eqref{eq:sbar-tensor}, so it has zero suspension. Since we have a canonical basis of chains, we can write down explicit representatives for the resulting homology classes in the various tensor and $\Tor$ terms; these are found in \zcref{app:cosuspension-on-generators} but will not be needed in the main text.

Using the values of $\sbar$ displayed above, we can read off the homology suspension. The $\BI$ case is immediate, so we begin with the $\BII$ case. The classes with potentially nonzero image are
\begin{align}
  \both{\qty[\gp_{p^i}(x)]_{\BII}}
  &\overset{\sigma}{\longmapsto}
  \both{\qty[\sigma \gp_{p^i}(x)]_{\AII}},
  \\
  \only{\qty[\frac{p^{\ell}}{\gcd(\eta, p^{\ell})} y]_{\BII}}
  &\overset{\sigma}{\longmapsto}
  \only{\qty[\frac{p^{\ell}}{\gcd(\eta, p^{\ell})} \sigma y]_{\AII}},
  \\
  \label{eq:BII-sigma-3}
  \only{\qty[\frac{p^{\ell}}{\gcd(\eta p^{i+1}, p^{\ell})} y \gp_{p^{i+1}-1}(x)]_{\BII}}
  &\overset{\sigma}{\longmapsto}
  \frac{1}{\rho_i} \frac{\eta p^i \gcd(\eta, p^\ell)}{\gcd(\eta p^{i+1}, p^\ell)} \only{\qty[\frac{p^{\ell}}{\gcd(\eta, p^{\ell})} \tp_p\!\gp_{p^i}\!x]_{\AII}}.
\end{align}
The latter class is trivial if $p^\ell \mid \eta p^i$, but otherwise it is a nontrivial class of order $p$.\footnote{Counting powers of $p$ with $f \defeq \nu_p(\eta)$, the RHS of \eqref{eq:BII-sigma-3} has order $p^{\max(f - \qty(f + i + \min(f, \ell) - \min(f + i + 1, \ell)), 0)}$, which is $p$ if $f + i < \ell$ but $1$ if $f + i \geq \ell$. Here $\nu_p(n)$ denotes the $p$-adic valuation: the largest $i$ such that $p^i$ divides $n$.}

In the $\AII$ case, there are some fortunate simplifications that make the expressions \eqref{eq:AII-sbar} simpler once we pass to homology. First, a necessary condition for the suspension $\sigma(\both{[x\gp_k(y)]_{\AII}}) = (-\eta)^k \both{[\gp_{k+1}(\sigma x)]_{\BII}}$ to be nonzero is that $k \nu_p(\eta) < \nu_p(k + 1) + \nu_p(\eta)$.\footnote{The RHS has order $p^{\max(\min(f + \nu_p(k + 1), \ell) - kf, 0)}$ with $f \defeq \nu_p(\eta)$. This is $1$ unless $f + \nu_p(k + 1) > kf$.} Since $\nu_p(\eta) \geq 1$ this implies $k - 1 < \nu_p(k + 1)$. This is impossible for $k \geq 2$, and for $k = 1$ it is possible only if $p = 2$. Therefore the suspension simplifies to
\begin{equation}
  \label{eq:AII-suspension}
  \both{\sigma[x\gp_k(y)]_{\AII}} =
  \begin{cases}
    \both{[\sigma x]_{\BII}} & k = 0 \\
    \eta \both{[\gp_2(\sigma x)]_{\BII}} & k = 1 \text{ and } p = 2 \\
    0 & \text{otherwise}.
  \end{cases}
\end{equation}
The class $\eta \both{[\gp_2(\sigma x)]_{\BII}} = -\eta \both{[\gp_2(\sigma x)]_{\BII}} = \frac{\eta}{2} (\both{[\sigma x]_{\BII}})^2$ has order $2$ (or is trivial when $\ell \leq f$) since $\eta\both{[\sigma x]_{\BII}} = 0$.

For the second class in \eqref{eq:AII-homology-classes}, the expression
\begin{equation}
  \sigma\only{\qty[\frac{p^\ell}{\gcd(\eta, p^\ell)} \gp_k(y)]_{\AII}}
  = \qty[\frac{p^\ell}{\gcd(\eta, p^\ell)} \sbar(\gp_k(y))]
\end{equation}
is zero mod $p^\ell$ whenever the expression for $\sbar(\gp_{k}(y))$ in \eqref{eq:AII-sbar} is a multiple of $\eta$. This is true whenever $k$ is not a power of $p$, so that $k - p^i \neq 0$. Furthermore, $\theta_i$ is divisible by $\eta$ whenever $p^i (p - 1) \geq 2$, that is, when $p \geq 3$ or $i \geq 1$; thus any term containing a $\theta_i$ also becomes zero in homology, except possibly $\theta_0 = -\frac{\eta}{2}$ for $p = 2$. The only terms of \eqref{eq:AII-sbar} that contribute are therefore those with $j = 0$, and the term with $j = 1$ in the case $p = 2$, $i = 1$ (that is, in the expression for $\sbar(\gamma_2(y))$).

Since $\eps_1 y = \sigma y$ by definition, this all means that the suspension simplifies to
\begin{IEEEeqnarray}{r}
  \sigma\only{\qty[\frac{p^\ell}{\gcd(\eta, p^\ell)} \gp_k(y)]_{\AII}} =
  \qty[\frac{p^\ell}{\gcd(\eta, p^\ell)}
  \begin{cases}
    \sigma y & k = 1 \\
    \eps_2 y - \frac{\eta}{2} \sigma x \cdot \sigma y & p = k = 2 \\
    \eps_{p^i} y & k = p^i \text{ ($i > 1$ or $p > 2$)} \\
    0 & \text{$k$ not a power of $p$}
  \end{cases}] \nonumber \\
  = \begin{cases}
    \only{\qty[\frac{p^\ell}{\gcd(\eta, p^\ell)} \sigma y]_{\BII}} & k = 1 \\
    \frac{2^\ell}{\gcd(\eta, 2^\ell)}
    \both{[\eps_2 y]_{\AII}} - \frac{\eta}{2} \frac{\gcd(2 \eta, 2^\ell)}{\gcd(\eta, 2^\ell)} \both{\qty[\frac{2^\ell}{\gcd(2 \eta, 2^\ell)} \sigma x \cdot \sigma y]_{\BII}}
                                                                   & p = k = 2 \\
    \frac{p^\ell}{\gcd(\eta, p^\ell)} \both{\qty[\eps_{p^i} y]_{\AII}} & k = p^i \text{ ($i > 1$ or $p > 2$)} \\
    0 & \text{$k$ not a power of $p$}. \IEEEeqnarraynumspace
  \end{cases}
\end{IEEEeqnarray}
Moreover, since $p\both{[\eps_{p^i} y]_{\AII}} = 0$, the terms with $\eps_{p^i} y$ disappear unless the prefactor is $1$; thus they appear only if $\eta = 0 \bmod{p^\ell}$. In this case the differential in the original $\AII$ complex vanishes, and it decomposes into a true product of $\AI$ and $\BI$ complexes, so it would have been unnecessary to consider $\AII$ in the first place. In summary, as long as $\eta \neq 0 \bmod{p^\ell}$ (but $p \mid \eta$ as always), the result is simply
\begin{equation}
  \sigma\only{\qty[\frac{p^\ell}{\gcd(\eta, p^\ell)} \gp_k(y)]_{\AII}}
  = \begin{cases}
    \only{\qty[\frac{p^\ell}{\gcd(\eta, p^\ell)} \sigma y]_{\BII}} & k = 1 \\
    - \eta \both{\qty[\frac{2^\ell}{\gcd(2 \eta, 2^\ell)} \sigma x \cdot \sigma y]_{\BII}}
                                                                   & p = k = 2 \\
    0 & \text{otherwise}. \IEEEeqnarraynumspace
  \end{cases}
\end{equation}

Now we consider cohomology with $A = \Z_{(p)}$ or $A = \Z_{p^k}$ coefficients. Let us first notice that the chains with $A$ coefficients determine the cochains with $A$ coefficients. These particular groups have $\Hom(A, A) \cong A$, so that
\begin{equation}
  \Hom(C_*(X; A), A) = \Hom(C_*(X) \otimes A, A) \cong
  \Hom(C_*(X), \Hom(A,A)) \cong C^*(X; A)
\end{equation}
by tensor-hom adjunction.

We have a canonical basis $\{x_{D,i}\}$ of $C_D(X; A)$, namely the one provided by Cartan's words, so we get the canonical dual basis $\{x_i^*\}$ of $C^{\bullet}(X; A)$, with $x_i^*(x_j) = \delta_{ij}$. Since the boundary map $\partial\colon C_D(X; A) \lto C_{D-1}(X; A)$ is rectangular diagonal in the $\{x_{D,i}\}$ basis (it takes the form $\partial x_{D,i} = n_i x_{D-1, i}$), its transpose $\dd$, the co-differential, is also diagonal, with $\dd x_{D-1,i}^* = n_i x_{D,i}^*$. 

In $\AII(x,y,n,\eta)$, the dual differential is $\dd(x\gp_k(y))^* = \eta \gp_{k+1}(y)^*$, so the cohomology classes are (using the same colour coding as before)
\begin{equation}
  \label{eq:AII-cohomology-classes}
  \both{\qty\Big[\gp_{k+1}(y)^*]_{\AII}}\quad\qq{and}\quad
  \only{\qty[\frac{p^\ell}{\gcd(\eta, p^\ell)} \qty(x \gp_k(y))^*]_{\AII}}
  \qquad\text{of order } \gcd(\eta, p^\ell).
\end{equation}
In $\BII(x,y,n,\eta)$, the dual differential is $\dd\gp_k(x)^* = \eta k
(\gp_{k-1}(x)\, y)^*$, so the cohomology classes are
\begin{equation}
  \label{eq:BII-cohomology-classes}
  \both{\qty\Big[\qty(\gp_{k-1}(x)\, y)^*]_{\BII}} \quad\qq{and}\quad
  \only{\qty[\frac{p^\ell}{\gcd(\eta k, p^\ell)} \gp_k(x)^*]_{\BII}}
  \qquad\text{of order } \gcd(\eta k, p^\ell).
\end{equation}

Now, we calculate the cosuspension $\Omega$ from $\sbar$. In the $\BII$ case \eqref{eq:BII-sbar}, we have
\begin{IEEEeqnarray}{rCl}
  \Omega\both{\qty\big[(\sigma y)^*]_{\AII}} &=& \both{\qty\big[y^*]_{\BII}} \\
  \label{eq:BII-Omega-2}
  \Omega\both{\qty\big[(\tp_p\! \gp_{p^i}\! x)^*]_{\AII}}
  &=& \frac{1}{\rho_i} \eta p^i \both{\qty\big[(y \gp_{p^{i+1} - 1}\! x)^*]_{\BII}} \\
  \label{eq:BII-Omega-3}
  \Omega\only{\qty\big[p^{\ell - 1} (\sigma\! \gp_{p^i}\! x)^*]_{\AII}}
  &=& \only{\qty\big[p^{\ell - 1} (\gp_{p^i}\! x)^*]_{\BII}}
  = \frac{\gcd(\eta p^i, p^\ell)}{p} \only{\qty[\frac{p^\ell}{\gcd(\eta p^i, p^\ell)} \gp_{p^i}(x)^*]_{\BII}}.
\end{IEEEeqnarray}
The class $\both{\qty\big[(y \gp_{p^{i+1} - 1}\! x)^*]_{\BII}}$ has order $\gcd(\eta p^{i+1}, p^\ell)$, from which we find that \eqref{eq:BII-Omega-2} is trivial if $p^\ell \mid \eta p^i$, but is a nontrivial element of order $p$ otherwise. Meanwhile, \eqref{eq:BII-Omega-3} always has order $p$.

For the $\AII$ case, since $\eta \both{[\gp_k(y)^*]_{\AII}} = 0$, any term with a factor of $\eta$ in the expression for $\sbar\qty(\gp_{p^i}(y)^h)$ will have zero cosuspension. As for homology, these are the terms with any $\theta_i$, with the exception of $\theta_0$ for $p = 2$. The remaining terms give rise to the following cosuspensions:
\begin{IEEEeqnarray}{lCl}
  \Omega\both{\qty[\qty(\sigma y)^*]_{\BII}}
  &=& \both{[y^*]_{\AII}}, \\
  \label{eq:AII-omega-2}
  \Omega\both{\qty[\qty(\sigma x \cdot \sigma y)^*]_{\BII}}
  &=& -\frac{\eta}{2} \both{\qty[\gp_2(y)^*]_{\AII}} \qquad (p = 2) \\
  \label{eq:AII-omega-3}
  \Omega\only{\qty[\frac{p^\ell}{\gcd(\eta, p^\ell)} \qty(\eps_{p^i}y)^*]_{\AII}}
  &=& \frac{p^\ell}{\gcd(\eta, p^\ell)} \both{\qty[\gp_{p^i}(y)^*]_{\AII}}.
\end{IEEEeqnarray}
Since $\eta \both{\qty[\gp_2(y)^*]_{\AII}} = 0$, \eqref{eq:AII-omega-2} has order at most $2$. Meanwhile, the order of \eqref{eq:AII-omega-3} can be a power of $p$. Moving on, from the expression for $\sbar(x\gp_k(y))$ in \eqref{eq:AII-sbar}, we read $\sbar^*\qty(\gp_{k+1}(\sigma x)^*) = (-\eta)^k \qty(x\gp_k(y))^*$. Since any factor of $\eta$ annihilates the second class in \eqref{eq:AII-cohomology-classes}, the cosuspension is simply
\begin{equation}
  \Omega\only{\qty[\frac{p^\ell}{\gcd(\eta,p^\ell)} \gp_k(\sigma x)^*]_{\BII}} = \begin{cases}
    \only{\qty[\frac{p^\ell}{\gcd(\eta,p^\ell)} x^*]_{\AII}} & k = 1 \\
    0 & k \neq 1.
  \end{cases}
\end{equation}
Cohomology in a tensor product of elementary complexes is completely analogous to the homology case described above. Any cohomology class mixing different complexes has zero cosuspension.

Let us also remark that the cosuspension $H^1(BG; \Lambda) \lto H^0(\Lambda[G]; \Lambda)$ is the only one that maps a single generator to a sum; we have $\sbar^*(\sigma a)^* = \sum_{n \in \Z} n(a^n)^*$ for $G = \Z$, and similarly for $G = \Z_N$.

Finally, the case of $\R/\Z$ coefficients is particularly easy to deal with because the cosuspension is simply the adjoint of the homology suspension with $\Z$ coefficients: $\Omega_{\R/\Z} = (\sigma_\Z)^*$, as mentioned in \eqref{eq:UCT-Omega-U1}. For example, from \eqref{eq:AII-suspension} we obtain, for generators $x, y$ of an $\AII$ complex,
\begin{equation}
  \Omega\qty(\frac{1}{\eta} [\gp_k(\sigma x)]^*) = \begin{cases}
    \frac{1}{\eta} [x]^* & k = 1 \\
    \frac{1}{2}[xy]^* & p = k = 2 \\
    0 & \text{otherwise},
  \end{cases}
\end{equation}
as we found for \eqref{eq:B2Zpf-Omega}.

\zcref[cap]{app:tables} contains several tables showing the behaviour of $\Omega$ for various coefficient groups. \zcref[cap]{tab:Omega-B2Z2-small} shows a small extract for $H^4(B^2 \Z_2; A) \lto H^3(B \Z_2; A)$ and $H^5(B^2 \Z_2; A) \lto H^4(B \Z_2; A)$, illustrating the difference between $A = \R/\Z$, $\Z$ and $\Z_2$ (the order of a generator is shown in angle brackets). The difference between $\Z$ and $\Z_2$ coefficients is particularly interesting: With $\Z$ coefficients, the generator $(\sigma^{2} u_{2} \cdot \sigma \psi_{2} u_{2})^*$ of $H^5$ has nonzero suspension according to \eqref{eq:AII-omega-2}. With $\Z_2$ coefficients, however, the same generator is now a tensor product $(\sigma^{2} u_{2} \otimes \sigma \psi_{2} u_{2})^*$ of generators from different elementary complexes (of type $\BI$ and $\AI$, respectively), so its cosuspension is zero. However, there is now another generator $(\sigma \gamma_{2} \psi_{2} u_{2})^*$, closed over $\Z_2$ but not over $\Z$, whose cosuspension is nonzero. Thus, $\Omega\colon H^5(B^2 \Z_2) \lto H^4(B\Z_2)$ is surjective in both cases.

\begin{table}[h]
  \centering
  {\footnotesize \input{table-omega-B2Z2-small.tex}}
  \caption{$\Omega\colon H^{D+1}(B^2 \Z_2; A) \lto H^D(B \Z_2; A)$.}
  \label{tab:Omega-B2Z2-small}
\end{table}

\section{More structures on Moore's constructions}

% Yes, this is a pun.
\label{sec:more-structures}
As mentioned in the introduction, the Cartan--Moore constructions are ``homology-first'', based on the algebra structure of chains, not cochains. It would be beneficial to translate this language into that of cochain operations such as cup products and Steenrod's higher cup products \cite{steenrodProductsCocyclesExtensions1947}, as this is the more widespread formulation. This is not least true in condensed matter physics, where explicit cocycle representatives are ubiquitous, such as
\begin{equation}
  \label{eq:pontryagin-square}
  \mathcal{P}(b) = b \cupp b + b \cupp_1 \dd{b}
\end{equation}
for the Pontryagin square $\mathcal{P}\colon H^2(\blank; \Z_2) \lto H^4(\blank; \Z_4)$ (see appendix C of \cite{Benini:2018reh} for a review). In this section, we enable this translation by giving an algorithm to compute the $\cupp$ and certain $\cupp_i$ products, and give a few examples of its use.

\subsection{Cup products}
\label{sec:cup-products}
As we have already stated, the product in the DGA-algebras discussed above corresponds to the Pontryagin product in homology, but the cup product in cohomology is extra data which we have not included. However, it is possible to reconstruct it, as follows. The cup product in a space $X$ is determined by a diagonal map $\Delta_0\colon C_{\bullet}(X) \lto C_{\bullet}(X) \otimes C_{\bullet}(X)$ (the subscript $0$ anticipates the notation of the next section). In a special construction $\cA \lto \cM \lto \cN$ where $\cA$ is commutative, a diagonal map $\Delta_0\colon \cA \lto \cA \otimes \cA$ can be extended uniquely to a good $\Delta_0\colon \cM \lto \cM \otimes \cM$, by recursively applying the following properties \cite[exposé~4, thm.~5]{cartanAlgebreDEilenbergMacLane1954}:
\begin{itemize}
\item $\Delta_0$ is a morphism of $\Lambda$-DGA algebras (in particular, it is multiplicative, commutes with $\partial$, preserves degree and $\Delta_0(1) = 1$).
\item For $n \in \cN$ of positive degree, $\Delta_0(n) = s\Delta_0(\partial n)$.\footnote{This property follows from $\Delta_0$ being a \emph{special homomorphism}, meaning that $\Delta_0(\cN) \subseteq \Lambda \oplus \im s$. For any $m \in \im s$, we have $m = sm' = (\partial s + s\partial)sm' = s \partial s m' = s\partial m$, so $\Delta_0(n) = s\partial\Delta_0(n) = s\Delta_0(\partial n)$.}
\end{itemize}
In particular, $\Delta_0$ acts on basis elements as $\Delta_0(a \otimes n) = \Delta_0(a)s\Delta_0(\partial n)$. The diagonal map thus obtained is guaranteed to agree with the one of the bar construction \cite[exposé~5, prop.~4]{cartanAlgebreDEilenbergMacLane1954}, which \textcite[sec.~7]{eilenbergGroupsHPII1954} showed agrees with the diagonal map defining the topological cup product.

Passing through the quotient gives a diagonal map $\Deltabar_0\colon \cN \lto \cN \otimes \cN$ of the delooping $\cN$. By starting with the map $\Delta_0\colon \Lambda[G] \lto \Lambda[G] \otimes \Lambda[G]$ given by $\Delta_0(g) = g \otimes g$ and performing this lifting $h$ times, one obtains the diagonal map on $C_{\bullet}(B^h G; \Lambda)$.

For example, in the construction \eqref{eq:LambdaZN-delooping} over $\Lambda[\Z_N]$, after a calculation using \eqref{eq:LambdaZN-s} and \eqref{eq:tensor-construction} (detailed in \zcref{app:diagonal}), we obtain
\begin{equation}
  \label{eq:Deltabar-BZN}
  \eqbox{c}{rCl}{
    \Deltabar_0(\sigma u) &=& \sigma u \otimes 1 + 1 \otimes \sigma u, \\
    \Deltabar_0(\tpsi_N\! u) &=&
    \tpsi_N\! u \otimes 1 + {N \choose 2} \sigma u \otimes \sigma u
    + 1 \otimes \tpsi_N\! u.}
\end{equation}
Because $\Delta_0$ is an algebra morphism, we can exploit exponentials to calculate
\begin{IEEEeqnarray*}{rCl}
  \sum_k \Deltabar_0 \qty(\gamma_k(\tpsi_N\! u))
  &=& \Deltabar_0 \qty(\e^{\tpsi_N\! u})
  = \e^{\Deltabar_0 (\tpsi_N\! u)} \\
  &=& \e^{\tpsi_N\! u \otimes 1 + 1 \otimes \tpsi_N\! u}
  \qty(1 + {N \choose 2} \sigma u \otimes \sigma u) \\
  &=& \sum_{i,j} \qty(\gamma_i(\tpsi_N\! u) \otimes \gamma_j(\tpsi_N\! u))
  \qty(1 + {N \choose 2} \sigma u \otimes \sigma u)
  \IEEEyesnumber
\end{IEEEeqnarray*}
and
\begin{IEEEeqnarray*}{l}
  \sum_k \Deltabar_0 \qty(\sigma u \cdot \gamma_k(\tpsi_N\! u))
  = \Deltabar_0 \qty(\sigma u \cdot \e^{\tpsi_N\! u})
  = \Deltabar_0(\sigma u) \e^{\Deltabar_0 (\tpsi_N\! u)} \\
  {} = \sum_{i,j} \qty\Big(
  \qty(\sigma u \cdot \gamma_i(\tpsi_N\! u)) \otimes \gamma_j(\tpsi_N\! u)
  + \gamma_i(\tpsi_N\! u) \otimes \qty(\sigma u \cdot \gamma_j(\tpsi_N\! u))).
  \IEEEyesnumber
\end{IEEEeqnarray*}
Dualising, we obtain the complete $\cupp$-product structure
\begin{equation}
  \eqbox{c}{rCl}{
    \gp_i(\tpsi_N\! u)^* \cupp \gp_j(\tpsi_N\! u)^*
    &=& \gp_{i+j}(\tpsi_N\! u)^*, \\
    (\sigma u)^* \cupp \gp_k(\tpsi_N\! u)^*
    = \gp_k(\tpsi_N\! u)^* \cupp (\sigma u)^*
    &=& \qty\big(\sigma u \cdot \gp_k(\tpsi_N\! u))^*, \\
    (\sigma u)^* \cupp (\sigma u)^*
    &=& {N \choose 2} (\tpsi_N\! u)^*.}
\end{equation}
Passing to cohomology, where $\dd(\sigma u \cdot \gp_k(\tpsi_N\! u))^* = N\gp_{k+1}(\tpsi_N\! u)^*$, gives the well-known cup product structure on the lens space $B\Z_N$. Notice especially how $[(\sigma u)^*] \cupp [(\sigma u)^*]$ becomes $0$ for odd $N$ (because then ${N \choose 2}$ is divisible by $N$), but $[(\tpsi_2\! u)^*]$ for $N = 2$.

There is no obstacle, except for the effort it takes, to pushing the calculation further to compute the cup product in $B^2 \Z_N$, $B^3 \Z_N$ and so on. In \zcref{app:diagonal}, we compute low-dimensional cup products in $B^2 \Z_N$ using the same method for the construction \eqref{eq:AII-delooping} with $\AII(\sigma u, \tpsi_N\! u, 1, N)$ as initial algebra. We find
\begin{IEEEeqnarray}{rCl}
  \Deltabar_0(\sigma^2 u) &=& \sigma^2 u \otimes 1 + 1 \otimes \sigma^2 u, \\
  \Deltabar_0(\sigma\!\tpsi_N\! u) &=& \sigma\!\tpsi_N\! u \otimes 1 + 1 \otimes \sigma\!\tpsi_N\! u
\end{IEEEeqnarray}
and, for $p = 2$ ($N$ even),
\begin{equation}
  \Deltabar_0(\eps_2\!\tpsi_N\! u)
  = \eps_2\!\tpsi_N\! u \otimes 1 + \frac{N}{2}\qty(\sigma^2 u \otimes \sigma\!\tpsi_N\! u - \sigma\!\tpsi_N\! u \otimes \sigma^2 u) + 1 \otimes \eps_2\!\tpsi_N\! u.
\end{equation}
Together with
\begin{IEEEeqnarray*}{l}
  \Deltabar_0(\sigma^2 u \cdot \sigma\!\tpsi_N\! u)
  = \Deltabar_0(\sigma^2 u) \Deltabar_0(\sigma\!\tpsi_N\! u) \\
  {} = (\sigma^2 u \cdot \sigma\!\tpsi_N\! u) \otimes 1
  + \sigma^2 u \otimes \sigma\!\tpsi_N\! u
  + \sigma\!\tpsi_N\! u \otimes \sigma^2 u
  + 1 \otimes (\sigma^2 u \cdot \sigma\!\tpsi_N\! u),
  \IEEEyesnumber
\end{IEEEeqnarray*}
this gives the cup products
\begin{IEEEeqnarray}{rCl}
  (\sigma^2 u)^* \smile (\sigma\!\tpsi_N\! u)^*
  &=& (\sigma^2 u \cdot \sigma\!\tpsi_N\! u)^*
  + \frac{N}{2} (\eps_2\!\tpsi_2\! u)^*, \\
  (\sigma\!\tpsi_N\! u)^* \smile (\sigma^2 u)^*
  &=& (\sigma^2 u \cdot \sigma\!\tpsi_N\! u)^*
  - \frac{N}{2} (\eps_2\!\tpsi_2\! u)^*,
\end{IEEEeqnarray}
exemplifying the nontrivial cup product structure in $B^2\Z_N$.

\subsection{Steenrod's cup-$i$ products}
We can reconstruct even more topological information about the Eilenberg--MacLane spaces in the form of Steenrod's $\cupp_i$ products \cite{steenrodProductsCocyclesExtensions1947}. Calculating these in general turns out to be quite hard. We will calculate the full $\cupp_i$ structure only for $BG$, where it can be done in the same way as $\cupp_0$. However, the resulting maps $\Deltabar_i$ do not satisfy the multiplicativity property required to iterate the algorithm to $B^2 G$, $B^3 G$ and so on.

We rely on a number of elementary manipulations of chain complexes and maps between them; these are collected in \zcref{app:koszul}. Essentially all signs follow from the Koszul sign convention.

Following \cite[chap.~2]{Mosher2008}, let $\tau$ be a generator of $\Z_2$ and let $W$ be a contractible chain complex with a free $\tau$-action. Let $\tau$ act on $\cM \otimes \cM$ as the braiding $T$ given by $T(m_1 \otimes m_2) = (-1)^{\abs{m_1} \abs{m_2}} (m_2 \otimes m_1)$. Then a cup-$i$ product structure on $\cM$ is a $\tau$-equivariant chain map
\begin{equation}
  \label{eq:Delta-signature}
  \Delta\colon W \otimes \cM \lto \cM \otimes \cM.
\end{equation}
Here the $\tau$-action on $W \otimes \cM$ is just $\tau(w \otimes m) = \tau w \otimes m$, so $\tau$-equivariance means that $\Delta(\tau w \otimes m) = T\Delta(w \otimes m)$.

For a concrete choice of $W$, we take the total algebra for the delooping of $\Lambda[\Z_N]$ discussed in \zcref{sec:simple-constructions}, with $N = 2$. Renaming $u$ to $\tau$, $\gamma_k(y)$ to $e_{2k}$ and $x\gamma_k(y)$ to $e_{2k+1}$, the boundary map \eqref{eq:LambdaZN-partial} simplifies to
\begin{equation}
  \partial e_i = \qty(\tau + (-1)^i) e_{i-1}, \qquad \partial \tau = 0.
\end{equation}
We define $\Delta_i (m) = \Delta(e_i \otimes m)$; then the fact that $\Delta$ is a chain map translates to
\begin{align}
  \label{eq:Delta-chain-map}
  \partial \Delta (e_i \otimes m) &= \Delta(\partial(e_i \otimes m)) = \Delta((\tau + (-1)^i) e_{i-1} \otimes m) + (-1)^i \Delta(e_i \otimes \partial m) \Rightarrow  \nonumber \\
  &\Rightarrow \partial \Delta_i(m) - (-1)^i \Delta_i (\partial m) = (T + (-1)^i) \Delta_{i-1}(m),
\end{align}
or more compactly $\delta \Delta_i = \qty(T + (-1)^i) \Delta_{i-1}$, where $\delta$ is the differential \eqref{eq:kos:delta} of the function complex. Dualising to cochains $\Hom(\cM, A)$, we get products $\cupp_i$ via\footnote{The sign comes from exchanging $\Delta_i$ and $\alpha \otimes \beta$ according to \eqref{eq:kos:pullback}; a sign-free expression is $(\mu_A)_* \Delta_i^* (\alpha \otimes \beta)$.}
\begin{equation}
  \label{eq:cup-i-def}
  \alpha \cupp_i \beta = (-1)^{(\abs{\alpha} + \abs{\beta})i} \mu_A \circ (\alpha \otimes \beta) \circ \Delta_i,
\end{equation}
and they satisfy the coboundary formula \eqref{eq:kos:coboundary-formula} (\cite[p.~16]{Mosher2008} modulo sign conventions):
\begin{equation}
  (-1)^{\abs{\alpha}\abs{\beta}} \beta \cupp_{i-1} \alpha + (-1)^i \alpha \cupp_{i-1} \beta
  = \dd(\alpha \cupp_i \beta)
  - (-1)^i \qty\Big( \dd\alpha \cupp_i \beta + (-1)^{\abs{\alpha}} \alpha \cupp_i \dd\beta).
\end{equation}

To calculate geometric $\cupp_i$-products, it is not enough to find any such chain map $\Delta$. When $\cM = C_\bullet(X)$ for a space $X$, the correct $\Delta$ is defined with reference to the geometric diagonal map $X \lto X \times X$, e.g.~via the acyclic carrier method \cite{Mosher2008}. In our case, we only have access to its chain approximation $\Delta_0\colon \cM \lto \cM \otimes \cM$, which is not enough to calculate $\Delta$ directly in this way. Instead, we will use the acyclicity of the total space of the fibration $B^h G \lto * \lto B^{h+1} G$ to transport $\Delta$ from the fibre to the base, as in the previous section: Given a special construction $\cA \lto \cM \overset{\pi}{\lto} \cN$ and a chain map $\Delta_\cA\colon W \otimes \cA \lto \cA \otimes \cA$, we would like to find an extension $\Delta\colon W \otimes \cM \lto \cM \otimes \cM$ such that it passes to a quotient map $\Deltabar\colon W \otimes \cN \lto \cN \otimes \cN$. This last condition means that $\Delta_i(\ker \pi) \subset \ker (\pi \otimes \pi)$ for all $i$. Using contractibility of $\cM$ one can show (similarly to \cite[exposé~2, thm.~1]{cartanAlgebreDEilenbergMacLane1954}) that the homotopy class of $\Deltabar$ is uniquely determined by that of $\Delta_\cA$, and is independent of the choice of $\Delta$. Thus, to derive $\Deltabar$ from $\Delta_\cA$, it suffices to find any such $\Delta$.
 
In the case of $\Delta_0$ of the previous section, compatibility with the quotient is ensured by the property that $\Delta_0$ is a DGA-module morphism compatible with $\Delta_{0,\cA}$, i.e.~$\Delta_0(an) = \Delta_0(a) \Delta_0(n)$. Indeed, the quotient map is $\pi = \eps_\cA \otimes \id_\cN$, and if $\pi(an) = (\eps_\cA a) n = 0$, then $(\pi \otimes \pi)\Delta_0(a \otimes n) = (\eps_\cA \Delta_0(a)) \Delta_0(n) = (\eps_\cA a) \Delta_0(n) = 0$. We can achieve the same compatibility condition for $\Delta_i$ for $BG$ as follows: $W \otimes \cM$ is an $\cA$-module and $\cM \otimes \cM$ is a contractible $(\cA \otimes \cA)$-module, and the standard $\Delta_{0,\cA}\colon \cA \lto \cA \otimes \cA$ for $\cA = \Lambda[G]$ is an algebra morphism. Therefore, Theorem~1 of \cite[exposé~2]{cartanAlgebreDEilenbergMacLane1954} ensures the existence of a DGA-module morphism $\Delta$ compatible with $\Delta_{0,\cA}$, unique up to homotopy respecting $\Delta_{0,\cA}$. The contraction $s$ on $\cM \otimes \cM$ provides a canonical choice if one requires $\Delta_i(n) \in \im s$ for $n \in \cN$. This gives the recursive recipe
\begin{IEEEeqnarray}{l}
  \label{eq:Deltai-an}
  \Delta_i(an) = \Delta_0(a) \Delta_i(n), \\
  \label{eq:Deltai-n}
  \eqbox{c}{rCl}{
    \Delta_i(n) = s\partial\Delta_i (n)
    &\overset{\eqref{eq:Delta-chain-map}}{=}&
    s\qty\Big[\qty(T + (-1)^i) \Delta_{i-1} (n) + (-1)^i \Delta_i(\partial n)] \\
    &\overset{s^2 = 0}{=}&
    s\qty\Big[T \Delta_{i-1} (n) + (-1)^i \Delta_i(\partial n)].
  }
\end{IEEEeqnarray}
Because $\Delta_i(a) = 0$ for $i \geq 1$, the resulting $\Delta\colon W \otimes \cM \lto \cM \otimes \cM$ is compatible with the known $\Delta_\cA\colon W \otimes \cA \lto \cA \otimes \cA$, and can therefore be used to compute $\cupp_i$-products in $BG$.

As an example, let us use this recipe to calculate the $\cupp_i$-products in $B\Z_2$. We have the delooping of $\cA (u) = \Z[u]/(u^2 - 1)$ given in \zcref{sec:simple-constructions}, with the acyclic algebra $\cM = \cA (u) \otimes A_{II}(x,y ,1,2)$, with $x = \sigma u$ and $y = \tpsi_2\!u$ in the systematic naming scheme. The differential is $\partial u = 0$, $\partial x = u - 1$ and $\partial y = (u + 1)x$. The contraction \eqref{eq:LambdaZN-s} simplifies to
\begin{equation}
  s(\gamma_k(y)) = s(x\gamma_k(y)) = 0,
  \qquad s(u\gamma_k(y)) = x\gamma_k(y), \qquad s(ux\gamma_k(y)) = \gamma_{k+1}(y).
\end{equation}
Starting from $\Delta_0(u) = u \otimes u$ and using the recursive definition \eqref{eq:Deltai-n}, we obtain the following values for $\Delta_i$ in low dimension:
\begin{equation}
  \Delta_0(x) = x \otimes u + 1 \otimes x, \qquad
  \Delta_1(x) = x \otimes x.
\end{equation}
\begin{equation}
  \Delta_0(y) = y \otimes 1 + x \otimes u x + 1 \otimes y, \qquad
  \Delta_1(y) = -y \otimes (1+u) x, \qquad
  \Delta_2(y) = -y \otimes y.
\end{equation}
\begin{equation}
  \eqbox{c}{rCl}{
    \Delta_0 (xy) &=& xy \otimes u + y \otimes x + x \otimes u y + 1 \otimes x y, \\
    \Delta_1(xy) &=& x \otimes x y + x y \otimes (2+u)x, \\
    \Delta_2(xy) &=& - y \otimes x y - x y \otimes u y, \\
    \Delta_3(xy) &=& - x y \otimes x y.
  }
\end{equation}
\begin{equation}
  \eqbox{c}{rCl}{
    \Delta_0(\gamma_2(y)) &=& \gamma_2(y) \otimes 1 + 1 \otimes \gamma_2(y) + x y \otimes u x + y \otimes y + x \otimes u x y,\\
    \Delta_1(\gamma_2(y)) &=& -y\otimes (u + 1)xy - 2 \gamma_2(y) \otimes (u + 1)x, \\
    \Delta_2(\gamma_2(y)) &=& -2\gamma_2(y) \otimes y - 2y \otimes \gamma_2(y) - xy \otimes uxy, \\
    \Delta_3(\gamma_2(y)) &=& \gamma_2(y) \otimes (u + 1)xy, \\
    \Delta_4(\gamma_2(y)) &=& \gamma_2(y) \otimes \gamma_2(y).
  }
\end{equation}
We then project onto the final algebra (set $u = 1$) to get $\Deltabar_i$, and dualise via \eqref{eq:cup-i-def} to obtain the $\cupp_i$-products of \zcref{tab:BZ2-cupi}.\footnote{The coefficients in the expressions for $\Delta_i$ acquire the sign $(-1)^{\abs{\alpha} + \abs{\beta}) i}$ from \eqref{eq:cup-i-def}, as well as a sign $(-1)^{\abs{\alpha} \abs{\beta}}$ from $(a^* \otimes b^*)(a \otimes b) = (-1)^{\abs{a} \abs{b}}$.} If we reduce mod $2$, we can verify well-known properties of the Steenrod squares $\Sq^i \alpha^{(j)} = \alpha \cupp_{j-i} \alpha$, for instance $\Sq^0 = \id$ and $\Sq^1 \circ \Sq^1 = 0$. Since $\dd(\gamma_k(y))^* = 0$, the Pontryagin square \eqref{eq:pontryagin-square} reduces to the cup square in this model for $BG$.
\begin{table}
  \centering
  \begin{subtable}[h]{.55\textwidth}
    \begin{tabular}{c|cccc}
      \diagbox[height=2em]{$\alpha$}{$\beta$} & $x^*$ & $y^*$ & $(xy)^*$ & $(\gamma_2 y)^*$ \\
      \hline
      $x^*$ & $-y^*$ & $(xy)^*$ & $-(\gamma_2 y)^*$ \\
      $y^*$ & $(xy)^*$ & $(\gamma_2 y)^*$ \\
      $(xy)^*$ & $-(\gamma_2 y)^*$ \\
      $(\gamma_2 y)^*$
    \end{tabular}
    \caption{$\alpha \cupp_0 \beta$.}
    \label{tab:BZ2-cup0}
  \end{subtable}
  \begin{subtable}[h]{.4\textwidth}
    \begin{tabular}{c|cccc}
      \diagbox[height=2em]{$\alpha$}{$\beta$} & $x^*$ & $y^*$ & $(xy)^*$ & $(\gamma_2 y)^*$ \\
      \hline
      $x^*$ & $-x^*$ & $0$ & $-(xy)^*$ & $0$ \\
      $y^*$ & $2y^*$ & $0$ & $2(\gamma_2 y)^*$ \\
      $(xy)^*$ & $-3(xy)^*$ & $0$ \\
      $(\gamma_2 y)^*$ & $4(\gamma_2 y)^*$
    \end{tabular}
    \caption{$\alpha \cupp_1 \beta$.}
    \label{tab:BZ2-cup1}
  \end{subtable}
  \\[1em]
  \begin{subtable}[h]{.55\textwidth}
    \begin{tabular}{c|cccc}
      \diagbox[height=2em]{$\alpha$}{$\beta$} & $x^*$ & $y^*$ & $(xy)^*$ & $(\gamma_2 y)^*$ \\
      \hline
      $x^*$ & $0$ & $0$ & $0$ & $0$ \\
      $y^*$ & $0$ & $-y^*$ & $-(xy)^*$ & $-2(\gamma_2 y)^*$ \\
      $(xy)^*$ & $0$ & $-(xy)^*$ & $(\gamma_2 y)^*$ \\
      $(\gamma_2 y)^*$ & $0$ & $-2(\gamma_2 y)^*$
    \end{tabular}
    \caption{$\alpha \cupp_2 \beta$.}
    \label{tab:BZ2-cup2}
  \end{subtable}
  \begin{subtable}[h]{.4\textwidth}
    \begin{tabular}{c|cccc}
      \diagbox[height=2em]{$\alpha$}{$\beta$} & $x^*$ & $y^*$ & $(xy)^*$ & $(\gamma_2 y)^*$ \\
      \hline
      $x^*$ & $0$ & $0$ & $0$ & $0$ \\
      $y^*$ & $0$ & $0$ & $0$ & $0$ \\
      $(xy)^*$ & $0$ & $0$ & $(xy)^*$ & $0$ \\
      $(\gamma_2 y)^*$ & $0$ & $0$ & $-2(\gamma_2 y)^*$
    \end{tabular}
    \caption{$\alpha \cupp_3 \beta$.}
    \label{tab:BZ2-cup3}
  \end{subtable}
  \\[1em]
  \begin{subtable}[h]{.49\textwidth}
    \begin{tabular}{c|cccc}
      \diagbox[height=2em]{$\alpha$}{$\beta$} & $x^*$ & $y^*$ & $(xy)^*$ & $(\gamma_2 y)^*$ \\
      \hline
      $x^*$ & $0$ & $0$ & $0$ & $0$ \\
      $y^*$ & $0$ & $0$ & $0$ & $0$ \\
      $(xy)^*$ & $0$ & $0$ & $0$ & $0$ \\
      $(\gamma_2 y)^*$ & $0$ & $0$ & $0$ & $(\gamma_2 y)^*$
    \end{tabular}
    \caption{$\alpha \cupp_4 \beta$.}
    \label{tab:BZ2-cup4}
  \end{subtable}
  \caption{Cup-$i$ products in $B\Z_2$ of degree $\leq 4$.}
  \label{tab:BZ2-cupi}
\end{table}

In order to compute $\cupp_i$-products in $B^2 G$, $B^3 G$ and so on, the above method needs to be generalised. In order to apply the crucial theorem \cite[exposé~2, thm.~1]{cartanAlgebreDEilenbergMacLane1954}, one needs to start with an algebra morphism. While $\Delta_0\colon \cA \lto \cA \otimes \cA$ remains an algebra morphism even when $\cA$ is an iterated delooping of $\Lambda[G]$, using \eqref{eq:Deltai-an} would forget all the nonzero values of $\Delta_i(a)$ for $i \geq 1$, and hence break compatibility with the known $\Delta_\cA\colon W \otimes \cA \lto \cA \otimes \cA$. We may attempt to obtain a better notion of multiplicativity as follows: Using the isomorphism
\begin{equation}
  \eqbox{c}{rCl}{
    \Hom_\tau(W \otimes \cM, \cM \otimes \cM)
    &\cong& \Hom(\cM, \Hom_\tau(W, \cM \otimes \cM)) \\
    f &\mapsto& \qty(m \mapsto \qty(w \mapsto (-1)^{\abs{m}\abs{w}} f(w \otimes m)))
  }
\end{equation}
(braiding \eqref{eq:kos:braiding} followed by the tensor-hom adjunction \eqref{eq:kos:tensor-hom}), where the subscript means $\tau$-equivariance, we can express $\Delta$ as a chain map
\begin{equation}
  \label{eq:Delta-signature-new}
  \Delta\colon \cM \lto \Hom_\tau(W, \cM \otimes \cM)
\end{equation}
with $\Delta(m)(e_i) = (-1)^{\abs{m}i} \Delta_i(m)$. The right hand side of \eqref{eq:Delta-signature-new} has a natural algebra structure given by the convolution $\star$ with respect to the coproduct $\Delta_{0,W}$ on $W$, which takes the form\footnote{This is computed as in \zcref{sec:cup-products}. The contraction \eqref{eq:LambdaZN-s} of $W$ becomes
\begin{equation}
  s(\e_i) = 0, \qquad s(\tau e_i) = e_{i+1},
\end{equation}
using which we can calculate $\Delta_{0,W}$.}
\begin{equation}
  \label{eq:r-def}
  \Delta_{0,W}(e_i) = \sum_{j+k=i} e_j \otimes \tau^j e_k, \qquad \Delta_{0,W}(\tau e_i) = (\tau \otimes \tau) \Delta_{0,W}(e_i).
\end{equation}
The convolution is
\begin{equation}
  (f \star g)(e_i) = (f \otimes g)r(e_i) = \sum_{j+k=i} (-1)^{\abs{g}j} f(e_j) g(\tau^j e_k),
\end{equation}
and requiring $\Delta$ to be a morphism of algebras amounts to the condition
\begin{equation}
  \label{eq:multiplicativity}
  \Delta_i(xy) = \sum_{j+k=i} (-1)^{\abs{x} k} \Delta_j(x) T^j \Delta_k(y).
\end{equation}
If $\star$ were commutative, \cite[exposé~4, thm.~5]{cartanAlgebreDEilenbergMacLane1954} would guarantee that a multiplicative $\Delta_\cA\colon \cA \lto \Hom_\tau(W, \cA \otimes \cA)$ gives rise to a multiplicative $\Delta\colon \cM \lto \Hom_\tau(W, \cM \otimes \cM)$, descending to a multiplicative $\Deltabar$, allowing us to iterate the process indefinitely. However, $\star$ is not commutative, and indeed one can check explicitly that the property $2\Delta(\gamma_2(y)) = \Delta(y) \star \Delta(y)$ fails in the above construction for $B\Z_2$, as
\begin{IEEEeqnarray*}{Cl}
  & -4\gamma_2(y) \otimes y - 4y \otimes \gamma_2(y) - 2xy \otimes uxy \\
  \neq &
  -4\gamma_2(y) \otimes y - 4y \otimes \gamma_2(y) - (u + 1)xy \otimes (u + 1)xy. \IEEEyesnumber
\end{IEEEeqnarray*}
Despite this, $\star$ is commutative up to coherent homotopy, because the $\Delta_{i,W}$ that we have described how to construct witness the homotopy cocommutativity of $\Delta_{0,W}$. In principle, we could use this to iteratively construct maps $\Delta$ that are multiplicative only up to homotopy, together with explicit coherent towers of homotopies witnessing this fact. However, the amount of extra data contained in these homotopies somewhat negates the computational simplicity of Moore's small models, and a calculation of $\cupp_i$-products for $B^h G$ along these lines falls beyond the scope of this work.

\section*{Acknowledgements}
\addcontentsline{toc}{section}{Acknowledgements}
We would like to thank Matteo Dell'Acqua, Domenico Fiorenza, Daniele Migliorati, Shu-Heng Shao, Hanyu Xue and Xiao-Gang Wen for illuminating discussions. We also extend special thanks to Arun Debray, Domenico Fiorenza and Andrea Grigoletto for helpful and insightful comments on an earlier draft.

The work of SNM is supported by DOE (HEP) Award DE-SC0013528, BSF grant 2022100, and a University Research Foundation grant at the University of Pennsylvania. SNM also thanks the Simons Center for Geometry and Physics for the hospitality during the ``23rd Simons Physics Summer Workshop: Theory, Experiment and the Emerging New Physics". ERG gratefully acknowledges the support of the Knut and Alice Wallenberg Foundation.

\appendix

\section{$\Z_p$ coefficients via Steenrod operations}
\label{app:Zp}
The cohomology groups $H^D(B^h \Z_p; \Z_p)$, and even the ring structure given by the cup product, can be expressed simply in terms of the Steenrod squares ($p = 2$) and powers ($p > 2$), and their cup products (see e.g.~\cite[sec.~4.L]{Hatcher2002}). The squares and powers are \emph{stable} operations, which means that they commute with the cohomology suspension, and the suspension of any cup product is zero. This is enough to fully determine the suspension. For instance, consider $\Omega\colon H^{D+1}(B^2 \Z_2; \Z_2) \lto H^D(B \Z_2; \Z_2)$. Letting $\iota_h$ be the fundamental class in $H^h(B^h \Z_2; \Z_2)$, each $H^D(B \Z_2; \Z_2)$ is generated by $(\iota_1)^D$. Writing down the generators of $H^{D+1}(B^2 \Z_2; \Z_2)$, their suspensions follow immediately:
\begin{equation}
  \label{eq:Z2-coeff-suspension}
  \eqbox{c}{llrCl}{
    H^2(B^2\Z_2; \Z_2) &\colon\ \ & \iota_2 &\ \longmapsto\ & \iota_1 \\
    H^3(B^2\Z_2; \Z_2) &\colon& \Sq^1\iota_2
    &\longmapsto& \Sq^1\iota_1 = (\iota_1)^2 \\
    H^4(B^2\Z_2; \Z_2) &\colon& (\iota_2)^2 &\longmapsto& 0 \\
    H^5(B^2\Z_2; \Z_2) &\colon&
       \Sq^2\Sq^1\iota_2 &\longmapsto& \Sq^2\Sq^1\iota_1 = (\iota_1)^4 \\
    && \iota_2 \cupp \Sq^1 \iota_2 &\longmapsto& 0 \\
    H^6(B^2\Z_2; \Z_2) &\colon&
       (\Sq^1\iota_2)^2 &\longmapsto& 0 \\
    && (\iota_2)^3 &\longmapsto& 0 \\
    H^7(B^2\Z_2; \Z_2) &\colon&
       (\iota_2)^2 \cupp \Sq^1 \iota_2 &\longmapsto& 0 \\
    && \iota_2 \cupp \Sq^2 \Sq^1 \iota_2 &\longmapsto& 0 \\
    H^8(B^2\Z_2; \Z_2) &\colon&
       \Sq^1 \iota_2 \cupp \Sq^2 \Sq^1 \iota_2 &\longmapsto& 0 \\
    && \iota_2 \cupp (\Sq^1 \iota_2)^2 &\longmapsto& 0 \\
    && (\iota_2)^4 &\longmapsto& 0 \\
    H^9(B^2\Z_2; \Z_2) &\colon&
       \Sq^4 \Sq^2 \Sq^1 \iota_2
       &\longmapsto& \Sq^4 \Sq^2 \Sq^1 \iota_1 = (\iota_1)^8 \\
    && (\Sq^1 \iota_2)^3 &\longmapsto& 0 \\
    && (\iota_2)^2 \cupp \Sq^2 \Sq^1 \iota_2 &\longmapsto& 0 \\
    && (\iota_2)^3 \cupp \Sq^1 \iota_2 &\longmapsto& 0 \\
    &&&\vdots}
\end{equation}
These results agree with the ones obtained using Moore's constructions in \zcref{tab:Omega-B2Z2-Z2}. A facility for this computation of $\Omega$ with $\Z_p$ coefficients is included in \texttt{emcm} \cite{riedelgardingEmcm2026}.

\section{Proof that
  $H^{\bullet}(X; \Z) \otimes \Z_{(p)} \cong H^{\bullet}(X; \Z_{(p)})$}
\label{app:p-local-cohomology}
To obtain the isomorphism \eqref{eq:p-local-cohomology-isomorphism} when $H_{\bullet}(X)$ is degreewise finitely generated, note that we have a chain map
\begin{equation}
  \label{eq:mult-chain-map}
  \eqbox{c}{rCl}{
    \Hom(K_{\bullet}, \Z) \otimes A
    &\xlongrightarrow{\mu}& \Hom(K_{\bullet}, A) \\
    \omega \otimes a &\longmapsto& (x \mapsto a \omega(x))
  }
\end{equation}
for any chain complex $K_{\bullet}$, and in particular
\begin{equation}
  \label{eq:mu-on-cochains}
  C^{\bullet}(X; \Z) \otimes A = \Hom(C_{\bullet}(X), \Z) \otimes A
  \xlongrightarrow{\mu} \Hom(C_{\bullet}(X), A) = C^{\bullet}(X; A),
\end{equation}
natural in $X$ and $A$. Now assume that $A$ is torsion-free; we then have $H^D(C^{\bullet}(X; \Z) \otimes A) \cong H^D(X; \Z) \otimes A$ (since $\blank \otimes A$ commutes with taking homology, as $A$ is a flat $\Z$-module), so that \eqref{eq:mu-on-cochains} induces a map $\mu_*\colon H^D(X; \Z) \otimes A \lto H^D(X; A)$ in cohomology.

Whenever $K_{\bullet}$ is a complex of finitely generated groups, the chain map \eqref{eq:mult-chain-map} is an isomorphism: It is injective since $A$ is torsion-free, and it is surjective since we can exhibit a right inverse:
\begin{equation}
  \Hom(K_D, A) \ni \eta \mapsto \mu^{-1}(\eta) = \sum_i \delta_{x_i} \otimes \eta(x_i),
\end{equation}
where $\{x_i, t_j\}$ is some basis for $K_D$, the $x_i$ and $t_j$ being free and torsional generators respectively, and $\delta_{x_i}(x_j) = \delta_{ij}$, $\delta_{x_i}(t_j) = 0$.

Now, any complex $K_{\bullet}$ of free abelian groups with finitely generated homology can be replaced by a chain homotopy equivalent one with finitely generated chain groups. Let $Z_{\bullet}$ denote the cycles and $B_{\bullet}$ the boundaries. Choosing a free resolution of finitely generated groups $0 \lto B_D' \lto Z_D' \lto H_D(K) \lto 0$, we obtain a chain map $f$
\begin{equation}
  \begin{tikzcd}
    0 \ar[r] & B_D \ar[r,"i"] \ar[d,"f_B"] & Z_D \ar[r] \ar[d,"f_Z"] & H_D(K) \ar[r] \ar[d,equal] & 0 \\
    0 \ar[r] & B_D' \ar[r,"i'"] & Z_D' \ar[r] & H_D(K) \ar[r] & 0
  \end{tikzcd}
\end{equation}
unique up to chain homotopy. The exact sequence $0 \lto Z_D \lto K_D \overset{\partial}{\lto} B_{D-1} \lto 0$ splits since $B_{D-1}$ is free, so we can identify $K_D \cong Z_D \oplus B_{D-1}$ and $\partial(z, b) = (ib, 0)$.
We analogously define $K_D' = Z_D' \oplus B_{D-1}'$ with $\partial'(z', b') = (i' b', 0)$. Then $f = f_Z \oplus f_B$ becomes a chain map $K \lto K'$ that induces the identity map on homology (a quasi-isomorphism). In fact, $f$ is a chain homotopy equivalence: The mapping cone of a quasi-isomorphism is acyclic \cite[cor.~1.5.4]{weibelIntroductionHomologicalAlgebra1994}, and since it has free chain groups, it admits a chain contraction $S$ \cite[exercise~1.4.1]{weibelIntroductionHomologicalAlgebra1994}. Writing the boundary map $D$
\cite[§1.5.1]{weibelIntroductionHomologicalAlgebra1994} and $S$ on $\cone(f)_D = K_{D-1} \oplus K'_D$ as
\begin{equation}
  D = \mqty[-\partial & 0 \\ -f & \partial'], \qquad
  S = \mqty[s & -g \\ h & -s']
\end{equation}
and expanding $DS + SD = 1$, we find that $g\colon K' \lto K$ is a chain map such that $fg - 1 = \partial' s' + s' \partial'$ and $gf - 1 = \partial s + s \partial$.

In conclusion, for $K_{\bullet} = C_{\bullet}(X)$ where $X$ is a space with finitely generated homology groups, we can find a chain homotopy equivalence $K \simeq K'$ where $K'$ is degreewise finitely generated. The chain map \eqref{eq:mult-chain-map} for $K'$ is then an isomorphism and induces the isomorphism $H^D(X; \Z) \otimes A \cong H^D(X; A)$ whenever $A$ is torsion-free.

\section{The explicit constructions}
\label{app:cartan-moore}

\subsection{Cartan's constructions}
\label{app:cartan-constructions}
Here we review the constructions of \cite{cartanAlgebreDEilenbergMacLane1954}. We also write down contractions $s$, showing that they are \emph{special} constructions; this was not explicitly done in the original seminars, but is quite important for calculating the (co)suspension.

In \zcref{sec:simple-constructions}, we have already found simple constructions over $\Lambda[G]$ and $\AI(x, n) = E(x, n)$. It is not as easy to find a small construction on the divided polynomial algebra $\BI(x, n) = \Gamma(x, n)$; this is the source of complexity in $B^h \Z$ for $h \geq 3$ and $B^h \Z_N$ for $h \geq 2$. However, Cartan noted that the problem simplifies over the base field $\Lambda = \Z_p$ for a prime $p$. In this case, $\gp_k(x) = x^k / k!$ for $0 \leq k < p$ while $x^p = p! \gp_p(x) = 0$, and there is no further relation between $x$ and $\gamma_p(x)$. In fact, for any $k$, $\gp_k(x)$ can be written as a product of factors $\gamma_{p^i}(x)$ as follows. Write $k = \dots k_2 k_1 k_0$ in base $p$; then
\begin{equation}
  \label{eq:decompose-gp}
  \underbrace{x \cdots x}_{k_0\text{ times}}
  \underbrace{\gp_p(x)  \cdots \gp_p(x)}_{k_1\text{ times}}
  \underbrace{\gp_{p^2}(x)  \cdots \gp_{p^2}(x)}_{k_2\text{ times}}
  \cdots
  = n_k \gp_k(x)
\end{equation}
for some integer $n_k$ according to \eqref{eq:gp-product}. Now, \emph{Kummer's theorem} \cite{kummerUberErganzungssatzeAllgemeinen1852} states that the $p$-adic valuation $\nu_p{k + l \choose k}$ is equal to the number of carries when $k$ and $l$ are added in base $p$. Adding the exponents in \eqref{eq:decompose-gp} produces no carry, so $n_k$ is not divisible by $p$. Therefore it can be inverted in $\Z_p$, and \eqref{eq:decompose-gp} can be solved for $\gp_k(x)$.

This all means that we have an isomorphism
\begin{equation}
  \label{eq:gamma-isomorphism-mod-p}
  \Gamma(x, n) \cong \bigotimes_{i \geq 0} Q_p\qty(\gp_{p^i}(x), p^i n)
\end{equation}
where $Q_p(x, n)$ denotes the truncated polynomial algebra $\Z_p[x] / (x^p)$ (with a generator $x$ of even degree $n$ and $\partial x = 0$). On each factor, we have the following special construction:
\begin{itemize}
\item For $\cA = Q_p(x, n)$ over $\Lambda = \Z_p$,
  let $\cN = E(a, n + 1) \otimes \Gamma(b, pn + 2)$, with the differential on $\cM$ given by
  \begin{equation}
    \partial a = x, \qquad \partial b = x^{p-1} a.
  \end{equation}
  A contraction is given by
  \begin{equation}
    \eqbox{c}{rCl}{
      s\qty\big(x^h \gp_k(b)) &=& \begin{cases}
        0 & h = 0 \\
        x^{h-1}\, a \gp_k(b) & 0 < h < p,
      \end{cases} \\
      s\qty\big(x^h\, a \gp_k(b)) &=& \begin{cases}
        0 & 0 \leq h < p - 1 \\
        \gp_{k+1}(b) & h = p - 1.
      \end{cases}}
  \end{equation}
  Clearly $\tildeN = \Z_p + \im s$ is a subalgebra containing $\cN$, so this defines a special construction. The induced differential is $\partialbar a = \partialbar b = 0$, and the suspension is
  \begin{equation}
    \sbar(x^h) = \begin{cases}
      a & h = 1 \\
      0 & h \neq 1
    \end{cases}
    \qquad (0 \leq h < p).
  \end{equation}
  For the general naming scheme used in \zcref{sec:general-Omega}, we rename $a$ to $\sigma x$ and $b$ to $\tp_p x$.
\end{itemize}
Taking the tensor product of all these constructions, we obtain a construction over $\Gamma(x, n)$, which we write out explicitly:
\begin{itemize}
\item For $\cA = \BI(x, n) = \Gamma(x, n)$ over $\Lambda = \Z_p$, let
  \begin{equation}
    \label{eq:Gamma-Zp-N}
    \cN = \bigotimes_{i \geq 0} \qty\Big(
    E(a_i, p^i n + 1) \otimes \Gamma(b_i, p^{i+1} n + 2))
  \end{equation}
  with differential on $\cM$ given by
  \begin{equation}
    \label{eq:Gamma-Zp-d}
    \partial a_i = \gp_{p^i}(x), \qquad
    \partial b_i = \gp_{p^i}(x)^{p-1}\, a_i.
  \end{equation}
  The formula \eqref{eq:tensor-construction} for the contraction in a tensor product means that to evaluate $s$ on a product of positive-degree elements from different tensor factors in \eqref{eq:gamma-isomorphism-mod-p} and \eqref{eq:Gamma-Zp-N}, we just apply $s$ to the element with the lowest index $i$. For example, $s\qty\big(x \gp_k(b_0)\, a_1 \gp_p(x)\, b_2) = s\qty\big(x \gp_k(b_0))\, a_1 \gp_p(x)\, b_2$ but $s\qty\big(a_1 \gp_p(x)\, b_2) = s\qty\big(a_1 \gp_p(x))\, b_2$. The contraction in a single tensor factor is
  \begin{equation}
    \label{eq:Gamma-Zp-s}
    \eqbox{c}{rCl}{
      s\qty\big(\gp_{p^i}(x)^h \gp_k(b_i)) &=& \begin{cases}
        0 & h = 0 \\
        \gp_{p^i}(x)^{h-1}\, a_i \gp_k(b_i) & 0 < h < p,
      \end{cases} \\
      s\qty\big(\gp_{p^i}(x)^h\, a_i \gp_k(b_i)) &=& \begin{cases}
        0 & 0 \leq h < p - 1 \\
        \gp_{k+1}(b_i) & h = p - 1.
    \end{cases}}
  \end{equation}
  The induced differential is $\partialbar = 0$, and the suspension becomes
  \begin{equation}
    \label{eq:Gamma-Zp-sbar}
    \sbar(\gp_k(x)) = \begin{cases}
      a_i & k = p^i \\
      0 & \text{$k$ not a power of $p$}.
    \end{cases}
  \end{equation}
  We rename $a_i$ to $\sigma\!\gp_{p^i}\!x$ and $b_i$ to $\tp_p\!\gp_{p^i}\!x$
  (omitting $\gamma_1$ when $i = 0$).
\end{itemize}

These constructions are enough for a complete calculation of $H^{\bullet}(B^h G; \Z_p)$ and its suspension maps. Since $B\Z_N$ is contractible over $\Z_p$ when $p \nmid N$ (see \eqref{eq:BZN-CW}), only the summands of $\Z$ and $\Z_{p^k}$ in $G$ contribute. For these cases, all the differentials $\partialbar$ in the final algebras are zero over $\Lambda = \Z_p$, so they decompose as tensor products of algebras $E(x, n)$ and $\Gamma(x, n)$, which can be used as initial algebras for constructions of the same type. Since the differentials are zero, each generator of an $E(x, D)$ or $\Gamma(x, D)$ contributes a homology class to $H_D(B^h G; \Z_p)$ and a cohomology class to $H^D(B^h G; \Z_p)$ (these groups are therefore isomorphic), and the suspensions $\sigma$ and $\Omega$ are simply transposes of each other. In the naming scheme we have described, the generators correspond to the ``admissible words'' made from the letters $\sigma$, $\tpsi_N$, $\tp_p$ and $\gp_k$.\footnote{$\tpsi_N$ and $\varphi_p$ are certain operations defined by Cartan for arbitrary acyclic constructions, but we only have need of them as labels for generators in our specific constructions. $\sigma$ is the homology suspension and $\gamma_k$ is the divided power.}

For instance, the suspension $\Omega\colon H^{D+1}(B^2 \Z_p; \Z_p) \lto H^D(B \Z_p; \Z_p)$ is the transpose of $\sigma\colon H_D(B \Z_p; \Z_p) \lto H_{D+1}(B^2 \Z_p; \Z_p)$, which is given by
\begin{equation}
  \eqbox{c}{rCll}{
    \sigma([\gp_k\!\tpsi_p\! u]) &=& \begin{cases}
      [\sigma\!\gp_{p^i}\!\tpsi_p\! u] & k = p^i \\
      0 & \text{$k$ not a power of $p$}
    \end{cases}
    \qquad & (D = 2k) \\
    \sigma([\sigma u] [\gp_k\!\tpsi_p\! u])
    &=& \begin{cases}
      [\sigma^2 u] & k = 0 \\
      0 & k > 0
    \end{cases}
    & (D = 2k + 1).
  }
\end{equation}
The map is nonzero only for $D = 1$ and $D = 2p^i$, a pattern which is visible in \eqref{eq:Z2-coeff-suspension}.

\subsection{Moore's constructions}
\label{app:moore-constructions}
We would now like to investigate the suspension maps with coefficients other than $\Z_p$. \Textcite{doi:10.1073/pnas.43.5.409} extended the above constructions over $\Z_p$ to one over the ring $\Lambda = \Z_{(p)}$ of integers localised at $p$, that is, the rational numbers with denominators not divisible by $p$.\footnote{The original constructions were over the $p$-adic integers, but \textcite[p.~207]{AST_1976__32-33__173_0} later noted that they work over $\Z_{(p)}$ (and in fact any ring such that any integer not divisible by $p$ is invertible).} Moore's paper is extremely short; it only lists the constructions without any additional explanation. In particular, the contractions $s$ are not discussed. Here, we give more details and write down $s$ explicitly.\footnote{Several other authors have also elaborated on Moore's paper \cite{armarioPminimalHomologicalModels1999,zakharovCohomologyEilenbergMacLaneSpaces}.}

For the initial algebra $\cA = \BI(x, n) = \Gamma(x, n)$, let us take the same final algebra \eqref{eq:Gamma-Zp-N} as for $\Lambda = \Z_p$. The differential \eqref{eq:Gamma-Zp-d} no longer satisfies $\partial^2 = 0$, since
\begin{equation}
  \partial^2 b_i = \gp_{p^i}(x)^p
  = \frac{(p^{i+1})!}{((p^i)!)^p} \gp_{p^{i+1}}(x)
  = p\, \zeta_i \gp_{p^{i+1}}(x),
\end{equation}
where $\zeta_i \defeq \frac{(p^{i+1})!}{p((p^i)!)^p}$ is an integer not divisible by $p$,\footnote{This can be seen using Kummer's theorem, or Legendre's formula $\nu_p(n!) = \sum_{k = 1}^\infty \left\lfloor\frac{n}{p^k}\right\rfloor$.} and therefore invertible. We therefore modify the differential by setting
\begin{equation}
  \partial b_i = \gp_{p^i}(x)^{p-1}\, a_i - p\, \zeta_i\, a_{i+1},
\end{equation}
which restores $\partial^2 = 0$.

When reduced modulo $p$, this construction agrees with the one of the previous section. Defining $s_1$ by the same expression \eqref{eq:Gamma-Zp-s}, now interpreted over $\Z_{(p)}$, we have $\partial s_1 + s_1 \partial = \id - \eps \bmod{p}$, so we can improve it to a true contraction $s$ by \eqref{eq:p-local-improvement}. It is impractical to write down a closed expression for $s$, but we can do so for $\sbar$. The initial algebra is generated by elements of the form $\gamma_k(x)$. Up to a unit, we can write this as $\gamma_{p^i}(x)^h \gamma_n(x)$, where $k = p^i h + n$ with $0 < h < p$ and $p^{i+1} \mid n$. Then
\begin{equation}
  s_1\qty(\gamma_{p^i}(x)^h \gamma_n(x)) = a_i \gamma_{p^i}(x)^{h-1} \gamma_n(x).
\end{equation}
Using $\partial x = 0$, we quickly see that the right hand side is in the kernel of $1 - s_1 \partial$, which means that the iteration \eqref{eq:p-local-improvement} terminates immediately; we have $s = s_1$ when restricted to $\cA$. Therefore $\sbar$ is given by the same formula \eqref{eq:Gamma-Zp-sbar} as over $\Z_p$. We summarise:
\begin{itemize}
\item For $\cA = \BI(x, n) = \Gamma(x, n)$ over $\Lambda = \Z_{(p)}$, let
  \begin{equation}
    \label{eq:Gamma-plocal-N}
    \cN = \bigotimes_{i \geq 0} \qty\Big(
    E(a_i, p^i n + 1) \otimes \Gamma(b_i, p^{i+1} n + 2))
  \end{equation}
  with differential on $\cM$ given by
  \begin{equation}
    \label{eq:Gamma-plocal-d}
    \partial a_i = \gp_{p^i}(x), \qquad
    \partial b_i = \gp_{p^i}(x)^{p-1}\, a_i - p\, \zeta_i\, a_{i+1}.
  \end{equation}
  The contraction $s$ is given by \eqref{eq:p-local-improvement}, where $s_1$ is the tensor product contraction \eqref{eq:tensor-construction} given by
  \begin{equation}
    \label{eq:Gamma-plocal-s1}
    \eqbox{c}{rCl}{
      s_1\qty\big(\gp_{p^i}(x)^h \gp_k(b_i)) &=& \begin{cases}
        0 & h = 0 \\
        \gp_{p^i}(x)^{h-1}\, a_i \gp_k(b_i) & 0 < h < p,
      \end{cases} \\
      s_1\qty\big(\gp_{p^i}(x)^h\, a_i \gp_k(b_i)) &=& \begin{cases}
        0 & 0 \leq h < p - 1 \\
        \gp_{k+1}(b_i) & h = p - 1.
    \end{cases}}
  \end{equation}
  The induced differential is $\partialbar b_i = -p\, \zeta_i\, a_{i+1}$, and the suspension becomes
  \begin{equation}
    \label{eq:Gamma-plocal-sbar}
    \sbar(\gp_k(x)) = \begin{cases}
      a_i & k = p^i \\
      0 & \text{$k$ not a power of $p$}.
    \end{cases}
  \end{equation}
  The resulting construction is a tensor product of DGAs
  \begin{equation}
    \cN = \AI(a_0, n + 1) \otimes \bigotimes_{i \geq 0} \AII(a_{i+1}, b_i, p^i n + 1, -p\, \zeta_i).
  \end{equation}
  We rename $a_i$ to $\sigma\!\gp_{p^i}\!x$ and $b_i$ to $\tp_p\!\gp_{p^i}\!x$.
\end{itemize}
Note that this does not give a construction over the integers. Over $\Z_{(p)}$, we can decompose any $\gamma_k(x)$ according to \eqref{eq:decompose-gp}, with an invertible $n_k$ just as over $\Z_p$. Over $\Z$, this is no longer true, so the contraction \eqref{eq:Gamma-plocal-s1} is not fully specified.

We proceed in the same way for the constructions over $\BII(x, y, n, \eta)$ and $\AII(x, y, n, \eta)$. We begin with the case $\BII$, since it turns out to be somewhat easier.
\begin{itemize}
\item For $\cA = \BII(x, y, n, \eta)$ over $\Lambda = \Z_{(p)}$, Moore gives the construction
  \begin{equation}
    \cN = \Gamma(z, n + 2) \otimes \bigotimes_{i \geq 0} \qty\Big(
    E(a_i, p^i n + 1) \otimes \Gamma(b_i, p^{i+1} n + 2))
  \end{equation}
  with
  \begin{equation}
    \eqbox{c}{c}{
      \partial y = \eta x, \qquad
      \partial z = y - \eta a_0, \\
      \partial x = 0, \qquad
      \partial a_i = \gp_{p^i}(x), \qquad
      \partial b_i = \gp_{p^i}(x)^{p-1}\, a_i - p\, \zeta_i\, a_{i+1}.}
  \end{equation}
  The induced differential on $\cN$ becomes
  \begin{equation}
    \partialbar z = -\eta a_0, \qquad
    \partialbar a_i = 0, \qquad
    \partialbar b_i = -p\, \zeta_i\, a_{i+1}.
  \end{equation}
  Thus $\cN$ can be written as a tensor product of DGA-algebras
  \begin{equation}
    \cN = \AII(a_0, z, n + 1, -\eta) \otimes
    \bigotimes_{i \geq 0} \AII(a_{i+1}, b_i, p^{i+1} n + 1, -p\zeta_i).
  \end{equation}

  We can view $\partial$ as the tensor product differential on $\AI(y, n + 1) \otimes \BI(x, n)$, perturbed by the parameter $\eta$ (once $\partial y = \eta x$ is fixed, the change in $\partial z$ becomes necessary to maintain $\partial^2 = 0$).\footnote{Hence, an alternative way to obtain an $s$ would be to use the homological perturbation lemma \eqref{eq:pert-lemma} starting from $s^{\AI} \otimes 1^{\BI} + \eps^{\AI} \otimes s^{\BI}$, where $s^{\BI}$ is the $p$-local contraction just derived.} Indeed, reduced modulo $p$, we obtain exactly $\AI(y, n + 1) \otimes \BI(x, n)$ since $\eta = 0$. Thus we can take $s_1$ as the tensor product contraction $s^{\AI} \otimes 1^{\BI} + \eps^{\AI} \otimes s_1^{\BI}$, where $s^{\AI}$ is given in \eqref{eq:AI-s} and $s_1^{\BI}$ is the mod $p$ contraction \eqref{eq:Gamma-Zp-s}.\footnote{We choose $\AI(y, n + 1)$ as the first tensor factor to simplify calculations (recall that the first factor takes priority when acting with the tensor product contraction \eqref{eq:tensor-construction}). Homotopy-invariant data, such as the (co)homology suspensions, does not depend on the choice of contraction.} Explicitly,
  \begin{equation}
    \eqbox{c}{rCl}{
      s_1\qty\big(\gp_{p^i}(x)^h \gp_k(b_i)) &=& \begin{cases}
        0 & h = 0 \\
        \gp_{p^i}(x)^{h-1}\, a_i \gp_k(b_i) & 0 < h < p,
      \end{cases} \\
      s_1\qty\big(\gp_{p^i}(x)^h\, a_i \gp_k(b_i)) &=& \begin{cases}
        0 & 0 \leq h < p - 1 \\
        \gp_{k+1}(b_i) & h = p - 1,
      \end{cases} \\
      s_1(y \gp_k(z)) &=& \gp_{k+1}(z), \\
      s_1(\gp_k(z)) &=& 0.
    }
  \end{equation}
  Then $s$ is defined by \eqref{eq:p-local-improvement}; we will compute the sum explicitly on the elements of $\cA$, and see that it terminates there. Again, we do not write down $s$ in closed form, only $\sbar$. First, on the elements $\gp_k(x) \in \BI(x, n) \subset \cA$, we already have the result \eqref{eq:Gamma-plocal-sbar}. On elements of the form $y \gp_k(x)$, we have to work a little harder. Factor this element as $y \gp_{p^{i_0}}(x)^{h_0} \gp_{p^{i_1}}(x)^{h_1} \cdots \gp_{p^{i_n}}(x)^{h_n}$ up to a unit, where $k = h_0 p^{i_0} + h_1 p^{i_1} + \cdots + h_n p^{i_n}$ with $i_0 < i_1 < \cdots < i_n$ and $0 < h_j < p$ (i.e.~the $h_j$ are the nonzero digits of $k$ in base $p$). We write out the terms $t_\ell = (1 - s_1 \partial)^\ell s_1(y \gp_{p^{i_0}}(x)^{h_0} \cdots)$ of the sum \eqref{eq:p-local-improvement}:\footnote{We write $\delta_{i_0,i_1,\dots}^{0,1,\dots}$ for the Kronecker delta $\delta_{i_0}^0 \delta_{i_1}^1 \dots$, and abbreviate $\delta_{h_0,h_1,\dots}^{p-1} = \delta_{h_0}^{p-1} \delta_{h_1}^{p-1} \dots$.}
  \begin{equation}
    \small
    \begin{tikzcd}
      y \gp_{p^{i_0}}(x)^{h_0} \cdots \ar[d,"s_1"] \\
      t_0 = z \gp_{p^{i_0}}(x)^{h_0} \cdots
      \ar[dd,"1 - s_1 \partial"] \ar[r,"\partial"]
      & (y - \eta a_0) \gp_{p^{i_0}}(x)^{h_0} \cdots
      \ar[d,"s_1"] \\
      & t_0 - \eta \delta_{i_0}^{0} \delta_{h_0}^{p-1} b_0 \gp_{p^{i_1}}(x)^{h_1} \cdots
      \\
      t_1 = \eta \delta_{i_0}^{0} \delta_{h_0}^{p-1} b_0 \gp_{p^{i_1}}(x)^{h_1} \cdots
      \ar[dd,"1 - s_1 \partial"] \ar[r,"\partial"]
      & \eta \delta_{i_0}^{0} \delta_{h_0}^{p-1} (x^{p-1} a_0 - pa_1) \gp_{p^{i_1}}(x)^{h_1} \cdots
      \ar[d,"s_1"] \\
      & t_1 - \eta p \delta_{i_0,i_1}^{0,1} \delta_{h_0,h_1}^{p-1} b_1 \gp_{p^{i_2}}(x)^{h_2} \cdots
      \\
      t_2 = \eta p \delta_{i_0,i_1}^{0,1} \delta_{h_0,h_1}^{p-1} b_1 \gp_{p^{i_2}}(x)^{h_2} \cdots
      \ar[dd,"1 - s_1 \partial"] \ar[r,"\partial"]
      & \eta p \delta_{i_0,i_1}^{0,1} \delta_{h_0,h_1}^{p-1} (\gp_p(x)^{p-1} a_1 - p a_2) \gp_{p^{i_2}}(x)^{h_2} \cdots
      \ar[d,"s_1"] \\
      & t_2 - \eta p^2 \delta_{i_0,i_1,i_2}^{0,1,2} \delta_{h_0,h_1,h_2}^{p-1} b_2 \gp_{p^{i_3}}(x)^{h_3} \cdots
      \\
      t_3 = \eta p^2 \delta_{i_0,i_1,i_2}^{0,1,2} \delta_{h_0,h_1,h_2}^{p-1} b_2 \gp_{p^{i_3}}(x)^{h_3} \cdots
      \ar[d] \\
      \vdots
    \end{tikzcd}
  \end{equation}
  The pattern is hopefully clear. The sum terminates whenever the next digit of $k$ is not equal to $p - 1$ (including if it is zero). Furthermore, each term $t_\ell$ contains a factor of $x$, which is annihilated by the quotient $\cM \lto \cN$, except for possibly the last term. Thus the only nonzero values of $\sbar$ are
  \begin{equation}
    \sbar(y) = z, \qquad \sbar\qty(y x^{p-1} \gp_p(x)^{p-1} \cdots \gp_{p^i}(x)^{p-1}) = \eta p^i b_i.
  \end{equation}
  For compactness, we write
  \begin{equation}
    x^{p-1} \gp_p(x)^{p-1} \cdots \gp_{p^i}(x)^{p-1} = \rho_i \gp_{p^{i+1} - 1}(x)
  \end{equation}
  where the coefficient
  \begin{equation}
    \rho_i = \frac{(p^{i+1} - 1)!}{\qty\big(p! (p^2)! \cdots (p^i)!)^{p-1}}
  \end{equation}
  is an integer not divisible by $p$ (hence a unit in $\Z_{(p)}$).

  In summary, we have found that
  \begin{equation}
    \eqbox{c}{rCl}{
      \sbar(\gp_k(x)) &=& \begin{cases}
        a_i & k = p^i \\
        0 & \text{$k$ not a power of $p$},
      \end{cases} \\
      \sbar(y \gp_k(x)) &=& \begin{cases}
        z & k = 0 \\
        \frac{1}{\rho_i} \eta p^i b_i & k = p^{i+1} - 1 \\
        0 & \text{otherwise}.
      \end{cases}}
  \end{equation}

  Finally, we again rename $z$ to $\sigma y$, $a_i$ to $\sigma\!\gp_{p^i}\!x$ and $b_i$ to $\varphi_p\!\gp_{p^i}\!x$.
\end{itemize}

We can find a delooping of $\AII$ in the same way, but the result is more complicated.

\begin{itemize}
\item For $\cA = \AII(x, y, n, \eta)$ over $\Lambda = \Z_{(p)}$, Moore gives the construction
  \begin{equation}
    \label{eq:AII-N}
    \cN = \Gamma(z, n + 1) \otimes
    \bigotimes_{i \geq 0} \qty\Big(
        E(a_i, p^i (n + 1) + 1) \otimes \Gamma(b_i, p^{i+1} (n + 1) + 2))
  \end{equation}
  with
  \begin{equation}
    \eqbox{c}{c}{
      \partial x = 0, \qquad
      \partial y = \eta x, \\
      \partial z = x, \qquad
      \partial a_i = \gp_{p^i}(y - \eta z), \qquad
      \partial b_i = \gp_{p^i}(y - \eta z)^{p-1} a_i - p\, \zeta_i\, a_{i+1}.
    }
  \end{equation}
  Again, this is a perturbation of the tensor product $\AI(x, n) \otimes \BI(y, n + 1)$, where the differential is found by imposing $\partial y = \eta x$ and $\partial^2 = 0$.

  The induced differential on $\cN$ is
  \begin{equation}
    \partialbar z = 0, \qquad
    \partialbar a_i = \gp_{p^i}(-\eta z), \qquad
    \partialbar b_i = \gp_{p^i}(-\eta z)^{p-1} a_i - p\, \zeta_i\, a_{i+1}.
  \end{equation}
  To write $\cN$ as a tensor product of DGAs, we make a basis change. For $i
  \geq 0$, define
  \begin{equation}
    \label{eq:ci-definition}
    c_{i+1} = a_{i+1} - \theta_i \gp_{p^i}(z)^{p-1}\, a_i,
    \qq{where} \theta_i = \frac{(-\eta)^{p^i(p - 1)}}{p} \zeta_i^{-1}.
  \end{equation}
  Here $\theta_i \zeta_i$ is an integer since $\eta$ is assumed to be divisible
  by $p$, and we can divide by $\zeta_i \in \Z$ since it is invertible in
  $\Z_{(p)}$; this means that we have an invertible basis change
  \begin{equation}
    (a_0, a_1, a_2, a_3, \dots) \longleftrightarrow
    (a_0, c_1, c_2, c_3, \dots);
  \end{equation}
  explicitly
  \begin{equation}
    \label{eq:AII-basis-change}
    a_i = \sum_{j = 0}^i
    \qty(\prod_{k = 1}^j \theta_{i-k} \gp_{p^{i-k}}(z)^{p-1}) c_{i-j}.
  \end{equation}
  The basis change is chosen such that the induced differential splits as
  \begin{equation}
    \partialbar z = 0, \qquad
    \partialbar a_0 = -\eta z, \qquad
    \partialbar c_i = 0, \qquad
    \partialbar b_i = -p\, \zeta_i\, c_{i+1};
  \end{equation}
  therefore we can write $\cN$ as the tensor product
  \begin{equation}
    \label{eq:AII-delooping-as-tensor}
    \cN = \BII(z, a_0, n + 1, -\eta) \otimes
    \bigotimes_{i \geq 0} \AII(c_{i+1}, b_i, p^{i+1} (n + 1) + 1, -p \zeta_i).
  \end{equation}

  Like for $\BII$, we have the mod $p$ tensor product contraction
  \begin{equation}
    \eqbox{c}{rCl}{
      s_1(x \gp_k(z)) &=& \gp_{k+1}(z), \\
      s_1(\gp_k(z)) &=& 0, \\
      s_1\qty\big(\gp_{p^i}(y)^h \gp_k(b_i)) &=& \begin{cases}
        0 & h = 0 \\
        \gp_{p^i}(y)^{h-1}\, a_i \gp_k(b_i) & 0 < h < p,
      \end{cases} \\
      s_1\qty\big(\gp_{p^i}(y)^h\, a_i \gp_k(b_i)) &=& \begin{cases}
        0 & 0 \leq h < p - 1 \\
        \gp_{k+1}(b_i) & h = p - 1.
      \end{cases}
    }
  \end{equation}

  Finally, we calculate the suspension $\sbar$. As for the $\BII$ case, we write out the terms of the sum \eqref{eq:p-local-improvement} when applied to the elements of $\cA$, and see explicitly that it terminates. For an element $x \gp_k(y) \in \cA$, we get
  \begin{equation}
    \small
    \begin{tikzcd}
      x \gp_k(y) \ar[d,"s_1"] \\
      t_0 = z \gp_k(y)
      \ar[dd,"1 - s_1 \partial"] \ar[r,"\partial"]
      & x \gp_k(y) + \eta z x \gp_{k-1}(y)
      \ar[d,"s_1"] \\
      & t_0 + \eta \gp_2(z) \gp_{k-1}(y)
      \\
      t_1 = -\eta \gp_2(z) \gp_{k-1}(y)
      \ar[dd,"1 - s_1 \partial"] \ar[r,"\partial"]
      & -\eta \qty(x z \gp_{k-1}(y) + \eta x \gp_2(z) \gp_{k-2}(y))
      \ar[d,"s_1"] \\
      & t_1 - \eta^2 \gp_3(z) \gp_{k-2}(y) \\
      t_2 = (-\eta)^2 \gp_3(z) \gp_{k-2}(y)
      \ar[d] \\
      \vdots
    \end{tikzcd}
  \end{equation}
  In general, $t_\ell = (-\eta)^\ell \gp_{\ell+1}(z) \gp_{k-\ell}(y)$. The quotient $\cM \lto \cN$ annihilates all terms with a $y$, so only the last term $t_k$ remains, and we find
  \begin{equation}
    \label{eq:AII-sbar-appendix}
    \sbar(x\gp_k(y)) = (-\eta)^k \gp_{k+1}(z).
  \end{equation}

  Similarly, for an element of the form $\gp_k(y) \in \cA$, factor it as $\gp_{p^i}(y)^h \gp_n(y)$ up to a unit, where $0 < h < p$ and $p^{i+1} \mid n$. Then the first term is
  \begin{equation}
    \label{eq:AII-t0-factor}
    t_0 = s_1\qty(\gp_{p^i}(y)^h \gp_n(y)) = a_i \gp_{p^i}(y)^{h-1} \gp_n(y) =
    \alpha a_i \gp_\kappa(y),
  \end{equation}
  where $\kappa = k - p^i$ and $\alpha$ is a unit. We can now proceed:
  \begin{equation}
    \small
    \begin{tikzcd}
      t_0 = \alpha a_i \gp_\kappa(y)
      \ar[dd,"1 - s_1 \partial"] \ar[r,"\partial"]
      & \alpha \qty(\gp_{p^i}(y - \eta z) \gp_\kappa(y) + \eta x a_i \gp_{\kappa-1}(y))
      \ar[d,"s_1"] \\
      & t_0 + \alpha \eta z a_i \gp_{\kappa-1}(y)
      \\
      t_1 = -\alpha \eta z a_i \gp_{\kappa-1}(y)
      \ar[dd,"1 - s_1 \partial"] \ar[r,"\partial"]
      & -\alpha \eta \qty\bigg(
      \qty\Big(x a_i + z \gp_{p^i}(y - \eta z)) \gp_{\kappa-1}(y)
      + \eta x z a_i \gp_{\kappa-2}(y))
      \ar[d,"s_1"] \\
      & t_1 - \alpha \eta^2 \gp_2(z) a_i \gp_{\kappa - 2}(y)
      \\
      t_2 = \alpha \eta^2 \gp_2(z) a_i \gp_{\kappa - 2}(y)
      \ar[d] \\
      \vdots
    \end{tikzcd}
  \end{equation}
  Some explanations are in order here. When we apply $s_1$, we always do so to terms containing at least one $x$ or $z$; hence the result is simple. The one exception to this is the first application; here we have used \eqref{eq:AII-t0-factor} as follows:
  \begin{equation}
    s_1\qty(\alpha\gp_{p^i}(y - \eta z) \gp_\kappa(y))
    = s_1\qty(\alpha\gp_{p^i}(y) \gp_\kappa(y))
    = s_1\qty(\gp_{p^i}(y)^h \gp_n(y)) = t_0.
  \end{equation}
  We identify the pattern $t_\ell = \alpha \gp_\ell(-\eta z) a_i \gp_{\kappa - \ell}(y)$. As before, all terms but the last contain a $y$ and disappear in the quotient $\cM \lto \cN$, so we have
  \begin{equation}
    \sbar(\gp_{p^i}(y)^h \gp_n(y)) = \alpha \gp_\kappa(-\eta z) a_i
    = \gp_{p^i}(-\eta z)^{h-1} \gp_n(-\eta z) a_i.
  \end{equation}
  Writing everything in terms of single divided powers, we can reformulate this as
  \begin{equation}
    \sbar(\gamma_k(y)) = {k \choose {p^i}}^{-1} \gp_{k - p^i}(-\eta z) a_i \qq{where} i = \nu_p(k).
  \end{equation}
  Rewriting $a_i$ in terms of $c_i$ according to \eqref{eq:AII-basis-change}, this becomes
  \begin{equation}
    \sbar(\gamma_k(y)) = {k \choose {p^i}}^{-1} \gp_{k - p^i}(-\eta z) \sum_{j = 0}^i
    \qty(\prod_{l = 1}^j \theta_{i-l} \gp_{p^{i-l}}(z)^{p-1}) c_{i-j} \qq{where} i = \nu_p(k).
  \end{equation}

  Finally, we rename $z$ to $\sigma x$, $a_0$ to $\sigma y$, $b_i$ to
  $\tp_p\!\gp_{p^i}\!y$ and $c_i$ to $\eps_{p^i} y$.
\end{itemize}

\section{Differential algebra and Koszul signs}
\label{app:koszul}
Here, we review some elementary constructions in differential algebra, with particular emphasis on the Koszul signs that make them consistent. We pay pedantic attention to the signs because they need to be correct for our computations in \zcref{sec:more-structures} to work. All the stated formulae can be verified by simply unfolding the definitions of both sides.

For two chain complexes $(X, \partial)$ and $(Y, \partial)$, we can form the \emph{function complex} $(\Hom(X, Y), \delta)$. A linear map $f\colon X \lto Y$ has degree $i = \abs{f}$ if it raises the degree of a chain by $i$, and the differential is
\begin{equation}
  \label{eq:kos:delta}
  \delta f = \partial \circ f - (-1)^\abs{f} f \circ \partial.
\end{equation}
We have the graded Leibniz rule
\begin{equation}
  \label{eq:kos:delta-leibniz}
  \delta(f \circ g) = \delta f \circ g + (-1)^{\abs{f}} f \circ \delta g.
\end{equation}
A \emph{chain map} is a degree-0 function with $\delta f = 0$, i.e.~$\partial \circ f = f \circ \partial$. A \emph{homotopy} $h\colon f \sim g$ is a map $h$ such that $\delta h = g - f$.

For the pushforward (postcomposition), we have a straightforward map
\begin{equation}
  \label{eq:kos:pushforward}
  \eqbox{c}{rl}{
    (\blank)_*\colon &\Hom(X, Y) \lto \Hom(\Hom(Z, X), \Hom(Z, Y)) \\
    & f_* g = f \circ g.
  }
\end{equation}
This is a chain map: If we call the differential on the right hand side $D$, to avoid confusion with $\delta$, defined such that $D\varphi = \delta \circ \varphi - (-1)^{\abs{\varphi}} \varphi \circ \delta$, we have $D(f_*) = (\delta f)_*$. However, for the pullback (precomposition), a sign is necessary to ensure $D(f^*) = (\delta f)^*$:
\begin{equation}
  \label{eq:kos:pullback}
  \eqbox{c}{rl}{
    (\blank)^*\colon &\Hom(X, Y) \lto \Hom(\Hom(Y, Z), \Hom(X, Z)) \\
    & f^* g = (-1)^{\abs{f} \abs{g}} g \circ f.
  }
\end{equation}
By definition, we have
\begin{equation}
  \label{eq:kos:pushforward-pullback}
  f^* g = (-1)^{\abs{f} \abs{g}} g_* f.
\end{equation}

In the tensor product $X \otimes Y$, we have the differential $\partial(x \otimes y) = \partial x \otimes y + (-1)^{\abs{x}} x \otimes \partial y$. We have a natural inclusion
\begin{equation}
  \label{eq:kos:tensor-inc}
  \eqbox{c}{rCl}{
    \theta\colon \Hom(X, Z) \otimes \Hom(Y, W) &\lto& \Hom(X \otimes Y, Z \otimes W) \\
    f \otimes g &\mapsto& \qty(x \otimes y \mapsto (-1)^{\abs{g}\abs{x}} fx \otimes gy)
  }
\end{equation}
which is a chain map, meaning that $\delta(\theta(f \otimes g)) = \theta(\delta f \otimes g + (-1)^{\abs{f}} f \otimes \delta g)$. We typically omit $\theta$, and just write $(f \otimes g)(x \otimes y) = (-1)^{\abs{g} \abs{x}} fx \otimes gy$. Then the fact that $\theta$ is a chain map just means that the two interpretations of $\delta(f \otimes g)$ ($\delta$ being the differential of functions or of tensor products) agree.

The tensor-hom adjunction
\begin{equation}
  \label{eq:kos:tensor-hom}
  \eqbox{c}{rCl}{
    \Hom(X \otimes Y, Z) &\lto& \Hom(X, \Hom(Y, Z)) \\
    f &\mapsto& \qty(x \mapsto \qty(y \mapsto f(x \otimes y)))
  }
\end{equation}
is a chain map without additional signs.

The \emph{braiding} is the natural isomorphism
\begin{equation}
  \label{eq:kos:braiding}
  \eqbox{c}{rCl}{
    T\colon X \otimes Y &\lto& Y \otimes X \\
    x \otimes y &\mapsto& (-1)^{\abs{x} \abs{y}} y \otimes x
  }
\end{equation}
with $T^2 = \id$. For $f\colon X \to Z$ and $g \colon Y \to W$, it is simple to show that
\begin{equation}
  \label{eq:kos:swap-conjugate}
  T_{Z \otimes W} \circ (f \otimes g) \circ T_{X \otimes Y} = (-1)^{\abs{f}\abs{g}} (g \otimes f),
\end{equation}
which can also be phrased in terms of the inclusion \eqref{eq:kos:tensor-inc} as
\begin{equation}
  \label{eq:kos:swap-conjugate-2}
  (T_{Z \otimes W})_* \circ (T_{X \otimes Y})^* \circ \theta = \theta \circ T_{\Hom(X, Z) \otimes \Hom(Y, W)}.
\end{equation}

A \emph{product} on $X$ is a map $\mu\colon X \otimes X \lto X$, written as $x_1 \otimes x_2 \mapsto x_1 x_2$, that is a chain map ($\delta\mu = 0$, which unfolds to the Leibniz rule) and \emph{associative}, meaning that $\mu \circ (\mu \otimes \id) = \mu \circ (\id \otimes \mu)\colon X \otimes X \otimes X \lto X$. It is \emph{commutative} if $\mu \circ T = \mu$ (i.e.~$(-1)^{\abs{x_2} \abs{x_1}} x_2 x_1 = x_1 x_2$). Dually, a \emph{coproduct} is a chain map $\Delta_0\colon X \lto X \otimes X$ that is \emph{coassociative} ($(\Delta_0 \otimes \id) \circ \Delta_0 = (\id \otimes \Delta_0) \circ \Delta_0$). It is \emph{cocommutative} if $T \circ \Delta_0 = \Delta_0$.

When $X$ has a coproduct $\Delta_0$ (coassociative) and $Y$ has a product $\mu$ (associative), there is an associative\footnote{$(f \cupp g) \cupp h = \mu \circ (\mu \otimes \id) \circ (f \otimes g \otimes h) \circ (\Delta_0 \otimes \id) \circ \Delta_0 = \mu \circ (\id \otimes \mu) \circ (f \otimes g \otimes h) \circ (\id \otimes \Delta_0) \circ \Delta_0 = f \cupp (g \cupp h)$.} product on $\Hom(X, Y)$, the \emph{convolution} with respect to $\Delta_0$:
\begin{equation}
  \label{eq:kos:convolution}
  f \cupp g = \mu_* \Delta_0^* (f \otimes g) = \mu \circ (f \otimes g) \circ \Delta_0.
\end{equation}
When $\mu$ is commutative and $\Delta_0$ is cocommutative, $\cupp$ is commutative:
\begin{equation}
  f \cupp g = \mu \circ (f \otimes g) \circ \Delta_0 =
  \mu \circ T \circ (f \otimes g) \circ T \circ \Delta_0
  \overset{\eqref{eq:kos:swap-conjugate}}{=}
  (-1)^{\abs{f} \abs{g}} \mu \circ (g \otimes f) \circ \Delta_0
  = (-1)^{\abs{f} \abs{g}} g \cupp f.
\end{equation}
However, in \zcref{sec:more-structures}, we deal with a $\Delta_0$ that is only cocommutative up to homotopy (while $\mu$ is still strictly commutative), specifically, we have a sequence of degree $i$ maps $\Delta_i\colon X \lto X \otimes X$ with $\delta \Delta_i = \qty(T + (-1)^i) \circ \Delta_{i-1}$. This makes the $\cupp$ operation commutative up to homotopy: we can define corresponding degree $-i$ maps $\cupp_i\colon \Hom(X, Y)^{\otimes 2} \lto \Hom(X, Y)$ by
\begin{equation}
  \label{eq:kos:cupp_i}
  f \cupp_i g = \mu_* \Delta_i^* (f \otimes g)
  = (-1)^{(\abs{f} + \abs{g})i} \mu \circ (f \otimes g) \circ \Delta_i,
\end{equation}
so that ${\cupp_0} = {\cupp}$ and $D(\cupp_i) = {\cupp_{i-1}} \circ \qty(T_{\Hom(X, Y)^{\otimes 2}} + (-1)^i)$.\footnote{Again, with $D\varphi = \delta \circ \varphi - (-1)^{\abs{\varphi}} \varphi \circ \delta$.} The last equation expands to the coboundary formula
\begin{equation}
  \label{eq:kos:coboundary-formula}
  \delta(f \cupp_i g) - (-1)^i \qty(\delta f \cupp_i g + (-1)^{\abs{f}} f \cupp_i \delta g)
  = (-1)^{\abs{f} \abs{g}} g \cupp_{i-1} f + (-1)^i f \cupp_{i-1} g.
\end{equation}
The proof is simple with the machinery we have built up. Using \eqref{eq:kos:tensor-inc} to write ${\cupp_i} = \mu_* \circ \Delta_i^* \circ \theta$, we have $D\theta \overset{\eqref{eq:kos:tensor-inc}}{=} 0$, $D(\mu_*) \overset{\eqref{eq:kos:pushforward}}{=} (\delta \mu)_* = 0$ and $D(\Delta_i^*) \overset{\eqref{eq:kos:pullback}}{=} (\delta \Delta_i)^*$ as $\theta$, $(\blank)_*$ and $(\blank)^*$ are chain maps. Then
\begin{IEEEeqnarray*}{rCl}
  D(\cupp_i) &=& D(\mu_* \circ \Delta_i^j* \circ \theta)
  \overset{\eqref{eq:kos:delta-leibniz}}{=}
  \mu_* \circ D(\Delta_i^*) \circ \theta
  = \mu_* \circ (\delta \Delta_i)^* \circ \theta \\
  &=& \mu_* \circ \Delta_{i-1}^* \circ (T_{X \otimes X} + (-1)^i)^* \circ \theta \\
  &=& \mu_* \circ \Delta_{i-1}^* \circ (T_{X \otimes X})^* \circ \theta + (-1)^i {\cupp_{i-1}} \\
  &\overset{\eqref{eq:kos:swap-conjugate-2}}{=}&
  \mu_* \circ \Delta_{i-1}^* \circ (T_{Y \otimes Y})_* \circ \theta \circ T_{\Hom(X,Y)^{\otimes 2}}
  + (-1)^i {\cupp_{i-1}} \\
  &\overset{\eqref{eq:kos:pushforward-pullback}}{=}& (\mu \circ T_{Y \otimes Y})_* \circ \Delta_{i-1}^* \circ \theta \circ T_{\Hom(X,Y)^{\otimes 2}}
  + (-1)^i {\cupp_{i-1}} \\
  &=& \mu_* \circ \Delta_{i-1}^* \circ \theta \circ T_{\Hom(X,Y)^{\otimes 2}}
  + (-1)^i {\cupp_{i-1}} \\
  &=& {\cupp_{i-1}} \circ \qty(T_{\Hom(X,Y)^{\otimes 2}} + (-1)^i),
  \IEEEyesnumber
\end{IEEEeqnarray*}
having used commutativity $\mu \circ T = \mu$.

\section{Computation of diagonal maps}
\label{app:diagonal}
In this appendix, we calculate the diagonal maps exhibited in \zcref{sec:more-structures}.

We begin with the derivation of \eqref{eq:Deltabar-BZN}. We start from the diagonal map $\Delta_0(u) = u \otimes u$ of $\Lambda[\Z_N]$ and follow the method of \zcref{sec:cup-products} to extend it to the construction \eqref{eq:LambdaZN-delooping}. Unlike when calculating the (co)suspension, we need the full map $s$, given in \eqref{eq:LambdaZN-s}; knowledge of only $\sbar$ is not sufficient. By multiplicativity, we only need to find $\Delta_0(x)$ and $\Delta_0(y)$ (where $x = \sigma u$ and $y = \tpsi_N\! u$). First, we compute
\begin{equation}
  \label{eq:BZN-Delta0x}
  \Delta_0(x) = s\Delta_0(\partial x) = s\Delta_0(u) = s(u \otimes u)
  = x \otimes u + 1 \otimes x.
\end{equation}
The next case is more complicated:
\begin{IEEEeqnarray*}{rCl}
  \Delta_0(y) &=& s\Delta_0(\partial y) = s\Delta_0\qty(\frac{u^N - 1}{u - 1} x)
  = \sum_{i = 0}^{N-1} s\qty(\Delta_0(u)^i \Delta_0(x)) \\
  &\overset{\eqref{eq:BZN-Delta0x}}{=}&
  \sum_{i = 0}^{N-1} s\qty(u^i x \otimes u^{i+1} + u^i \otimes u^i x) \\
  &\overset{\eqref{eq:tensor-construction}}{=}&
  \sum_{i = 0}^{N-1} \qty\Big(s(u^i x) \otimes u^{i+1}
  + s(u^i) \otimes u^i x + 1 \otimes s(u^i x)).
  \IEEEyesnumber
\end{IEEEeqnarray*}
According to \eqref{eq:LambdaZN-s}, $s(u^i x)$ is nonzero only for $i = N - 1$, but $s(u^i)$ is always nonzero. We obtain
\begin{IEEEeqnarray*}{rCl}
  \label{eq:BZN-Delta0y}
  \Delta_0(y)
  &=& y \otimes 1
  + \sum_{i = 0}^{N-1} \frac{u^i - 1}{u - 1} x \otimes u^i x
  + 1 \otimes y \\
  &=& y \otimes 1
  + \sum_{0 \leq i < j < N} u^i x \otimes u^j x
  + 1 \otimes y.
  \IEEEyesnumber
\end{IEEEeqnarray*}
Taking the quotient $\cM \lto \cN$ (setting $u = 1$), we get \eqref{eq:Deltabar-BZN}, with the coefficient ${N \choose 2}$ coming from $\sum_{0 \leq i < j < N} 1 = {N \choose 2}$.

Next, we compute some values of $\Delta_0$ in $B^2 \Z_N$, given by the construction \eqref{eq:AII-N}. Beginning with \eqref{eq:Deltabar-BZN}, written as
\begin{equation}
  \Delta_0(x) = x \otimes 1 + 1 \otimes x, \qquad
  \Delta_0(y) = y \otimes 1 + {N \choose 2} x \otimes x + 1 \otimes y,
\end{equation}
we again use multiplicativity of $\Delta_0$ and $\Delta_0(n) = s\Delta_0(\partial n)$. It is straightforward to find
\begin{equation}
  \Delta_0(z) = z \otimes 1 + 1 \otimes z
  \qq{and}
  \Delta_0(a_0) = a_0 \otimes 1 + {N \choose 2} x \otimes x
  + 1 \otimes a_0.
\end{equation}
We will also compute $\Delta_0(a_1)$, restricting to the case $p = 2$ (and therefore $N$ even) for simplicity. We have
\begin{IEEEeqnarray*}{rCl}
  \Delta_0(a_1) &=& s\Delta_0\gamma_2(y - Nz) = s\gamma_2\Delta_0(y - Nz) \\
  &=& s\gamma_2\qty((y - Nz) \otimes 1 + {N \choose 2} x \otimes x + 1 \otimes (y - Nz)) \\
  &=& s\Bigg[\gamma_2(y - Nz) \otimes 1 + (y - Nz) \otimes (y - Nz) + 1 \otimes \gamma_2(y - Nz) \\
  && \quad {} + {N \choose 2}\qty\Big((xy - Nxz) \otimes x + x \otimes (xy - Nxz))\Bigg].
  \IEEEyesnumber
\end{IEEEeqnarray*}
Several terms disappear since $s(z) = s(zy) = 0$. We have $s(xz) = \gamma_2(z)$, $s(y) = a_0$ and $s(\gamma_2(y)) = a_1$. For $xy$, we do not simply have $s(xy) = s(x)y$; instead, the iteration \eqref{eq:p-local-improvement} gives a correction $s(xy) = \qty(1 + (1 - s_1 \partial)) s_1(xy) = zy - N\gamma_2(z)$. Thus, we compute
\begin{IEEEeqnarray*}{rCl}
  \Delta_0(a_1) &=& a_1 \otimes 1 + a_0 \otimes (y - Nz) + 1 \otimes a_1 \\
  && {} + {N \choose 2} \qty\Big((zy - 2N\gamma_2(z)) \otimes x + z \otimes (xy - Nxz)).
  \IEEEyesnumber
\end{IEEEeqnarray*}
Passing to $\cN$ by setting $x = y = 0$, we obtain
\begin{IEEEeqnarray}{rCl}
  \Deltabar_0(z) &=& z \otimes 1 + 1 \otimes z, \\
  \Deltabar_0(a_0) &=& a_0 \otimes 1 + 1 \otimes a_0, \\
  \Deltabar_0(a_1) &=& a_1 \otimes 1 - N a_0 \otimes z + 1 \otimes a_1.
\end{IEEEeqnarray}
Finally, we write this in terms of the $c$-basis \eqref{eq:ci-definition} $c_0 = a_0$ and $c_1 = a_1 + \frac{N}{2} z a_0$:
\begin{equation}
  \Deltabar_0(c_1) = c_1 \otimes 1 + \frac{N}{2} \qty(z \otimes c_0 - c_0 \otimes z) + 1 \otimes c_1.
\end{equation}

\section{Tables}
\label{app:tables}

This appendix lists tables of the cohomology groups of Eilenberg--MacLane spaces and their cosuspensions with various coefficients. The tables are generated by \texttt{emcm} \cite{riedelgardingEmcm2026}.

\subsection{Integral stable and unstable groups}
\zcref[cap]{tab:cohomology-BhZ-Z} shows some integral cohomology groups that do not fit in later tables. \zcref[cap]{tab:stable-homology} shows the stable homology groups, calculated by \texttt{emcm} essentially as outlined by \textcite[exposé~11, §6]{cartanAlgebreDEilenbergMacLane1954}. The notation $_N G$ means the $N$-torsion subgroup of $G$.

\begin{sidewaystable}[h]
  {\tiny \input{tables/cohomology-BhZ-Z.tex}}
  \caption{The cohomology groups $H^D(B^h \Z; \Z)$.}
  \label{tab:cohomology-BhZ-Z}
\end{sidewaystable}

\label{app:stable-groups}
\begin{table}[h]
  \centering
  {\footnotesize \input{tables/stable-homology.tex}}
  \caption{The stable homology groups $H^s_{h-1}(G) = H_{2h-1+k}(B^{h+k} G)$, independent of $k \geq 0$.}
  \label{tab:stable-homology}
\end{table}

\clearpage
\subsection{Cosuspension on generators}
\label{app:cosuspension-on-generators}
The following tables display the generators of the groups $H^D(B^h G; A)$ for $h = 2, 3$, $G = \Z_2, \Z_3$ and $A = \R/\Z, \Z, G$, together with their images under the cosuspension $\Omega$. The order of a generator is written next to it in angle brackets.

The generators for cohomology classes mixing different elementary complexes (as in the Künneth formula) are written using the tensor product $\otimes$ as well as the letters $\tau$ and $\kappa$, which we fix for definiteness:
\begin{itemize}
\item If $\partial x_1 = \partial x_2 = 0$, then $\partial(x_1 x_2) = 0$.
\item If $\partial x_1 = 0$ and $n_2 x_2 = \partial y_2$, then
  $(-1)^{\abs{x_1}} n_2 x_1 x_2 = \partial\qty(x_1 y_2)$.
\item If $n_1 x_1 = \partial y_1$ and $\partial x_2 = 0$, then
  $n_1 x_1 x_2 = \partial\qty(y_1 x_2)$.
\item If $n_1 x_1 = \partial y_1$ and $n_2 x_2 = \partial y_2$, let
  \begin{equation}
    \label{eq:tau}
    \tau(x_1, y_1, x_2, y_2) =
    \frac{n_1}{\gcd(n_1, n_2)} x_1 y_2
    + (-1)^{\abs{y_1}} \frac{n_2}{\gcd(n_1, n_2)} y_1 x_2
  \end{equation}
  and
  \begin{equation}
    \kappa(x_1, y_1, x_2, y_2) = a y_1 x_2 + (-1)^{\abs{x_1}} b x_1 y_2
  \end{equation}
  where $a$ and $b$ are Bézout coefficients such that
  $a n_1 + b n_2 = \gcd(n_1, n_2)$. Then
  \begin{equation}
    \gcd(n_1, n_2) x_1 x_2 = \partial \kappa
    \qq{and}
    \gcd(n_1, n_2) \tau = \partial (y_1 y_2).
  \end{equation}
\end{itemize}

\clearpage
\subsubsection{$B^2\Z_2$}
\begin{table}[h]
  {\footnotesize \input{tables/omega-B2Z2-RZ.tex}}
  \caption{$\Omega\colon H^{D+1}(B^2 \Z_2; \R/\Z) \lto H^D(B \Z_2; \R/\Z)$.}
  \label{tab:Omega-B2Z2-RZ}
\end{table}
\begin{table}[h]
  {\footnotesize \input{tables/omega-B2Z2-Z.tex}}
  \caption{$\Omega\colon H^{D+1}(B^2 \Z_2; \Z) \lto H^D(B \Z_2; \Z)$.}
  \label{tab:Omega-B2Z2-Z}
\end{table}
\begin{table}[h]
  {\footnotesize \input{tables/omega-B2Z2-Z2.tex}}
  \caption{$\Omega\colon H^{D+1}(B^2 \Z_2; \Z_2) \lto H^D(B \Z_2; \Z_2)$.}
  \label{tab:Omega-B2Z2-Z2}
\end{table}
\begin{table}[h]
  {\footnotesize \input{tables/omega-B2Z2-Z4.tex}}
  \caption{$\Omega\colon H^{D+1}(B^2 \Z_2; \Z_4) \lto H^D(B \Z_2; \Z_4)$.}
  \label{tab:Omega-B2Z2-Z4}
\end{table}

\clearpage
\subsubsection{$B^3\Z_2$}
\begin{table}[h]
  {\footnotesize \input{tables/omega-B3Z2-RZ.tex}}
  \caption{$\Omega\colon H^{D+1}(B^3 \Z_2; \R/\Z) \lto H^D(B^2 \Z_2; \R/\Z)$.}
  \label{tab:Omega-B3Z2-RZ}
\end{table}
\begin{table}[h]
  {\footnotesize \input{tables/omega-B3Z2-Z.tex}}
  \caption{$\Omega\colon H^{D+1}(B^3 \Z_2; \Z) \lto H^D(B^2 \Z_2; \Z)$.}
  \label{tab:Omega-B3Z2-Z}
\end{table}
\begin{table}[h]
  {\footnotesize \input{tables/omega-B3Z2-Z2.tex}}
  \caption{$\Omega\colon H^{D+1}(B^3 \Z_2; \Z_2) \lto H^D(B^2 \Z_2; \Z_2)$.}
  \label{tab:Omega-B3Z2-Z2}
\end{table}
\begin{table}[h]
  {\footnotesize \input{tables/omega-B3Z2-Z4.tex}}
  \caption{$\Omega\colon H^{D+1}(B^3 \Z_2; \Z_4) \lto H^D(B^2 \Z_2; \Z_4)$.}
  \label{tab:Omega-B3Z2-Z4}
\end{table}

\clearpage
\subsubsection{$B^2\Z_3$}
\begin{table}[h]
  {\footnotesize \input{tables/omega-B2Z3-RZ.tex}}
  \caption{$\Omega\colon H^{D+1}(B^2 \Z_3; \R/\Z) \lto H^D(B \Z_3; \R/\Z)$.}
  \label{tab:Omega-B2Z3-RZ}
\end{table}
\begin{table}[h]
  {\footnotesize \input{tables/omega-B2Z3-Z.tex}}
  \caption{$\Omega\colon H^{D+1}(B^2 \Z_3; \Z) \lto H^D(B \Z_3; \Z)$.}
  \label{tab:Omega-B2Z3-Z}
\end{table}
\begin{table}[h]
  {\footnotesize \input{tables/omega-B2Z3-Z3.tex}}
  \caption{$\Omega\colon H^{D+1}(B^2 \Z_3; \Z_3) \lto H^D(B \Z_3; \Z_3)$.}
  \label{tab:Omega-B2Z3-Z3}
\end{table}

\clearpage
\subsubsection{$B^3\Z_3$}
\begin{table}[h]
  {\footnotesize \input{tables/omega-B3Z3-RZ.tex}}
  \caption{$\Omega\colon H^{D+1}(B^3 \Z_3; \R/\Z) \lto H^D(B^2 \Z_3; \R/\Z)$.}
  \label{tab:Omega-B3Z3-RZ}
\end{table}
\begin{table}[h]
  {\footnotesize \input{tables/omega-B3Z3-Z.tex}}
  \caption{$\Omega\colon H^{D+1}(B^3 \Z_3; \Z) \lto H^D(B^2 \Z_3; \Z)$.}
  \label{tab:Omega-B3Z3-Z}
\end{table}
\begin{table}[h]
  {\footnotesize \input{tables/omega-B3Z3-Z3.tex}}
  \caption{$\Omega\colon H^{D+1}(B^3 \Z_3; \Z_3) \lto H^D(B^2 \Z_3; \Z_3)$.}
  \label{tab:Omega-B3Z3-Z3}
\end{table}

\FloatBarrier
\subsection{Cosuspension in high dimension}
\label{app:high-dimension}
The following tables display the cosuspension $\Omega\colon H^{D+1}(B^{h+1} G; A) \lto H^D(B^h G; A)$ for $G = \Z, \Z_2, \Z_4, \Z_3, \Z_5$ and various $A$. When $G$ is finite, the case $A = \Z$ is omitted, as it is the same as $A = \R/\Z$ by \eqref{eq:connecting-isomorphism}. Arrows are drawn as injections and surjections in the standard way, and labelled with the order of the generators with nonzero images, and the order of those images. Repeated orders are written with an exponent; for instance, an arrow like
$\begin{tikzcd}
  (\Z_2)^{86} \oplus (\Z_4)^{2}
  \ar[r,"{2,4^2}" {gray,scale=0.8,pos=0.2},%
  "{2^3}" {gray,scale=0.8,pos=0.8}]
  & (\Z_2)^{102}
\end{tikzcd}$
means that $\Omega$ sends three generators with orders $(2, 4, 4)$ to elements of order $(2, 2, 2)$ respectively. The absence of an arrow means that $\Omega$ is zero. The stable range is highlighted in red.

\clearpage
\subsubsection{$\Z$}
\begin{center}
  {\tiny\input{tables/omega-Z-RZ.tex}}
  \captionof{table}{$\Omega\colon
    H^{D+1}(B^{h+1} \Z; \R/\Z) \lto H^D(B^h \Z; \R/\Z)$.}
    \label{tab:Omega-BZ-R/Z}
\end{center}

\clearpage
\begin{center}
  {\tiny\input{tables/omega-Z-Z.tex}}
  \captionof{table}{$\Omega\colon
    H^{D+1}(B^{h+1} \Z; \Z) \lto H^D(B^h \Z; \Z)$.}
    \label{tab:Omega-BZ-Z}
\end{center}

\clearpage
\subsubsection{$\Z_2$}
\vspace{-1.5em}
\begin{center}
  {\tiny\input{tables/omega-Z2-RZ.tex}}
  \captionof{table}{$\Omega\colon
    H^{D+1}(B^{h+1} \Z_2; \R/\Z) \lto H^D(B^h \Z_2; \R/\Z)$.}
\end{center}

\clearpage
\begin{center}
  {\tiny\input{tables/omega-Z2-Z2.tex}}
  \captionof{table}{$\Omega\colon
    H^{D+1}(B^{h+1} \Z_2; \Z_2) \lto H^D(B^h \Z_2; \Z_2)$.}
\end{center}

\clearpage
\begin{center}
  {\tiny\input{tables/omega-Z2-Z4.tex}}
  \captionof{table}{$\Omega\colon
    H^{D+1}(B^{h+1} \Z_2; \Z_4) \lto H^D(B^h \Z_2; \Z_4)$.}
\end{center}

\clearpage
\subsubsection{$\Z_4$}
\vspace{-1.5em}
\begin{center}
  {\tiny\input{tables/omega-Z4-RZ.tex}}
  \captionof{table}{$\Omega\colon
    H^{D+1}(B^{h+1} \Z_4; \R/\Z) \lto H^D(B^h \Z_4; \R/\Z)$.}
\end{center}

\clearpage
\begin{center}
  {\tiny\input{tables/omega-Z4-Z4.tex}}
  \captionof{table}{$\Omega\colon
    H^{D+1}(B^{h+1} \Z_4; \Z_4) \lto H^D(B^h \Z_4; \Z_4)$.}
\end{center}

\clearpage
\subsubsection{$\Z_3$}
\begin{center}
  {\tiny\input{tables/omega-Z3-RZ.tex}}
  \captionof{table}{$\Omega\colon
    H^{D+1}(B^{h+1} \Z_3; \R/\Z) \lto H^D(B^h \Z_3; \R/\Z)$.}
\end{center}

\clearpage
\begin{center}
  {\tiny\input{tables/omega-Z3-Z3.tex}}
  \captionof{table}{$\Omega\colon
    H^{D+1}(B^{h+1} \Z_3; \Z_3) \lto H^D(B^h \Z_3; \Z_3)$.}
\end{center}

\clearpage
\begin{center}
  {\tiny\input{tables/omega-Z3-Z9.tex}}
  \captionof{table}{$\Omega\colon
    H^{D+1}(B^{h+1} \Z_3; \Z_9) \lto H^D(B^h \Z_3; \Z_9)$.}
\end{center}

\clearpage
\subsubsection{$\Z_5$}
\begin{center}
  {\tiny\input{tables/omega-Z5-RZ.tex}}
  \captionof{table}{$\Omega\colon
    H^{D+1}(B^{h+1} \Z_5; \R/\Z) \lto H^D(B^h \Z_5; \R/\Z)$.}
\end{center}

\clearpage
\begin{center}
  {\tiny\input{tables/omega-Z5-Z5.tex}}
  \captionof{table}{$\Omega\colon
    H^{D+1}(B^{h+1} \Z_5; \Z_5) \lto H^D(B^h \Z_5; \Z_5)$.}
\end{center}

\clearpage
\begin{center}
  {\tiny\input{tables/omega-Z5-Z25.tex}}
  \captionof{table}{$\Omega\colon
    H^{D+1}(B^{h+1} \Z_5; \Z_{25}) \lto H^D(B^h \Z_5; \Z_{25})$.}
\end{center}

\clearpage
\printbibliography
\end{document}

%% file: preamble.tex
\usepackage[UKenglish]{babel}

\usepackage{IEEEtrantools}
\usepackage{jheppub}
\usepackage{subcaption}
\usepackage{physics}
\usepackage{amssymb}
\usepackage{mathtools}
\usepackage[new]{old-arrows}
\usepackage{mathrsfs}
\usepackage[dvipsnames]{xcolor}
\usepackage{graphicx}
\usepackage{caption}
\usepackage{placeins}
\usepackage{rotating}
\usepackage{diagbox}
\usepackage{zref-clever}
\zcsetup{nameinlink=false}
\usepackage{csquotes}
\usepackage{xparse}
\usepackage{extarrows}
\usepackage[backend=biber,style=numeric-comp,sorting=none,backref,maxnames=10]{biblatex}
\usepackage{etoolbox}
\patchcmd{\abstract}{\small}{}{}{}

\usepackage{tikz}
\usetikzlibrary{3d}
\usetikzlibrary{calc}
\usetikzlibrary{decorations.pathreplacing}
\usetikzlibrary{decorations.pathmorphing}
\usetikzlibrary{decorations.markings}
\usetikzlibrary{arrows.meta}
\usetikzlibrary{knots}
\usetikzlibrary{topaths}
\usetikzlibrary{perspective}
\usetikzlibrary{shapes.multipart}
\usepackage{tikz-3dplot}

\tikzset{dotnode/.style={inner sep=0pt,outer sep=0pt,
  circle,fill,minimum size=5pt}}

\usepackage{tikz-cd}

\usepackage[status=draft]{fixme}
\fxusetheme{color}
\FXRegisterAuthor{erg}{aerg}{\textbf{Elias}}
\FXRegisterAuthor{sm}{asm}{\textbf{Shani}}

\newcommand{\eqbox}[3]{\begin{IEEEeqnarraybox}[][#1]{#2}#3\end{IEEEeqnarraybox}}

\DeclareMathOperator{\im}{im}
\DeclareMathOperator{\coim}{coim}
\DeclareMathOperator{\Hom}{Hom}

\DeclareMathOperator{\Maps}{Maps}
\DeclareMathOperator{\Tor}{Tor}
\DeclareMathOperator{\Ext}{Ext}

\DeclareMathOperator{\Nat}{Nat}
\DeclareMathOperator{\evMap}{ev}
\DeclareMathOperator{\Sq}{Sq}
\DeclareMathOperator{\ord}{ord}
\DeclareMathOperator{\cone}{cone}

\DeclareMathOperator{\gp}{\gamma} % divided power
\DeclareMathOperator{\tp}{\varphi} % transpotence
\DeclareMathOperator{\tpsi}{\psi} % transpotence into H_2

\newcommand{\id}{\mathrm{id}}

\newcommand{\defeq}{\coloneqq}

\newcommand{\blank}{-}
\newcommand{\cupp}{\smile}
\newcommand{\lto}{\longrightarrow}

\newcommand{\eps}{\varepsilon}

\newcommand{\e}{\mathrm{e}}
\newcommand{\ii}{\mathrm{i}}

\newcommand{\Z}{\mathbb{Z}}

\newcommand{\R}{\mathbb{R}}
\newcommand{\C}{\mathbb{C}}
\newcommand{\bbG}{\mathbb{G}}

\newcommand{\cA}{\mathcal{A}}
\newcommand{\cC}{\mathcal{C}}
\newcommand{\cM}{\mathcal{M}}
\newcommand{\cN}{\mathcal{N}}
\newcommand{\cZ}{\mathcal{Z}}
\newcommand{\tildeN}{\widetilde{\cN}}
\newcommand{\partialbar}{\bar{\partial}}
\newcommand{\sbar}{\bar{s}}
\newcommand{\Deltabar}{\bar{\Delta}}

\newcommand{\UU}{\mathrm{U}}

\newcommand{\AI}{A_{\mathrm{I}}}
\newcommand{\BI}{B_{\mathrm{I}}}
\newcommand{\AII}{A_{\mathrm{II}}}
\newcommand{\BII}{B_{\mathrm{II}}}

\newcommand{\both}[1]{\textcolor{NavyBlue}{#1}}
\newcommand{\only}[1]{\textcolor{OrangeRed}{#1}}

%% file: table-omega-B2Z2-small.tex
\begin{IEEEeqnarraybox}[][c]{rClrCl}
\mathbb{Z}_{4} &\cong& H^{4}(B^{2}\mathbb{Z}_{2}; \mathbb{R}/\mathbb{Z}) \colon & \frac{1}{4} (\gamma_{2} \sigma^{2} u_{2})^*\langle 4 \rangle &\longmapsto& -\frac{1}{2} (\sigma u_{2} \cdot \psi_{2} u_{2})^*\langle 2 \rangle \\
\mathbb{Z}_{2} &\cong& H^{5}(B^{2}\mathbb{Z}_{2}; \mathbb{R}/\mathbb{Z}) \colon & \frac{1}{2} (\varepsilon_{2} \psi_{2} u_{2})^*\langle 2 \rangle &\longmapsto& 0
\\[1em]
0 &\cong& H^{4}(B^{2}\mathbb{Z}_{2}; \mathbb{Z}) \colon &  \\
\mathbb{Z}_{4} &\cong& H^{5}(B^{2}\mathbb{Z}_{2}; \mathbb{Z}) \colon & (\sigma^{2} u_{2} \cdot \sigma \psi_{2} u_{2})^*\langle 4 \rangle &\longmapsto& -(\gamma_{2} \psi_{2} u_{2})^*\langle 2 \rangle
\\[1em]
\mathbb{Z}_{2} &\cong& H^{4}(B^{2}\mathbb{Z}_{2}; \mathbb{Z}_{2}) \colon & (\gamma_{2} \sigma^{2} u_{2})^*\langle 2 \rangle &\longmapsto& 0 \\
(\mathbb{Z}_{2})^{2} &\cong& H^{5}(B^{2}\mathbb{Z}_{2}; \mathbb{Z}_{2}) \colon & (\sigma \gamma_{2} \psi_{2} u_{2})^*\langle 2 \rangle &\longmapsto& (\gamma_{2} \psi_{2} u_{2})^*\langle 2 \rangle \\
&&& (\sigma^{2} u_{2} \otimes \sigma \psi_{2} u_{2})^*\langle 2 \rangle &\longmapsto& 0
\end{IEEEeqnarraybox}

%% file: tables/cohomology-BhZ-Z.tex
\begin{tabular}{l|ccccccccc}
\diagbox[width=4em]{$D$}{$h$} & 1 & 2 & 3 & 4 & 5 & 6 & 7 & 8 & 9 \\
\hline
0 & $\mathbb{Z}$ & $\mathbb{Z}$ & $\mathbb{Z}$ & $\mathbb{Z}$ & $\mathbb{Z}$ & $\mathbb{Z}$ & $\mathbb{Z}$ & $\mathbb{Z}$ & $\mathbb{Z}$ \\
1 & $\mathbb{Z}$ & $0$ & $0$ & $0$ & $0$ & $0$ & $0$ & $0$ & $0$ \\
2 & $0$ & $\mathbb{Z}$ & $0$ & $0$ & $0$ & $0$ & $0$ & $0$ & $0$ \\
3 & $0$ & $0$ & $\mathbb{Z}$ & $0$ & $0$ & $0$ & $0$ & $0$ & $0$ \\
4 & $0$ & $\mathbb{Z}$ & $0$ & $\mathbb{Z}$ & $0$ & $0$ & $0$ & $0$ & $0$ \\
5 & $0$ & $0$ & $0$ & $0$ & $\mathbb{Z}$ & $0$ & $0$ & $0$ & $0$ \\
6 & $0$ & $\mathbb{Z}$ & $\mathbb{Z}_{2}$ & $0$ & $0$ & $\mathbb{Z}$ & $0$ & $0$ & $0$ \\
7 & $0$ & $0$ & $0$ & $\mathbb{Z}_{2}$ & $0$ & $0$ & $\mathbb{Z}$ & $0$ & $0$ \\
8 & $0$ & $\mathbb{Z}$ & $\mathbb{Z}_{3}$ & $\mathbb{Z}$ & $\mathbb{Z}_{2}$ & $0$ & $0$ & $\mathbb{Z}$ & $0$ \\
9 & $0$ & $0$ & $\mathbb{Z}_{2}$ & $\mathbb{Z}_{3}$ & $0$ & $\mathbb{Z}_{2}$ & $0$ & $0$ & $\mathbb{Z}$ \\
10 & $0$ & $\mathbb{Z}$ & $\mathbb{Z}_{2}$ & $0$ & $\mathbb{Z}_{2} \oplus \mathbb{Z}_{3}$ & $0$ & $\mathbb{Z}_{2}$ & $0$ & $0$ \\
11 & $0$ & $0$ & $\mathbb{Z}_{3}$ & $(\mathbb{Z}_{2})^{2}$ & $0$ & $\mathbb{Z}_{2} \oplus \mathbb{Z}_{3}$ & $0$ & $\mathbb{Z}_{2}$ & $0$ \\
12 & $0$ & $\mathbb{Z}$ & $\mathbb{Z}_{2} \oplus \mathbb{Z}_{5}$ & $\mathbb{Z}$ & $\mathbb{Z}_{2}$ & $\mathbb{Z}$ & $\mathbb{Z}_{2} \oplus \mathbb{Z}_{3}$ & $0$ & $\mathbb{Z}_{2}$ \\
13 & $0$ & $0$ & $\mathbb{Z}_{2}$ & $\mathbb{Z}_{3} \oplus \mathbb{Z}_{4} \oplus \mathbb{Z}_{5}$ & $\mathbb{Z}_{2}$ & $\mathbb{Z}_{2}$ & $0$ & $\mathbb{Z}_{2} \oplus \mathbb{Z}_{3}$ & $0$ \\
14 & $0$ & $\mathbb{Z}$ & $0$ & $\mathbb{Z}_{2}$ & $\mathbb{Z}_{2} \oplus \mathbb{Z}_{3} \oplus \mathbb{Z}_{5}$ & $0$ & $(\mathbb{Z}_{2})^{2}$ & $0$ & $\mathbb{Z}_{2} \oplus \mathbb{Z}_{3}$ \\
15 & $0$ & $0$ & $(\mathbb{Z}_{2})^{2} \oplus \mathbb{Z}_{5}$ & $(\mathbb{Z}_{2})^{2}$ & $(\mathbb{Z}_{2})^{2} \oplus \mathbb{Z}_{3}$ & $(\mathbb{Z}_{2})^{2} \oplus \mathbb{Z}_{3} \oplus \mathbb{Z}_{5}$ & $0$ & $(\mathbb{Z}_{2})^{2}$ & $0$ \\
16 & $0$ & $\mathbb{Z}$ & $\mathbb{Z}_{2} \oplus \mathbb{Z}_{3} \oplus \mathbb{Z}_{7}$ & $\mathbb{Z}$ & $\mathbb{Z}_{2}$ & $\mathbb{Z}_{2}$ & $\mathbb{Z}_{2} \oplus \mathbb{Z}_{3} \oplus \mathbb{Z}_{5}$ & $\mathbb{Z}$ & $(\mathbb{Z}_{2})^{2}$ \\
17 & $0$ & $0$ & $0$ & $\mathbb{Z}_{2} \oplus (\mathbb{Z}_{3})^{2} \oplus \mathbb{Z}_{4} \oplus \mathbb{Z}_{5} \oplus \mathbb{Z}_{7}$ & $(\mathbb{Z}_{2})^{2}$ & $\mathbb{Z}_{2} \oplus \mathbb{Z}_{3} \oplus \mathbb{Z}_{4}$ & $(\mathbb{Z}_{2})^{2}$ & $\mathbb{Z}_{2} \oplus \mathbb{Z}_{3} \oplus \mathbb{Z}_{5}$ & $0$ \\
18 & $0$ & $\mathbb{Z}$ & $(\mathbb{Z}_{2})^{3}$ & $(\mathbb{Z}_{2})^{2}$ & $(\mathbb{Z}_{2})^{2} \oplus \mathbb{Z}_{7}$ & $\mathbb{Z} \oplus \mathbb{Z}_{2}$ & $\mathbb{Z}_{2}$ & $\mathbb{Z}_{2}$ & $(\mathbb{Z}_{2})^{2} \oplus \mathbb{Z}_{3} \oplus \mathbb{Z}_{5}$ \\
19 & $0$ & $0$ & $\mathbb{Z}_{2} \oplus \mathbb{Z}_{3} \oplus \mathbb{Z}_{7}$ & $(\mathbb{Z}_{2})^{4}$ & $(\mathbb{Z}_{2})^{2} \oplus \mathbb{Z}_{3} \oplus \mathbb{Z}_{5}$ & $(\mathbb{Z}_{2})^{3} \oplus \mathbb{Z}_{7}$ & $(\mathbb{Z}_{2})^{2} \oplus \mathbb{Z}_{3}$ & $(\mathbb{Z}_{2})^{2}$ & $\mathbb{Z}_{2}$ \\
20 & $0$ & $\mathbb{Z}$ & $\mathbb{Z}_{2} \oplus \mathbb{Z}_{3}$ & $\mathbb{Z} \oplus \mathbb{Z}_{2}$ & $(\mathbb{Z}_{2})^{4} \oplus \mathbb{Z}_{3}$ & $\mathbb{Z}_{2}$ & $(\mathbb{Z}_{2})^{2} \oplus \mathbb{Z}_{3} \oplus \mathbb{Z}_{7}$ & $\mathbb{Z}_{2}$ & $\mathbb{Z}_{2}$ \\
21 & $0$ & $0$ & $(\mathbb{Z}_{2})^{3}$ & $(\mathbb{Z}_{2})^{2} \oplus (\mathbb{Z}_{3})^{3} \oplus (\mathbb{Z}_{4})^{2} \oplus \mathbb{Z}_{5} \oplus \mathbb{Z}_{7}$ & $(\mathbb{Z}_{2})^{3}$ & $(\mathbb{Z}_{2})^{4} \oplus (\mathbb{Z}_{3})^{2} \oplus \mathbb{Z}_{4} \oplus \mathbb{Z}_{5}$ & $(\mathbb{Z}_{2})^{3}$ & $(\mathbb{Z}_{2})^{2} \oplus (\mathbb{Z}_{3})^{2} \oplus \mathbb{Z}_{4} \oplus \mathbb{Z}_{7}$ & $(\mathbb{Z}_{2})^{2}$ \\
22 & $0$ & $\mathbb{Z}$ & $\mathbb{Z}_{2}$ & $(\mathbb{Z}_{2})^{3}$ & $(\mathbb{Z}_{2})^{5} \oplus \mathbb{Z}_{3} \oplus \mathbb{Z}_{5}$ & $(\mathbb{Z}_{2})^{3}$ & $(\mathbb{Z}_{2})^{3}$ & $\mathbb{Z}_{2}$ & $(\mathbb{Z}_{2})^{2} \oplus \mathbb{Z}_{3} \oplus \mathbb{Z}_{7}$ \\
23 & $0$ & $0$ & $(\mathbb{Z}_{2})^{2} \oplus \mathbb{Z}_{3}$ & $(\mathbb{Z}_{2})^{5}$ & $(\mathbb{Z}_{2})^{5} \oplus \mathbb{Z}_{3} \oplus \mathbb{Z}_{7}$ & $(\mathbb{Z}_{2})^{4} \oplus (\mathbb{Z}_{3})^{2} \oplus \mathbb{Z}_{4} \oplus \mathbb{Z}_{5}$ & $(\mathbb{Z}_{2})^{4} \oplus \mathbb{Z}_{3} \oplus \mathbb{Z}_{5}$ & $(\mathbb{Z}_{2})^{5}$ & $(\mathbb{Z}_{2})^{2} \oplus \mathbb{Z}_{3}$ \\
24 & $0$ & $\mathbb{Z}$ & $(\mathbb{Z}_{2})^{3} \oplus \mathbb{Z}_{3} \oplus \mathbb{Z}_{5} \oplus \mathbb{Z}_{11}$ & $\mathbb{Z} \oplus (\mathbb{Z}_{2})^{3}$ & $(\mathbb{Z}_{2})^{5} \oplus \mathbb{Z}_{3}$ & $\mathbb{Z} \oplus (\mathbb{Z}_{2})^{5}$ & $(\mathbb{Z}_{2})^{5} \oplus (\mathbb{Z}_{3})^{2} \oplus \mathbb{Z}_{5}$ & $\mathbb{Z} \oplus (\mathbb{Z}_{2})^{2}$ & $(\mathbb{Z}_{2})^{3}$ \\
\end{tabular}

%% file: tables/stable-homology.tex
\begin{tabular}{l|l}
$H^s_{0}(G; \mathbb{Z}) = H_{1 + \bullet}(B^{1 + \bullet}G; \mathbb{Z})$ & $G$ \\
\hline$H^s_{1}(G; \mathbb{Z}) = H_{3 + \bullet}(B^{2 + \bullet}G; \mathbb{Z})$ & $0$ \\
\hline$H^s_{2}(G; \mathbb{Z}) = H_{5 + \bullet}(B^{3 + \bullet}G; \mathbb{Z})$ & $G/2G$ \\
\hline$H^s_{3}(G; \mathbb{Z}) = H_{7 + \bullet}(B^{4 + \bullet}G; \mathbb{Z})$ & ${_{2}}G$ \\
\hline$H^s_{4}(G; \mathbb{Z}) = H_{9 + \bullet}(B^{5 + \bullet}G; \mathbb{Z})$ & $G/2G \oplus G/3G$ \\
\hline$H^s_{5}(G; \mathbb{Z}) = H_{11 + \bullet}(B^{6 + \bullet}G; \mathbb{Z})$ & ${_{2}}G \oplus {_{3}}G$ \\
\hline$H^s_{6}(G; \mathbb{Z}) = H_{13 + \bullet}(B^{7 + \bullet}G; \mathbb{Z})$ & $(G/2G)^{2}$ \\
\hline$H^s_{7}(G; \mathbb{Z}) = H_{15 + \bullet}(B^{8 + \bullet}G; \mathbb{Z})$ & $({_{2}}G)^{2}$ \\
\hline$H^s_{8}(G; \mathbb{Z}) = H_{17 + \bullet}(B^{9 + \bullet}G; \mathbb{Z})$ & $(G/2G)^{2} \oplus G/3G \oplus G/5G$ \\
\hline$H^s_{9}(G; \mathbb{Z}) = H_{19 + \bullet}(B^{10 + \bullet}G; \mathbb{Z})$ & $G/2G \oplus ({_{2}}G)^{2} \oplus {_{3}}G \oplus {_{5}}G$ \\
\hline$H^s_{10}(G; \mathbb{Z}) = H_{21 + \bullet}(B^{11 + \bullet}G; \mathbb{Z})$ & $(G/2G)^{2} \oplus {_{2}}G$ \\
\hline$H^s_{11}(G; \mathbb{Z}) = H_{23 + \bullet}(B^{12 + \bullet}G; \mathbb{Z})$ & $G/2G \oplus ({_{2}}G)^{2}$ \\
\hline$H^s_{12}(G; \mathbb{Z}) = H_{25 + \bullet}(B^{13 + \bullet}G; \mathbb{Z})$ & $(G/2G)^{3} \oplus {_{2}}G \oplus G/3G \oplus G/7G$ \\
\hline$H^s_{13}(G; \mathbb{Z}) = H_{27 + \bullet}(B^{14 + \bullet}G; \mathbb{Z})$ & $G/2G \oplus ({_{2}}G)^{3} \oplus {_{3}}G \oplus {_{7}}G$ \\
\hline$H^s_{14}(G; \mathbb{Z}) = H_{29 + \bullet}(B^{15 + \bullet}G; \mathbb{Z})$ & $(G/2G)^{4} \oplus {_{2}}G$ \\
\hline$H^s_{15}(G; \mathbb{Z}) = H_{31 + \bullet}(B^{16 + \bullet}G; \mathbb{Z})$ & $(G/2G)^{2} \oplus ({_{2}}G)^{4}$ \\
\hline$H^s_{16}(G; \mathbb{Z}) = H_{33 + \bullet}(B^{17 + \bullet}G; \mathbb{Z})$ & $(G/2G)^{4} \oplus ({_{2}}G)^{2} \oplus (G/3G)^{2} \oplus G/5G$ \\
\hline$H^s_{17}(G; \mathbb{Z}) = H_{35 + \bullet}(B^{18 + \bullet}G; \mathbb{Z})$ & $(G/2G)^{3} \oplus ({_{2}}G)^{4} \oplus ({_{3}}G)^{2} \oplus {_{5}}G$ \\
\hline$H^s_{18}(G; \mathbb{Z}) = H_{37 + \bullet}(B^{19 + \bullet}G; \mathbb{Z})$ & $(G/2G)^{5} \oplus ({_{2}}G)^{3}$ \\
\hline$H^s_{19}(G; \mathbb{Z}) = H_{39 + \bullet}(B^{20 + \bullet}G; \mathbb{Z})$ & $(G/2G)^{3} \oplus ({_{2}}G)^{5}$ \\
\hline$H^s_{20}(G; \mathbb{Z}) = H_{41 + \bullet}(B^{21 + \bullet}G; \mathbb{Z})$ & $(G/2G)^{6} \oplus ({_{2}}G)^{3} \oplus (G/3G)^{2} \oplus G/11G$ \\
\hline$H^s_{21}(G; \mathbb{Z}) = H_{43 + \bullet}(B^{22 + \bullet}G; \mathbb{Z})$ & $(G/2G)^{5} \oplus ({_{2}}G)^{6} \oplus G/3G \oplus ({_{3}}G)^{2} \oplus {_{11}}G$ \\
\hline$H^s_{22}(G; \mathbb{Z}) = H_{45 + \bullet}(B^{23 + \bullet}G; \mathbb{Z})$ & $(G/2G)^{6} \oplus ({_{2}}G)^{5} \oplus {_{3}}G$ \\
\hline$H^s_{23}(G; \mathbb{Z}) = H_{47 + \bullet}(B^{24 + \bullet}G; \mathbb{Z})$ & $(G/2G)^{6} \oplus ({_{2}}G)^{6}$ \\
\hline$H^s_{24}(G; \mathbb{Z}) = H_{49 + \bullet}(B^{25 + \bullet}G; \mathbb{Z})$ & $(G/2G)^{8} \oplus ({_{2}}G)^{6} \oplus (G/3G)^{2} \oplus G/5G \oplus G/7G \oplus G/13G$ \\
\hline$H^s_{25}(G; \mathbb{Z}) = H_{51 + \bullet}(B^{26 + \bullet}G; \mathbb{Z})$ & $(G/2G)^{6} \oplus ({_{2}}G)^{8} \oplus G/3G \oplus ({_{3}}G)^{2} \oplus {_{5}}G \oplus {_{7}}G \oplus {_{13}}G$ \\
\hline$H^s_{26}(G; \mathbb{Z}) = H_{53 + \bullet}(B^{27 + \bullet}G; \mathbb{Z})$ & $(G/2G)^{9} \oplus ({_{2}}G)^{6} \oplus {_{3}}G$ \\
\hline$H^s_{27}(G; \mathbb{Z}) = H_{55 + \bullet}(B^{28 + \bullet}G; \mathbb{Z})$ & $(G/2G)^{8} \oplus ({_{2}}G)^{9}$ \\
\hline$H^s_{28}(G; \mathbb{Z}) = H_{57 + \bullet}(B^{29 + \bullet}G; \mathbb{Z})$ & $(G/2G)^{10} \oplus ({_{2}}G)^{8} \oplus (G/3G)^{2}$ \\
\hline$H^s_{29}(G; \mathbb{Z}) = H_{59 + \bullet}(B^{30 + \bullet}G; \mathbb{Z})$ & $(G/2G)^{9} \oplus ({_{2}}G)^{10} \oplus G/3G \oplus ({_{3}}G)^{2}$ \\
\hline$H^s_{30}(G; \mathbb{Z}) = H_{61 + \bullet}(B^{31 + \bullet}G; \mathbb{Z})$ & $(G/2G)^{13} \oplus ({_{2}}G)^{9} \oplus {_{3}}G$ \\
\hline$H^s_{31}(G; \mathbb{Z}) = H_{63 + \bullet}(B^{32 + \bullet}G; \mathbb{Z})$ & $(G/2G)^{10} \oplus ({_{2}}G)^{13}$ \\
\hline$H^s_{32}(G; \mathbb{Z}) = H_{65 + \bullet}(B^{33 + \bullet}G; \mathbb{Z})$ & $(G/2G)^{14} \oplus ({_{2}}G)^{10} \oplus (G/3G)^{3} \oplus G/5G \oplus G/17G$ \\
\hline$H^s_{33}(G; \mathbb{Z}) = H_{67 + \bullet}(B^{34 + \bullet}G; \mathbb{Z})$ & $(G/2G)^{13} \oplus ({_{2}}G)^{14} \oplus G/3G \oplus ({_{3}}G)^{3} \oplus {_{5}}G \oplus {_{17}}G$ \\
\hline$H^s_{34}(G; \mathbb{Z}) = H_{69 + \bullet}(B^{35 + \bullet}G; \mathbb{Z})$ & $(G/2G)^{15} \oplus ({_{2}}G)^{13} \oplus {_{3}}G$ \\
\hline$H^s_{35}(G; \mathbb{Z}) = H_{71 + \bullet}(B^{36 + \bullet}G; \mathbb{Z})$ & $(G/2G)^{15} \oplus ({_{2}}G)^{15}$ \\
\hline$H^s_{36}(G; \mathbb{Z}) = H_{73 + \bullet}(B^{37 + \bullet}G; \mathbb{Z})$ & $(G/2G)^{18} \oplus ({_{2}}G)^{15} \oplus (G/3G)^{3} \oplus G/7G \oplus G/19G$ \\
\hline$H^s_{37}(G; \mathbb{Z}) = H_{75 + \bullet}(B^{38 + \bullet}G; \mathbb{Z})$ & $(G/2G)^{17} \oplus ({_{2}}G)^{18} \oplus (G/3G)^{2} \oplus ({_{3}}G)^{3} \oplus {_{7}}G \oplus {_{19}}G$ \\
\hline$H^s_{38}(G; \mathbb{Z}) = H_{77 + \bullet}(B^{39 + \bullet}G; \mathbb{Z})$ & $(G/2G)^{20} \oplus ({_{2}}G)^{17} \oplus ({_{3}}G)^{2}$ \\
\hline$H^s_{39}(G; \mathbb{Z}) = H_{79 + \bullet}(B^{40 + \bullet}G; \mathbb{Z})$ & $(G/2G)^{20} \oplus ({_{2}}G)^{20}$ \\
\hline$H^s_{40}(G; \mathbb{Z}) = H_{81 + \bullet}(B^{41 + \bullet}G; \mathbb{Z})$ & $(G/2G)^{22} \oplus ({_{2}}G)^{20} \oplus (G/3G)^{3} \oplus G/5G \oplus G/11G$ \\
\hline$H^s_{41}(G; \mathbb{Z}) = H_{83 + \bullet}(B^{42 + \bullet}G; \mathbb{Z})$ & $(G/2G)^{22} \oplus ({_{2}}G)^{22} \oplus (G/3G)^{2} \oplus ({_{3}}G)^{3} \oplus {_{5}}G \oplus {_{11}}G$ \\
\hline$H^s_{42}(G; \mathbb{Z}) = H_{85 + \bullet}(B^{43 + \bullet}G; \mathbb{Z})$ & $(G/2G)^{26} \oplus ({_{2}}G)^{22} \oplus ({_{3}}G)^{2}$ \\
\hline$H^s_{43}(G; \mathbb{Z}) = H_{87 + \bullet}(B^{44 + \bullet}G; \mathbb{Z})$ & $(G/2G)^{24} \oplus ({_{2}}G)^{26}$ \\
\hline$H^s_{44}(G; \mathbb{Z}) = H_{89 + \bullet}(B^{45 + \bullet}G; \mathbb{Z})$ & $(G/2G)^{29} \oplus ({_{2}}G)^{24} \oplus (G/3G)^{3} \oplus G/23G$ \\
\hline$H^s_{45}(G; \mathbb{Z}) = H_{91 + \bullet}(B^{46 + \bullet}G; \mathbb{Z})$ & $(G/2G)^{29} \oplus ({_{2}}G)^{29} \oplus (G/3G)^{2} \oplus ({_{3}}G)^{3} \oplus {_{23}}G$ \\
\hline$H^s_{46}(G; \mathbb{Z}) = H_{93 + \bullet}(B^{47 + \bullet}G; \mathbb{Z})$ & $(G/2G)^{31} \oplus ({_{2}}G)^{29} \oplus ({_{3}}G)^{2}$ \\
\hline$H^s_{47}(G; \mathbb{Z}) = H_{95 + \bullet}(B^{48 + \bullet}G; \mathbb{Z})$ & $(G/2G)^{32} \oplus ({_{2}}G)^{31}$ \\
\hline$H^s_{48}(G; \mathbb{Z}) = H_{97 + \bullet}(B^{49 + \bullet}G; \mathbb{Z})$ & $(G/2G)^{36} \oplus ({_{2}}G)^{32} \oplus (G/3G)^{4} \oplus (G/5G)^{2} \oplus G/7G \oplus G/13G$ \\
\hline$H^s_{49}(G; \mathbb{Z}) = H_{99 + \bullet}(B^{50 + \bullet}G; \mathbb{Z})$ & $(G/2G)^{35} \oplus ({_{2}}G)^{36} \oplus (G/3G)^{2} \oplus ({_{3}}G)^{4} \oplus ({_{5}}G)^{2} \oplus {_{7}}G \oplus {_{13}}G$ \\
\hline$H^s_{50}(G; \mathbb{Z}) = H_{101 + \bullet}(B^{51 + \bullet}G; \mathbb{Z})$ & $(G/2G)^{39} \oplus ({_{2}}G)^{35} \oplus ({_{3}}G)^{2}$ \\

\end{tabular}

%% file: tables/omega-B2Z2-RZ.tex
\begin{IEEEeqnarraybox}[][c]{rClrCl}
\mathbb{Z}_{2} &\cong& H^{2}(B^{2}\mathbb{Z}_{2}; \mathbb{R}/\mathbb{Z}) \colon & \frac{1}{2} (\sigma^{2} u_{2})^*\langle 2 \rangle &\longmapsto& \frac{1}{2} (\sigma u_{2})^*\langle 2 \rangle \\
0 &\cong& H^{3}(B^{2}\mathbb{Z}_{2}; \mathbb{R}/\mathbb{Z}) \colon &  \\
\mathbb{Z}_{4} &\cong& H^{4}(B^{2}\mathbb{Z}_{2}; \mathbb{R}/\mathbb{Z}) \colon & \frac{1}{4} (\gamma_{2} \sigma^{2} u_{2})^*\langle 4 \rangle &\longmapsto& -\frac{1}{2} (\sigma u_{2} \cdot \psi_{2} u_{2})^*\langle 2 \rangle \\
\mathbb{Z}_{2} &\cong& H^{5}(B^{2}\mathbb{Z}_{2}; \mathbb{R}/\mathbb{Z}) \colon & \frac{1}{2} (\varepsilon_{2} \psi_{2} u_{2})^*\langle 2 \rangle &\longmapsto& 0 \\
\mathbb{Z}_{2} &\cong& H^{6}(B^{2}\mathbb{Z}_{2}; \mathbb{R}/\mathbb{Z}) \colon & \frac{1}{2} (\gamma_{3} \sigma^{2} u_{2})^*\langle 2 \rangle &\longmapsto& 0 \\
\mathbb{Z}_{2} &\cong& H^{7}(B^{2}\mathbb{Z}_{2}; \mathbb{R}/\mathbb{Z}) \colon & \frac{1}{2} (\sigma^{2} u_{2} \otimes \varepsilon_{2} \psi_{2} u_{2})^*\langle 2 \rangle &\longmapsto& 0 \\
\mathbb{Z}_{2} \oplus \mathbb{Z}_{8} &\cong& H^{8}(B^{2}\mathbb{Z}_{2}; \mathbb{R}/\mathbb{Z}) \colon & \frac{1}{2} (\tau(\sigma^{2} u_{2},\sigma \psi_{2} u_{2},\varepsilon_{2} \psi_{2} u_{2},\varphi_{2} \psi_{2} u_{2}))^*\langle 2 \rangle &\longmapsto& 0 \\
&&& \frac{1}{8} (\gamma_{4} \sigma^{2} u_{2})^*\langle 8 \rangle &\longmapsto& 0 \\
(\mathbb{Z}_{2})^{2} &\cong& H^{9}(B^{2}\mathbb{Z}_{2}; \mathbb{R}/\mathbb{Z}) \colon & \frac{1}{2} (\varepsilon_{4} \psi_{2} u_{2})^*\langle 2 \rangle &\longmapsto& 0 \\
&&& \frac{1}{2} (\gamma_{2} \sigma^{2} u_{2} \otimes \varepsilon_{2} \psi_{2} u_{2})^*\langle 2 \rangle &\longmapsto& 0 \\
(\mathbb{Z}_{2})^{2} &\cong& H^{10}(B^{2}\mathbb{Z}_{2}; \mathbb{R}/\mathbb{Z}) \colon & \frac{1}{2} (\tau(\gamma_{2} \sigma^{2} u_{2},\sigma^{2} u_{2} \cdot \sigma \psi_{2} u_{2},\varepsilon_{2} \psi_{2} u_{2},\varphi_{2} \psi_{2} u_{2}))^*\langle 2 \rangle &\longmapsto& 0 \\
&&& \frac{1}{2} (\gamma_{5} \sigma^{2} u_{2})^*\langle 2 \rangle &\longmapsto& 0 \\
(\mathbb{Z}_{2})^{3} &\cong& H^{11}(B^{2}\mathbb{Z}_{2}; \mathbb{R}/\mathbb{Z}) \colon & \frac{1}{2} (\varepsilon_{2} \psi_{2} u_{2} \cdot \varphi_{2} \psi_{2} u_{2})^*\langle 2 \rangle &\longmapsto& 0 \\
&&& \frac{1}{2} (\sigma^{2} u_{2} \otimes \varepsilon_{4} \psi_{2} u_{2})^*\langle 2 \rangle &\longmapsto& 0 \\
&&& \frac{1}{2} (\gamma_{3} \sigma^{2} u_{2} \otimes \varepsilon_{2} \psi_{2} u_{2})^*\langle 2 \rangle &\longmapsto& 0 \\
(\mathbb{Z}_{2})^{2} \oplus \mathbb{Z}_{4} &\cong& H^{12}(B^{2}\mathbb{Z}_{2}; \mathbb{R}/\mathbb{Z}) \colon & \frac{1}{2} (\tau(\sigma^{2} u_{2},\sigma \psi_{2} u_{2},\varepsilon_{4} \psi_{2} u_{2},\varphi_{2} \gamma_{2} \psi_{2} u_{2}))^*\langle 2 \rangle &\longmapsto& 0 \\
&&& \frac{1}{2} (\tau(\gamma_{3} \sigma^{2} u_{2},\gamma_{2} \sigma^{2} u_{2} \cdot \sigma \psi_{2} u_{2},\varepsilon_{2} \psi_{2} u_{2},\varphi_{2} \psi_{2} u_{2}))^*\langle 2 \rangle &\longmapsto& 0 \\
&&& \frac{1}{4} (\gamma_{6} \sigma^{2} u_{2})^*\langle 4 \rangle &\longmapsto& 0 \\
(\mathbb{Z}_{2})^{3} &\cong& H^{13}(B^{2}\mathbb{Z}_{2}; \mathbb{R}/\mathbb{Z}) \colon & \frac{1}{2} (\sigma^{2} u_{2} \otimes \varepsilon_{2} \psi_{2} u_{2} \cdot \varphi_{2} \psi_{2} u_{2})^*\langle 2 \rangle &\longmapsto& 0 \\
&&& \frac{1}{2} (\gamma_{2} \sigma^{2} u_{2} \otimes \varepsilon_{4} \psi_{2} u_{2})^*\langle 2 \rangle &\longmapsto& 0 \\
&&& \frac{1}{2} (\gamma_{4} \sigma^{2} u_{2} \otimes \varepsilon_{2} \psi_{2} u_{2})^*\langle 2 \rangle &\longmapsto& 0 \\
(\mathbb{Z}_{2})^{5} &\cong& H^{14}(B^{2}\mathbb{Z}_{2}; \mathbb{R}/\mathbb{Z}) \colon & \frac{1}{2} (\varepsilon_{2} \psi_{2} u_{2} \otimes \varepsilon_{4} \psi_{2} u_{2})^*\langle 2 \rangle &\longmapsto& 0 \\
&&& \frac{1}{2} (\tau(\sigma^{2} u_{2},\sigma \psi_{2} u_{2},\varepsilon_{2} \psi_{2} u_{2} \cdot \varphi_{2} \psi_{2} u_{2},\gamma_{2} \varphi_{2} \psi_{2} u_{2}))^*\langle 2 \rangle &\longmapsto& 0 \\
&&& \frac{1}{2} (\tau(\gamma_{2} \sigma^{2} u_{2},\sigma^{2} u_{2} \cdot \sigma \psi_{2} u_{2},\varepsilon_{4} \psi_{2} u_{2},\varphi_{2} \gamma_{2} \psi_{2} u_{2}))^*\langle 2 \rangle &\longmapsto& 0 \\
&&& \frac{1}{2} (\tau(\gamma_{4} \sigma^{2} u_{2},\gamma_{3} \sigma^{2} u_{2} \cdot \sigma \psi_{2} u_{2},\varepsilon_{2} \psi_{2} u_{2},\varphi_{2} \psi_{2} u_{2}))^*\langle 2 \rangle &\longmapsto& 0 \\
&&& \frac{1}{2} (\gamma_{7} \sigma^{2} u_{2})^*\langle 2 \rangle &\longmapsto& 0
\end{IEEEeqnarraybox}

%% file: tables/omega-B2Z2-Z.tex
\begin{IEEEeqnarraybox}[][c]{rClrCl}
\mathbb{Z}_{2} &\cong& H^{3}(B^{2}\mathbb{Z}_{2}; \mathbb{Z}) \colon & (\sigma \psi_{2} u_{2})^*\langle 2 \rangle &\longmapsto& (\psi_{2} u_{2})^*\langle 2 \rangle \\
0 &\cong& H^{4}(B^{2}\mathbb{Z}_{2}; \mathbb{Z}) \colon &  \\
\mathbb{Z}_{4} &\cong& H^{5}(B^{2}\mathbb{Z}_{2}; \mathbb{Z}) \colon & (\sigma^{2} u_{2} \cdot \sigma \psi_{2} u_{2})^*\langle 4 \rangle &\longmapsto& -(\gamma_{2} \psi_{2} u_{2})^*\langle 2 \rangle \\
\mathbb{Z}_{2} &\cong& H^{6}(B^{2}\mathbb{Z}_{2}; \mathbb{Z}) \colon & (\varphi_{2} \psi_{2} u_{2})^*\langle 2 \rangle &\longmapsto& 0 \\
\mathbb{Z}_{2} &\cong& H^{7}(B^{2}\mathbb{Z}_{2}; \mathbb{Z}) \colon & (\gamma_{2} \sigma^{2} u_{2} \cdot \sigma \psi_{2} u_{2})^*\langle 2 \rangle &\longmapsto& 0 \\
\mathbb{Z}_{2} &\cong& H^{8}(B^{2}\mathbb{Z}_{2}; \mathbb{Z}) \colon & (\kappa(\sigma^{2} u_{2},\sigma \psi_{2} u_{2},\varepsilon_{2} \psi_{2} u_{2},\varphi_{2} \psi_{2} u_{2}))^*\langle 2 \rangle &\longmapsto& 0 \\
\mathbb{Z}_{2} \oplus \mathbb{Z}_{8} &\cong& H^{9}(B^{2}\mathbb{Z}_{2}; \mathbb{Z}) \colon & (\sigma \psi_{2} u_{2} \otimes \varphi_{2} \psi_{2} u_{2})^*\langle 2 \rangle &\longmapsto& 0 \\
&&& (\gamma_{3} \sigma^{2} u_{2} \cdot \sigma \psi_{2} u_{2})^*\langle 8 \rangle &\longmapsto& 0 \\
(\mathbb{Z}_{2})^{2} &\cong& H^{10}(B^{2}\mathbb{Z}_{2}; \mathbb{Z}) \colon & (\varphi_{2} \gamma_{2} \psi_{2} u_{2})^*\langle 2 \rangle &\longmapsto& 0 \\
&&& (\kappa(\gamma_{2} \sigma^{2} u_{2},\sigma^{2} u_{2} \cdot \sigma \psi_{2} u_{2},\varepsilon_{2} \psi_{2} u_{2},\varphi_{2} \psi_{2} u_{2}))^*\langle 2 \rangle &\longmapsto& 0 \\
(\mathbb{Z}_{2})^{2} &\cong& H^{11}(B^{2}\mathbb{Z}_{2}; \mathbb{Z}) \colon & (\sigma^{2} u_{2} \cdot \sigma \psi_{2} u_{2} \otimes \varphi_{2} \psi_{2} u_{2})^*\langle 2 \rangle &\longmapsto& 0 \\
&&& (\gamma_{4} \sigma^{2} u_{2} \cdot \sigma \psi_{2} u_{2})^*\langle 2 \rangle &\longmapsto& 0 \\
(\mathbb{Z}_{2})^{3} &\cong& H^{12}(B^{2}\mathbb{Z}_{2}; \mathbb{Z}) \colon & (\gamma_{2} \varphi_{2} \psi_{2} u_{2})^*\langle 2 \rangle &\longmapsto& 0 \\
&&& (\kappa(\sigma^{2} u_{2},\sigma \psi_{2} u_{2},\varepsilon_{4} \psi_{2} u_{2},\varphi_{2} \gamma_{2} \psi_{2} u_{2}))^*\langle 2 \rangle &\longmapsto& 0 \\
&&& (\kappa(\gamma_{3} \sigma^{2} u_{2},\gamma_{2} \sigma^{2} u_{2} \cdot \sigma \psi_{2} u_{2},\varepsilon_{2} \psi_{2} u_{2},\varphi_{2} \psi_{2} u_{2}))^*\langle 2 \rangle &\longmapsto& 0 \\
(\mathbb{Z}_{2})^{2} \oplus \mathbb{Z}_{4} &\cong& H^{13}(B^{2}\mathbb{Z}_{2}; \mathbb{Z}) \colon & (\sigma \psi_{2} u_{2} \otimes \varphi_{2} \gamma_{2} \psi_{2} u_{2})^*\langle 2 \rangle &\longmapsto& 0 \\
&&& (\gamma_{2} \sigma^{2} u_{2} \cdot \sigma \psi_{2} u_{2} \otimes \varphi_{2} \psi_{2} u_{2})^*\langle 2 \rangle &\longmapsto& 0 \\
&&& (\gamma_{5} \sigma^{2} u_{2} \cdot \sigma \psi_{2} u_{2})^*\langle 4 \rangle &\longmapsto& 0 \\
(\mathbb{Z}_{2})^{3} &\cong& H^{14}(B^{2}\mathbb{Z}_{2}; \mathbb{Z}) \colon & (\kappa(\sigma^{2} u_{2},\sigma \psi_{2} u_{2},\varepsilon_{2} \psi_{2} u_{2} \cdot \varphi_{2} \psi_{2} u_{2},\gamma_{2} \varphi_{2} \psi_{2} u_{2}))^*\langle 2 \rangle &\longmapsto& 0 \\
&&& (\kappa(\gamma_{2} \sigma^{2} u_{2},\sigma^{2} u_{2} \cdot \sigma \psi_{2} u_{2},\varepsilon_{4} \psi_{2} u_{2},\varphi_{2} \gamma_{2} \psi_{2} u_{2}))^*\langle 2 \rangle &\longmapsto& 0 \\
&&& (\kappa(\gamma_{4} \sigma^{2} u_{2},\gamma_{3} \sigma^{2} u_{2} \cdot \sigma \psi_{2} u_{2},\varepsilon_{2} \psi_{2} u_{2},\varphi_{2} \psi_{2} u_{2}))^*\langle 2 \rangle &\longmapsto& 0 \\
(\mathbb{Z}_{2})^{5} &\cong& H^{15}(B^{2}\mathbb{Z}_{2}; \mathbb{Z}) \colon & (\kappa(\varepsilon_{2} \psi_{2} u_{2},\varphi_{2} \psi_{2} u_{2},\varepsilon_{4} \psi_{2} u_{2},\varphi_{2} \gamma_{2} \psi_{2} u_{2}))^*\langle 2 \rangle &\longmapsto& 0 \\
&&& (\sigma \psi_{2} u_{2} \otimes \gamma_{2} \varphi_{2} \psi_{2} u_{2})^*\langle 2 \rangle &\longmapsto& 0 \\
&&& (\sigma^{2} u_{2} \cdot \sigma \psi_{2} u_{2} \otimes \varphi_{2} \gamma_{2} \psi_{2} u_{2})^*\langle 2 \rangle &\longmapsto& 0 \\
&&& (\gamma_{3} \sigma^{2} u_{2} \cdot \sigma \psi_{2} u_{2} \otimes \varphi_{2} \psi_{2} u_{2})^*\langle 2 \rangle &\longmapsto& 0 \\
&&& (\gamma_{6} \sigma^{2} u_{2} \cdot \sigma \psi_{2} u_{2})^*\langle 2 \rangle &\longmapsto& 0
\end{IEEEeqnarraybox}

%% file: tables/omega-B2Z2-Z2.tex
\begin{IEEEeqnarraybox}[][c]{rClrCl}
\mathbb{Z}_{2} &\cong& H^{2}(B^{2}\mathbb{Z}_{2}; \mathbb{Z}_{2}) \colon & (\sigma^{2} u_{2})^*\langle 2 \rangle &\longmapsto& (\sigma u_{2})^*\langle 2 \rangle \\
\mathbb{Z}_{2} &\cong& H^{3}(B^{2}\mathbb{Z}_{2}; \mathbb{Z}_{2}) \colon & (\sigma \psi_{2} u_{2})^*\langle 2 \rangle &\longmapsto& (\psi_{2} u_{2})^*\langle 2 \rangle \\
\mathbb{Z}_{2} &\cong& H^{4}(B^{2}\mathbb{Z}_{2}; \mathbb{Z}_{2}) \colon & (\gamma_{2} \sigma^{2} u_{2})^*\langle 2 \rangle &\longmapsto& 0 \\
(\mathbb{Z}_{2})^{2} &\cong& H^{5}(B^{2}\mathbb{Z}_{2}; \mathbb{Z}_{2}) \colon & (\sigma \gamma_{2} \psi_{2} u_{2})^*\langle 2 \rangle &\longmapsto& (\gamma_{2} \psi_{2} u_{2})^*\langle 2 \rangle \\
&&& (\sigma^{2} u_{2} \otimes \sigma \psi_{2} u_{2})^*\langle 2 \rangle &\longmapsto& 0 \\
(\mathbb{Z}_{2})^{2} &\cong& H^{6}(B^{2}\mathbb{Z}_{2}; \mathbb{Z}_{2}) \colon & (\varphi_{2} \psi_{2} u_{2})^*\langle 2 \rangle &\longmapsto& 0 \\
&&& (\gamma_{3} \sigma^{2} u_{2})^*\langle 2 \rangle &\longmapsto& 0 \\
(\mathbb{Z}_{2})^{2} &\cong& H^{7}(B^{2}\mathbb{Z}_{2}; \mathbb{Z}_{2}) \colon & (\sigma^{2} u_{2} \otimes \sigma \gamma_{2} \psi_{2} u_{2})^*\langle 2 \rangle &\longmapsto& 0 \\
&&& (\gamma_{2} \sigma^{2} u_{2} \otimes \sigma \psi_{2} u_{2})^*\langle 2 \rangle &\longmapsto& 0 \\
(\mathbb{Z}_{2})^{3} &\cong& H^{8}(B^{2}\mathbb{Z}_{2}; \mathbb{Z}_{2}) \colon & (\sigma \psi_{2} u_{2} \otimes \sigma \gamma_{2} \psi_{2} u_{2})^*\langle 2 \rangle &\longmapsto& 0 \\
&&& (\sigma^{2} u_{2} \otimes \varphi_{2} \psi_{2} u_{2})^*\langle 2 \rangle &\longmapsto& 0 \\
&&& (\gamma_{4} \sigma^{2} u_{2})^*\langle 2 \rangle &\longmapsto& 0 \\
(\mathbb{Z}_{2})^{4} &\cong& H^{9}(B^{2}\mathbb{Z}_{2}; \mathbb{Z}_{2}) \colon & (\sigma \gamma_{4} \psi_{2} u_{2})^*\langle 2 \rangle &\longmapsto& (\gamma_{4} \psi_{2} u_{2})^*\langle 2 \rangle \\
&&& (\sigma \psi_{2} u_{2} \otimes \varphi_{2} \psi_{2} u_{2})^*\langle 2 \rangle &\longmapsto& 0 \\
&&& (\gamma_{2} \sigma^{2} u_{2} \otimes \sigma \gamma_{2} \psi_{2} u_{2})^*\langle 2 \rangle &\longmapsto& 0 \\
&&& (\gamma_{3} \sigma^{2} u_{2} \otimes \sigma \psi_{2} u_{2})^*\langle 2 \rangle &\longmapsto& 0 \\
(\mathbb{Z}_{2})^{4} &\cong& H^{10}(B^{2}\mathbb{Z}_{2}; \mathbb{Z}_{2}) \colon & (\varphi_{2} \gamma_{2} \psi_{2} u_{2})^*\langle 2 \rangle &\longmapsto& 0 \\
&&& (\sigma^{2} u_{2} \otimes \sigma \psi_{2} u_{2} \otimes \sigma \gamma_{2} \psi_{2} u_{2})^*\langle 2 \rangle &\longmapsto& 0 \\
&&& (\gamma_{2} \sigma^{2} u_{2} \otimes \varphi_{2} \psi_{2} u_{2})^*\langle 2 \rangle &\longmapsto& 0 \\
&&& (\gamma_{5} \sigma^{2} u_{2})^*\langle 2 \rangle &\longmapsto& 0 \\
(\mathbb{Z}_{2})^{5} &\cong& H^{11}(B^{2}\mathbb{Z}_{2}; \mathbb{Z}_{2}) \colon & (\sigma \gamma_{2} \psi_{2} u_{2} \otimes \varphi_{2} \psi_{2} u_{2})^*\langle 2 \rangle &\longmapsto& 0 \\
&&& (\sigma^{2} u_{2} \otimes \sigma \gamma_{4} \psi_{2} u_{2})^*\langle 2 \rangle &\longmapsto& 0 \\
&&& (\sigma^{2} u_{2} \otimes \sigma \psi_{2} u_{2} \otimes \varphi_{2} \psi_{2} u_{2})^*\langle 2 \rangle &\longmapsto& 0 \\
&&& (\gamma_{3} \sigma^{2} u_{2} \otimes \sigma \gamma_{2} \psi_{2} u_{2})^*\langle 2 \rangle &\longmapsto& 0 \\
&&& (\gamma_{4} \sigma^{2} u_{2} \otimes \sigma \psi_{2} u_{2})^*\langle 2 \rangle &\longmapsto& 0 \\
(\mathbb{Z}_{2})^{6} &\cong& H^{12}(B^{2}\mathbb{Z}_{2}; \mathbb{Z}_{2}) \colon & (\gamma_{2} \varphi_{2} \psi_{2} u_{2})^*\langle 2 \rangle &\longmapsto& 0 \\
&&& (\sigma \psi_{2} u_{2} \otimes \sigma \gamma_{4} \psi_{2} u_{2})^*\langle 2 \rangle &\longmapsto& 0 \\
&&& (\sigma^{2} u_{2} \otimes \varphi_{2} \gamma_{2} \psi_{2} u_{2})^*\langle 2 \rangle &\longmapsto& 0 \\
&&& (\gamma_{2} \sigma^{2} u_{2} \otimes \sigma \psi_{2} u_{2} \otimes \sigma \gamma_{2} \psi_{2} u_{2})^*\langle 2 \rangle &\longmapsto& 0 \\
&&& (\gamma_{3} \sigma^{2} u_{2} \otimes \varphi_{2} \psi_{2} u_{2})^*\langle 2 \rangle &\longmapsto& 0 \\
&&& (\gamma_{6} \sigma^{2} u_{2})^*\langle 2 \rangle &\longmapsto& 0 \\
(\mathbb{Z}_{2})^{6} &\cong& H^{13}(B^{2}\mathbb{Z}_{2}; \mathbb{Z}_{2}) \colon & (\sigma \psi_{2} u_{2} \otimes \varphi_{2} \gamma_{2} \psi_{2} u_{2})^*\langle 2 \rangle &\longmapsto& 0 \\
&&& (\sigma^{2} u_{2} \otimes \sigma \gamma_{2} \psi_{2} u_{2} \otimes \varphi_{2} \psi_{2} u_{2})^*\langle 2 \rangle &\longmapsto& 0 \\
&&& (\gamma_{2} \sigma^{2} u_{2} \otimes \sigma \gamma_{4} \psi_{2} u_{2})^*\langle 2 \rangle &\longmapsto& 0 \\
&&& (\gamma_{2} \sigma^{2} u_{2} \otimes \sigma \psi_{2} u_{2} \otimes \varphi_{2} \psi_{2} u_{2})^*\langle 2 \rangle &\longmapsto& 0 \\
&&& (\gamma_{4} \sigma^{2} u_{2} \otimes \sigma \gamma_{2} \psi_{2} u_{2})^*\langle 2 \rangle &\longmapsto& 0 \\
&&& (\gamma_{5} \sigma^{2} u_{2} \otimes \sigma \psi_{2} u_{2})^*\langle 2 \rangle &\longmapsto& 0
\end{IEEEeqnarraybox}

%% file: tables/omega-B2Z2-Z4.tex
\begin{IEEEeqnarraybox}[][c]{rClrCl}
\mathbb{Z}_{2} &\cong& H^{2}(B^{2}\mathbb{Z}_{2}; \mathbb{Z}_{4}) \colon & 2 (\sigma^{2} u_{2})^*\langle 2 \rangle &\longmapsto& 2 (\sigma u_{2})^*\langle 2 \rangle \\
\mathbb{Z}_{2} &\cong& H^{3}(B^{2}\mathbb{Z}_{2}; \mathbb{Z}_{4}) \colon & (\sigma \psi_{2} u_{2})^*\langle 2 \rangle &\longmapsto& (\psi_{2} u_{2})^*\langle 2 \rangle \\
\mathbb{Z}_{4} &\cong& H^{4}(B^{2}\mathbb{Z}_{2}; \mathbb{Z}_{4}) \colon & (\gamma_{2} \sigma^{2} u_{2})^*\langle 4 \rangle &\longmapsto& 2 (\sigma u_{2} \cdot \psi_{2} u_{2})^*\langle 2 \rangle \\
\mathbb{Z}_{2} \oplus \mathbb{Z}_{4} &\cong& H^{5}(B^{2}\mathbb{Z}_{2}; \mathbb{Z}_{4}) \colon & 2 (\varepsilon_{2} \psi_{2} u_{2})^*\langle 2 \rangle &\longmapsto& 0 \\
&&& (\sigma^{2} u_{2} \cdot \sigma \psi_{2} u_{2})^*\langle 4 \rangle &\longmapsto& -(\gamma_{2} \psi_{2} u_{2})^*\langle 2 \rangle \\
(\mathbb{Z}_{2})^{2} &\cong& H^{6}(B^{2}\mathbb{Z}_{2}; \mathbb{Z}_{4}) \colon & (\varphi_{2} \psi_{2} u_{2})^*\langle 2 \rangle &\longmapsto& 0 \\
&&& 2 (\gamma_{3} \sigma^{2} u_{2})^*\langle 2 \rangle &\longmapsto& 0 \\
(\mathbb{Z}_{2})^{2} &\cong& H^{7}(B^{2}\mathbb{Z}_{2}; \mathbb{Z}_{4}) \colon & 2 (\sigma^{2} u_{2} \otimes \varepsilon_{2} \psi_{2} u_{2})^*\langle 2 \rangle &\longmapsto& 0 \\
&&& (\gamma_{2} \sigma^{2} u_{2} \cdot \sigma \psi_{2} u_{2})^*\langle 2 \rangle &\longmapsto& 0 \\
(\mathbb{Z}_{2})^{2} \oplus \mathbb{Z}_{4} &\cong& H^{8}(B^{2}\mathbb{Z}_{2}; \mathbb{Z}_{4}) \colon & (\kappa(\sigma^{2} u_{2},\sigma \psi_{2} u_{2},\varepsilon_{2} \psi_{2} u_{2},\varphi_{2} \psi_{2} u_{2}))^*\langle 2 \rangle &\longmapsto& 0 \\
&&& 2 (\tau(\sigma^{2} u_{2},\sigma \psi_{2} u_{2},\varepsilon_{2} \psi_{2} u_{2},\varphi_{2} \psi_{2} u_{2}))^*\langle 2 \rangle &\longmapsto& 0 \\
&&& (\gamma_{4} \sigma^{2} u_{2})^*\langle 4 \rangle &\longmapsto& 0 \\
(\mathbb{Z}_{2})^{3} \oplus \mathbb{Z}_{4} &\cong& H^{9}(B^{2}\mathbb{Z}_{2}; \mathbb{Z}_{4}) \colon & 2 (\varepsilon_{4} \psi_{2} u_{2})^*\langle 2 \rangle &\longmapsto& 0 \\
&&& (\sigma \psi_{2} u_{2} \otimes \varphi_{2} \psi_{2} u_{2})^*\langle 2 \rangle &\longmapsto& 0 \\
&&& 2 (\gamma_{2} \sigma^{2} u_{2} \otimes \varepsilon_{2} \psi_{2} u_{2})^*\langle 2 \rangle &\longmapsto& 0 \\
&&& (\gamma_{3} \sigma^{2} u_{2} \cdot \sigma \psi_{2} u_{2})^*\langle 4 \rangle &\longmapsto& 0 \\
(\mathbb{Z}_{2})^{4} &\cong& H^{10}(B^{2}\mathbb{Z}_{2}; \mathbb{Z}_{4}) \colon & (\varphi_{2} \gamma_{2} \psi_{2} u_{2})^*\langle 2 \rangle &\longmapsto& 0 \\
&&& (\gamma_{2} \sigma^{2} u_{2} \otimes \varphi_{2} \psi_{2} u_{2})^*\langle 2 \rangle &\longmapsto& 0 \\
&&& 2 (\sigma^{2} u_{2} \cdot \sigma \psi_{2} u_{2} \otimes \varepsilon_{2} \psi_{2} u_{2})^*\langle 2 \rangle &\longmapsto& 0 \\
&&& 2 (\gamma_{5} \sigma^{2} u_{2})^*\langle 2 \rangle &\longmapsto& 0 \\
(\mathbb{Z}_{2})^{5} &\cong& H^{11}(B^{2}\mathbb{Z}_{2}; \mathbb{Z}_{4}) \colon & 2 (\varepsilon_{2} \psi_{2} u_{2} \cdot \varphi_{2} \psi_{2} u_{2})^*\langle 2 \rangle &\longmapsto& 0 \\
&&& 2 (\sigma^{2} u_{2} \otimes \varepsilon_{4} \psi_{2} u_{2})^*\langle 2 \rangle &\longmapsto& 0 \\
&&& (\sigma^{2} u_{2} \cdot \sigma \psi_{2} u_{2} \otimes \varphi_{2} \psi_{2} u_{2})^*\langle 2 \rangle &\longmapsto& 0 \\
&&& 2 (\gamma_{3} \sigma^{2} u_{2} \otimes \varepsilon_{2} \psi_{2} u_{2})^*\langle 2 \rangle &\longmapsto& 0 \\
&&& (\gamma_{4} \sigma^{2} u_{2} \cdot \sigma \psi_{2} u_{2})^*\langle 2 \rangle &\longmapsto& 0 \\
(\mathbb{Z}_{2})^{5} \oplus \mathbb{Z}_{4} &\cong& H^{12}(B^{2}\mathbb{Z}_{2}; \mathbb{Z}_{4}) \colon & (\gamma_{2} \varphi_{2} \psi_{2} u_{2})^*\langle 2 \rangle &\longmapsto& 0 \\
&&& (\kappa(\sigma^{2} u_{2},\sigma \psi_{2} u_{2},\varepsilon_{4} \psi_{2} u_{2},\varphi_{2} \gamma_{2} \psi_{2} u_{2}))^*\langle 2 \rangle &\longmapsto& 0 \\
&&& 2 (\tau(\sigma^{2} u_{2},\sigma \psi_{2} u_{2},\varepsilon_{4} \psi_{2} u_{2},\varphi_{2} \gamma_{2} \psi_{2} u_{2}))^*\langle 2 \rangle &\longmapsto& 0 \\
&&& (\kappa(\gamma_{3} \sigma^{2} u_{2},\gamma_{2} \sigma^{2} u_{2} \cdot \sigma \psi_{2} u_{2},\varepsilon_{2} \psi_{2} u_{2},\varphi_{2} \psi_{2} u_{2}))^*\langle 2 \rangle &\longmapsto& 0 \\
&&& 2 (\tau(\gamma_{3} \sigma^{2} u_{2},\gamma_{2} \sigma^{2} u_{2} \cdot \sigma \psi_{2} u_{2},\varepsilon_{2} \psi_{2} u_{2},\varphi_{2} \psi_{2} u_{2}))^*\langle 2 \rangle &\longmapsto& 0 \\
&&& (\gamma_{6} \sigma^{2} u_{2})^*\langle 4 \rangle &\longmapsto& 0 \\
(\mathbb{Z}_{2})^{5} \oplus \mathbb{Z}_{4} &\cong& H^{13}(B^{2}\mathbb{Z}_{2}; \mathbb{Z}_{4}) \colon & (\sigma \psi_{2} u_{2} \otimes \varphi_{2} \gamma_{2} \psi_{2} u_{2})^*\langle 2 \rangle &\longmapsto& 0 \\
&&& 2 (\sigma^{2} u_{2} \otimes \varepsilon_{2} \psi_{2} u_{2} \cdot \varphi_{2} \psi_{2} u_{2})^*\langle 2 \rangle &\longmapsto& 0 \\
&&& 2 (\gamma_{2} \sigma^{2} u_{2} \otimes \varepsilon_{4} \psi_{2} u_{2})^*\langle 2 \rangle &\longmapsto& 0 \\
&&& (\gamma_{2} \sigma^{2} u_{2} \cdot \sigma \psi_{2} u_{2} \otimes \varphi_{2} \psi_{2} u_{2})^*\langle 2 \rangle &\longmapsto& 0 \\
&&& 2 (\gamma_{4} \sigma^{2} u_{2} \otimes \varepsilon_{2} \psi_{2} u_{2})^*\langle 2 \rangle &\longmapsto& 0 \\
&&& (\gamma_{5} \sigma^{2} u_{2} \cdot \sigma \psi_{2} u_{2})^*\langle 4 \rangle &\longmapsto& 0
\end{IEEEeqnarraybox}

%% file: tables/omega-B3Z2-RZ.tex
\begin{IEEEeqnarraybox}[][c]{rClrCl}
\mathbb{Z}_{2} &\cong& H^{3}(B^{3}\mathbb{Z}_{2}; \mathbb{R}/\mathbb{Z}) \colon & \frac{1}{2} (\sigma^{3} u_{2})^*\langle 2 \rangle &\longmapsto& \frac{1}{2} (\sigma^{2} u_{2})^*\langle 2 \rangle \\
0 &\cong& H^{4}(B^{3}\mathbb{Z}_{2}; \mathbb{R}/\mathbb{Z}) \colon &  \\
\mathbb{Z}_{2} &\cong& H^{5}(B^{3}\mathbb{Z}_{2}; \mathbb{R}/\mathbb{Z}) \colon & \frac{1}{2} (\sigma \gamma_{2} \sigma^{2} u_{2})^*\langle 2 \rangle &\longmapsto& \frac{1}{2} (\gamma_{2} \sigma^{2} u_{2})^*\langle 2 \rangle \\
\mathbb{Z}_{2} &\cong& H^{6}(B^{3}\mathbb{Z}_{2}; \mathbb{R}/\mathbb{Z}) \colon & \frac{1}{2} (\sigma \varepsilon_{2} \psi_{2} u_{2})^*\langle 2 \rangle &\longmapsto& \frac{1}{2} (\varepsilon_{2} \psi_{2} u_{2})^*\langle 2 \rangle \\
\mathbb{Z}_{2} &\cong& H^{7}(B^{3}\mathbb{Z}_{2}; \mathbb{R}/\mathbb{Z}) \colon & \frac{1}{2} (\sigma^{3} u_{2} \cdot \sigma^{2} \psi_{2} u_{2})^*\langle 2 \rangle &\longmapsto& 0 \\
\mathbb{Z}_{2} &\cong& H^{8}(B^{3}\mathbb{Z}_{2}; \mathbb{R}/\mathbb{Z}) \colon & \frac{1}{2} (\sigma^{3} u_{2} \otimes \sigma \gamma_{2} \sigma^{2} u_{2})^*\langle 2 \rangle &\longmapsto& 0 \\
(\mathbb{Z}_{2})^{3} &\cong& H^{9}(B^{3}\mathbb{Z}_{2}; \mathbb{R}/\mathbb{Z}) \colon & \frac{1}{2} (\sigma \gamma_{4} \sigma^{2} u_{2})^*\langle 2 \rangle &\longmapsto& \frac{1}{2} (\gamma_{4} \sigma^{2} u_{2})^*\langle 2 \rangle \\
&&& \frac{1}{2} (\sigma^{3} u_{2} \otimes \sigma \varepsilon_{2} \psi_{2} u_{2})^*\langle 2 \rangle &\longmapsto& 0 \\
&&& \frac{1}{2} (\tau(\sigma^{3} u_{2},\sigma^{2} \psi_{2} u_{2},\sigma \gamma_{2} \sigma^{2} u_{2},\varphi_{2} \sigma^{2} u_{2}))^*\langle 2 \rangle &\longmapsto& 0 \\
(\mathbb{Z}_{2})^{2} &\cong& H^{10}(B^{3}\mathbb{Z}_{2}; \mathbb{R}/\mathbb{Z}) \colon & \frac{1}{2} (\sigma \varepsilon_{4} \psi_{2} u_{2})^*\langle 2 \rangle &\longmapsto& \frac{1}{2} (\varepsilon_{4} \psi_{2} u_{2})^*\langle 2 \rangle \\
&&& \frac{1}{2} (\tau(\sigma^{3} u_{2},\sigma^{2} \psi_{2} u_{2},\sigma \varepsilon_{2} \psi_{2} u_{2},\sigma \varphi_{2} \psi_{2} u_{2}))^*\langle 2 \rangle &\longmapsto& 0 \\
(\mathbb{Z}_{2})^{3} &\cong& H^{11}(B^{3}\mathbb{Z}_{2}; \mathbb{R}/\mathbb{Z}) \colon & \frac{1}{2} (\sigma \gamma_{2} \sigma^{2} u_{2} \otimes \sigma \varepsilon_{2} \psi_{2} u_{2})^*\langle 2 \rangle &\longmapsto& 0 \\
&&& \frac{1}{2} (\sigma \gamma_{2} \sigma^{2} u_{2} \cdot \varphi_{2} \sigma^{2} u_{2})^*\langle 2 \rangle &\longmapsto& 0 \\
&&& \frac{1}{2} (\sigma^{3} u_{2} \cdot \gamma_{2} \sigma^{2} \psi_{2} u_{2})^*\langle 2 \rangle &\longmapsto& 0 \\
(\mathbb{Z}_{2})^{3} \oplus \mathbb{Z}_{4} &\cong& H^{12}(B^{3}\mathbb{Z}_{2}; \mathbb{R}/\mathbb{Z}) \colon & \frac{1}{4} (\gamma_{2} \sigma \varepsilon_{2} \psi_{2} u_{2})^*\langle 4 \rangle &\longmapsto& \frac{1}{2} (\varepsilon_{2} \psi_{2} u_{2} \cdot \varphi_{2} \psi_{2} u_{2})^*\langle 2 \rangle \\
&&& \frac{1}{2} (\tau(\sigma \gamma_{2} \sigma^{2} u_{2},\varphi_{2} \sigma^{2} u_{2},\sigma \varepsilon_{2} \psi_{2} u_{2},\sigma \varphi_{2} \psi_{2} u_{2}))^*\langle 2 \rangle &\longmapsto& 0 \\
&&& \frac{1}{2} (\sigma^{3} u_{2} \otimes \sigma \gamma_{4} \sigma^{2} u_{2})^*\langle 2 \rangle &\longmapsto& 0 \\
&&& \frac{1}{2} (\sigma^{3} u_{2} \cdot \sigma^{2} \psi_{2} u_{2} \otimes \sigma \gamma_{2} \sigma^{2} u_{2})^*\langle 2 \rangle &\longmapsto& 0 \\
(\mathbb{Z}_{2})^{5} &\cong& H^{13}(B^{3}\mathbb{Z}_{2}; \mathbb{R}/\mathbb{Z}) \colon & \frac{1}{2} (\varepsilon_{2} \varphi_{2} \psi_{2} u_{2})^*\langle 2 \rangle &\longmapsto& 0 \\
&&& \frac{1}{2} (\sigma^{3} u_{2} \otimes \sigma \varepsilon_{4} \psi_{2} u_{2})^*\langle 2 \rangle &\longmapsto& 0 \\
&&& \frac{1}{2} (\tau(\sigma^{3} u_{2},\sigma^{2} \psi_{2} u_{2},\sigma \gamma_{4} \sigma^{2} u_{2},\varphi_{2} \gamma_{2} \sigma^{2} u_{2}))^*\langle 2 \rangle &\longmapsto& 0 \\
&&& \frac{1}{2} (\sigma^{3} u_{2} \cdot \sigma^{2} \psi_{2} u_{2} \otimes \sigma \varepsilon_{2} \psi_{2} u_{2})^*\langle 2 \rangle &\longmapsto& 0 \\
&&& \frac{1}{2} (\tau(\sigma^{3} u_{2} \cdot \sigma^{2} \psi_{2} u_{2},\gamma_{2} \sigma^{2} \psi_{2} u_{2},\sigma \gamma_{2} \sigma^{2} u_{2},\varphi_{2} \sigma^{2} u_{2}))^*\langle 2 \rangle &\longmapsto& 0 \\
(\mathbb{Z}_{2})^{5} &\cong& H^{14}(B^{3}\mathbb{Z}_{2}; \mathbb{R}/\mathbb{Z}) \colon & \frac{1}{2} (\sigma \gamma_{2} \sigma^{2} u_{2} \otimes \sigma \gamma_{4} \sigma^{2} u_{2})^*\langle 2 \rangle &\longmapsto& 0 \\
&&& \frac{1}{2} (\tau(\sigma^{3} u_{2},\sigma^{2} \psi_{2} u_{2},\sigma \varepsilon_{4} \psi_{2} u_{2},\sigma \varphi_{2} \gamma_{2} \psi_{2} u_{2}))^*\langle 2 \rangle &\longmapsto& 0 \\
&&& \frac{1}{2} (\sigma^{3} u_{2} \otimes \sigma \gamma_{2} \sigma^{2} u_{2} \otimes \sigma \varepsilon_{2} \psi_{2} u_{2})^*\langle 2 \rangle &\longmapsto& 0 \\
&&& \frac{1}{2} (\sigma^{3} u_{2} \otimes \sigma \gamma_{2} \sigma^{2} u_{2} \cdot \varphi_{2} \sigma^{2} u_{2})^*\langle 2 \rangle &\longmapsto& 0 \\
&&& \frac{1}{2} (\tau(\sigma^{3} u_{2} \cdot \sigma^{2} \psi_{2} u_{2},\gamma_{2} \sigma^{2} \psi_{2} u_{2},\sigma \varepsilon_{2} \psi_{2} u_{2},\sigma \varphi_{2} \psi_{2} u_{2}))^*\langle 2 \rangle &\longmapsto& 0
\end{IEEEeqnarraybox}

%% file: tables/omega-B3Z2-Z.tex
\begin{IEEEeqnarraybox}[][c]{rClrCl}
\mathbb{Z}_{2} &\cong& H^{4}(B^{3}\mathbb{Z}_{2}; \mathbb{Z}) \colon & (\sigma^{2} \psi_{2} u_{2})^*\langle 2 \rangle &\longmapsto& (\sigma \psi_{2} u_{2})^*\langle 2 \rangle \\
0 &\cong& H^{5}(B^{3}\mathbb{Z}_{2}; \mathbb{Z}) \colon &  \\
\mathbb{Z}_{2} &\cong& H^{6}(B^{3}\mathbb{Z}_{2}; \mathbb{Z}) \colon & (\varphi_{2} \sigma^{2} u_{2})^*\langle 2 \rangle &\longmapsto& -2 (\sigma^{2} u_{2} \cdot \sigma \psi_{2} u_{2})^*\langle 2 \rangle \\
\mathbb{Z}_{2} &\cong& H^{7}(B^{3}\mathbb{Z}_{2}; \mathbb{Z}) \colon & (\sigma \varphi_{2} \psi_{2} u_{2})^*\langle 2 \rangle &\longmapsto& (\varphi_{2} \psi_{2} u_{2})^*\langle 2 \rangle \\
\mathbb{Z}_{2} &\cong& H^{8}(B^{3}\mathbb{Z}_{2}; \mathbb{Z}) \colon & (\gamma_{2} \sigma^{2} \psi_{2} u_{2})^*\langle 2 \rangle &\longmapsto& 0 \\
\mathbb{Z}_{2} &\cong& H^{9}(B^{3}\mathbb{Z}_{2}; \mathbb{Z}) \colon & (\kappa(\sigma^{3} u_{2},\sigma^{2} \psi_{2} u_{2},\sigma \gamma_{2} \sigma^{2} u_{2},\varphi_{2} \sigma^{2} u_{2}))^*\langle 2 \rangle &\longmapsto& 0 \\
(\mathbb{Z}_{2})^{3} &\cong& H^{10}(B^{3}\mathbb{Z}_{2}; \mathbb{Z}) \colon & (\varphi_{2} \gamma_{2} \sigma^{2} u_{2})^*\langle 2 \rangle &\longmapsto& -4 (\gamma_{3} \sigma^{2} u_{2} \cdot \sigma \psi_{2} u_{2})^*\langle 2 \rangle \\
&&& (\kappa(\sigma^{3} u_{2},\sigma^{2} \psi_{2} u_{2},\sigma \varepsilon_{2} \psi_{2} u_{2},\sigma \varphi_{2} \psi_{2} u_{2}))^*\langle 2 \rangle &\longmapsto& 0 \\
&&& (\sigma^{2} \psi_{2} u_{2} \otimes \varphi_{2} \sigma^{2} u_{2})^*\langle 2 \rangle &\longmapsto& 0 \\
(\mathbb{Z}_{2})^{2} &\cong& H^{11}(B^{3}\mathbb{Z}_{2}; \mathbb{Z}) \colon & (\sigma \varphi_{2} \gamma_{2} \psi_{2} u_{2})^*\langle 2 \rangle &\longmapsto& (\varphi_{2} \gamma_{2} \psi_{2} u_{2})^*\langle 2 \rangle \\
&&& (\sigma^{2} \psi_{2} u_{2} \otimes \sigma \varphi_{2} \psi_{2} u_{2})^*\langle 2 \rangle &\longmapsto& 0 \\
(\mathbb{Z}_{2})^{3} &\cong& H^{12}(B^{3}\mathbb{Z}_{2}; \mathbb{Z}) \colon & (\kappa(\sigma \gamma_{2} \sigma^{2} u_{2},\varphi_{2} \sigma^{2} u_{2},\sigma \varepsilon_{2} \psi_{2} u_{2},\sigma \varphi_{2} \psi_{2} u_{2}))^*\langle 2 \rangle &\longmapsto& 0 \\
&&& (\gamma_{2} \varphi_{2} \sigma^{2} u_{2})^*\langle 2 \rangle &\longmapsto& 0 \\
&&& (\gamma_{3} \sigma^{2} \psi_{2} u_{2})^*\langle 2 \rangle &\longmapsto& 0 \\
(\mathbb{Z}_{2})^{3} \oplus \mathbb{Z}_{4} &\cong& H^{13}(B^{3}\mathbb{Z}_{2}; \mathbb{Z}) \colon & (\sigma \varepsilon_{2} \psi_{2} u_{2} \cdot \sigma \varphi_{2} \psi_{2} u_{2})^*\langle 4 \rangle &\longmapsto& (\gamma_{2} \varphi_{2} \psi_{2} u_{2})^*\langle 2 \rangle \\
&&& (\varphi_{2} \sigma^{2} u_{2} \otimes \sigma \varphi_{2} \psi_{2} u_{2})^*\langle 2 \rangle &\longmapsto& 0 \\
&&& (\kappa(\sigma^{3} u_{2},\sigma^{2} \psi_{2} u_{2},\sigma \gamma_{4} \sigma^{2} u_{2},\varphi_{2} \gamma_{2} \sigma^{2} u_{2}))^*\langle 2 \rangle &\longmapsto& 0 \\
&&& (\kappa(\sigma^{3} u_{2} \cdot \sigma^{2} \psi_{2} u_{2},\gamma_{2} \sigma^{2} \psi_{2} u_{2},\sigma \gamma_{2} \sigma^{2} u_{2},\varphi_{2} \sigma^{2} u_{2}))^*\langle 2 \rangle &\longmapsto& 0 \\
(\mathbb{Z}_{2})^{5} &\cong& H^{14}(B^{3}\mathbb{Z}_{2}; \mathbb{Z}) \colon & (\varphi_{2}^{2} \psi_{2} u_{2})^*\langle 2 \rangle &\longmapsto& 0 \\
&&& (\kappa(\sigma^{3} u_{2},\sigma^{2} \psi_{2} u_{2},\sigma \varepsilon_{4} \psi_{2} u_{2},\sigma \varphi_{2} \gamma_{2} \psi_{2} u_{2}))^*\langle 2 \rangle &\longmapsto& 0 \\
&&& (\sigma^{2} \psi_{2} u_{2} \otimes \varphi_{2} \gamma_{2} \sigma^{2} u_{2})^*\langle 2 \rangle &\longmapsto& 0 \\
&&& (\kappa(\sigma^{3} u_{2} \cdot \sigma^{2} \psi_{2} u_{2},\gamma_{2} \sigma^{2} \psi_{2} u_{2},\sigma \varepsilon_{2} \psi_{2} u_{2},\sigma \varphi_{2} \psi_{2} u_{2}))^*\langle 2 \rangle &\longmapsto& 0 \\
&&& (\gamma_{2} \sigma^{2} \psi_{2} u_{2} \otimes \varphi_{2} \sigma^{2} u_{2})^*\langle 2 \rangle &\longmapsto& 0
\end{IEEEeqnarraybox}

%% file: tables/omega-B3Z2-Z2.tex
\begin{IEEEeqnarraybox}[][c]{rClrCl}
\mathbb{Z}_{2} &\cong& H^{3}(B^{3}\mathbb{Z}_{2}; \mathbb{Z}_{2}) \colon & (\sigma^{3} u_{2})^*\langle 2 \rangle &\longmapsto& (\sigma^{2} u_{2})^*\langle 2 \rangle \\
\mathbb{Z}_{2} &\cong& H^{4}(B^{3}\mathbb{Z}_{2}; \mathbb{Z}_{2}) \colon & (\sigma^{2} \psi_{2} u_{2})^*\langle 2 \rangle &\longmapsto& (\sigma \psi_{2} u_{2})^*\langle 2 \rangle \\
\mathbb{Z}_{2} &\cong& H^{5}(B^{3}\mathbb{Z}_{2}; \mathbb{Z}_{2}) \colon & (\sigma \gamma_{2} \sigma^{2} u_{2})^*\langle 2 \rangle &\longmapsto& (\gamma_{2} \sigma^{2} u_{2})^*\langle 2 \rangle \\
(\mathbb{Z}_{2})^{2} &\cong& H^{6}(B^{3}\mathbb{Z}_{2}; \mathbb{Z}_{2}) \colon & (\sigma^{2} \gamma_{2} \psi_{2} u_{2})^*\langle 2 \rangle &\longmapsto& (\sigma \gamma_{2} \psi_{2} u_{2})^*\langle 2 \rangle \\
&&& (\varphi_{2} \sigma^{2} u_{2})^*\langle 2 \rangle &\longmapsto& 0 \\
(\mathbb{Z}_{2})^{2} &\cong& H^{7}(B^{3}\mathbb{Z}_{2}; \mathbb{Z}_{2}) \colon & (\sigma \varphi_{2} \psi_{2} u_{2})^*\langle 2 \rangle &\longmapsto& (\varphi_{2} \psi_{2} u_{2})^*\langle 2 \rangle \\
&&& (\sigma^{3} u_{2} \otimes \sigma^{2} \psi_{2} u_{2})^*\langle 2 \rangle &\longmapsto& 0 \\
(\mathbb{Z}_{2})^{2} &\cong& H^{8}(B^{3}\mathbb{Z}_{2}; \mathbb{Z}_{2}) \colon & (\gamma_{2} \sigma^{2} \psi_{2} u_{2})^*\langle 2 \rangle &\longmapsto& 0 \\
&&& (\sigma^{3} u_{2} \otimes \sigma \gamma_{2} \sigma^{2} u_{2})^*\langle 2 \rangle &\longmapsto& 0 \\
(\mathbb{Z}_{2})^{4} &\cong& H^{9}(B^{3}\mathbb{Z}_{2}; \mathbb{Z}_{2}) \colon & (\sigma \gamma_{4} \sigma^{2} u_{2})^*\langle 2 \rangle &\longmapsto& (\gamma_{4} \sigma^{2} u_{2})^*\langle 2 \rangle \\
&&& (\sigma \gamma_{2} \sigma^{2} u_{2} \otimes \sigma^{2} \psi_{2} u_{2})^*\langle 2 \rangle &\longmapsto& 0 \\
&&& (\sigma^{3} u_{2} \otimes \sigma^{2} \gamma_{2} \psi_{2} u_{2})^*\langle 2 \rangle &\longmapsto& 0 \\
&&& (\sigma^{3} u_{2} \otimes \varphi_{2} \sigma^{2} u_{2})^*\langle 2 \rangle &\longmapsto& 0 \\
(\mathbb{Z}_{2})^{5} &\cong& H^{10}(B^{3}\mathbb{Z}_{2}; \mathbb{Z}_{2}) \colon & (\sigma^{2} \gamma_{4} \psi_{2} u_{2})^*\langle 2 \rangle &\longmapsto& (\sigma \gamma_{4} \psi_{2} u_{2})^*\langle 2 \rangle \\
&&& (\sigma^{2} \psi_{2} u_{2} \otimes \sigma^{2} \gamma_{2} \psi_{2} u_{2})^*\langle 2 \rangle &\longmapsto& 0 \\
&&& (\varphi_{2} \gamma_{2} \sigma^{2} u_{2})^*\langle 2 \rangle &\longmapsto& 0 \\
&&& (\varphi_{2} \sigma^{2} u_{2} \otimes \sigma^{2} \psi_{2} u_{2})^*\langle 2 \rangle &\longmapsto& 0 \\
&&& (\sigma^{3} u_{2} \otimes \sigma \varphi_{2} \psi_{2} u_{2})^*\langle 2 \rangle &\longmapsto& 0 \\
(\mathbb{Z}_{2})^{5} &\cong& H^{11}(B^{3}\mathbb{Z}_{2}; \mathbb{Z}_{2}) \colon & (\sigma \varphi_{2} \gamma_{2} \psi_{2} u_{2})^*\langle 2 \rangle &\longmapsto& (\varphi_{2} \gamma_{2} \psi_{2} u_{2})^*\langle 2 \rangle \\
&&& (\sigma^{2} \psi_{2} u_{2} \otimes \sigma \varphi_{2} \psi_{2} u_{2})^*\langle 2 \rangle &\longmapsto& 0 \\
&&& (\sigma \gamma_{2} \sigma^{2} u_{2} \otimes \sigma^{2} \gamma_{2} \psi_{2} u_{2})^*\langle 2 \rangle &\longmapsto& 0 \\
&&& (\sigma \gamma_{2} \sigma^{2} u_{2} \otimes \varphi_{2} \sigma^{2} u_{2})^*\langle 2 \rangle &\longmapsto& 0 \\
&&& (\sigma^{3} u_{2} \otimes \gamma_{2} \sigma^{2} \psi_{2} u_{2})^*\langle 2 \rangle &\longmapsto& 0 \\
(\mathbb{Z}_{2})^{7} &\cong& H^{12}(B^{3}\mathbb{Z}_{2}; \mathbb{Z}_{2}) \colon & (\gamma_{2} \sigma^{2} \gamma_{2} \psi_{2} u_{2})^*\langle 2 \rangle &\longmapsto& 0 \\
&&& (\gamma_{3} \sigma^{2} \psi_{2} u_{2})^*\langle 2 \rangle &\longmapsto& 0 \\
&&& (\varphi_{2} \sigma^{2} u_{2} \otimes \sigma^{2} \gamma_{2} \psi_{2} u_{2})^*\langle 2 \rangle &\longmapsto& 0 \\
&&& (\gamma_{2} \varphi_{2} \sigma^{2} u_{2})^*\langle 2 \rangle &\longmapsto& 0 \\
&&& (\sigma \gamma_{2} \sigma^{2} u_{2} \otimes \sigma \varphi_{2} \psi_{2} u_{2})^*\langle 2 \rangle &\longmapsto& 0 \\
&&& (\sigma^{3} u_{2} \otimes \sigma \gamma_{4} \sigma^{2} u_{2})^*\langle 2 \rangle &\longmapsto& 0 \\
&&& (\sigma^{3} u_{2} \otimes \sigma \gamma_{2} \sigma^{2} u_{2} \otimes \sigma^{2} \psi_{2} u_{2})^*\langle 2 \rangle &\longmapsto& 0 \\
(\mathbb{Z}_{2})^{9} &\cong& H^{13}(B^{3}\mathbb{Z}_{2}; \mathbb{Z}_{2}) \colon & (\sigma \gamma_{2} \varphi_{2} \psi_{2} u_{2})^*\langle 2 \rangle &\longmapsto& (\gamma_{2} \varphi_{2} \psi_{2} u_{2})^*\langle 2 \rangle \\
&&& (\sigma^{2} \gamma_{2} \psi_{2} u_{2} \otimes \sigma \varphi_{2} \psi_{2} u_{2})^*\langle 2 \rangle &\longmapsto& 0 \\
&&& (\sigma \gamma_{4} \sigma^{2} u_{2} \otimes \sigma^{2} \psi_{2} u_{2})^*\langle 2 \rangle &\longmapsto& 0 \\
&&& (\varphi_{2} \sigma^{2} u_{2} \otimes \sigma \varphi_{2} \psi_{2} u_{2})^*\langle 2 \rangle &\longmapsto& 0 \\
&&& (\sigma \gamma_{2} \sigma^{2} u_{2} \otimes \gamma_{2} \sigma^{2} \psi_{2} u_{2})^*\langle 2 \rangle &\longmapsto& 0 \\
&&& (\sigma^{3} u_{2} \otimes \sigma^{2} \gamma_{4} \psi_{2} u_{2})^*\langle 2 \rangle &\longmapsto& 0 \\
&&& (\sigma^{3} u_{2} \otimes \sigma^{2} \psi_{2} u_{2} \otimes \sigma^{2} \gamma_{2} \psi_{2} u_{2})^*\langle 2 \rangle &\longmapsto& 0 \\
&&& (\sigma^{3} u_{2} \otimes \varphi_{2} \gamma_{2} \sigma^{2} u_{2})^*\langle 2 \rangle &\longmapsto& 0 \\
&&& (\sigma^{3} u_{2} \otimes \varphi_{2} \sigma^{2} u_{2} \otimes \sigma^{2} \psi_{2} u_{2})^*\langle 2 \rangle &\longmapsto& 0
\end{IEEEeqnarraybox}

%% file: tables/omega-B3Z2-Z4.tex
\begin{IEEEeqnarraybox}[][c]{rClrCl}
\mathbb{Z}_{2} &\cong& H^{3}(B^{3}\mathbb{Z}_{2}; \mathbb{Z}_{4}) \colon & 2 (\sigma^{3} u_{2})^*\langle 2 \rangle &\longmapsto& 2 (\sigma^{2} u_{2})^*\langle 2 \rangle \\
\mathbb{Z}_{2} &\cong& H^{4}(B^{3}\mathbb{Z}_{2}; \mathbb{Z}_{4}) \colon & (\sigma^{2} \psi_{2} u_{2})^*\langle 2 \rangle &\longmapsto& (\sigma \psi_{2} u_{2})^*\langle 2 \rangle \\
\mathbb{Z}_{2} &\cong& H^{5}(B^{3}\mathbb{Z}_{2}; \mathbb{Z}_{4}) \colon & 2 (\sigma \gamma_{2} \sigma^{2} u_{2})^*\langle 2 \rangle &\longmapsto& 2 (\gamma_{2} \sigma^{2} u_{2})^*\langle 2 \rangle \\
(\mathbb{Z}_{2})^{2} &\cong& H^{6}(B^{3}\mathbb{Z}_{2}; \mathbb{Z}_{4}) \colon & 2 (\sigma \varepsilon_{2} \psi_{2} u_{2})^*\langle 2 \rangle &\longmapsto& 2 (\varepsilon_{2} \psi_{2} u_{2})^*\langle 2 \rangle \\
&&& (\varphi_{2} \sigma^{2} u_{2})^*\langle 2 \rangle &\longmapsto& 2 (\sigma^{2} u_{2} \cdot \sigma \psi_{2} u_{2})^*\langle 2 \rangle \\
(\mathbb{Z}_{2})^{2} &\cong& H^{7}(B^{3}\mathbb{Z}_{2}; \mathbb{Z}_{4}) \colon & (\sigma \varphi_{2} \psi_{2} u_{2})^*\langle 2 \rangle &\longmapsto& (\varphi_{2} \psi_{2} u_{2})^*\langle 2 \rangle \\
&&& 2 (\sigma^{3} u_{2} \cdot \sigma^{2} \psi_{2} u_{2})^*\langle 2 \rangle &\longmapsto& 0 \\
(\mathbb{Z}_{2})^{2} &\cong& H^{8}(B^{3}\mathbb{Z}_{2}; \mathbb{Z}_{4}) \colon & 2 (\sigma^{3} u_{2} \otimes \sigma \gamma_{2} \sigma^{2} u_{2})^*\langle 2 \rangle &\longmapsto& 0 \\
&&& (\gamma_{2} \sigma^{2} \psi_{2} u_{2})^*\langle 2 \rangle &\longmapsto& 0 \\
(\mathbb{Z}_{2})^{4} &\cong& H^{9}(B^{3}\mathbb{Z}_{2}; \mathbb{Z}_{4}) \colon & 2 (\sigma \gamma_{4} \sigma^{2} u_{2})^*\langle 2 \rangle &\longmapsto& 2 (\gamma_{4} \sigma^{2} u_{2})^*\langle 2 \rangle \\
&&& 2 (\sigma^{3} u_{2} \otimes \sigma \varepsilon_{2} \psi_{2} u_{2})^*\langle 2 \rangle &\longmapsto& 0 \\
&&& (\kappa(\sigma^{3} u_{2},\sigma^{2} \psi_{2} u_{2},\sigma \gamma_{2} \sigma^{2} u_{2},\varphi_{2} \sigma^{2} u_{2}))^*\langle 2 \rangle &\longmapsto& 0 \\
&&& 2 (\tau(\sigma^{3} u_{2},\sigma^{2} \psi_{2} u_{2},\sigma \gamma_{2} \sigma^{2} u_{2},\varphi_{2} \sigma^{2} u_{2}))^*\langle 2 \rangle &\longmapsto& 0 \\
(\mathbb{Z}_{2})^{5} &\cong& H^{10}(B^{3}\mathbb{Z}_{2}; \mathbb{Z}_{4}) \colon & 2 (\sigma \varepsilon_{4} \psi_{2} u_{2})^*\langle 2 \rangle &\longmapsto& 2 (\varepsilon_{4} \psi_{2} u_{2})^*\langle 2 \rangle \\
&&& (\varphi_{2} \gamma_{2} \sigma^{2} u_{2})^*\langle 2 \rangle &\longmapsto& 0 \\
&&& (\kappa(\sigma^{3} u_{2},\sigma^{2} \psi_{2} u_{2},\sigma \varepsilon_{2} \psi_{2} u_{2},\sigma \varphi_{2} \psi_{2} u_{2}))^*\langle 2 \rangle &\longmapsto& 0 \\
&&& 2 (\tau(\sigma^{3} u_{2},\sigma^{2} \psi_{2} u_{2},\sigma \varepsilon_{2} \psi_{2} u_{2},\sigma \varphi_{2} \psi_{2} u_{2}))^*\langle 2 \rangle &\longmapsto& 0 \\
&&& (\sigma^{2} \psi_{2} u_{2} \otimes \varphi_{2} \sigma^{2} u_{2})^*\langle 2 \rangle &\longmapsto& 0 \\
(\mathbb{Z}_{2})^{5} &\cong& H^{11}(B^{3}\mathbb{Z}_{2}; \mathbb{Z}_{4}) \colon & (\sigma \varphi_{2} \gamma_{2} \psi_{2} u_{2})^*\langle 2 \rangle &\longmapsto& (\varphi_{2} \gamma_{2} \psi_{2} u_{2})^*\langle 2 \rangle \\
&&& 2 (\sigma \gamma_{2} \sigma^{2} u_{2} \otimes \sigma \varepsilon_{2} \psi_{2} u_{2})^*\langle 2 \rangle &\longmapsto& 0 \\
&&& 2 (\sigma \gamma_{2} \sigma^{2} u_{2} \cdot \varphi_{2} \sigma^{2} u_{2})^*\langle 2 \rangle &\longmapsto& 0 \\
&&& (\sigma^{2} \psi_{2} u_{2} \otimes \sigma \varphi_{2} \psi_{2} u_{2})^*\langle 2 \rangle &\longmapsto& 0 \\
&&& 2 (\sigma^{3} u_{2} \cdot \gamma_{2} \sigma^{2} \psi_{2} u_{2})^*\langle 2 \rangle &\longmapsto& 0 \\
(\mathbb{Z}_{2})^{6} \oplus \mathbb{Z}_{4} &\cong& H^{12}(B^{3}\mathbb{Z}_{2}; \mathbb{Z}_{4}) \colon & (\gamma_{2} \sigma \varepsilon_{2} \psi_{2} u_{2})^*\langle 4 \rangle &\longmapsto& 2 (\varepsilon_{2} \psi_{2} u_{2} \cdot \varphi_{2} \psi_{2} u_{2})^*\langle 2 \rangle \\
&&& (\kappa(\sigma \gamma_{2} \sigma^{2} u_{2},\varphi_{2} \sigma^{2} u_{2},\sigma \varepsilon_{2} \psi_{2} u_{2},\sigma \varphi_{2} \psi_{2} u_{2}))^*\langle 2 \rangle &\longmapsto& 0 \\
&&& 2 (\tau(\sigma \gamma_{2} \sigma^{2} u_{2},\varphi_{2} \sigma^{2} u_{2},\sigma \varepsilon_{2} \psi_{2} u_{2},\sigma \varphi_{2} \psi_{2} u_{2}))^*\langle 2 \rangle &\longmapsto& 0 \\
&&& (\gamma_{2} \varphi_{2} \sigma^{2} u_{2})^*\langle 2 \rangle &\longmapsto& 0 \\
&&& 2 (\sigma^{3} u_{2} \otimes \sigma \gamma_{4} \sigma^{2} u_{2})^*\langle 2 \rangle &\longmapsto& 0 \\
&&& 2 (\sigma^{3} u_{2} \cdot \sigma^{2} \psi_{2} u_{2} \otimes \sigma \gamma_{2} \sigma^{2} u_{2})^*\langle 2 \rangle &\longmapsto& 0 \\
&&& (\gamma_{3} \sigma^{2} \psi_{2} u_{2})^*\langle 2 \rangle &\longmapsto& 0 \\
(\mathbb{Z}_{2})^{8} \oplus \mathbb{Z}_{4} &\cong& H^{13}(B^{3}\mathbb{Z}_{2}; \mathbb{Z}_{4}) \colon & 2 (\varepsilon_{2} \varphi_{2} \psi_{2} u_{2})^*\langle 2 \rangle &\longmapsto& 0 \\
&&& (\sigma \varepsilon_{2} \psi_{2} u_{2} \cdot \sigma \varphi_{2} \psi_{2} u_{2})^*\langle 4 \rangle &\longmapsto& -(\gamma_{2} \varphi_{2} \psi_{2} u_{2})^*\langle 2 \rangle \\
&&& (\varphi_{2} \sigma^{2} u_{2} \otimes \sigma \varphi_{2} \psi_{2} u_{2})^*\langle 2 \rangle &\longmapsto& 0 \\
&&& 2 (\sigma^{3} u_{2} \otimes \sigma \varepsilon_{4} \psi_{2} u_{2})^*\langle 2 \rangle &\longmapsto& 0 \\
&&& (\kappa(\sigma^{3} u_{2},\sigma^{2} \psi_{2} u_{2},\sigma \gamma_{4} \sigma^{2} u_{2},\varphi_{2} \gamma_{2} \sigma^{2} u_{2}))^*\langle 2 \rangle &\longmapsto& 0 \\
&&& 2 (\tau(\sigma^{3} u_{2},\sigma^{2} \psi_{2} u_{2},\sigma \gamma_{4} \sigma^{2} u_{2},\varphi_{2} \gamma_{2} \sigma^{2} u_{2}))^*\langle 2 \rangle &\longmapsto& 0 \\
&&& 2 (\sigma^{3} u_{2} \cdot \sigma^{2} \psi_{2} u_{2} \otimes \sigma \varepsilon_{2} \psi_{2} u_{2})^*\langle 2 \rangle &\longmapsto& 0 \\
&&& (\kappa(\sigma^{3} u_{2} \cdot \sigma^{2} \psi_{2} u_{2},\gamma_{2} \sigma^{2} \psi_{2} u_{2},\sigma \gamma_{2} \sigma^{2} u_{2},\varphi_{2} \sigma^{2} u_{2}))^*\langle 2 \rangle &\longmapsto& 0 \\
&&& 2 (\tau(\sigma^{3} u_{2} \cdot \sigma^{2} \psi_{2} u_{2},\gamma_{2} \sigma^{2} \psi_{2} u_{2},\sigma \gamma_{2} \sigma^{2} u_{2},\varphi_{2} \sigma^{2} u_{2}))^*\langle 2 \rangle &\longmapsto& 0
\end{IEEEeqnarraybox}

%% file: tables/omega-B2Z3-RZ.tex
\begin{IEEEeqnarraybox}[][c]{rClrCl}
\mathbb{Z}_{3} &\cong& H^{2}(B^{2}\mathbb{Z}_{3}; \mathbb{R}/\mathbb{Z}) \colon & \frac{1}{3} (\sigma^{2} u_{3})^*\langle 3 \rangle &\longmapsto& \frac{1}{3} (\sigma u_{3})^*\langle 3 \rangle \\
0 &\cong& H^{3}(B^{2}\mathbb{Z}_{3}; \mathbb{R}/\mathbb{Z}) \colon &  \\
\mathbb{Z}_{3} &\cong& H^{4}(B^{2}\mathbb{Z}_{3}; \mathbb{R}/\mathbb{Z}) \colon & \frac{1}{3} (\gamma_{2} \sigma^{2} u_{3})^*\langle 3 \rangle &\longmapsto& 0 \\
0 &\cong& H^{5}(B^{2}\mathbb{Z}_{3}; \mathbb{R}/\mathbb{Z}) \colon &  \\
\mathbb{Z}_{9} &\cong& H^{6}(B^{2}\mathbb{Z}_{3}; \mathbb{R}/\mathbb{Z}) \colon & \frac{1}{9} (\gamma_{3} \sigma^{2} u_{3})^*\langle 9 \rangle &\longmapsto& 0 \\
\mathbb{Z}_{3} &\cong& H^{7}(B^{2}\mathbb{Z}_{3}; \mathbb{R}/\mathbb{Z}) \colon & \frac{1}{3} (\varepsilon_{3} \psi_{3} u_{3})^*\langle 3 \rangle &\longmapsto& 0 \\
\mathbb{Z}_{3} &\cong& H^{8}(B^{2}\mathbb{Z}_{3}; \mathbb{R}/\mathbb{Z}) \colon & \frac{1}{3} (\gamma_{4} \sigma^{2} u_{3})^*\langle 3 \rangle &\longmapsto& 0 \\
\mathbb{Z}_{3} &\cong& H^{9}(B^{2}\mathbb{Z}_{3}; \mathbb{R}/\mathbb{Z}) \colon & \frac{1}{3} (\sigma^{2} u_{3} \otimes \varepsilon_{3} \psi_{3} u_{3})^*\langle 3 \rangle &\longmapsto& 0 \\
(\mathbb{Z}_{3})^{2} &\cong& H^{10}(B^{2}\mathbb{Z}_{3}; \mathbb{R}/\mathbb{Z}) \colon & \frac{1}{3} (\tau(\sigma^{2} u_{3},\sigma \psi_{3} u_{3},\varepsilon_{3} \psi_{3} u_{3},\varphi_{3} \psi_{3} u_{3}))^*\langle 3 \rangle &\longmapsto& 0 \\
&&& \frac{1}{3} (\gamma_{5} \sigma^{2} u_{3})^*\langle 3 \rangle &\longmapsto& 0 \\
\mathbb{Z}_{3} &\cong& H^{11}(B^{2}\mathbb{Z}_{3}; \mathbb{R}/\mathbb{Z}) \colon & \frac{1}{3} (\gamma_{2} \sigma^{2} u_{3} \otimes \varepsilon_{3} \psi_{3} u_{3})^*\langle 3 \rangle &\longmapsto& 0 \\
\mathbb{Z}_{3} \oplus \mathbb{Z}_{9} &\cong& H^{12}(B^{2}\mathbb{Z}_{3}; \mathbb{R}/\mathbb{Z}) \colon & \frac{1}{3} (\tau(\gamma_{2} \sigma^{2} u_{3},\sigma^{2} u_{3} \cdot \sigma \psi_{3} u_{3},\varepsilon_{3} \psi_{3} u_{3},\varphi_{3} \psi_{3} u_{3}))^*\langle 3 \rangle &\longmapsto& 0 \\
&&& \frac{1}{9} (\gamma_{6} \sigma^{2} u_{3})^*\langle 9 \rangle &\longmapsto& 0 \\
\mathbb{Z}_{3} &\cong& H^{13}(B^{2}\mathbb{Z}_{3}; \mathbb{R}/\mathbb{Z}) \colon & \frac{1}{3} (\gamma_{3} \sigma^{2} u_{3} \otimes \varepsilon_{3} \psi_{3} u_{3})^*\langle 3 \rangle &\longmapsto& 0 \\
(\mathbb{Z}_{3})^{2} &\cong& H^{14}(B^{2}\mathbb{Z}_{3}; \mathbb{R}/\mathbb{Z}) \colon & \frac{1}{3} (\tau(\gamma_{3} \sigma^{2} u_{3},\gamma_{2} \sigma^{2} u_{3} \cdot \sigma \psi_{3} u_{3},\varepsilon_{3} \psi_{3} u_{3},\varphi_{3} \psi_{3} u_{3}))^*\langle 3 \rangle &\longmapsto& 0 \\
&&& \frac{1}{3} (\gamma_{7} \sigma^{2} u_{3})^*\langle 3 \rangle &\longmapsto& 0 \\
(\mathbb{Z}_{3})^{2} &\cong& H^{15}(B^{2}\mathbb{Z}_{3}; \mathbb{R}/\mathbb{Z}) \colon & \frac{1}{3} (\varepsilon_{3} \psi_{3} u_{3} \cdot \varphi_{3} \psi_{3} u_{3})^*\langle 3 \rangle &\longmapsto& 0 \\
&&& \frac{1}{3} (\gamma_{4} \sigma^{2} u_{3} \otimes \varepsilon_{3} \psi_{3} u_{3})^*\langle 3 \rangle &\longmapsto& 0
\end{IEEEeqnarraybox}

%% file: tables/omega-B2Z3-Z.tex
\begin{IEEEeqnarraybox}[][c]{rClrCl}
\mathbb{Z}_{3} &\cong& H^{3}(B^{2}\mathbb{Z}_{3}; \mathbb{Z}) \colon & (\sigma \psi_{3} u_{3})^*\langle 3 \rangle &\longmapsto& (\psi_{3} u_{3})^*\langle 3 \rangle \\
0 &\cong& H^{4}(B^{2}\mathbb{Z}_{3}; \mathbb{Z}) \colon &  \\
\mathbb{Z}_{3} &\cong& H^{5}(B^{2}\mathbb{Z}_{3}; \mathbb{Z}) \colon & (\sigma^{2} u_{3} \cdot \sigma \psi_{3} u_{3})^*\langle 3 \rangle &\longmapsto& 0 \\
0 &\cong& H^{6}(B^{2}\mathbb{Z}_{3}; \mathbb{Z}) \colon &  \\
\mathbb{Z}_{9} &\cong& H^{7}(B^{2}\mathbb{Z}_{3}; \mathbb{Z}) \colon & (\gamma_{2} \sigma^{2} u_{3} \cdot \sigma \psi_{3} u_{3})^*\langle 9 \rangle &\longmapsto& 0 \\
\mathbb{Z}_{3} &\cong& H^{8}(B^{2}\mathbb{Z}_{3}; \mathbb{Z}) \colon & (\varphi_{3} \psi_{3} u_{3})^*\langle 3 \rangle &\longmapsto& 0 \\
\mathbb{Z}_{3} &\cong& H^{9}(B^{2}\mathbb{Z}_{3}; \mathbb{Z}) \colon & (\gamma_{3} \sigma^{2} u_{3} \cdot \sigma \psi_{3} u_{3})^*\langle 3 \rangle &\longmapsto& 0 \\
\mathbb{Z}_{3} &\cong& H^{10}(B^{2}\mathbb{Z}_{3}; \mathbb{Z}) \colon & (\kappa(\sigma^{2} u_{3},\sigma \psi_{3} u_{3},\varepsilon_{3} \psi_{3} u_{3},\varphi_{3} \psi_{3} u_{3}))^*\langle 3 \rangle &\longmapsto& 0 \\
(\mathbb{Z}_{3})^{2} &\cong& H^{11}(B^{2}\mathbb{Z}_{3}; \mathbb{Z}) \colon & (\sigma \psi_{3} u_{3} \otimes \varphi_{3} \psi_{3} u_{3})^*\langle 3 \rangle &\longmapsto& 0 \\
&&& (\gamma_{4} \sigma^{2} u_{3} \cdot \sigma \psi_{3} u_{3})^*\langle 3 \rangle &\longmapsto& 0 \\
\mathbb{Z}_{3} &\cong& H^{12}(B^{2}\mathbb{Z}_{3}; \mathbb{Z}) \colon & (\kappa(\gamma_{2} \sigma^{2} u_{3},\sigma^{2} u_{3} \cdot \sigma \psi_{3} u_{3},\varepsilon_{3} \psi_{3} u_{3},\varphi_{3} \psi_{3} u_{3}))^*\langle 3 \rangle &\longmapsto& 0 \\
\mathbb{Z}_{3} \oplus \mathbb{Z}_{9} &\cong& H^{13}(B^{2}\mathbb{Z}_{3}; \mathbb{Z}) \colon & (\sigma^{2} u_{3} \cdot \sigma \psi_{3} u_{3} \otimes \varphi_{3} \psi_{3} u_{3})^*\langle 3 \rangle &\longmapsto& 0 \\
&&& (\gamma_{5} \sigma^{2} u_{3} \cdot \sigma \psi_{3} u_{3})^*\langle 9 \rangle &\longmapsto& 0 \\
\mathbb{Z}_{3} &\cong& H^{14}(B^{2}\mathbb{Z}_{3}; \mathbb{Z}) \colon & (\kappa(\gamma_{3} \sigma^{2} u_{3},\gamma_{2} \sigma^{2} u_{3} \cdot \sigma \psi_{3} u_{3},\varepsilon_{3} \psi_{3} u_{3},\varphi_{3} \psi_{3} u_{3}))^*\langle 3 \rangle &\longmapsto& 0 \\
(\mathbb{Z}_{3})^{2} &\cong& H^{15}(B^{2}\mathbb{Z}_{3}; \mathbb{Z}) \colon & (\gamma_{2} \sigma^{2} u_{3} \cdot \sigma \psi_{3} u_{3} \otimes \varphi_{3} \psi_{3} u_{3})^*\langle 3 \rangle &\longmapsto& 0 \\
&&& (\gamma_{6} \sigma^{2} u_{3} \cdot \sigma \psi_{3} u_{3})^*\langle 3 \rangle &\longmapsto& 0
\end{IEEEeqnarraybox}

%% file: tables/omega-B2Z3-Z3.tex
\begin{IEEEeqnarraybox}[][c]{rClrCl}
\mathbb{Z}_{3} &\cong& H^{2}(B^{2}\mathbb{Z}_{3}; \mathbb{Z}_{3}) \colon & (\sigma^{2} u_{3})^*\langle 3 \rangle &\longmapsto& (\sigma u_{3})^*\langle 3 \rangle \\
\mathbb{Z}_{3} &\cong& H^{3}(B^{2}\mathbb{Z}_{3}; \mathbb{Z}_{3}) \colon & (\sigma \psi_{3} u_{3})^*\langle 3 \rangle &\longmapsto& (\psi_{3} u_{3})^*\langle 3 \rangle \\
\mathbb{Z}_{3} &\cong& H^{4}(B^{2}\mathbb{Z}_{3}; \mathbb{Z}_{3}) \colon & (\gamma_{2} \sigma^{2} u_{3})^*\langle 3 \rangle &\longmapsto& 0 \\
\mathbb{Z}_{3} &\cong& H^{5}(B^{2}\mathbb{Z}_{3}; \mathbb{Z}_{3}) \colon & (\sigma^{2} u_{3} \otimes \sigma \psi_{3} u_{3})^*\langle 3 \rangle &\longmapsto& 0 \\
\mathbb{Z}_{3} &\cong& H^{6}(B^{2}\mathbb{Z}_{3}; \mathbb{Z}_{3}) \colon & (\gamma_{3} \sigma^{2} u_{3})^*\langle 3 \rangle &\longmapsto& 0 \\
(\mathbb{Z}_{3})^{2} &\cong& H^{7}(B^{2}\mathbb{Z}_{3}; \mathbb{Z}_{3}) \colon & (\sigma \gamma_{3} \psi_{3} u_{3})^*\langle 3 \rangle &\longmapsto& (\gamma_{3} \psi_{3} u_{3})^*\langle 3 \rangle \\
&&& (\gamma_{2} \sigma^{2} u_{3} \otimes \sigma \psi_{3} u_{3})^*\langle 3 \rangle &\longmapsto& 0 \\
(\mathbb{Z}_{3})^{2} &\cong& H^{8}(B^{2}\mathbb{Z}_{3}; \mathbb{Z}_{3}) \colon & (\varphi_{3} \psi_{3} u_{3})^*\langle 3 \rangle &\longmapsto& 0 \\
&&& (\gamma_{4} \sigma^{2} u_{3})^*\langle 3 \rangle &\longmapsto& 0 \\
(\mathbb{Z}_{3})^{2} &\cong& H^{9}(B^{2}\mathbb{Z}_{3}; \mathbb{Z}_{3}) \colon & (\sigma^{2} u_{3} \otimes \sigma \gamma_{3} \psi_{3} u_{3})^*\langle 3 \rangle &\longmapsto& 0 \\
&&& (\gamma_{3} \sigma^{2} u_{3} \otimes \sigma \psi_{3} u_{3})^*\langle 3 \rangle &\longmapsto& 0 \\
(\mathbb{Z}_{3})^{3} &\cong& H^{10}(B^{2}\mathbb{Z}_{3}; \mathbb{Z}_{3}) \colon & (\sigma \psi_{3} u_{3} \otimes \sigma \gamma_{3} \psi_{3} u_{3})^*\langle 3 \rangle &\longmapsto& 0 \\
&&& (\sigma^{2} u_{3} \otimes \varphi_{3} \psi_{3} u_{3})^*\langle 3 \rangle &\longmapsto& 0 \\
&&& (\gamma_{5} \sigma^{2} u_{3})^*\langle 3 \rangle &\longmapsto& 0 \\
(\mathbb{Z}_{3})^{3} &\cong& H^{11}(B^{2}\mathbb{Z}_{3}; \mathbb{Z}_{3}) \colon & (\sigma \psi_{3} u_{3} \otimes \varphi_{3} \psi_{3} u_{3})^*\langle 3 \rangle &\longmapsto& 0 \\
&&& (\gamma_{2} \sigma^{2} u_{3} \otimes \sigma \gamma_{3} \psi_{3} u_{3})^*\langle 3 \rangle &\longmapsto& 0 \\
&&& (\gamma_{4} \sigma^{2} u_{3} \otimes \sigma \psi_{3} u_{3})^*\langle 3 \rangle &\longmapsto& 0 \\
(\mathbb{Z}_{3})^{3} &\cong& H^{12}(B^{2}\mathbb{Z}_{3}; \mathbb{Z}_{3}) \colon & (\sigma^{2} u_{3} \otimes \sigma \psi_{3} u_{3} \otimes \sigma \gamma_{3} \psi_{3} u_{3})^*\langle 3 \rangle &\longmapsto& 0 \\
&&& (\gamma_{2} \sigma^{2} u_{3} \otimes \varphi_{3} \psi_{3} u_{3})^*\langle 3 \rangle &\longmapsto& 0 \\
&&& (\gamma_{6} \sigma^{2} u_{3})^*\langle 3 \rangle &\longmapsto& 0 \\
(\mathbb{Z}_{3})^{3} &\cong& H^{13}(B^{2}\mathbb{Z}_{3}; \mathbb{Z}_{3}) \colon & (\sigma^{2} u_{3} \otimes \sigma \psi_{3} u_{3} \otimes \varphi_{3} \psi_{3} u_{3})^*\langle 3 \rangle &\longmapsto& 0 \\
&&& (\gamma_{3} \sigma^{2} u_{3} \otimes \sigma \gamma_{3} \psi_{3} u_{3})^*\langle 3 \rangle &\longmapsto& 0 \\
&&& (\gamma_{5} \sigma^{2} u_{3} \otimes \sigma \psi_{3} u_{3})^*\langle 3 \rangle &\longmapsto& 0 \\
(\mathbb{Z}_{3})^{3} &\cong& H^{14}(B^{2}\mathbb{Z}_{3}; \mathbb{Z}_{3}) \colon & (\gamma_{2} \sigma^{2} u_{3} \otimes \sigma \psi_{3} u_{3} \otimes \sigma \gamma_{3} \psi_{3} u_{3})^*\langle 3 \rangle &\longmapsto& 0 \\
&&& (\gamma_{3} \sigma^{2} u_{3} \otimes \varphi_{3} \psi_{3} u_{3})^*\langle 3 \rangle &\longmapsto& 0 \\
&&& (\gamma_{7} \sigma^{2} u_{3})^*\langle 3 \rangle &\longmapsto& 0 \\
(\mathbb{Z}_{3})^{4} &\cong& H^{15}(B^{2}\mathbb{Z}_{3}; \mathbb{Z}_{3}) \colon & (\sigma \gamma_{3} \psi_{3} u_{3} \otimes \varphi_{3} \psi_{3} u_{3})^*\langle 3 \rangle &\longmapsto& 0 \\
&&& (\gamma_{2} \sigma^{2} u_{3} \otimes \sigma \psi_{3} u_{3} \otimes \varphi_{3} \psi_{3} u_{3})^*\langle 3 \rangle &\longmapsto& 0 \\
&&& (\gamma_{4} \sigma^{2} u_{3} \otimes \sigma \gamma_{3} \psi_{3} u_{3})^*\langle 3 \rangle &\longmapsto& 0 \\
&&& (\gamma_{6} \sigma^{2} u_{3} \otimes \sigma \psi_{3} u_{3})^*\langle 3 \rangle &\longmapsto& 0
\end{IEEEeqnarraybox}

%% file: tables/omega-B3Z3-RZ.tex
\begin{IEEEeqnarraybox}[][c]{rClrCl}
\mathbb{Z}_{3} &\cong& H^{3}(B^{3}\mathbb{Z}_{3}; \mathbb{R}/\mathbb{Z}) \colon & \frac{1}{3} (\sigma^{3} u_{3})^*\langle 3 \rangle &\longmapsto& \frac{1}{3} (\sigma^{2} u_{3})^*\langle 3 \rangle \\
0 &\cong& H^{4}(B^{3}\mathbb{Z}_{3}; \mathbb{R}/\mathbb{Z}) \colon &  \\
0 &\cong& H^{5}(B^{3}\mathbb{Z}_{3}; \mathbb{R}/\mathbb{Z}) \colon &  \\
0 &\cong& H^{6}(B^{3}\mathbb{Z}_{3}; \mathbb{R}/\mathbb{Z}) \colon &  \\
(\mathbb{Z}_{3})^{2} &\cong& H^{7}(B^{3}\mathbb{Z}_{3}; \mathbb{R}/\mathbb{Z}) \colon & \frac{1}{3} (\sigma \gamma_{3} \sigma^{2} u_{3})^*\langle 3 \rangle &\longmapsto& \frac{1}{3} (\gamma_{3} \sigma^{2} u_{3})^*\langle 3 \rangle \\
&&& \frac{1}{3} (\sigma^{3} u_{3} \cdot \sigma^{2} \psi_{3} u_{3})^*\langle 3 \rangle &\longmapsto& 0 \\
\mathbb{Z}_{3} &\cong& H^{8}(B^{3}\mathbb{Z}_{3}; \mathbb{R}/\mathbb{Z}) \colon & \frac{1}{3} (\sigma \varepsilon_{3} \psi_{3} u_{3})^*\langle 3 \rangle &\longmapsto& \frac{1}{3} (\varepsilon_{3} \psi_{3} u_{3})^*\langle 3 \rangle \\
0 &\cong& H^{9}(B^{3}\mathbb{Z}_{3}; \mathbb{R}/\mathbb{Z}) \colon &  \\
\mathbb{Z}_{3} &\cong& H^{10}(B^{3}\mathbb{Z}_{3}; \mathbb{R}/\mathbb{Z}) \colon & \frac{1}{3} (\sigma^{3} u_{3} \otimes \sigma \gamma_{3} \sigma^{2} u_{3})^*\langle 3 \rangle &\longmapsto& 0 \\
(\mathbb{Z}_{3})^{3} &\cong& H^{11}(B^{3}\mathbb{Z}_{3}; \mathbb{R}/\mathbb{Z}) \colon & \frac{1}{3} (\sigma^{3} u_{3} \otimes \sigma \varepsilon_{3} \psi_{3} u_{3})^*\langle 3 \rangle &\longmapsto& 0 \\
&&& \frac{1}{3} (\tau(\sigma^{3} u_{3},\sigma^{2} \psi_{3} u_{3},\sigma \gamma_{3} \sigma^{2} u_{3},\varphi_{3} \sigma^{2} u_{3}))^*\langle 3 \rangle &\longmapsto& 0 \\
&&& \frac{1}{3} (\sigma^{3} u_{3} \cdot \gamma_{2} \sigma^{2} \psi_{3} u_{3})^*\langle 3 \rangle &\longmapsto& 0 \\
\mathbb{Z}_{3} &\cong& H^{12}(B^{3}\mathbb{Z}_{3}; \mathbb{R}/\mathbb{Z}) \colon & \frac{1}{3} (\tau(\sigma^{3} u_{3},\sigma^{2} \psi_{3} u_{3},\sigma \varepsilon_{3} \psi_{3} u_{3},\sigma \varphi_{3} \psi_{3} u_{3}))^*\langle 3 \rangle &\longmapsto& 0 \\
0 &\cong& H^{13}(B^{3}\mathbb{Z}_{3}; \mathbb{R}/\mathbb{Z}) \colon &  \\
\mathbb{Z}_{3} &\cong& H^{14}(B^{3}\mathbb{Z}_{3}; \mathbb{R}/\mathbb{Z}) \colon & \frac{1}{3} (\sigma^{3} u_{3} \cdot \sigma^{2} \psi_{3} u_{3} \otimes \sigma \gamma_{3} \sigma^{2} u_{3})^*\langle 3 \rangle &\longmapsto& 0 \\
(\mathbb{Z}_{3})^{5} &\cong& H^{15}(B^{3}\mathbb{Z}_{3}; \mathbb{R}/\mathbb{Z}) \colon & \frac{1}{3} (\sigma \gamma_{3} \sigma^{2} u_{3} \otimes \sigma \varepsilon_{3} \psi_{3} u_{3})^*\langle 3 \rangle &\longmapsto& 0 \\
&&& \frac{1}{3} (\sigma \gamma_{3} \sigma^{2} u_{3} \cdot \varphi_{3} \sigma^{2} u_{3})^*\langle 3 \rangle &\longmapsto& 0 \\
&&& \frac{1}{3} (\sigma^{3} u_{3} \cdot \sigma^{2} \psi_{3} u_{3} \otimes \sigma \varepsilon_{3} \psi_{3} u_{3})^*\langle 3 \rangle &\longmapsto& 0 \\
&&& \frac{1}{3} (\tau(\sigma^{3} u_{3} \cdot \sigma^{2} \psi_{3} u_{3},\gamma_{2} \sigma^{2} \psi_{3} u_{3},\sigma \gamma_{3} \sigma^{2} u_{3},\varphi_{3} \sigma^{2} u_{3}))^*\langle 3 \rangle &\longmapsto& 0 \\
&&& \frac{1}{3} (\sigma^{3} u_{3} \cdot \gamma_{3} \sigma^{2} \psi_{3} u_{3})^*\langle 3 \rangle &\longmapsto& 0
\end{IEEEeqnarraybox}

%% file: tables/omega-B3Z3-Z.tex
\begin{IEEEeqnarraybox}[][c]{rClrCl}
\mathbb{Z}_{3} &\cong& H^{4}(B^{3}\mathbb{Z}_{3}; \mathbb{Z}) \colon & (\sigma^{2} \psi_{3} u_{3})^*\langle 3 \rangle &\longmapsto& (\sigma \psi_{3} u_{3})^*\langle 3 \rangle \\
0 &\cong& H^{5}(B^{3}\mathbb{Z}_{3}; \mathbb{Z}) \colon &  \\
0 &\cong& H^{6}(B^{3}\mathbb{Z}_{3}; \mathbb{Z}) \colon &  \\
0 &\cong& H^{7}(B^{3}\mathbb{Z}_{3}; \mathbb{Z}) \colon &  \\
(\mathbb{Z}_{3})^{2} &\cong& H^{8}(B^{3}\mathbb{Z}_{3}; \mathbb{Z}) \colon & (\varphi_{3} \sigma^{2} u_{3})^*\langle 3 \rangle &\longmapsto& -6 (\gamma_{2} \sigma^{2} u_{3} \cdot \sigma \psi_{3} u_{3})^*\langle 3 \rangle \\
&&& (\gamma_{2} \sigma^{2} \psi_{3} u_{3})^*\langle 3 \rangle &\longmapsto& 0 \\
\mathbb{Z}_{3} &\cong& H^{9}(B^{3}\mathbb{Z}_{3}; \mathbb{Z}) \colon & (\sigma \varphi_{3} \psi_{3} u_{3})^*\langle 3 \rangle &\longmapsto& (\varphi_{3} \psi_{3} u_{3})^*\langle 3 \rangle \\
0 &\cong& H^{10}(B^{3}\mathbb{Z}_{3}; \mathbb{Z}) \colon &  \\
\mathbb{Z}_{3} &\cong& H^{11}(B^{3}\mathbb{Z}_{3}; \mathbb{Z}) \colon & (\kappa(\sigma^{3} u_{3},\sigma^{2} \psi_{3} u_{3},\sigma \gamma_{3} \sigma^{2} u_{3},\varphi_{3} \sigma^{2} u_{3}))^*\langle 3 \rangle &\longmapsto& 0 \\
(\mathbb{Z}_{3})^{3} &\cong& H^{12}(B^{3}\mathbb{Z}_{3}; \mathbb{Z}) \colon & (\kappa(\sigma^{3} u_{3},\sigma^{2} \psi_{3} u_{3},\sigma \varepsilon_{3} \psi_{3} u_{3},\sigma \varphi_{3} \psi_{3} u_{3}))^*\langle 3 \rangle &\longmapsto& 0 \\
&&& (\sigma^{2} \psi_{3} u_{3} \otimes \varphi_{3} \sigma^{2} u_{3})^*\langle 3 \rangle &\longmapsto& 0 \\
&&& (\gamma_{3} \sigma^{2} \psi_{3} u_{3})^*\langle 3 \rangle &\longmapsto& 0 \\
\mathbb{Z}_{3} &\cong& H^{13}(B^{3}\mathbb{Z}_{3}; \mathbb{Z}) \colon & (\sigma^{2} \psi_{3} u_{3} \otimes \sigma \varphi_{3} \psi_{3} u_{3})^*\langle 3 \rangle &\longmapsto& 0 \\
0 &\cong& H^{14}(B^{3}\mathbb{Z}_{3}; \mathbb{Z}) \colon &  \\
\mathbb{Z}_{3} &\cong& H^{15}(B^{3}\mathbb{Z}_{3}; \mathbb{Z}) \colon & (\kappa(\sigma^{3} u_{3} \cdot \sigma^{2} \psi_{3} u_{3},\gamma_{2} \sigma^{2} \psi_{3} u_{3},\sigma \gamma_{3} \sigma^{2} u_{3},\varphi_{3} \sigma^{2} u_{3}))^*\langle 3 \rangle &\longmapsto& 0
\end{IEEEeqnarraybox}

%% file: tables/omega-B3Z3-Z3.tex
\begin{IEEEeqnarraybox}[][c]{rClrCl}
\mathbb{Z}_{3} &\cong& H^{3}(B^{3}\mathbb{Z}_{3}; \mathbb{Z}_{3}) \colon & (\sigma^{3} u_{3})^*\langle 3 \rangle &\longmapsto& (\sigma^{2} u_{3})^*\langle 3 \rangle \\
\mathbb{Z}_{3} &\cong& H^{4}(B^{3}\mathbb{Z}_{3}; \mathbb{Z}_{3}) \colon & (\sigma^{2} \psi_{3} u_{3})^*\langle 3 \rangle &\longmapsto& (\sigma \psi_{3} u_{3})^*\langle 3 \rangle \\
0 &\cong& H^{5}(B^{3}\mathbb{Z}_{3}; \mathbb{Z}_{3}) \colon &  \\
0 &\cong& H^{6}(B^{3}\mathbb{Z}_{3}; \mathbb{Z}_{3}) \colon &  \\
(\mathbb{Z}_{3})^{2} &\cong& H^{7}(B^{3}\mathbb{Z}_{3}; \mathbb{Z}_{3}) \colon & (\sigma \gamma_{3} \sigma^{2} u_{3})^*\langle 3 \rangle &\longmapsto& (\gamma_{3} \sigma^{2} u_{3})^*\langle 3 \rangle \\
&&& (\sigma^{3} u_{3} \otimes \sigma^{2} \psi_{3} u_{3})^*\langle 3 \rangle &\longmapsto& 0 \\
(\mathbb{Z}_{3})^{3} &\cong& H^{8}(B^{3}\mathbb{Z}_{3}; \mathbb{Z}_{3}) \colon & (\sigma^{2} \gamma_{3} \psi_{3} u_{3})^*\langle 3 \rangle &\longmapsto& (\sigma \gamma_{3} \psi_{3} u_{3})^*\langle 3 \rangle \\
&&& (\gamma_{2} \sigma^{2} \psi_{3} u_{3})^*\langle 3 \rangle &\longmapsto& 0 \\
&&& (\varphi_{3} \sigma^{2} u_{3})^*\langle 3 \rangle &\longmapsto& 0 \\
\mathbb{Z}_{3} &\cong& H^{9}(B^{3}\mathbb{Z}_{3}; \mathbb{Z}_{3}) \colon & (\sigma \varphi_{3} \psi_{3} u_{3})^*\langle 3 \rangle &\longmapsto& (\varphi_{3} \psi_{3} u_{3})^*\langle 3 \rangle \\
\mathbb{Z}_{3} &\cong& H^{10}(B^{3}\mathbb{Z}_{3}; \mathbb{Z}_{3}) \colon & (\sigma^{3} u_{3} \otimes \sigma \gamma_{3} \sigma^{2} u_{3})^*\langle 3 \rangle &\longmapsto& 0 \\
(\mathbb{Z}_{3})^{4} &\cong& H^{11}(B^{3}\mathbb{Z}_{3}; \mathbb{Z}_{3}) \colon & (\sigma \gamma_{3} \sigma^{2} u_{3} \otimes \sigma^{2} \psi_{3} u_{3})^*\langle 3 \rangle &\longmapsto& 0 \\
&&& (\sigma^{3} u_{3} \otimes \sigma^{2} \gamma_{3} \psi_{3} u_{3})^*\langle 3 \rangle &\longmapsto& 0 \\
&&& (\sigma^{3} u_{3} \otimes \gamma_{2} \sigma^{2} \psi_{3} u_{3})^*\langle 3 \rangle &\longmapsto& 0 \\
&&& (\sigma^{3} u_{3} \otimes \varphi_{3} \sigma^{2} u_{3})^*\langle 3 \rangle &\longmapsto& 0 \\
(\mathbb{Z}_{3})^{4} &\cong& H^{12}(B^{3}\mathbb{Z}_{3}; \mathbb{Z}_{3}) \colon & (\sigma^{2} \psi_{3} u_{3} \otimes \sigma^{2} \gamma_{3} \psi_{3} u_{3})^*\langle 3 \rangle &\longmapsto& 0 \\
&&& (\gamma_{3} \sigma^{2} \psi_{3} u_{3})^*\langle 3 \rangle &\longmapsto& 0 \\
&&& (\varphi_{3} \sigma^{2} u_{3} \otimes \sigma^{2} \psi_{3} u_{3})^*\langle 3 \rangle &\longmapsto& 0 \\
&&& (\sigma^{3} u_{3} \otimes \sigma \varphi_{3} \psi_{3} u_{3})^*\langle 3 \rangle &\longmapsto& 0 \\
\mathbb{Z}_{3} &\cong& H^{13}(B^{3}\mathbb{Z}_{3}; \mathbb{Z}_{3}) \colon & (\sigma^{2} \psi_{3} u_{3} \otimes \sigma \varphi_{3} \psi_{3} u_{3})^*\langle 3 \rangle &\longmapsto& 0 \\
\mathbb{Z}_{3} &\cong& H^{14}(B^{3}\mathbb{Z}_{3}; \mathbb{Z}_{3}) \colon & (\sigma^{3} u_{3} \otimes \sigma \gamma_{3} \sigma^{2} u_{3} \otimes \sigma^{2} \psi_{3} u_{3})^*\langle 3 \rangle &\longmapsto& 0 \\
(\mathbb{Z}_{3})^{6} &\cong& H^{15}(B^{3}\mathbb{Z}_{3}; \mathbb{Z}_{3}) \colon & (\sigma \gamma_{3} \sigma^{2} u_{3} \otimes \sigma^{2} \gamma_{3} \psi_{3} u_{3})^*\langle 3 \rangle &\longmapsto& 0 \\
&&& (\sigma \gamma_{3} \sigma^{2} u_{3} \otimes \gamma_{2} \sigma^{2} \psi_{3} u_{3})^*\langle 3 \rangle &\longmapsto& 0 \\
&&& (\sigma \gamma_{3} \sigma^{2} u_{3} \otimes \varphi_{3} \sigma^{2} u_{3})^*\langle 3 \rangle &\longmapsto& 0 \\
&&& (\sigma^{3} u_{3} \otimes \sigma^{2} \psi_{3} u_{3} \otimes \sigma^{2} \gamma_{3} \psi_{3} u_{3})^*\langle 3 \rangle &\longmapsto& 0 \\
&&& (\sigma^{3} u_{3} \otimes \gamma_{3} \sigma^{2} \psi_{3} u_{3})^*\langle 3 \rangle &\longmapsto& 0 \\
&&& (\sigma^{3} u_{3} \otimes \varphi_{3} \sigma^{2} u_{3} \otimes \sigma^{2} \psi_{3} u_{3})^*\langle 3 \rangle &\longmapsto& 0
\end{IEEEeqnarraybox}

%% file: tables/omega-Z-RZ.tex
\begin{tikzcd}[row sep=tiny,column sep=-.5em]
& B\mathbb{Z} & B^{2}\mathbb{Z} & B^{3}\mathbb{Z} & B^{4}\mathbb{Z} & B^{5}\mathbb{Z} \\
H^{0} & \mathbb{R}/\mathbb{Z} & \mathbb{R}/\mathbb{Z} & \mathbb{R}/\mathbb{Z} & \mathbb{R}/\mathbb{Z} & \mathbb{R}/\mathbb{Z} \\
H^{1} & \mathbb{R}/\mathbb{Z} & 0 & 0 & 0 & 0 \\
H^{2} & 0 & \mathbb{R}/\mathbb{Z} \ar[lu,BrickRed,hook,two heads,"{{\infty}}" {gray,pos=0.08,scale=0.8,below left},"{{\infty}}" {gray,pos=0.92,scale=0.8,above right}] & 0 & 0 & 0 \\
H^{3} & 0 & 0 & \mathbb{R}/\mathbb{Z} \ar[lu,BrickRed,hook,two heads,"{{\infty}}" {gray,pos=0.08,scale=0.8,below left},"{{\infty}}" {gray,pos=0.92,scale=0.8,above right}] & 0 & 0 \\
H^{4} & 0 & \mathbb{R}/\mathbb{Z} & 0 & \mathbb{R}/\mathbb{Z} \ar[lu,BrickRed,hook,two heads,"{{\infty}}" {gray,pos=0.08,scale=0.8,below left},"{{\infty}}" {gray,pos=0.92,scale=0.8,above right}] & 0 \\
H^{5} & 0 & 0 & \mathbb{Z}_{2} \ar[lu,hook,"{{2}}" {gray,pos=0.08,scale=0.8,below left},"{{2}}" {gray,pos=0.92,scale=0.8,above right}] & 0 & \mathbb{R}/\mathbb{Z} \ar[lu,BrickRed,hook,two heads,"{{\infty}}" {gray,pos=0.08,scale=0.8,below left},"{{\infty}}" {gray,pos=0.92,scale=0.8,above right}] \\
H^{6} & 0 & \mathbb{R}/\mathbb{Z} & 0 & \mathbb{Z}_{2} \ar[lu,BrickRed,hook,two heads,"{{2}}" {gray,pos=0.08,scale=0.8,below left},"{{2}}" {gray,pos=0.92,scale=0.8,above right}] & 0 \\
H^{7} & 0 & 0 & \mathbb{Z}_{3} \ar[lu,hook,"{{3}}" {gray,pos=0.08,scale=0.8,below left},"{{3}}" {gray,pos=0.92,scale=0.8,above right}] & 0 & \mathbb{Z}_{2} \ar[lu,BrickRed,hook,two heads,"{{2}}" {gray,pos=0.08,scale=0.8,below left},"{{2}}" {gray,pos=0.92,scale=0.8,above right}] \\
H^{8} & 0 & \mathbb{R}/\mathbb{Z} & \mathbb{Z}_{2} & \mathbb{R}/\mathbb{Z} \oplus \mathbb{Z}_{3} \ar[lu,two heads,"{{3}}" {gray,pos=0.08,scale=0.8,below left},"{{3}}" {gray,pos=0.92,scale=0.8,above right}] & 0 \\
H^{9} & 0 & 0 & \mathbb{Z}_{2} \ar[lu,hook,"{{2}}" {gray,pos=0.08,scale=0.8,below left},"{{2}}" {gray,pos=0.92,scale=0.8,above right}] & 0 & \mathbb{Z}_{2} \oplus \mathbb{Z}_{3} \ar[lu,hook,"{{2,3}}" {gray,pos=0.08,scale=0.8,below left},"{{2,3}}" {gray,pos=0.92,scale=0.8,above right}] \\
H^{10} & 0 & \mathbb{R}/\mathbb{Z} & \mathbb{Z}_{3} & (\mathbb{Z}_{2})^{2} \ar[lu,two heads,"{{2}}" {gray,pos=0.08,scale=0.8,below left},"{{2}}" {gray,pos=0.92,scale=0.8,above right}] & 0 \\
H^{11} & 0 & 0 & \mathbb{Z}_{2} \oplus \mathbb{Z}_{5} \ar[lu,"{{5}}" {gray,pos=0.08,scale=0.8,below left},"{{5}}" {gray,pos=0.92,scale=0.8,above right}] & 0 & \mathbb{Z}_{2} \ar[lu,hook,"{{2}}" {gray,pos=0.08,scale=0.8,below left},"{{2}}" {gray,pos=0.92,scale=0.8,above right}] \\
H^{12} & 0 & \mathbb{R}/\mathbb{Z} & \mathbb{Z}_{2} & \mathbb{R}/\mathbb{Z} \oplus \mathbb{Z}_{3} \oplus \mathbb{Z}_{4} \oplus \mathbb{Z}_{5} \ar[lu,two heads,"{{4,5}}" {gray,pos=0.08,scale=0.8,below left},"{{2,5}}" {gray,pos=0.92,scale=0.8,above right}] & \mathbb{Z}_{2} \\
H^{13} & 0 & 0 & 0 & \mathbb{Z}_{2} & \mathbb{Z}_{2} \oplus \mathbb{Z}_{3} \oplus \mathbb{Z}_{5} \ar[lu,hook,"{{2,3,5}}" {gray,pos=0.08,scale=0.8,below left},"{{2,3,5}}" {gray,pos=0.92,scale=0.8,above right}] \\
H^{14} & 0 & \mathbb{R}/\mathbb{Z} & (\mathbb{Z}_{2})^{2} \oplus \mathbb{Z}_{5} & (\mathbb{Z}_{2})^{2} & (\mathbb{Z}_{2})^{2} \oplus \mathbb{Z}_{3} \ar[lu,two heads,"{{2}}" {gray,pos=0.08,scale=0.8,below left},"{{2}}" {gray,pos=0.92,scale=0.8,above right}] \\
H^{15} & 0 & 0 & \mathbb{Z}_{2} \oplus \mathbb{Z}_{3} \oplus \mathbb{Z}_{7} \ar[lu,"{{7}}" {gray,pos=0.08,scale=0.8,below left},"{{7}}" {gray,pos=0.92,scale=0.8,above right}] & 0 & \mathbb{Z}_{2} \\
H^{16} & 0 & \mathbb{R}/\mathbb{Z} & 0 & \mathbb{R}/\mathbb{Z} \oplus \mathbb{Z}_{2} \oplus (\mathbb{Z}_{3})^{2} \oplus \mathbb{Z}_{4} \oplus \mathbb{Z}_{5} \oplus \mathbb{Z}_{7} \ar[lu,"{{7}}" {gray,pos=0.08,scale=0.8,below left},"{{7}}" {gray,pos=0.92,scale=0.8,above right}] & (\mathbb{Z}_{2})^{2} \\
H^{17} & 0 & 0 & (\mathbb{Z}_{2})^{3} \ar[lu,"{{2}}" {gray,pos=0.08,scale=0.8,below left},"{{2}}" {gray,pos=0.92,scale=0.8,above right}] & (\mathbb{Z}_{2})^{2} & (\mathbb{Z}_{2})^{2} \oplus \mathbb{Z}_{7} \ar[lu,"{{2,7}}" {gray,pos=0.08,scale=0.8,below left},"{{2,7}}" {gray,pos=0.92,scale=0.8,above right}] \\
H^{18} & 0 & \mathbb{R}/\mathbb{Z} & \mathbb{Z}_{2} \oplus \mathbb{Z}_{3} \oplus \mathbb{Z}_{7} & (\mathbb{Z}_{2})^{4} \ar[lu,"{{2}}" {gray,pos=0.08,scale=0.8,below left},"{{2}}" {gray,pos=0.92,scale=0.8,above right}] & (\mathbb{Z}_{2})^{2} \oplus \mathbb{Z}_{3} \oplus \mathbb{Z}_{5} \\
H^{19} & 0 & 0 & \mathbb{Z}_{2} \oplus \mathbb{Z}_{3} \ar[lu,"{{3}}" {gray,pos=0.08,scale=0.8,below left},"{{3}}" {gray,pos=0.92,scale=0.8,above right}] & \mathbb{Z}_{2} & (\mathbb{Z}_{2})^{4} \oplus \mathbb{Z}_{3} \ar[lu,"{{2}}" {gray,pos=0.08,scale=0.8,below left},"{{2}}" {gray,pos=0.92,scale=0.8,above right}] \\
H^{20} & 0 & \mathbb{R}/\mathbb{Z} & (\mathbb{Z}_{2})^{3} & \mathbb{R}/\mathbb{Z} \oplus (\mathbb{Z}_{2})^{2} \oplus (\mathbb{Z}_{3})^{3} \oplus (\mathbb{Z}_{4})^{2} \oplus \mathbb{Z}_{5} \oplus \mathbb{Z}_{7} \ar[lu,two heads,"{{3,4}}" {gray,pos=0.08,scale=0.8,below left},"{{3,2}}" {gray,pos=0.92,scale=0.8,above right}] & (\mathbb{Z}_{2})^{3} \\
H^{21} & 0 & 0 & \mathbb{Z}_{2} & (\mathbb{Z}_{2})^{3} & (\mathbb{Z}_{2})^{5} \oplus \mathbb{Z}_{3} \oplus \mathbb{Z}_{5} \ar[lu,"{{2,3,5}}" {gray,pos=0.08,scale=0.8,below left},"{{2,3,5}}" {gray,pos=0.92,scale=0.8,above right}] \\
H^{22} & 0 & \mathbb{R}/\mathbb{Z} & (\mathbb{Z}_{2})^{2} \oplus \mathbb{Z}_{3} & (\mathbb{Z}_{2})^{5} & (\mathbb{Z}_{2})^{5} \oplus \mathbb{Z}_{3} \oplus \mathbb{Z}_{7} \ar[lu,"{{2}}" {gray,pos=0.08,scale=0.8,below left},"{{2}}" {gray,pos=0.92,scale=0.8,above right}] \\
H^{23} & 0 & 0 & (\mathbb{Z}_{2})^{3} \oplus \mathbb{Z}_{3} \oplus \mathbb{Z}_{5} \oplus \mathbb{Z}_{11} \ar[lu,"{{11}}" {gray,pos=0.08,scale=0.8,below left},"{{11}}" {gray,pos=0.92,scale=0.8,above right}] & (\mathbb{Z}_{2})^{3} & (\mathbb{Z}_{2})^{5} \oplus \mathbb{Z}_{3} \\
H^{24} & 0 & \mathbb{R}/\mathbb{Z} & (\mathbb{Z}_{2})^{2} & \mathbb{R}/\mathbb{Z} \oplus (\mathbb{Z}_{2})^{4} \oplus (\mathbb{Z}_{3})^{3} \oplus (\mathbb{Z}_{4})^{2} \oplus (\mathbb{Z}_{5})^{2} \oplus \mathbb{Z}_{7} \oplus \mathbb{Z}_{8} \oplus \mathbb{Z}_{9} \oplus \mathbb{Z}_{11} \ar[lu,"{{11}}" {gray,pos=0.08,scale=0.8,below left},"{{11}}" {gray,pos=0.92,scale=0.8,above right}] & (\mathbb{Z}_{2})^{7} \oplus \mathbb{Z}_{3} \\
H^{25} & 0 & 0 & (\mathbb{Z}_{2})^{2} & (\mathbb{Z}_{2})^{6} \oplus \mathbb{Z}_{3} & (\mathbb{Z}_{2})^{7} \oplus \mathbb{Z}_{3} \oplus \mathbb{Z}_{11} \ar[lu,"{{2,3,11}}" {gray,pos=0.08,scale=0.8,below left},"{{2,3,11}}" {gray,pos=0.92,scale=0.8,above right}] \\
H^{26} & 0 & \mathbb{R}/\mathbb{Z} & (\mathbb{Z}_{2})^{4} \oplus (\mathbb{Z}_{3})^{2} \oplus \mathbb{Z}_{5} \oplus \mathbb{Z}_{11} & (\mathbb{Z}_{2})^{7} & (\mathbb{Z}_{2})^{10} \oplus (\mathbb{Z}_{3})^{2} \oplus \mathbb{Z}_{5} \ar[lu,"{{2,3}}" {gray,pos=0.08,scale=0.8,below left},"{{2,3}}" {gray,pos=0.92,scale=0.8,above right}] \\
H^{27} & 0 & 0 & (\mathbb{Z}_{2})^{3} \oplus \mathbb{Z}_{3} \oplus \mathbb{Z}_{13} \ar[lu,"{{13}}" {gray,pos=0.08,scale=0.8,below left},"{{13}}" {gray,pos=0.92,scale=0.8,above right}] & (\mathbb{Z}_{2})^{6} & (\mathbb{Z}_{2})^{10} \oplus (\mathbb{Z}_{3})^{2} \oplus \mathbb{Z}_{5} \\
H^{28} & 0 & \mathbb{R}/\mathbb{Z} & (\mathbb{Z}_{2})^{2} & \mathbb{R}/\mathbb{Z} \oplus (\mathbb{Z}_{2})^{7} \oplus (\mathbb{Z}_{3})^{4} \oplus (\mathbb{Z}_{4})^{2} \oplus (\mathbb{Z}_{5})^{2} \oplus \mathbb{Z}_{7} \oplus \mathbb{Z}_{8} \oplus \mathbb{Z}_{9} \oplus \mathbb{Z}_{11} \oplus \mathbb{Z}_{13} \ar[lu,"{{13}}" {gray,pos=0.08,scale=0.8,below left},"{{13}}" {gray,pos=0.92,scale=0.8,above right}] & (\mathbb{Z}_{2})^{11} \oplus \mathbb{Z}_{3} \oplus \mathbb{Z}_{4} \ar[lu,"{{4}}" {gray,pos=0.08,scale=0.8,below left},"{{2}}" {gray,pos=0.92,scale=0.8,above right}] \\
H^{29} & 0 & 0 & (\mathbb{Z}_{2})^{5} \oplus \mathbb{Z}_{3} & (\mathbb{Z}_{2})^{9} \oplus (\mathbb{Z}_{3})^{2} & (\mathbb{Z}_{2})^{14} \oplus \mathbb{Z}_{3} \oplus \mathbb{Z}_{7} \oplus \mathbb{Z}_{13} \ar[lu,"{{7,13}}" {gray,pos=0.08,scale=0.8,below left},"{{7,13}}" {gray,pos=0.92,scale=0.8,above right}] \\
H^{30} & 0 & \mathbb{R}/\mathbb{Z} & (\mathbb{Z}_{2})^{3} \oplus \mathbb{Z}_{3} \oplus \mathbb{Z}_{13} & (\mathbb{Z}_{2})^{12} & (\mathbb{Z}_{2})^{15} \oplus (\mathbb{Z}_{3})^{2} \oplus \mathbb{Z}_{11} \\
H^{31} & 0 & 0 & (\mathbb{Z}_{2})^{3} \oplus \mathbb{Z}_{3} \oplus \mathbb{Z}_{7} & (\mathbb{Z}_{2})^{12} & (\mathbb{Z}_{2})^{18} \oplus (\mathbb{Z}_{3})^{2} \\
H^{32} & 0 & \mathbb{R}/\mathbb{Z} & (\mathbb{Z}_{2})^{6} & \mathbb{R}/\mathbb{Z} \oplus (\mathbb{Z}_{2})^{11} \oplus (\mathbb{Z}_{3})^{5} \oplus (\mathbb{Z}_{4})^{3} \oplus (\mathbb{Z}_{5})^{2} \oplus (\mathbb{Z}_{7})^{2} \oplus \mathbb{Z}_{8} \oplus \mathbb{Z}_{9} \oplus \mathbb{Z}_{11} \oplus \mathbb{Z}_{13} & (\mathbb{Z}_{2})^{20} \oplus (\mathbb{Z}_{3})^{2} \oplus \mathbb{Z}_{5} \\
H^{33} & 0 & 0 & (\mathbb{Z}_{2})^{4} \ar[lu,"{{2}}" {gray,pos=0.08,scale=0.8,below left},"{{2}}" {gray,pos=0.92,scale=0.8,above right}] & (\mathbb{Z}_{2})^{12} \oplus (\mathbb{Z}_{3})^{3} \oplus \mathbb{Z}_{4} & (\mathbb{Z}_{2})^{24} \oplus \mathbb{Z}_{3} \oplus \mathbb{Z}_{4} \ar[lu,"{{2}}" {gray,pos=0.08,scale=0.8,below left},"{{2}}" {gray,pos=0.92,scale=0.8,above right}] \\
H^{34} & 0 & \mathbb{R}/\mathbb{Z} & (\mathbb{Z}_{2})^{4} \oplus (\mathbb{Z}_{3})^{2} \oplus \mathbb{Z}_{7} & (\mathbb{Z}_{2})^{19} \oplus \mathbb{Z}_{3} \ar[lu,"{{2}}" {gray,pos=0.08,scale=0.8,below left},"{{2}}" {gray,pos=0.92,scale=0.8,above right}] & (\mathbb{Z}_{2})^{27} \oplus (\mathbb{Z}_{3})^{3} \oplus \mathbb{Z}_{5} \oplus \mathbb{Z}_{7} \oplus \mathbb{Z}_{13} \\
H^{35} & 0 & 0 & (\mathbb{Z}_{2})^{7} \oplus \mathbb{Z}_{3} \oplus \mathbb{Z}_{5} \oplus \mathbb{Z}_{17} \ar[lu,"{{17}}" {gray,pos=0.08,scale=0.8,below left},"{{17}}" {gray,pos=0.92,scale=0.8,above right}] & (\mathbb{Z}_{2})^{18} & (\mathbb{Z}_{2})^{32} \oplus (\mathbb{Z}_{3})^{4} \oplus \mathbb{Z}_{5} \oplus \mathbb{Z}_{7} \ar[lu,"{{2}}" {gray,pos=0.08,scale=0.8,below left},"{{2}}" {gray,pos=0.92,scale=0.8,above right}] \\
\end{tikzcd}

%% file: tables/omega-Z-Z.tex
\begin{tikzcd}[row sep=tiny,column sep=-.5em]
& B\mathbb{Z} & B^{2}\mathbb{Z} & B^{3}\mathbb{Z} & B^{4}\mathbb{Z} & B^{5}\mathbb{Z} \\
H^{0} & \mathbb{Z} & \mathbb{Z} & \mathbb{Z} & \mathbb{Z} & \mathbb{Z} \\
H^{1} & \mathbb{Z} & 0 & 0 & 0 & 0 \\
H^{2} & 0 & \mathbb{Z} \ar[lu,BrickRed,hook,two heads,"{{\infty}}" {gray,pos=0.08,scale=0.8,below left},"{{\infty}}" {gray,pos=0.92,scale=0.8,above right}] & 0 & 0 & 0 \\
H^{3} & 0 & 0 & \mathbb{Z} \ar[lu,BrickRed,hook,two heads,"{{\infty}}" {gray,pos=0.08,scale=0.8,below left},"{{\infty}}" {gray,pos=0.92,scale=0.8,above right}] & 0 & 0 \\
H^{4} & 0 & \mathbb{Z} & 0 & \mathbb{Z} \ar[lu,BrickRed,hook,two heads,"{{\infty}}" {gray,pos=0.08,scale=0.8,below left},"{{\infty}}" {gray,pos=0.92,scale=0.8,above right}] & 0 \\
H^{5} & 0 & 0 & 0 & 0 & \mathbb{Z} \ar[lu,BrickRed,hook,two heads,"{{\infty}}" {gray,pos=0.08,scale=0.8,below left},"{{\infty}}" {gray,pos=0.92,scale=0.8,above right}] \\
H^{6} & 0 & \mathbb{Z} & \mathbb{Z}_{2} & 0 & 0 \\
H^{7} & 0 & 0 & 0 & \mathbb{Z}_{2} \ar[lu,hook,two heads,"{{2}}" {gray,pos=0.08,scale=0.8,below left},"{{2}}" {gray,pos=0.92,scale=0.8,above right}] & 0 \\
H^{8} & 0 & \mathbb{Z} & \mathbb{Z}_{3} & \mathbb{Z} & \mathbb{Z}_{2} \ar[lu,BrickRed,hook,two heads,"{{2}}" {gray,pos=0.08,scale=0.8,below left},"{{2}}" {gray,pos=0.92,scale=0.8,above right}] \\
H^{9} & 0 & 0 & \mathbb{Z}_{2} & \mathbb{Z}_{3} \ar[lu,hook,two heads,"{{3}}" {gray,pos=0.08,scale=0.8,below left},"{{3}}" {gray,pos=0.92,scale=0.8,above right}] & 0 \\
H^{10} & 0 & \mathbb{Z} & \mathbb{Z}_{2} & 0 & \mathbb{Z}_{2} \oplus \mathbb{Z}_{3} \ar[lu,two heads,"{{3}}" {gray,pos=0.08,scale=0.8,below left},"{{3}}" {gray,pos=0.92,scale=0.8,above right}] \\
H^{11} & 0 & 0 & \mathbb{Z}_{3} & (\mathbb{Z}_{2})^{2} \ar[lu,two heads,"{{2}}" {gray,pos=0.08,scale=0.8,below left},"{{2}}" {gray,pos=0.92,scale=0.8,above right}] & 0 \\
H^{12} & 0 & \mathbb{Z} & \mathbb{Z}_{2} \oplus \mathbb{Z}_{5} & \mathbb{Z} & \mathbb{Z}_{2} \ar[lu,hook,"{{2}}" {gray,pos=0.08,scale=0.8,below left},"{{2}}" {gray,pos=0.92,scale=0.8,above right}] \\
H^{13} & 0 & 0 & \mathbb{Z}_{2} & \mathbb{Z}_{3} \oplus \mathbb{Z}_{4} \oplus \mathbb{Z}_{5} \ar[lu,two heads,"{{4,5}}" {gray,pos=0.08,scale=0.8,below left},"{{2,5}}" {gray,pos=0.92,scale=0.8,above right}] & \mathbb{Z}_{2} \\
H^{14} & 0 & \mathbb{Z} & 0 & \mathbb{Z}_{2} & \mathbb{Z}_{2} \oplus \mathbb{Z}_{3} \oplus \mathbb{Z}_{5} \ar[lu,"{{2,5}}" {gray,pos=0.08,scale=0.8,below left},"{{2,5}}" {gray,pos=0.92,scale=0.8,above right}] \\
H^{15} & 0 & 0 & (\mathbb{Z}_{2})^{2} \oplus \mathbb{Z}_{5} & (\mathbb{Z}_{2})^{2} & (\mathbb{Z}_{2})^{2} \oplus \mathbb{Z}_{3} \ar[lu,two heads,"{{2}}" {gray,pos=0.08,scale=0.8,below left},"{{2}}" {gray,pos=0.92,scale=0.8,above right}] \\
H^{16} & 0 & \mathbb{Z} & \mathbb{Z}_{2} \oplus \mathbb{Z}_{3} \oplus \mathbb{Z}_{7} & \mathbb{Z} & \mathbb{Z}_{2} \\
H^{17} & 0 & 0 & 0 & \mathbb{Z}_{2} \oplus (\mathbb{Z}_{3})^{2} \oplus \mathbb{Z}_{4} \oplus \mathbb{Z}_{5} \oplus \mathbb{Z}_{7} \ar[lu,"{{7}}" {gray,pos=0.08,scale=0.8,below left},"{{7}}" {gray,pos=0.92,scale=0.8,above right}] & (\mathbb{Z}_{2})^{2} \\
H^{18} & 0 & \mathbb{Z} & (\mathbb{Z}_{2})^{3} & (\mathbb{Z}_{2})^{2} & (\mathbb{Z}_{2})^{2} \oplus \mathbb{Z}_{7} \ar[lu,"{{7}}" {gray,pos=0.08,scale=0.8,below left},"{{7}}" {gray,pos=0.92,scale=0.8,above right}] \\
H^{19} & 0 & 0 & \mathbb{Z}_{2} \oplus \mathbb{Z}_{3} \oplus \mathbb{Z}_{7} & (\mathbb{Z}_{2})^{4} \ar[lu,"{{2}}" {gray,pos=0.08,scale=0.8,below left},"{{2}}" {gray,pos=0.92,scale=0.8,above right}] & (\mathbb{Z}_{2})^{2} \oplus \mathbb{Z}_{3} \oplus \mathbb{Z}_{5} \\
H^{20} & 0 & \mathbb{Z} & \mathbb{Z}_{2} \oplus \mathbb{Z}_{3} & \mathbb{Z} \oplus \mathbb{Z}_{2} & (\mathbb{Z}_{2})^{4} \oplus \mathbb{Z}_{3} \ar[lu,"{{2}}" {gray,pos=0.08,scale=0.8,below left},"{{2}}" {gray,pos=0.92,scale=0.8,above right}] \\
H^{21} & 0 & 0 & (\mathbb{Z}_{2})^{3} & (\mathbb{Z}_{2})^{2} \oplus (\mathbb{Z}_{3})^{3} \oplus (\mathbb{Z}_{4})^{2} \oplus \mathbb{Z}_{5} \oplus \mathbb{Z}_{7} \ar[lu,two heads,"{{3,4}}" {gray,pos=0.08,scale=0.8,below left},"{{3,2}}" {gray,pos=0.92,scale=0.8,above right}] & (\mathbb{Z}_{2})^{3} \\
H^{22} & 0 & \mathbb{Z} & \mathbb{Z}_{2} & (\mathbb{Z}_{2})^{3} & (\mathbb{Z}_{2})^{5} \oplus \mathbb{Z}_{3} \oplus \mathbb{Z}_{5} \ar[lu,"{{2,3}}" {gray,pos=0.08,scale=0.8,below left},"{{2,3}}" {gray,pos=0.92,scale=0.8,above right}] \\
H^{23} & 0 & 0 & (\mathbb{Z}_{2})^{2} \oplus \mathbb{Z}_{3} & (\mathbb{Z}_{2})^{5} & (\mathbb{Z}_{2})^{5} \oplus \mathbb{Z}_{3} \oplus \mathbb{Z}_{7} \ar[lu,"{{2}}" {gray,pos=0.08,scale=0.8,below left},"{{2}}" {gray,pos=0.92,scale=0.8,above right}] \\
H^{24} & 0 & \mathbb{Z} & (\mathbb{Z}_{2})^{3} \oplus \mathbb{Z}_{3} \oplus \mathbb{Z}_{5} \oplus \mathbb{Z}_{11} & \mathbb{Z} \oplus (\mathbb{Z}_{2})^{3} & (\mathbb{Z}_{2})^{5} \oplus \mathbb{Z}_{3} \\
H^{25} & 0 & 0 & (\mathbb{Z}_{2})^{2} & (\mathbb{Z}_{2})^{4} \oplus (\mathbb{Z}_{3})^{3} \oplus (\mathbb{Z}_{4})^{2} \oplus (\mathbb{Z}_{5})^{2} \oplus \mathbb{Z}_{7} \oplus \mathbb{Z}_{8} \oplus \mathbb{Z}_{9} \oplus \mathbb{Z}_{11} \ar[lu,"{{11}}" {gray,pos=0.08,scale=0.8,below left},"{{11}}" {gray,pos=0.92,scale=0.8,above right}] & (\mathbb{Z}_{2})^{7} \oplus \mathbb{Z}_{3} \\
H^{26} & 0 & \mathbb{Z} & (\mathbb{Z}_{2})^{2} & (\mathbb{Z}_{2})^{6} \oplus \mathbb{Z}_{3} & (\mathbb{Z}_{2})^{7} \oplus \mathbb{Z}_{3} \oplus \mathbb{Z}_{11} \ar[lu,"{{2,3,11}}" {gray,pos=0.08,scale=0.8,below left},"{{2,3,11}}" {gray,pos=0.92,scale=0.8,above right}] \\
H^{27} & 0 & 0 & (\mathbb{Z}_{2})^{4} \oplus (\mathbb{Z}_{3})^{2} \oplus \mathbb{Z}_{5} \oplus \mathbb{Z}_{11} & (\mathbb{Z}_{2})^{7} & (\mathbb{Z}_{2})^{10} \oplus (\mathbb{Z}_{3})^{2} \oplus \mathbb{Z}_{5} \ar[lu,"{{2,3}}" {gray,pos=0.08,scale=0.8,below left},"{{2,3}}" {gray,pos=0.92,scale=0.8,above right}] \\
H^{28} & 0 & \mathbb{Z} & (\mathbb{Z}_{2})^{3} \oplus \mathbb{Z}_{3} \oplus \mathbb{Z}_{13} & \mathbb{Z} \oplus (\mathbb{Z}_{2})^{6} & (\mathbb{Z}_{2})^{10} \oplus (\mathbb{Z}_{3})^{2} \oplus \mathbb{Z}_{5} \\
H^{29} & 0 & 0 & (\mathbb{Z}_{2})^{2} & (\mathbb{Z}_{2})^{7} \oplus (\mathbb{Z}_{3})^{4} \oplus (\mathbb{Z}_{4})^{2} \oplus (\mathbb{Z}_{5})^{2} \oplus \mathbb{Z}_{7} \oplus \mathbb{Z}_{8} \oplus \mathbb{Z}_{9} \oplus \mathbb{Z}_{11} \oplus \mathbb{Z}_{13} \ar[lu,"{{13}}" {gray,pos=0.08,scale=0.8,below left},"{{13}}" {gray,pos=0.92,scale=0.8,above right}] & (\mathbb{Z}_{2})^{11} \oplus \mathbb{Z}_{3} \oplus \mathbb{Z}_{4} \ar[lu,"{{4}}" {gray,pos=0.08,scale=0.8,below left},"{{2}}" {gray,pos=0.92,scale=0.8,above right}] \\
H^{30} & 0 & \mathbb{Z} & (\mathbb{Z}_{2})^{5} \oplus \mathbb{Z}_{3} & (\mathbb{Z}_{2})^{9} \oplus (\mathbb{Z}_{3})^{2} & (\mathbb{Z}_{2})^{14} \oplus \mathbb{Z}_{3} \oplus \mathbb{Z}_{7} \oplus \mathbb{Z}_{13} \ar[lu,"{{13}}" {gray,pos=0.08,scale=0.8,below left},"{{13}}" {gray,pos=0.92,scale=0.8,above right}] \\
H^{31} & 0 & 0 & (\mathbb{Z}_{2})^{3} \oplus \mathbb{Z}_{3} \oplus \mathbb{Z}_{13} & (\mathbb{Z}_{2})^{12} & (\mathbb{Z}_{2})^{15} \oplus (\mathbb{Z}_{3})^{2} \oplus \mathbb{Z}_{11} \\
H^{32} & 0 & \mathbb{Z} & (\mathbb{Z}_{2})^{3} \oplus \mathbb{Z}_{3} \oplus \mathbb{Z}_{7} & \mathbb{Z} \oplus (\mathbb{Z}_{2})^{12} & (\mathbb{Z}_{2})^{18} \oplus (\mathbb{Z}_{3})^{2} \\
H^{33} & 0 & 0 & (\mathbb{Z}_{2})^{6} & (\mathbb{Z}_{2})^{11} \oplus (\mathbb{Z}_{3})^{5} \oplus (\mathbb{Z}_{4})^{3} \oplus (\mathbb{Z}_{5})^{2} \oplus (\mathbb{Z}_{7})^{2} \oplus \mathbb{Z}_{8} \oplus \mathbb{Z}_{9} \oplus \mathbb{Z}_{11} \oplus \mathbb{Z}_{13} & (\mathbb{Z}_{2})^{20} \oplus (\mathbb{Z}_{3})^{2} \oplus \mathbb{Z}_{5} \\
H^{34} & 0 & \mathbb{Z} & (\mathbb{Z}_{2})^{4} & (\mathbb{Z}_{2})^{12} \oplus (\mathbb{Z}_{3})^{3} \oplus \mathbb{Z}_{4} & (\mathbb{Z}_{2})^{24} \oplus \mathbb{Z}_{3} \oplus \mathbb{Z}_{4} \\
H^{35} & 0 & 0 & (\mathbb{Z}_{2})^{4} \oplus (\mathbb{Z}_{3})^{2} \oplus \mathbb{Z}_{7} & (\mathbb{Z}_{2})^{19} \oplus \mathbb{Z}_{3} \ar[lu,"{{2}}" {gray,pos=0.08,scale=0.8,below left},"{{2}}" {gray,pos=0.92,scale=0.8,above right}] & (\mathbb{Z}_{2})^{27} \oplus (\mathbb{Z}_{3})^{3} \oplus \mathbb{Z}_{5} \oplus \mathbb{Z}_{7} \oplus \mathbb{Z}_{13} \\
H^{36} & 0 & \mathbb{Z} & (\mathbb{Z}_{2})^{7} \oplus \mathbb{Z}_{3} \oplus \mathbb{Z}_{5} \oplus \mathbb{Z}_{17} & \mathbb{Z} \oplus (\mathbb{Z}_{2})^{18} & (\mathbb{Z}_{2})^{32} \oplus (\mathbb{Z}_{3})^{4} \oplus \mathbb{Z}_{5} \oplus \mathbb{Z}_{7} \ar[lu,"{{2}}" {gray,pos=0.08,scale=0.8,below left},"{{2}}" {gray,pos=0.92,scale=0.8,above right}] \\
H^{37} & 0 & 0 & (\mathbb{Z}_{2})^{5} & (\mathbb{Z}_{2})^{17} \oplus (\mathbb{Z}_{3})^{6} \oplus (\mathbb{Z}_{4})^{5} \oplus (\mathbb{Z}_{5})^{3} \oplus (\mathbb{Z}_{7})^{2} \oplus \mathbb{Z}_{8} \oplus \mathbb{Z}_{9} \oplus \mathbb{Z}_{11} \oplus \mathbb{Z}_{13} \oplus \mathbb{Z}_{17} \ar[lu,"{{4,17}}" {gray,pos=0.08,scale=0.8,below left},"{{2,17}}" {gray,pos=0.92,scale=0.8,above right}] & (\mathbb{Z}_{2})^{35} \oplus (\mathbb{Z}_{3})^{3} \\
H^{38} & 0 & \mathbb{Z} & (\mathbb{Z}_{2})^{5} \oplus \mathbb{Z}_{3} & (\mathbb{Z}_{2})^{21} \oplus (\mathbb{Z}_{3})^{4} \oplus \mathbb{Z}_{4} & (\mathbb{Z}_{2})^{40} \oplus (\mathbb{Z}_{3})^{3} \oplus \mathbb{Z}_{17} \ar[lu,"{{2,17}}" {gray,pos=0.08,scale=0.8,below left},"{{2,17}}" {gray,pos=0.92,scale=0.8,above right}] \\
H^{39} & 0 & 0 & (\mathbb{Z}_{2})^{8} \oplus \mathbb{Z}_{3} \oplus \mathbb{Z}_{5} \oplus \mathbb{Z}_{17} & (\mathbb{Z}_{2})^{29} \oplus \mathbb{Z}_{3} & (\mathbb{Z}_{2})^{46} \oplus (\mathbb{Z}_{3})^{2} \ar[lu,"{{2}}" {gray,pos=0.08,scale=0.8,below left},"{{2}}" {gray,pos=0.92,scale=0.8,above right}] \\
H^{40} & 0 & \mathbb{Z} & (\mathbb{Z}_{2})^{6} \oplus (\mathbb{Z}_{3})^{2} \oplus \mathbb{Z}_{19} & \mathbb{Z} \oplus (\mathbb{Z}_{2})^{27} & (\mathbb{Z}_{2})^{51} \oplus (\mathbb{Z}_{3})^{5} \oplus \mathbb{Z}_{5} \\
\end{tikzcd}

%% file: tables/omega-Z2-RZ.tex
\begin{tikzcd}[row sep=tiny,column sep=tiny]
& B\mathbb{Z}_{2} & B^{2}\mathbb{Z}_{2} & B^{3}\mathbb{Z}_{2} & B^{4}\mathbb{Z}_{2} & B^{5}\mathbb{Z}_{2} & B^{6}\mathbb{Z}_{2} & B^{7}\mathbb{Z}_{2} \\
H^{0} & \mathbb{R}/\mathbb{Z} & \mathbb{R}/\mathbb{Z} & \mathbb{R}/\mathbb{Z} & \mathbb{R}/\mathbb{Z} & \mathbb{R}/\mathbb{Z} & \mathbb{R}/\mathbb{Z} & \mathbb{R}/\mathbb{Z} \\
H^{1} & \mathbb{Z}_{2} & 0 & 0 & 0 & 0 & 0 & 0 \\
H^{2} & 0 & \mathbb{Z}_{2} \ar[lu,BrickRed,hook,two heads,"{{2}}" {gray,pos=0.08,scale=0.8,below left},"{{2}}" {gray,pos=0.92,scale=0.8,above right}] & 0 & 0 & 0 & 0 & 0 \\
H^{3} & \mathbb{Z}_{2} & 0 & \mathbb{Z}_{2} \ar[lu,BrickRed,hook,two heads,"{{2}}" {gray,pos=0.08,scale=0.8,below left},"{{2}}" {gray,pos=0.92,scale=0.8,above right}] & 0 & 0 & 0 & 0 \\
H^{4} & 0 & \mathbb{Z}_{4} \ar[lu,two heads,"{{4}}" {gray,pos=0.08,scale=0.8,below left},"{{2}}" {gray,pos=0.92,scale=0.8,above right}] & 0 & \mathbb{Z}_{2} \ar[lu,BrickRed,hook,two heads,"{{2}}" {gray,pos=0.08,scale=0.8,below left},"{{2}}" {gray,pos=0.92,scale=0.8,above right}] & 0 & 0 & 0 \\
H^{5} & \mathbb{Z}_{2} & \mathbb{Z}_{2} & \mathbb{Z}_{2} \ar[lu,hook,"{{2}}" {gray,pos=0.08,scale=0.8,below left},"{{2}}" {gray,pos=0.92,scale=0.8,above right}] & 0 & \mathbb{Z}_{2} \ar[lu,BrickRed,hook,two heads,"{{2}}" {gray,pos=0.08,scale=0.8,below left},"{{2}}" {gray,pos=0.92,scale=0.8,above right}] & 0 & 0 \\
H^{6} & 0 & \mathbb{Z}_{2} & \mathbb{Z}_{2} \ar[lu,hook,two heads,"{{2}}" {gray,pos=0.08,scale=0.8,below left},"{{2}}" {gray,pos=0.92,scale=0.8,above right}] & \mathbb{Z}_{2} \ar[lu,BrickRed,hook,two heads,"{{2}}" {gray,pos=0.08,scale=0.8,below left},"{{2}}" {gray,pos=0.92,scale=0.8,above right}] & 0 & \mathbb{Z}_{2} \ar[lu,BrickRed,hook,two heads,"{{2}}" {gray,pos=0.08,scale=0.8,below left},"{{2}}" {gray,pos=0.92,scale=0.8,above right}] & 0 \\
H^{7} & \mathbb{Z}_{2} & \mathbb{Z}_{2} & \mathbb{Z}_{2} & \mathbb{Z}_{2} \ar[lu,hook,two heads,"{{2}}" {gray,pos=0.08,scale=0.8,below left},"{{2}}" {gray,pos=0.92,scale=0.8,above right}] & \mathbb{Z}_{2} \ar[lu,BrickRed,hook,two heads,"{{2}}" {gray,pos=0.08,scale=0.8,below left},"{{2}}" {gray,pos=0.92,scale=0.8,above right}] & 0 & \mathbb{Z}_{2} \ar[lu,BrickRed,hook,two heads,"{{2}}" {gray,pos=0.08,scale=0.8,below left},"{{2}}" {gray,pos=0.92,scale=0.8,above right}] \\
H^{8} & 0 & \mathbb{Z}_{2} \oplus \mathbb{Z}_{8} & \mathbb{Z}_{2} & \mathbb{Z}_{4} \ar[lu,two heads,"{{4}}" {gray,pos=0.08,scale=0.8,below left},"{{2}}" {gray,pos=0.92,scale=0.8,above right}] & \mathbb{Z}_{2} \ar[lu,BrickRed,hook,two heads,"{{2}}" {gray,pos=0.08,scale=0.8,below left},"{{2}}" {gray,pos=0.92,scale=0.8,above right}] & \mathbb{Z}_{2} \ar[lu,BrickRed,hook,two heads,"{{2}}" {gray,pos=0.08,scale=0.8,below left},"{{2}}" {gray,pos=0.92,scale=0.8,above right}] & 0 \\
H^{9} & \mathbb{Z}_{2} & (\mathbb{Z}_{2})^{2} & (\mathbb{Z}_{2})^{3} \ar[lu,"{{2}}" {gray,pos=0.08,scale=0.8,below left},"{{2}}" {gray,pos=0.92,scale=0.8,above right}] & \mathbb{Z}_{2} & \mathbb{Z}_{2} \ar[lu,hook,"{{2}}" {gray,pos=0.08,scale=0.8,below left},"{{2}}" {gray,pos=0.92,scale=0.8,above right}] & \mathbb{Z}_{2} \ar[lu,BrickRed,hook,two heads,"{{2}}" {gray,pos=0.08,scale=0.8,below left},"{{2}}" {gray,pos=0.92,scale=0.8,above right}] & \mathbb{Z}_{2} \ar[lu,BrickRed,hook,two heads,"{{2}}" {gray,pos=0.08,scale=0.8,below left},"{{2}}" {gray,pos=0.92,scale=0.8,above right}] \\
H^{10} & 0 & (\mathbb{Z}_{2})^{2} & (\mathbb{Z}_{2})^{2} \ar[lu,"{{2}}" {gray,pos=0.08,scale=0.8,below left},"{{2}}" {gray,pos=0.92,scale=0.8,above right}] & (\mathbb{Z}_{2})^{2} \ar[lu,"{{2}}" {gray,pos=0.08,scale=0.8,below left},"{{2}}" {gray,pos=0.92,scale=0.8,above right}] & \mathbb{Z}_{2} \ar[lu,hook,two heads,"{{2}}" {gray,pos=0.08,scale=0.8,below left},"{{2}}" {gray,pos=0.92,scale=0.8,above right}] & \mathbb{Z}_{2} \ar[lu,BrickRed,hook,two heads,"{{2}}" {gray,pos=0.08,scale=0.8,below left},"{{2}}" {gray,pos=0.92,scale=0.8,above right}] & \mathbb{Z}_{2} \ar[lu,BrickRed,hook,two heads,"{{2}}" {gray,pos=0.08,scale=0.8,below left},"{{2}}" {gray,pos=0.92,scale=0.8,above right}] \\
H^{11} & \mathbb{Z}_{2} & (\mathbb{Z}_{2})^{3} & (\mathbb{Z}_{2})^{3} & (\mathbb{Z}_{2})^{3} \ar[lu,"{{2}}" {gray,pos=0.08,scale=0.8,below left},"{{2}}" {gray,pos=0.92,scale=0.8,above right}] & (\mathbb{Z}_{2})^{2} \ar[lu,"{{2}}" {gray,pos=0.08,scale=0.8,below left},"{{2}}" {gray,pos=0.92,scale=0.8,above right}] & \mathbb{Z}_{2} \ar[lu,hook,two heads,"{{2}}" {gray,pos=0.08,scale=0.8,below left},"{{2}}" {gray,pos=0.92,scale=0.8,above right}] & \mathbb{Z}_{2} \ar[lu,BrickRed,hook,two heads,"{{2}}" {gray,pos=0.08,scale=0.8,below left},"{{2}}" {gray,pos=0.92,scale=0.8,above right}] \\
H^{12} & 0 & (\mathbb{Z}_{2})^{2} \oplus \mathbb{Z}_{4} & (\mathbb{Z}_{2})^{3} \oplus \mathbb{Z}_{4} \ar[lu,"{{4}}" {gray,pos=0.08,scale=0.8,below left},"{{2}}" {gray,pos=0.92,scale=0.8,above right}] & (\mathbb{Z}_{2})^{2} \oplus \mathbb{Z}_{4} \ar[lu,"{{4}}" {gray,pos=0.08,scale=0.8,below left},"{{2}}" {gray,pos=0.92,scale=0.8,above right}] & (\mathbb{Z}_{2})^{2} \ar[lu,"{{2}}" {gray,pos=0.08,scale=0.8,below left},"{{2}}" {gray,pos=0.92,scale=0.8,above right}] & \mathbb{Z}_{2} \oplus \mathbb{Z}_{4} \ar[lu,two heads,"{{2,4}}" {gray,pos=0.08,scale=0.8,below left},"{{2^{2}}}" {gray,pos=0.92,scale=0.8,above right}] & \mathbb{Z}_{2} \ar[lu,BrickRed,hook,two heads,"{{2}}" {gray,pos=0.08,scale=0.8,below left},"{{2}}" {gray,pos=0.92,scale=0.8,above right}] \\
H^{13} & \mathbb{Z}_{2} & (\mathbb{Z}_{2})^{3} & (\mathbb{Z}_{2})^{5} & (\mathbb{Z}_{2})^{4} \ar[lu,"{{2}}" {gray,pos=0.08,scale=0.8,below left},"{{2}}" {gray,pos=0.92,scale=0.8,above right}] & (\mathbb{Z}_{2})^{3} \ar[lu,"{{2}}" {gray,pos=0.08,scale=0.8,below left},"{{2}}" {gray,pos=0.92,scale=0.8,above right}] & (\mathbb{Z}_{2})^{2} \ar[lu,"{{2}}" {gray,pos=0.08,scale=0.8,below left},"{{2}}" {gray,pos=0.92,scale=0.8,above right}] & (\mathbb{Z}_{2})^{2} \ar[lu,hook,"{{2^{2}}}" {gray,pos=0.08,scale=0.8,below left},"{{2^{2}}}" {gray,pos=0.92,scale=0.8,above right}] \\
H^{14} & 0 & (\mathbb{Z}_{2})^{5} & (\mathbb{Z}_{2})^{5} & (\mathbb{Z}_{2})^{5} \ar[lu,"{{2}}" {gray,pos=0.08,scale=0.8,below left},"{{2}}" {gray,pos=0.92,scale=0.8,above right}] & (\mathbb{Z}_{2})^{4} \ar[lu,"{{2^{2}}}" {gray,pos=0.08,scale=0.8,below left},"{{2^{2}}}" {gray,pos=0.92,scale=0.8,above right}] & (\mathbb{Z}_{2})^{2} \ar[lu,"{{2}}" {gray,pos=0.08,scale=0.8,below left},"{{2}}" {gray,pos=0.92,scale=0.8,above right}] & (\mathbb{Z}_{2})^{2} \ar[lu,hook,two heads,"{{2^{2}}}" {gray,pos=0.08,scale=0.8,below left},"{{2^{2}}}" {gray,pos=0.92,scale=0.8,above right}] \\
H^{15} & \mathbb{Z}_{2} & (\mathbb{Z}_{2})^{4} & (\mathbb{Z}_{2})^{8} & (\mathbb{Z}_{2})^{6} & (\mathbb{Z}_{2})^{5} \ar[lu,"{{2}}" {gray,pos=0.08,scale=0.8,below left},"{{2}}" {gray,pos=0.92,scale=0.8,above right}] & (\mathbb{Z}_{2})^{4} \ar[lu,"{{2^{2}}}" {gray,pos=0.08,scale=0.8,below left},"{{2^{2}}}" {gray,pos=0.92,scale=0.8,above right}] & (\mathbb{Z}_{2})^{2} \ar[lu,"{{2}}" {gray,pos=0.08,scale=0.8,below left},"{{2}}" {gray,pos=0.92,scale=0.8,above right}] \\
H^{16} & 0 & (\mathbb{Z}_{2})^{4} \oplus \mathbb{Z}_{16} & (\mathbb{Z}_{2})^{8} & (\mathbb{Z}_{2})^{6} \oplus \mathbb{Z}_{8} & (\mathbb{Z}_{2})^{4} \oplus \mathbb{Z}_{4} \ar[lu,"{{4}}" {gray,pos=0.08,scale=0.8,below left},"{{2}}" {gray,pos=0.92,scale=0.8,above right}] & (\mathbb{Z}_{2})^{3} \oplus \mathbb{Z}_{4} \ar[lu,"{{2,4}}" {gray,pos=0.08,scale=0.8,below left},"{{2^{2}}}" {gray,pos=0.92,scale=0.8,above right}] & (\mathbb{Z}_{2})^{3} \ar[lu,"{{2^{2}}}" {gray,pos=0.08,scale=0.8,below left},"{{2^{2}}}" {gray,pos=0.92,scale=0.8,above right}] \\
H^{17} & \mathbb{Z}_{2} & (\mathbb{Z}_{2})^{7} & (\mathbb{Z}_{2})^{11} \ar[lu,"{{2}}" {gray,pos=0.08,scale=0.8,below left},"{{2}}" {gray,pos=0.92,scale=0.8,above right}] & (\mathbb{Z}_{2})^{10} & (\mathbb{Z}_{2})^{8} \ar[lu,"{{2}}" {gray,pos=0.08,scale=0.8,below left},"{{2}}" {gray,pos=0.92,scale=0.8,above right}] & (\mathbb{Z}_{2})^{5} \ar[lu,"{{2}}" {gray,pos=0.08,scale=0.8,below left},"{{2}}" {gray,pos=0.92,scale=0.8,above right}] & (\mathbb{Z}_{2})^{4} \ar[lu,"{{2^{2}}}" {gray,pos=0.08,scale=0.8,below left},"{{2^{2}}}" {gray,pos=0.92,scale=0.8,above right}] \\
H^{18} & 0 & (\mathbb{Z}_{2})^{6} & (\mathbb{Z}_{2})^{13} \ar[lu,"{{2}}" {gray,pos=0.08,scale=0.8,below left},"{{2}}" {gray,pos=0.92,scale=0.8,above right}] & (\mathbb{Z}_{2})^{13} \ar[lu,"{{2}}" {gray,pos=0.08,scale=0.8,below left},"{{2}}" {gray,pos=0.92,scale=0.8,above right}] & (\mathbb{Z}_{2})^{8} \ar[lu,"{{2}}" {gray,pos=0.08,scale=0.8,below left},"{{2}}" {gray,pos=0.92,scale=0.8,above right}] & (\mathbb{Z}_{2})^{7} \ar[lu,"{{2^{2}}}" {gray,pos=0.08,scale=0.8,below left},"{{2^{2}}}" {gray,pos=0.92,scale=0.8,above right}] & (\mathbb{Z}_{2})^{4} \ar[lu,"{{2^{2}}}" {gray,pos=0.08,scale=0.8,below left},"{{2^{2}}}" {gray,pos=0.92,scale=0.8,above right}] \\
H^{19} & \mathbb{Z}_{2} & (\mathbb{Z}_{2})^{8} & (\mathbb{Z}_{2})^{15} & (\mathbb{Z}_{2})^{15} \ar[lu,"{{2}}" {gray,pos=0.08,scale=0.8,below left},"{{2}}" {gray,pos=0.92,scale=0.8,above right}] & (\mathbb{Z}_{2})^{12} \ar[lu,"{{2}}" {gray,pos=0.08,scale=0.8,below left},"{{2}}" {gray,pos=0.92,scale=0.8,above right}] & (\mathbb{Z}_{2})^{8} \ar[lu,"{{2}}" {gray,pos=0.08,scale=0.8,below left},"{{2}}" {gray,pos=0.92,scale=0.8,above right}] & (\mathbb{Z}_{2})^{6} \ar[lu,"{{2^{2}}}" {gray,pos=0.08,scale=0.8,below left},"{{2^{2}}}" {gray,pos=0.92,scale=0.8,above right}] \\
H^{20} & 0 & (\mathbb{Z}_{2})^{8} \oplus \mathbb{Z}_{4} & (\mathbb{Z}_{2})^{17} \oplus \mathbb{Z}_{4} \ar[lu,"{{4}}" {gray,pos=0.08,scale=0.8,below left},"{{2}}" {gray,pos=0.92,scale=0.8,above right}] & (\mathbb{Z}_{2})^{16} \oplus (\mathbb{Z}_{4})^{2} \ar[lu,"{{4}}" {gray,pos=0.08,scale=0.8,below left},"{{2}}" {gray,pos=0.92,scale=0.8,above right}] & (\mathbb{Z}_{2})^{14} \oplus \mathbb{Z}_{4} \ar[lu,"{{2,4}}" {gray,pos=0.08,scale=0.8,below left},"{{2^{2}}}" {gray,pos=0.92,scale=0.8,above right}] & (\mathbb{Z}_{2})^{9} \oplus \mathbb{Z}_{4} \ar[lu,"{{2,4}}" {gray,pos=0.08,scale=0.8,below left},"{{2^{2}}}" {gray,pos=0.92,scale=0.8,above right}] & (\mathbb{Z}_{2})^{6} \oplus \mathbb{Z}_{4} \ar[lu,"{{2,4}}" {gray,pos=0.08,scale=0.8,below left},"{{2^{2}}}" {gray,pos=0.92,scale=0.8,above right}] \\
H^{21} & \mathbb{Z}_{2} & (\mathbb{Z}_{2})^{9} & (\mathbb{Z}_{2})^{23} & (\mathbb{Z}_{2})^{22} \oplus \mathbb{Z}_{4} \ar[lu,"{{2}}" {gray,pos=0.08,scale=0.8,below left},"{{2}}" {gray,pos=0.92,scale=0.8,above right}] & (\mathbb{Z}_{2})^{18} \ar[lu,"{{2}}" {gray,pos=0.08,scale=0.8,below left},"{{2}}" {gray,pos=0.92,scale=0.8,above right}] & (\mathbb{Z}_{2})^{14} \ar[lu,"{{2^{2}}}" {gray,pos=0.08,scale=0.8,below left},"{{2^{2}}}" {gray,pos=0.92,scale=0.8,above right}] & (\mathbb{Z}_{2})^{10} \ar[lu,"{{2^{2}}}" {gray,pos=0.08,scale=0.8,below left},"{{2^{2}}}" {gray,pos=0.92,scale=0.8,above right}] \\
H^{22} & 0 & (\mathbb{Z}_{2})^{11} & (\mathbb{Z}_{2})^{25} & (\mathbb{Z}_{2})^{27} \ar[lu,"{{2}}" {gray,pos=0.08,scale=0.8,below left},"{{2}}" {gray,pos=0.92,scale=0.8,above right}] & (\mathbb{Z}_{2})^{22} \ar[lu,"{{2^{2}}}" {gray,pos=0.08,scale=0.8,below left},"{{2^{2}}}" {gray,pos=0.92,scale=0.8,above right}] & (\mathbb{Z}_{2})^{16} \ar[lu,"{{2^{2}}}" {gray,pos=0.08,scale=0.8,below left},"{{2^{2}}}" {gray,pos=0.92,scale=0.8,above right}] & (\mathbb{Z}_{2})^{11} \ar[lu,"{{2^{3}}}" {gray,pos=0.08,scale=0.8,below left},"{{2^{3}}}" {gray,pos=0.92,scale=0.8,above right}] \\
H^{23} & \mathbb{Z}_{2} & (\mathbb{Z}_{2})^{12} & (\mathbb{Z}_{2})^{31} & (\mathbb{Z}_{2})^{33} & (\mathbb{Z}_{2})^{28} \ar[lu,"{{2}}" {gray,pos=0.08,scale=0.8,below left},"{{2}}" {gray,pos=0.92,scale=0.8,above right}] & (\mathbb{Z}_{2})^{20} \ar[lu,"{{2^{2}}}" {gray,pos=0.08,scale=0.8,below left},"{{2^{2}}}" {gray,pos=0.92,scale=0.8,above right}] & (\mathbb{Z}_{2})^{15} \ar[lu,"{{2^{2}}}" {gray,pos=0.08,scale=0.8,below left},"{{2^{2}}}" {gray,pos=0.92,scale=0.8,above right}] \\
H^{24} & 0 & (\mathbb{Z}_{2})^{12} \oplus \mathbb{Z}_{8} & (\mathbb{Z}_{2})^{36} \oplus \mathbb{Z}_{8} & (\mathbb{Z}_{2})^{39} \oplus \mathbb{Z}_{4} \oplus \mathbb{Z}_{8} & (\mathbb{Z}_{2})^{31} \oplus \mathbb{Z}_{4} \ar[lu,"{{4}}" {gray,pos=0.08,scale=0.8,below left},"{{2}}" {gray,pos=0.92,scale=0.8,above right}] & (\mathbb{Z}_{2})^{23} \oplus \mathbb{Z}_{4} \oplus \mathbb{Z}_{8} \ar[lu,"{{2,4}}" {gray,pos=0.08,scale=0.8,below left},"{{2^{2}}}" {gray,pos=0.92,scale=0.8,above right}] & (\mathbb{Z}_{2})^{17} \oplus \mathbb{Z}_{4} \ar[lu,"{{2^{2},4}}" {gray,pos=0.08,scale=0.8,below left},"{{2^{3}}}" {gray,pos=0.92,scale=0.8,above right}] \\
H^{25} & \mathbb{Z}_{2} & (\mathbb{Z}_{2})^{14} & (\mathbb{Z}_{2})^{43} & (\mathbb{Z}_{2})^{52} \ar[lu,"{{2}}" {gray,pos=0.08,scale=0.8,below left},"{{2}}" {gray,pos=0.92,scale=0.8,above right}] & (\mathbb{Z}_{2})^{41} \ar[lu,"{{2}}" {gray,pos=0.08,scale=0.8,below left},"{{2}}" {gray,pos=0.92,scale=0.8,above right}] & (\mathbb{Z}_{2})^{30} \ar[lu,"{{2}}" {gray,pos=0.08,scale=0.8,below left},"{{2}}" {gray,pos=0.92,scale=0.8,above right}] & (\mathbb{Z}_{2})^{22} \ar[lu,"{{2^{3}}}" {gray,pos=0.08,scale=0.8,below left},"{{2^{3}}}" {gray,pos=0.92,scale=0.8,above right}] \\
H^{26} & 0 & (\mathbb{Z}_{2})^{17} & (\mathbb{Z}_{2})^{49} & (\mathbb{Z}_{2})^{60} \ar[lu,"{{2}}" {gray,pos=0.08,scale=0.8,below left},"{{2}}" {gray,pos=0.92,scale=0.8,above right}] & (\mathbb{Z}_{2})^{51} \ar[lu,"{{2^{2}}}" {gray,pos=0.08,scale=0.8,below left},"{{2^{2}}}" {gray,pos=0.92,scale=0.8,above right}] & (\mathbb{Z}_{2})^{36} \ar[lu,"{{2^{2}}}" {gray,pos=0.08,scale=0.8,below left},"{{2^{2}}}" {gray,pos=0.92,scale=0.8,above right}] & (\mathbb{Z}_{2})^{26} \ar[lu,"{{2^{3}}}" {gray,pos=0.08,scale=0.8,below left},"{{2^{3}}}" {gray,pos=0.92,scale=0.8,above right}] \\
H^{27} & \mathbb{Z}_{2} & (\mathbb{Z}_{2})^{17} & (\mathbb{Z}_{2})^{61} & (\mathbb{Z}_{2})^{72} & (\mathbb{Z}_{2})^{62} \ar[lu,"{{2}}" {gray,pos=0.08,scale=0.8,below left},"{{2}}" {gray,pos=0.92,scale=0.8,above right}] & (\mathbb{Z}_{2})^{47} \ar[lu,"{{2^{2}}}" {gray,pos=0.08,scale=0.8,below left},"{{2^{2}}}" {gray,pos=0.92,scale=0.8,above right}] & (\mathbb{Z}_{2})^{33} \ar[lu,"{{2^{2}}}" {gray,pos=0.08,scale=0.8,below left},"{{2^{2}}}" {gray,pos=0.92,scale=0.8,above right}] \\
H^{28} & 0 & (\mathbb{Z}_{2})^{18} \oplus \mathbb{Z}_{4} & (\mathbb{Z}_{2})^{68} & (\mathbb{Z}_{2})^{84} \oplus (\mathbb{Z}_{4})^{3} \ar[lu,"{{4}}" {gray,pos=0.08,scale=0.8,below left},"{{2}}" {gray,pos=0.92,scale=0.8,above right}] & (\mathbb{Z}_{2})^{73} \oplus (\mathbb{Z}_{4})^{2} \ar[lu,"{{4^{2}}}" {gray,pos=0.08,scale=0.8,below left},"{{2^{2}}}" {gray,pos=0.92,scale=0.8,above right}] & (\mathbb{Z}_{2})^{53} \oplus (\mathbb{Z}_{4})^{2} \ar[lu,"{{2,4}}" {gray,pos=0.08,scale=0.8,below left},"{{2^{2}}}" {gray,pos=0.92,scale=0.8,above right}] & (\mathbb{Z}_{2})^{37} \oplus (\mathbb{Z}_{4})^{2} \ar[lu,"{{2^{2},4^{2}}}" {gray,pos=0.08,scale=0.8,below left},"{{2^{4}}}" {gray,pos=0.92,scale=0.8,above right}] \\
H^{29} & \mathbb{Z}_{2} & (\mathbb{Z}_{2})^{22} & (\mathbb{Z}_{2})^{80} & (\mathbb{Z}_{2})^{104} \oplus (\mathbb{Z}_{4})^{2} & (\mathbb{Z}_{2})^{94} \ar[lu,"{{2}}" {gray,pos=0.08,scale=0.8,below left},"{{2}}" {gray,pos=0.92,scale=0.8,above right}] & (\mathbb{Z}_{2})^{68} \oplus \mathbb{Z}_{4} \ar[lu,"{{2^{2}}}" {gray,pos=0.08,scale=0.8,below left},"{{2^{2}}}" {gray,pos=0.92,scale=0.8,above right}] & (\mathbb{Z}_{2})^{49} \ar[lu,"{{2^{2}}}" {gray,pos=0.08,scale=0.8,below left},"{{2^{2}}}" {gray,pos=0.92,scale=0.8,above right}] \\
H^{30} & 0 & (\mathbb{Z}_{2})^{22} & (\mathbb{Z}_{2})^{95} & (\mathbb{Z}_{2})^{126} & (\mathbb{Z}_{2})^{110} \ar[lu,"{{2}}" {gray,pos=0.08,scale=0.8,below left},"{{2}}" {gray,pos=0.92,scale=0.8,above right}] & (\mathbb{Z}_{2})^{85} \ar[lu,"{{2^{3}}}" {gray,pos=0.08,scale=0.8,below left},"{{2^{3}}}" {gray,pos=0.92,scale=0.8,above right}] & (\mathbb{Z}_{2})^{59} \ar[lu,"{{2^{3}}}" {gray,pos=0.08,scale=0.8,below left},"{{2^{3}}}" {gray,pos=0.92,scale=0.8,above right}] \\
H^{31} & \mathbb{Z}_{2} & (\mathbb{Z}_{2})^{25} & (\mathbb{Z}_{2})^{108} & (\mathbb{Z}_{2})^{151} & (\mathbb{Z}_{2})^{135} & (\mathbb{Z}_{2})^{102} \ar[lu,"{{2}}" {gray,pos=0.08,scale=0.8,below left},"{{2}}" {gray,pos=0.92,scale=0.8,above right}] & (\mathbb{Z}_{2})^{73} \ar[lu,"{{2^{3}}}" {gray,pos=0.08,scale=0.8,below left},"{{2^{3}}}" {gray,pos=0.92,scale=0.8,above right}] \\
H^{32} & 0 & (\mathbb{Z}_{2})^{27} \oplus \mathbb{Z}_{32} & (\mathbb{Z}_{2})^{123} \oplus \mathbb{Z}_{4} & (\mathbb{Z}_{2})^{180} \oplus (\mathbb{Z}_{4})^{2} \oplus \mathbb{Z}_{16} & (\mathbb{Z}_{2})^{163} \oplus \mathbb{Z}_{8} & (\mathbb{Z}_{2})^{120} \oplus (\mathbb{Z}_{4})^{2} \oplus \mathbb{Z}_{8} \ar[lu,"{{4}}" {gray,pos=0.08,scale=0.8,below left},"{{2}}" {gray,pos=0.92,scale=0.8,above right}] & (\mathbb{Z}_{2})^{86} \oplus (\mathbb{Z}_{4})^{2} \ar[lu,"{{2,4^{2}}}" {gray,pos=0.08,scale=0.8,below left},"{{2^{3}}}" {gray,pos=0.92,scale=0.8,above right}] \\
H^{33} & \mathbb{Z}_{2} & (\mathbb{Z}_{2})^{29} & (\mathbb{Z}_{2})^{146} \oplus \mathbb{Z}_{4} \ar[lu,"{{2}}" {gray,pos=0.08,scale=0.8,below left},"{{2}}" {gray,pos=0.92,scale=0.8,above right}] & (\mathbb{Z}_{2})^{214} \oplus (\mathbb{Z}_{4})^{2} & (\mathbb{Z}_{2})^{198} \ar[lu,"{{2}}" {gray,pos=0.08,scale=0.8,below left},"{{2}}" {gray,pos=0.92,scale=0.8,above right}] & (\mathbb{Z}_{2})^{152} \oplus \mathbb{Z}_{4} \ar[lu,"{{2}}" {gray,pos=0.08,scale=0.8,below left},"{{2}}" {gray,pos=0.92,scale=0.8,above right}] & (\mathbb{Z}_{2})^{108} \ar[lu,"{{2^{2}}}" {gray,pos=0.08,scale=0.8,below left},"{{2^{2}}}" {gray,pos=0.92,scale=0.8,above right}] \\
H^{34} & 0 & (\mathbb{Z}_{2})^{32} & (\mathbb{Z}_{2})^{167} \ar[lu,"{{2}}" {gray,pos=0.08,scale=0.8,below left},"{{2}}" {gray,pos=0.92,scale=0.8,above right}] & (\mathbb{Z}_{2})^{255} \ar[lu,"{{2}}" {gray,pos=0.08,scale=0.8,below left},"{{2}}" {gray,pos=0.92,scale=0.8,above right}] & (\mathbb{Z}_{2})^{240} \ar[lu,"{{2}}" {gray,pos=0.08,scale=0.8,below left},"{{2}}" {gray,pos=0.92,scale=0.8,above right}] & (\mathbb{Z}_{2})^{183} \ar[lu,"{{2^{2}}}" {gray,pos=0.08,scale=0.8,below left},"{{2^{2}}}" {gray,pos=0.92,scale=0.8,above right}] & (\mathbb{Z}_{2})^{130} \ar[lu,"{{2^{3}}}" {gray,pos=0.08,scale=0.8,below left},"{{2^{3}}}" {gray,pos=0.92,scale=0.8,above right}] \\
H^{35} & \mathbb{Z}_{2} & (\mathbb{Z}_{2})^{36} & (\mathbb{Z}_{2})^{192} & (\mathbb{Z}_{2})^{304} \ar[lu,"{{2}}" {gray,pos=0.08,scale=0.8,below left},"{{2}}" {gray,pos=0.92,scale=0.8,above right}] & (\mathbb{Z}_{2})^{292} \ar[lu,"{{2}}" {gray,pos=0.08,scale=0.8,below left},"{{2}}" {gray,pos=0.92,scale=0.8,above right}] & (\mathbb{Z}_{2})^{222} \ar[lu,"{{2}}" {gray,pos=0.08,scale=0.8,below left},"{{2}}" {gray,pos=0.92,scale=0.8,above right}] & (\mathbb{Z}_{2})^{160} \ar[lu,"{{2^{2}}}" {gray,pos=0.08,scale=0.8,below left},"{{2^{2}}}" {gray,pos=0.92,scale=0.8,above right}] \\
H^{36} & 0 & (\mathbb{Z}_{2})^{36} \oplus \mathbb{Z}_{4} & (\mathbb{Z}_{2})^{220} \oplus (\mathbb{Z}_{4})^{2} \ar[lu,"{{4}}" {gray,pos=0.08,scale=0.8,below left},"{{2}}" {gray,pos=0.92,scale=0.8,above right}] & (\mathbb{Z}_{2})^{356} \oplus (\mathbb{Z}_{4})^{5} \ar[lu,"{{4}}" {gray,pos=0.08,scale=0.8,below left},"{{2}}" {gray,pos=0.92,scale=0.8,above right}] & (\mathbb{Z}_{2})^{344} \oplus (\mathbb{Z}_{4})^{2} \ar[lu,"{{2,4}}" {gray,pos=0.08,scale=0.8,below left},"{{2^{2}}}" {gray,pos=0.92,scale=0.8,above right}] & (\mathbb{Z}_{2})^{267} \oplus (\mathbb{Z}_{4})^{5} \ar[lu,"{{2,4^{2}}}" {gray,pos=0.08,scale=0.8,below left},"{{2^{3}}}" {gray,pos=0.92,scale=0.8,above right}] & (\mathbb{Z}_{2})^{190} \oplus (\mathbb{Z}_{4})^{2} \ar[lu,"{{2,4^{2}}}" {gray,pos=0.08,scale=0.8,below left},"{{2^{3}}}" {gray,pos=0.92,scale=0.8,above right}] \\
H^{37} & \mathbb{Z}_{2} & (\mathbb{Z}_{2})^{41} & (\mathbb{Z}_{2})^{254} & (\mathbb{Z}_{2})^{425} \oplus (\mathbb{Z}_{4})^{3} \ar[lu,"{{2}}" {gray,pos=0.08,scale=0.8,below left},"{{2}}" {gray,pos=0.92,scale=0.8,above right}] & (\mathbb{Z}_{2})^{417} \oplus \mathbb{Z}_{4} \ar[lu,"{{2}}" {gray,pos=0.08,scale=0.8,below left},"{{2}}" {gray,pos=0.92,scale=0.8,above right}] & (\mathbb{Z}_{2})^{325} \oplus (\mathbb{Z}_{4})^{2} \ar[lu,"{{2^{2}}}" {gray,pos=0.08,scale=0.8,below left},"{{2^{2}}}" {gray,pos=0.92,scale=0.8,above right}] & (\mathbb{Z}_{2})^{235} \ar[lu,"{{2^{3}}}" {gray,pos=0.08,scale=0.8,below left},"{{2^{3}}}" {gray,pos=0.92,scale=0.8,above right}] \\
H^{38} & 0 & (\mathbb{Z}_{2})^{45} & (\mathbb{Z}_{2})^{289} & (\mathbb{Z}_{2})^{504} \ar[lu,"{{2}}" {gray,pos=0.08,scale=0.8,below left},"{{2}}" {gray,pos=0.92,scale=0.8,above right}] & (\mathbb{Z}_{2})^{501} \ar[lu,"{{2^{2}}}" {gray,pos=0.08,scale=0.8,below left},"{{2^{2}}}" {gray,pos=0.92,scale=0.8,above right}] & (\mathbb{Z}_{2})^{392} \ar[lu,"{{2^{2}}}" {gray,pos=0.08,scale=0.8,below left},"{{2^{2}}}" {gray,pos=0.92,scale=0.8,above right}] & (\mathbb{Z}_{2})^{283} \ar[lu,"{{2^{4}}}" {gray,pos=0.08,scale=0.8,below left},"{{2^{4}}}" {gray,pos=0.92,scale=0.8,above right}] \\
H^{39} & \mathbb{Z}_{2} & (\mathbb{Z}_{2})^{47} & (\mathbb{Z}_{2})^{334} & (\mathbb{Z}_{2})^{598} & (\mathbb{Z}_{2})^{597} \ar[lu,"{{2}}" {gray,pos=0.08,scale=0.8,below left},"{{2}}" {gray,pos=0.92,scale=0.8,above right}] & (\mathbb{Z}_{2})^{477} \ar[lu,"{{2^{2}}}" {gray,pos=0.08,scale=0.8,below left},"{{2^{2}}}" {gray,pos=0.92,scale=0.8,above right}] & (\mathbb{Z}_{2})^{342} \ar[lu,"{{2^{2}}}" {gray,pos=0.08,scale=0.8,below left},"{{2^{2}}}" {gray,pos=0.92,scale=0.8,above right}] \\
H^{40} & 0 & (\mathbb{Z}_{2})^{50} \oplus \mathbb{Z}_{8} & (\mathbb{Z}_{2})^{377} \oplus \mathbb{Z}_{8} & (\mathbb{Z}_{2})^{696} \oplus (\mathbb{Z}_{4})^{3} \oplus (\mathbb{Z}_{8})^{2} & (\mathbb{Z}_{2})^{713} \oplus (\mathbb{Z}_{4})^{2} \oplus \mathbb{Z}_{8} \ar[lu,"{{4}}" {gray,pos=0.08,scale=0.8,below left},"{{2}}" {gray,pos=0.92,scale=0.8,above right}] & (\mathbb{Z}_{2})^{565} \oplus (\mathbb{Z}_{4})^{4} \oplus \mathbb{Z}_{8} \ar[lu,"{{2,4}}" {gray,pos=0.08,scale=0.8,below left},"{{2^{2}}}" {gray,pos=0.92,scale=0.8,above right}] & (\mathbb{Z}_{2})^{408} \oplus \mathbb{Z}_{4} \oplus \mathbb{Z}_{8} \ar[lu,"{{2^{2},4}}" {gray,pos=0.08,scale=0.8,below left},"{{2^{3}}}" {gray,pos=0.92,scale=0.8,above right}] \\
\end{tikzcd}

%% file: tables/omega-Z2-Z2.tex
\begin{tikzcd}[row sep=tiny,column sep=large]
& B\mathbb{Z}_{2} & B^{2}\mathbb{Z}_{2} & B^{3}\mathbb{Z}_{2} & B^{4}\mathbb{Z}_{2} & B^{5}\mathbb{Z}_{2} & B^{6}\mathbb{Z}_{2} & B^{7}\mathbb{Z}_{2} & B^{8}\mathbb{Z}_{2} \\
H^{0} & \mathbb{Z}_{2} & \mathbb{Z}_{2} & \mathbb{Z}_{2} & \mathbb{Z}_{2} & \mathbb{Z}_{2} & \mathbb{Z}_{2} & \mathbb{Z}_{2} & \mathbb{Z}_{2} \\
H^{1} & \mathbb{Z}_{2} & 0 & 0 & 0 & 0 & 0 & 0 & 0 \\
H^{2} & \mathbb{Z}_{2} & \mathbb{Z}_{2} \ar[lu,BrickRed,hook,two heads,"{{2}}" {gray,pos=0.08,scale=0.8,below left},"{{2}}" {gray,pos=0.92,scale=0.8,above right}] & 0 & 0 & 0 & 0 & 0 & 0 \\
H^{3} & \mathbb{Z}_{2} & \mathbb{Z}_{2} \ar[lu,hook,two heads,"{{2}}" {gray,pos=0.08,scale=0.8,below left},"{{2}}" {gray,pos=0.92,scale=0.8,above right}] & \mathbb{Z}_{2} \ar[lu,BrickRed,hook,two heads,"{{2}}" {gray,pos=0.08,scale=0.8,below left},"{{2}}" {gray,pos=0.92,scale=0.8,above right}] & 0 & 0 & 0 & 0 & 0 \\
H^{4} & \mathbb{Z}_{2} & \mathbb{Z}_{2} & \mathbb{Z}_{2} \ar[lu,BrickRed,hook,two heads,"{{2}}" {gray,pos=0.08,scale=0.8,below left},"{{2}}" {gray,pos=0.92,scale=0.8,above right}] & \mathbb{Z}_{2} \ar[lu,BrickRed,hook,two heads,"{{2}}" {gray,pos=0.08,scale=0.8,below left},"{{2}}" {gray,pos=0.92,scale=0.8,above right}] & 0 & 0 & 0 & 0 \\
H^{5} & \mathbb{Z}_{2} & (\mathbb{Z}_{2})^{2} \ar[lu,two heads,"{{2}}" {gray,pos=0.08,scale=0.8,below left},"{{2}}" {gray,pos=0.92,scale=0.8,above right}] & \mathbb{Z}_{2} \ar[lu,hook,two heads,"{{2}}" {gray,pos=0.08,scale=0.8,below left},"{{2}}" {gray,pos=0.92,scale=0.8,above right}] & \mathbb{Z}_{2} \ar[lu,BrickRed,hook,two heads,"{{2}}" {gray,pos=0.08,scale=0.8,below left},"{{2}}" {gray,pos=0.92,scale=0.8,above right}] & \mathbb{Z}_{2} \ar[lu,BrickRed,hook,two heads,"{{2}}" {gray,pos=0.08,scale=0.8,below left},"{{2}}" {gray,pos=0.92,scale=0.8,above right}] & 0 & 0 & 0 \\
H^{6} & \mathbb{Z}_{2} & (\mathbb{Z}_{2})^{2} & (\mathbb{Z}_{2})^{2} \ar[lu,"{{2}}" {gray,pos=0.08,scale=0.8,below left},"{{2}}" {gray,pos=0.92,scale=0.8,above right}] & \mathbb{Z}_{2} \ar[lu,BrickRed,hook,two heads,"{{2}}" {gray,pos=0.08,scale=0.8,below left},"{{2}}" {gray,pos=0.92,scale=0.8,above right}] & \mathbb{Z}_{2} \ar[lu,BrickRed,hook,two heads,"{{2}}" {gray,pos=0.08,scale=0.8,below left},"{{2}}" {gray,pos=0.92,scale=0.8,above right}] & \mathbb{Z}_{2} \ar[lu,BrickRed,hook,two heads,"{{2}}" {gray,pos=0.08,scale=0.8,below left},"{{2}}" {gray,pos=0.92,scale=0.8,above right}] & 0 & 0 \\
H^{7} & \mathbb{Z}_{2} & (\mathbb{Z}_{2})^{2} & (\mathbb{Z}_{2})^{2} \ar[lu,"{{2}}" {gray,pos=0.08,scale=0.8,below left},"{{2}}" {gray,pos=0.92,scale=0.8,above right}] & (\mathbb{Z}_{2})^{2} \ar[lu,hook,two heads,"{{2^{2}}}" {gray,pos=0.08,scale=0.8,below left},"{{2^{2}}}" {gray,pos=0.92,scale=0.8,above right}] & \mathbb{Z}_{2} \ar[lu,BrickRed,hook,two heads,"{{2}}" {gray,pos=0.08,scale=0.8,below left},"{{2}}" {gray,pos=0.92,scale=0.8,above right}] & \mathbb{Z}_{2} \ar[lu,BrickRed,hook,two heads,"{{2}}" {gray,pos=0.08,scale=0.8,below left},"{{2}}" {gray,pos=0.92,scale=0.8,above right}] & \mathbb{Z}_{2} \ar[lu,BrickRed,hook,two heads,"{{2}}" {gray,pos=0.08,scale=0.8,below left},"{{2}}" {gray,pos=0.92,scale=0.8,above right}] & 0 \\
H^{8} & \mathbb{Z}_{2} & (\mathbb{Z}_{2})^{3} & (\mathbb{Z}_{2})^{2} & (\mathbb{Z}_{2})^{2} \ar[lu,"{{2}}" {gray,pos=0.08,scale=0.8,below left},"{{2}}" {gray,pos=0.92,scale=0.8,above right}] & (\mathbb{Z}_{2})^{2} \ar[lu,BrickRed,hook,two heads,"{{2^{2}}}" {gray,pos=0.08,scale=0.8,below left},"{{2^{2}}}" {gray,pos=0.92,scale=0.8,above right}] & \mathbb{Z}_{2} \ar[lu,BrickRed,hook,two heads,"{{2}}" {gray,pos=0.08,scale=0.8,below left},"{{2}}" {gray,pos=0.92,scale=0.8,above right}] & \mathbb{Z}_{2} \ar[lu,BrickRed,hook,two heads,"{{2}}" {gray,pos=0.08,scale=0.8,below left},"{{2}}" {gray,pos=0.92,scale=0.8,above right}] & \mathbb{Z}_{2} \ar[lu,BrickRed,hook,two heads,"{{2}}" {gray,pos=0.08,scale=0.8,below left},"{{2}}" {gray,pos=0.92,scale=0.8,above right}] \\
H^{9} & \mathbb{Z}_{2} & (\mathbb{Z}_{2})^{4} \ar[lu,two heads,"{{2}}" {gray,pos=0.08,scale=0.8,below left},"{{2}}" {gray,pos=0.92,scale=0.8,above right}] & (\mathbb{Z}_{2})^{4} \ar[lu,"{{2}}" {gray,pos=0.08,scale=0.8,below left},"{{2}}" {gray,pos=0.92,scale=0.8,above right}] & (\mathbb{Z}_{2})^{2} \ar[lu,"{{2}}" {gray,pos=0.08,scale=0.8,below left},"{{2}}" {gray,pos=0.92,scale=0.8,above right}] & (\mathbb{Z}_{2})^{2} \ar[lu,hook,two heads,"{{2^{2}}}" {gray,pos=0.08,scale=0.8,below left},"{{2^{2}}}" {gray,pos=0.92,scale=0.8,above right}] & (\mathbb{Z}_{2})^{2} \ar[lu,BrickRed,hook,two heads,"{{2^{2}}}" {gray,pos=0.08,scale=0.8,below left},"{{2^{2}}}" {gray,pos=0.92,scale=0.8,above right}] & \mathbb{Z}_{2} \ar[lu,BrickRed,hook,two heads,"{{2}}" {gray,pos=0.08,scale=0.8,below left},"{{2}}" {gray,pos=0.92,scale=0.8,above right}] & \mathbb{Z}_{2} \ar[lu,BrickRed,hook,two heads,"{{2}}" {gray,pos=0.08,scale=0.8,below left},"{{2}}" {gray,pos=0.92,scale=0.8,above right}] \\
H^{10} & \mathbb{Z}_{2} & (\mathbb{Z}_{2})^{4} & (\mathbb{Z}_{2})^{5} \ar[lu,"{{2}}" {gray,pos=0.08,scale=0.8,below left},"{{2}}" {gray,pos=0.92,scale=0.8,above right}] & (\mathbb{Z}_{2})^{3} \ar[lu,"{{2}}" {gray,pos=0.08,scale=0.8,below left},"{{2}}" {gray,pos=0.92,scale=0.8,above right}] & (\mathbb{Z}_{2})^{2} \ar[lu,"{{2}}" {gray,pos=0.08,scale=0.8,below left},"{{2}}" {gray,pos=0.92,scale=0.8,above right}] & (\mathbb{Z}_{2})^{2} \ar[lu,BrickRed,hook,two heads,"{{2^{2}}}" {gray,pos=0.08,scale=0.8,below left},"{{2^{2}}}" {gray,pos=0.92,scale=0.8,above right}] & (\mathbb{Z}_{2})^{2} \ar[lu,BrickRed,hook,two heads,"{{2^{2}}}" {gray,pos=0.08,scale=0.8,below left},"{{2^{2}}}" {gray,pos=0.92,scale=0.8,above right}] & \mathbb{Z}_{2} \ar[lu,BrickRed,hook,two heads,"{{2}}" {gray,pos=0.08,scale=0.8,below left},"{{2}}" {gray,pos=0.92,scale=0.8,above right}] \\
H^{11} & \mathbb{Z}_{2} & (\mathbb{Z}_{2})^{5} & (\mathbb{Z}_{2})^{5} \ar[lu,"{{2}}" {gray,pos=0.08,scale=0.8,below left},"{{2}}" {gray,pos=0.92,scale=0.8,above right}] & (\mathbb{Z}_{2})^{5} \ar[lu,"{{2^{2}}}" {gray,pos=0.08,scale=0.8,below left},"{{2^{2}}}" {gray,pos=0.92,scale=0.8,above right}] & (\mathbb{Z}_{2})^{3} \ar[lu,"{{2^{2}}}" {gray,pos=0.08,scale=0.8,below left},"{{2^{2}}}" {gray,pos=0.92,scale=0.8,above right}] & (\mathbb{Z}_{2})^{2} \ar[lu,hook,two heads,"{{2^{2}}}" {gray,pos=0.08,scale=0.8,below left},"{{2^{2}}}" {gray,pos=0.92,scale=0.8,above right}] & (\mathbb{Z}_{2})^{2} \ar[lu,BrickRed,hook,two heads,"{{2^{2}}}" {gray,pos=0.08,scale=0.8,below left},"{{2^{2}}}" {gray,pos=0.92,scale=0.8,above right}] & (\mathbb{Z}_{2})^{2} \ar[lu,BrickRed,hook,two heads,"{{2^{2}}}" {gray,pos=0.08,scale=0.8,below left},"{{2^{2}}}" {gray,pos=0.92,scale=0.8,above right}] \\
H^{12} & \mathbb{Z}_{2} & (\mathbb{Z}_{2})^{6} & (\mathbb{Z}_{2})^{7} & (\mathbb{Z}_{2})^{6} \ar[lu,"{{2}}" {gray,pos=0.08,scale=0.8,below left},"{{2}}" {gray,pos=0.92,scale=0.8,above right}] & (\mathbb{Z}_{2})^{4} \ar[lu,"{{2^{2}}}" {gray,pos=0.08,scale=0.8,below left},"{{2^{2}}}" {gray,pos=0.92,scale=0.8,above right}] & (\mathbb{Z}_{2})^{3} \ar[lu,"{{2^{2}}}" {gray,pos=0.08,scale=0.8,below left},"{{2^{2}}}" {gray,pos=0.92,scale=0.8,above right}] & (\mathbb{Z}_{2})^{2} \ar[lu,BrickRed,hook,two heads,"{{2^{2}}}" {gray,pos=0.08,scale=0.8,below left},"{{2^{2}}}" {gray,pos=0.92,scale=0.8,above right}] & (\mathbb{Z}_{2})^{2} \ar[lu,BrickRed,hook,two heads,"{{2^{2}}}" {gray,pos=0.08,scale=0.8,below left},"{{2^{2}}}" {gray,pos=0.92,scale=0.8,above right}] \\
H^{13} & \mathbb{Z}_{2} & (\mathbb{Z}_{2})^{6} & (\mathbb{Z}_{2})^{9} \ar[lu,"{{2}}" {gray,pos=0.08,scale=0.8,below left},"{{2}}" {gray,pos=0.92,scale=0.8,above right}] & (\mathbb{Z}_{2})^{7} \ar[lu,"{{2^{2}}}" {gray,pos=0.08,scale=0.8,below left},"{{2^{2}}}" {gray,pos=0.92,scale=0.8,above right}] & (\mathbb{Z}_{2})^{5} \ar[lu,"{{2^{2}}}" {gray,pos=0.08,scale=0.8,below left},"{{2^{2}}}" {gray,pos=0.92,scale=0.8,above right}] & (\mathbb{Z}_{2})^{4} \ar[lu,"{{2^{3}}}" {gray,pos=0.08,scale=0.8,below left},"{{2^{3}}}" {gray,pos=0.92,scale=0.8,above right}] & (\mathbb{Z}_{2})^{3} \ar[lu,hook,two heads,"{{2^{3}}}" {gray,pos=0.08,scale=0.8,below left},"{{2^{3}}}" {gray,pos=0.92,scale=0.8,above right}] & (\mathbb{Z}_{2})^{2} \ar[lu,BrickRed,hook,two heads,"{{2^{2}}}" {gray,pos=0.08,scale=0.8,below left},"{{2^{2}}}" {gray,pos=0.92,scale=0.8,above right}] \\
H^{14} & \mathbb{Z}_{2} & (\mathbb{Z}_{2})^{8} & (\mathbb{Z}_{2})^{10} & (\mathbb{Z}_{2})^{9} \ar[lu,"{{2}}" {gray,pos=0.08,scale=0.8,below left},"{{2}}" {gray,pos=0.92,scale=0.8,above right}] & (\mathbb{Z}_{2})^{7} \ar[lu,"{{2^{2}}}" {gray,pos=0.08,scale=0.8,below left},"{{2^{2}}}" {gray,pos=0.92,scale=0.8,above right}] & (\mathbb{Z}_{2})^{4} \ar[lu,"{{2^{2}}}" {gray,pos=0.08,scale=0.8,below left},"{{2^{2}}}" {gray,pos=0.92,scale=0.8,above right}] & (\mathbb{Z}_{2})^{4} \ar[lu,"{{2^{3}}}" {gray,pos=0.08,scale=0.8,below left},"{{2^{3}}}" {gray,pos=0.92,scale=0.8,above right}] & (\mathbb{Z}_{2})^{3} \ar[lu,BrickRed,hook,two heads,"{{2^{3}}}" {gray,pos=0.08,scale=0.8,below left},"{{2^{3}}}" {gray,pos=0.92,scale=0.8,above right}] \\
H^{15} & \mathbb{Z}_{2} & (\mathbb{Z}_{2})^{9} & (\mathbb{Z}_{2})^{13} & (\mathbb{Z}_{2})^{11} \ar[lu,"{{2}}" {gray,pos=0.08,scale=0.8,below left},"{{2}}" {gray,pos=0.92,scale=0.8,above right}] & (\mathbb{Z}_{2})^{9} \ar[lu,"{{2^{3}}}" {gray,pos=0.08,scale=0.8,below left},"{{2^{3}}}" {gray,pos=0.92,scale=0.8,above right}] & (\mathbb{Z}_{2})^{6} \ar[lu,"{{2^{3}}}" {gray,pos=0.08,scale=0.8,below left},"{{2^{3}}}" {gray,pos=0.92,scale=0.8,above right}] & (\mathbb{Z}_{2})^{4} \ar[lu,"{{2^{3}}}" {gray,pos=0.08,scale=0.8,below left},"{{2^{3}}}" {gray,pos=0.92,scale=0.8,above right}] & (\mathbb{Z}_{2})^{4} \ar[lu,hook,two heads,"{{2^{4}}}" {gray,pos=0.08,scale=0.8,below left},"{{2^{4}}}" {gray,pos=0.92,scale=0.8,above right}] \\
H^{16} & \mathbb{Z}_{2} & (\mathbb{Z}_{2})^{9} & (\mathbb{Z}_{2})^{16} & (\mathbb{Z}_{2})^{13} & (\mathbb{Z}_{2})^{10} \ar[lu,"{{2}}" {gray,pos=0.08,scale=0.8,below left},"{{2}}" {gray,pos=0.92,scale=0.8,above right}] & (\mathbb{Z}_{2})^{8} \ar[lu,"{{2^{3}}}" {gray,pos=0.08,scale=0.8,below left},"{{2^{3}}}" {gray,pos=0.92,scale=0.8,above right}] & (\mathbb{Z}_{2})^{5} \ar[lu,"{{2^{3}}}" {gray,pos=0.08,scale=0.8,below left},"{{2^{3}}}" {gray,pos=0.92,scale=0.8,above right}] & (\mathbb{Z}_{2})^{4} \ar[lu,"{{2^{3}}}" {gray,pos=0.08,scale=0.8,below left},"{{2^{3}}}" {gray,pos=0.92,scale=0.8,above right}] \\
H^{17} & \mathbb{Z}_{2} & (\mathbb{Z}_{2})^{12} \ar[lu,two heads,"{{2}}" {gray,pos=0.08,scale=0.8,below left},"{{2}}" {gray,pos=0.92,scale=0.8,above right}] & (\mathbb{Z}_{2})^{19} \ar[lu,"{{2}}" {gray,pos=0.08,scale=0.8,below left},"{{2}}" {gray,pos=0.92,scale=0.8,above right}] & (\mathbb{Z}_{2})^{17} \ar[lu,"{{2}}" {gray,pos=0.08,scale=0.8,below left},"{{2}}" {gray,pos=0.92,scale=0.8,above right}] & (\mathbb{Z}_{2})^{13} \ar[lu,"{{2^{2}}}" {gray,pos=0.08,scale=0.8,below left},"{{2^{2}}}" {gray,pos=0.92,scale=0.8,above right}] & (\mathbb{Z}_{2})^{9} \ar[lu,"{{2^{3}}}" {gray,pos=0.08,scale=0.8,below left},"{{2^{3}}}" {gray,pos=0.92,scale=0.8,above right}] & (\mathbb{Z}_{2})^{7} \ar[lu,"{{2^{4}}}" {gray,pos=0.08,scale=0.8,below left},"{{2^{4}}}" {gray,pos=0.92,scale=0.8,above right}] & (\mathbb{Z}_{2})^{5} \ar[lu,"{{2^{4}}}" {gray,pos=0.08,scale=0.8,below left},"{{2^{4}}}" {gray,pos=0.92,scale=0.8,above right}] \\
H^{18} & \mathbb{Z}_{2} & (\mathbb{Z}_{2})^{13} & (\mathbb{Z}_{2})^{24} \ar[lu,"{{2}}" {gray,pos=0.08,scale=0.8,below left},"{{2}}" {gray,pos=0.92,scale=0.8,above right}] & (\mathbb{Z}_{2})^{23} \ar[lu,"{{2}}" {gray,pos=0.08,scale=0.8,below left},"{{2}}" {gray,pos=0.92,scale=0.8,above right}] & (\mathbb{Z}_{2})^{16} \ar[lu,"{{2}}" {gray,pos=0.08,scale=0.8,below left},"{{2}}" {gray,pos=0.92,scale=0.8,above right}] & (\mathbb{Z}_{2})^{12} \ar[lu,"{{2^{2}}}" {gray,pos=0.08,scale=0.8,below left},"{{2^{2}}}" {gray,pos=0.92,scale=0.8,above right}] & (\mathbb{Z}_{2})^{8} \ar[lu,"{{2^{3}}}" {gray,pos=0.08,scale=0.8,below left},"{{2^{3}}}" {gray,pos=0.92,scale=0.8,above right}] & (\mathbb{Z}_{2})^{6} \ar[lu,"{{2^{4}}}" {gray,pos=0.08,scale=0.8,below left},"{{2^{4}}}" {gray,pos=0.92,scale=0.8,above right}] \\
H^{19} & \mathbb{Z}_{2} & (\mathbb{Z}_{2})^{14} & (\mathbb{Z}_{2})^{28} \ar[lu,"{{2}}" {gray,pos=0.08,scale=0.8,below left},"{{2}}" {gray,pos=0.92,scale=0.8,above right}] & (\mathbb{Z}_{2})^{28} \ar[lu,"{{2^{2}}}" {gray,pos=0.08,scale=0.8,below left},"{{2^{2}}}" {gray,pos=0.92,scale=0.8,above right}] & (\mathbb{Z}_{2})^{20} \ar[lu,"{{2^{2}}}" {gray,pos=0.08,scale=0.8,below left},"{{2^{2}}}" {gray,pos=0.92,scale=0.8,above right}] & (\mathbb{Z}_{2})^{15} \ar[lu,"{{2^{3}}}" {gray,pos=0.08,scale=0.8,below left},"{{2^{3}}}" {gray,pos=0.92,scale=0.8,above right}] & (\mathbb{Z}_{2})^{10} \ar[lu,"{{2^{4}}}" {gray,pos=0.08,scale=0.8,below left},"{{2^{4}}}" {gray,pos=0.92,scale=0.8,above right}] & (\mathbb{Z}_{2})^{7} \ar[lu,"{{2^{4}}}" {gray,pos=0.08,scale=0.8,below left},"{{2^{4}}}" {gray,pos=0.92,scale=0.8,above right}] \\
H^{20} & \mathbb{Z}_{2} & (\mathbb{Z}_{2})^{17} & (\mathbb{Z}_{2})^{33} & (\mathbb{Z}_{2})^{33} \ar[lu,"{{2}}" {gray,pos=0.08,scale=0.8,below left},"{{2}}" {gray,pos=0.92,scale=0.8,above right}] & (\mathbb{Z}_{2})^{27} \ar[lu,"{{2^{2}}}" {gray,pos=0.08,scale=0.8,below left},"{{2^{2}}}" {gray,pos=0.92,scale=0.8,above right}] & (\mathbb{Z}_{2})^{18} \ar[lu,"{{2^{2}}}" {gray,pos=0.08,scale=0.8,below left},"{{2^{2}}}" {gray,pos=0.92,scale=0.8,above right}] & (\mathbb{Z}_{2})^{13} \ar[lu,"{{2^{3}}}" {gray,pos=0.08,scale=0.8,below left},"{{2^{3}}}" {gray,pos=0.92,scale=0.8,above right}] & (\mathbb{Z}_{2})^{9} \ar[lu,"{{2^{4}}}" {gray,pos=0.08,scale=0.8,below left},"{{2^{4}}}" {gray,pos=0.92,scale=0.8,above right}] \\
H^{21} & \mathbb{Z}_{2} & (\mathbb{Z}_{2})^{18} & (\mathbb{Z}_{2})^{41} \ar[lu,"{{2}}" {gray,pos=0.08,scale=0.8,below left},"{{2}}" {gray,pos=0.92,scale=0.8,above right}] & (\mathbb{Z}_{2})^{41} \ar[lu,"{{2^{2}}}" {gray,pos=0.08,scale=0.8,below left},"{{2^{2}}}" {gray,pos=0.92,scale=0.8,above right}] & (\mathbb{Z}_{2})^{33} \ar[lu,"{{2^{3}}}" {gray,pos=0.08,scale=0.8,below left},"{{2^{3}}}" {gray,pos=0.92,scale=0.8,above right}] & (\mathbb{Z}_{2})^{24} \ar[lu,"{{2^{4}}}" {gray,pos=0.08,scale=0.8,below left},"{{2^{4}}}" {gray,pos=0.92,scale=0.8,above right}] & (\mathbb{Z}_{2})^{17} \ar[lu,"{{2^{4}}}" {gray,pos=0.08,scale=0.8,below left},"{{2^{4}}}" {gray,pos=0.92,scale=0.8,above right}] & (\mathbb{Z}_{2})^{11} \ar[lu,"{{2^{5}}}" {gray,pos=0.08,scale=0.8,below left},"{{2^{5}}}" {gray,pos=0.92,scale=0.8,above right}] \\
H^{22} & \mathbb{Z}_{2} & (\mathbb{Z}_{2})^{20} & (\mathbb{Z}_{2})^{48} & (\mathbb{Z}_{2})^{50} \ar[lu,"{{2}}" {gray,pos=0.08,scale=0.8,below left},"{{2}}" {gray,pos=0.92,scale=0.8,above right}] & (\mathbb{Z}_{2})^{40} \ar[lu,"{{2^{2}}}" {gray,pos=0.08,scale=0.8,below left},"{{2^{2}}}" {gray,pos=0.92,scale=0.8,above right}] & (\mathbb{Z}_{2})^{30} \ar[lu,"{{2^{3}}}" {gray,pos=0.08,scale=0.8,below left},"{{2^{3}}}" {gray,pos=0.92,scale=0.8,above right}] & (\mathbb{Z}_{2})^{21} \ar[lu,"{{2^{4}}}" {gray,pos=0.08,scale=0.8,below left},"{{2^{4}}}" {gray,pos=0.92,scale=0.8,above right}] & (\mathbb{Z}_{2})^{14} \ar[lu,"{{2^{4}}}" {gray,pos=0.08,scale=0.8,below left},"{{2^{4}}}" {gray,pos=0.92,scale=0.8,above right}] \\
H^{23} & \mathbb{Z}_{2} & (\mathbb{Z}_{2})^{23} & (\mathbb{Z}_{2})^{56} & (\mathbb{Z}_{2})^{60} \ar[lu,"{{2}}" {gray,pos=0.08,scale=0.8,below left},"{{2}}" {gray,pos=0.92,scale=0.8,above right}] & (\mathbb{Z}_{2})^{50} \ar[lu,"{{2^{3}}}" {gray,pos=0.08,scale=0.8,below left},"{{2^{3}}}" {gray,pos=0.92,scale=0.8,above right}] & (\mathbb{Z}_{2})^{36} \ar[lu,"{{2^{4}}}" {gray,pos=0.08,scale=0.8,below left},"{{2^{4}}}" {gray,pos=0.92,scale=0.8,above right}] & (\mathbb{Z}_{2})^{26} \ar[lu,"{{2^{5}}}" {gray,pos=0.08,scale=0.8,below left},"{{2^{5}}}" {gray,pos=0.92,scale=0.8,above right}] & (\mathbb{Z}_{2})^{19} \ar[lu,"{{2^{6}}}" {gray,pos=0.08,scale=0.8,below left},"{{2^{6}}}" {gray,pos=0.92,scale=0.8,above right}] \\
H^{24} & \mathbb{Z}_{2} & (\mathbb{Z}_{2})^{25} & (\mathbb{Z}_{2})^{68} & (\mathbb{Z}_{2})^{74} & (\mathbb{Z}_{2})^{60} \ar[lu,"{{2}}" {gray,pos=0.08,scale=0.8,below left},"{{2}}" {gray,pos=0.92,scale=0.8,above right}] & (\mathbb{Z}_{2})^{45} \ar[lu,"{{2^{3}}}" {gray,pos=0.08,scale=0.8,below left},"{{2^{3}}}" {gray,pos=0.92,scale=0.8,above right}] & (\mathbb{Z}_{2})^{33} \ar[lu,"{{2^{4}}}" {gray,pos=0.08,scale=0.8,below left},"{{2^{4}}}" {gray,pos=0.92,scale=0.8,above right}] & (\mathbb{Z}_{2})^{23} \ar[lu,"{{2^{5}}}" {gray,pos=0.08,scale=0.8,below left},"{{2^{5}}}" {gray,pos=0.92,scale=0.8,above right}] \\
H^{25} & \mathbb{Z}_{2} & (\mathbb{Z}_{2})^{27} & (\mathbb{Z}_{2})^{80} \ar[lu,"{{2}}" {gray,pos=0.08,scale=0.8,below left},"{{2}}" {gray,pos=0.92,scale=0.8,above right}] & (\mathbb{Z}_{2})^{93} \ar[lu,"{{2^{2}}}" {gray,pos=0.08,scale=0.8,below left},"{{2^{2}}}" {gray,pos=0.92,scale=0.8,above right}] & (\mathbb{Z}_{2})^{73} \ar[lu,"{{2^{2}}}" {gray,pos=0.08,scale=0.8,below left},"{{2^{2}}}" {gray,pos=0.92,scale=0.8,above right}] & (\mathbb{Z}_{2})^{55} \ar[lu,"{{2^{4}}}" {gray,pos=0.08,scale=0.8,below left},"{{2^{4}}}" {gray,pos=0.92,scale=0.8,above right}] & (\mathbb{Z}_{2})^{40} \ar[lu,"{{2^{6}}}" {gray,pos=0.08,scale=0.8,below left},"{{2^{6}}}" {gray,pos=0.92,scale=0.8,above right}] & (\mathbb{Z}_{2})^{28} \ar[lu,"{{2^{6}}}" {gray,pos=0.08,scale=0.8,below left},"{{2^{6}}}" {gray,pos=0.92,scale=0.8,above right}] \\
H^{26} & \mathbb{Z}_{2} & (\mathbb{Z}_{2})^{31} & (\mathbb{Z}_{2})^{92} & (\mathbb{Z}_{2})^{112} \ar[lu,"{{2}}" {gray,pos=0.08,scale=0.8,below left},"{{2}}" {gray,pos=0.92,scale=0.8,above right}] & (\mathbb{Z}_{2})^{92} \ar[lu,"{{2^{2}}}" {gray,pos=0.08,scale=0.8,below left},"{{2^{2}}}" {gray,pos=0.92,scale=0.8,above right}] & (\mathbb{Z}_{2})^{66} \ar[lu,"{{2^{2}}}" {gray,pos=0.08,scale=0.8,below left},"{{2^{2}}}" {gray,pos=0.92,scale=0.8,above right}] & (\mathbb{Z}_{2})^{48} \ar[lu,"{{2^{4}}}" {gray,pos=0.08,scale=0.8,below left},"{{2^{4}}}" {gray,pos=0.92,scale=0.8,above right}] & (\mathbb{Z}_{2})^{36} \ar[lu,"{{2^{6}}}" {gray,pos=0.08,scale=0.8,below left},"{{2^{6}}}" {gray,pos=0.92,scale=0.8,above right}] \\
H^{27} & \mathbb{Z}_{2} & (\mathbb{Z}_{2})^{34} & (\mathbb{Z}_{2})^{110} & (\mathbb{Z}_{2})^{132} \ar[lu,"{{2}}" {gray,pos=0.08,scale=0.8,below left},"{{2}}" {gray,pos=0.92,scale=0.8,above right}] & (\mathbb{Z}_{2})^{113} \ar[lu,"{{2^{3}}}" {gray,pos=0.08,scale=0.8,below left},"{{2^{3}}}" {gray,pos=0.92,scale=0.8,above right}] & (\mathbb{Z}_{2})^{83} \ar[lu,"{{2^{4}}}" {gray,pos=0.08,scale=0.8,below left},"{{2^{4}}}" {gray,pos=0.92,scale=0.8,above right}] & (\mathbb{Z}_{2})^{59} \ar[lu,"{{2^{5}}}" {gray,pos=0.08,scale=0.8,below left},"{{2^{5}}}" {gray,pos=0.92,scale=0.8,above right}] & (\mathbb{Z}_{2})^{43} \ar[lu,"{{2^{7}}}" {gray,pos=0.08,scale=0.8,below left},"{{2^{7}}}" {gray,pos=0.92,scale=0.8,above right}] \\
H^{28} & \mathbb{Z}_{2} & (\mathbb{Z}_{2})^{36} & (\mathbb{Z}_{2})^{129} & (\mathbb{Z}_{2})^{159} & (\mathbb{Z}_{2})^{137} \ar[lu,"{{2}}" {gray,pos=0.08,scale=0.8,below left},"{{2}}" {gray,pos=0.92,scale=0.8,above right}] & (\mathbb{Z}_{2})^{102} \ar[lu,"{{2^{3}}}" {gray,pos=0.08,scale=0.8,below left},"{{2^{3}}}" {gray,pos=0.92,scale=0.8,above right}] & (\mathbb{Z}_{2})^{72} \ar[lu,"{{2^{4}}}" {gray,pos=0.08,scale=0.8,below left},"{{2^{4}}}" {gray,pos=0.92,scale=0.8,above right}] & (\mathbb{Z}_{2})^{51} \ar[lu,"{{2^{5}}}" {gray,pos=0.08,scale=0.8,below left},"{{2^{5}}}" {gray,pos=0.92,scale=0.8,above right}] \\
H^{29} & \mathbb{Z}_{2} & (\mathbb{Z}_{2})^{41} & (\mathbb{Z}_{2})^{148} & (\mathbb{Z}_{2})^{193} \ar[lu,"{{2}}" {gray,pos=0.08,scale=0.8,below left},"{{2}}" {gray,pos=0.92,scale=0.8,above right}] & (\mathbb{Z}_{2})^{169} \ar[lu,"{{2^{3}}}" {gray,pos=0.08,scale=0.8,below left},"{{2^{3}}}" {gray,pos=0.92,scale=0.8,above right}] & (\mathbb{Z}_{2})^{124} \ar[lu,"{{2^{4}}}" {gray,pos=0.08,scale=0.8,below left},"{{2^{4}}}" {gray,pos=0.92,scale=0.8,above right}] & (\mathbb{Z}_{2})^{88} \ar[lu,"{{2^{6}}}" {gray,pos=0.08,scale=0.8,below left},"{{2^{6}}}" {gray,pos=0.92,scale=0.8,above right}] & (\mathbb{Z}_{2})^{64} \ar[lu,"{{2^{8}}}" {gray,pos=0.08,scale=0.8,below left},"{{2^{8}}}" {gray,pos=0.92,scale=0.8,above right}] \\
H^{30} & \mathbb{Z}_{2} & (\mathbb{Z}_{2})^{44} & (\mathbb{Z}_{2})^{175} & (\mathbb{Z}_{2})^{232} & (\mathbb{Z}_{2})^{204} \ar[lu,"{{2}}" {gray,pos=0.08,scale=0.8,below left},"{{2}}" {gray,pos=0.92,scale=0.8,above right}] & (\mathbb{Z}_{2})^{154} \ar[lu,"{{2^{3}}}" {gray,pos=0.08,scale=0.8,below left},"{{2^{3}}}" {gray,pos=0.92,scale=0.8,above right}] & (\mathbb{Z}_{2})^{108} \ar[lu,"{{2^{4}}}" {gray,pos=0.08,scale=0.8,below left},"{{2^{4}}}" {gray,pos=0.92,scale=0.8,above right}] & (\mathbb{Z}_{2})^{77} \ar[lu,"{{2^{6}}}" {gray,pos=0.08,scale=0.8,below left},"{{2^{6}}}" {gray,pos=0.92,scale=0.8,above right}] \\
H^{31} & \mathbb{Z}_{2} & (\mathbb{Z}_{2})^{47} & (\mathbb{Z}_{2})^{203} & (\mathbb{Z}_{2})^{277} & (\mathbb{Z}_{2})^{245} \ar[lu,"{{2}}" {gray,pos=0.08,scale=0.8,below left},"{{2}}" {gray,pos=0.92,scale=0.8,above right}] & (\mathbb{Z}_{2})^{187} \ar[lu,"{{2^{4}}}" {gray,pos=0.08,scale=0.8,below left},"{{2^{4}}}" {gray,pos=0.92,scale=0.8,above right}] & (\mathbb{Z}_{2})^{132} \ar[lu,"{{2^{6}}}" {gray,pos=0.08,scale=0.8,below left},"{{2^{6}}}" {gray,pos=0.92,scale=0.8,above right}] & (\mathbb{Z}_{2})^{92} \ar[lu,"{{2^{7}}}" {gray,pos=0.08,scale=0.8,below left},"{{2^{7}}}" {gray,pos=0.92,scale=0.8,above right}] \\
H^{32} & \mathbb{Z}_{2} & (\mathbb{Z}_{2})^{53} & (\mathbb{Z}_{2})^{232} & (\mathbb{Z}_{2})^{334} & (\mathbb{Z}_{2})^{299} & (\mathbb{Z}_{2})^{225} \ar[lu,"{{2}}" {gray,pos=0.08,scale=0.8,below left},"{{2}}" {gray,pos=0.92,scale=0.8,above right}] & (\mathbb{Z}_{2})^{161} \ar[lu,"{{2^{4}}}" {gray,pos=0.08,scale=0.8,below left},"{{2^{4}}}" {gray,pos=0.92,scale=0.8,above right}] & (\mathbb{Z}_{2})^{113} \ar[lu,"{{2^{6}}}" {gray,pos=0.08,scale=0.8,below left},"{{2^{6}}}" {gray,pos=0.92,scale=0.8,above right}] \\
H^{33} & \mathbb{Z}_{2} & (\mathbb{Z}_{2})^{57} \ar[lu,two heads,"{{2}}" {gray,pos=0.08,scale=0.8,below left},"{{2}}" {gray,pos=0.92,scale=0.8,above right}] & (\mathbb{Z}_{2})^{271} \ar[lu,"{{2}}" {gray,pos=0.08,scale=0.8,below left},"{{2}}" {gray,pos=0.92,scale=0.8,above right}] & (\mathbb{Z}_{2})^{399} \ar[lu,"{{2}}" {gray,pos=0.08,scale=0.8,below left},"{{2}}" {gray,pos=0.92,scale=0.8,above right}] & (\mathbb{Z}_{2})^{362} \ar[lu,"{{2^{2}}}" {gray,pos=0.08,scale=0.8,below left},"{{2^{2}}}" {gray,pos=0.92,scale=0.8,above right}] & (\mathbb{Z}_{2})^{276} \ar[lu,"{{2^{3}}}" {gray,pos=0.08,scale=0.8,below left},"{{2^{3}}}" {gray,pos=0.92,scale=0.8,above right}] & (\mathbb{Z}_{2})^{196} \ar[lu,"{{2^{5}}}" {gray,pos=0.08,scale=0.8,below left},"{{2^{5}}}" {gray,pos=0.92,scale=0.8,above right}] & (\mathbb{Z}_{2})^{137} \ar[lu,"{{2^{8}}}" {gray,pos=0.08,scale=0.8,below left},"{{2^{8}}}" {gray,pos=0.92,scale=0.8,above right}] \\
H^{34} & \mathbb{Z}_{2} & (\mathbb{Z}_{2})^{61} & (\mathbb{Z}_{2})^{314} \ar[lu,"{{2}}" {gray,pos=0.08,scale=0.8,below left},"{{2}}" {gray,pos=0.92,scale=0.8,above right}] & (\mathbb{Z}_{2})^{471} \ar[lu,"{{2}}" {gray,pos=0.08,scale=0.8,below left},"{{2}}" {gray,pos=0.92,scale=0.8,above right}] & (\mathbb{Z}_{2})^{438} \ar[lu,"{{2}}" {gray,pos=0.08,scale=0.8,below left},"{{2}}" {gray,pos=0.92,scale=0.8,above right}] & (\mathbb{Z}_{2})^{336} \ar[lu,"{{2^{2}}}" {gray,pos=0.08,scale=0.8,below left},"{{2^{2}}}" {gray,pos=0.92,scale=0.8,above right}] & (\mathbb{Z}_{2})^{238} \ar[lu,"{{2^{3}}}" {gray,pos=0.08,scale=0.8,below left},"{{2^{3}}}" {gray,pos=0.92,scale=0.8,above right}] & (\mathbb{Z}_{2})^{165} \ar[lu,"{{2^{5}}}" {gray,pos=0.08,scale=0.8,below left},"{{2^{5}}}" {gray,pos=0.92,scale=0.8,above right}] \\
H^{35} & \mathbb{Z}_{2} & (\mathbb{Z}_{2})^{68} & (\mathbb{Z}_{2})^{359} \ar[lu,"{{2}}" {gray,pos=0.08,scale=0.8,below left},"{{2}}" {gray,pos=0.92,scale=0.8,above right}] & (\mathbb{Z}_{2})^{559} \ar[lu,"{{2^{2}}}" {gray,pos=0.08,scale=0.8,below left},"{{2^{2}}}" {gray,pos=0.92,scale=0.8,above right}] & (\mathbb{Z}_{2})^{532} \ar[lu,"{{2^{2}}}" {gray,pos=0.08,scale=0.8,below left},"{{2^{2}}}" {gray,pos=0.92,scale=0.8,above right}] & (\mathbb{Z}_{2})^{405} \ar[lu,"{{2^{3}}}" {gray,pos=0.08,scale=0.8,below left},"{{2^{3}}}" {gray,pos=0.92,scale=0.8,above right}] & (\mathbb{Z}_{2})^{290} \ar[lu,"{{2^{5}}}" {gray,pos=0.08,scale=0.8,below left},"{{2^{5}}}" {gray,pos=0.92,scale=0.8,above right}] & (\mathbb{Z}_{2})^{201} \ar[lu,"{{2^{7}}}" {gray,pos=0.08,scale=0.8,below left},"{{2^{7}}}" {gray,pos=0.92,scale=0.8,above right}] \\
H^{36} & \mathbb{Z}_{2} & (\mathbb{Z}_{2})^{73} & (\mathbb{Z}_{2})^{414} & (\mathbb{Z}_{2})^{665} \ar[lu,"{{2}}" {gray,pos=0.08,scale=0.8,below left},"{{2}}" {gray,pos=0.92,scale=0.8,above right}] & (\mathbb{Z}_{2})^{638} \ar[lu,"{{2^{2}}}" {gray,pos=0.08,scale=0.8,below left},"{{2^{2}}}" {gray,pos=0.92,scale=0.8,above right}] & (\mathbb{Z}_{2})^{494} \ar[lu,"{{2^{2}}}" {gray,pos=0.08,scale=0.8,below left},"{{2^{2}}}" {gray,pos=0.92,scale=0.8,above right}] & (\mathbb{Z}_{2})^{352} \ar[lu,"{{2^{3}}}" {gray,pos=0.08,scale=0.8,below left},"{{2^{3}}}" {gray,pos=0.92,scale=0.8,above right}] & (\mathbb{Z}_{2})^{243} \ar[lu,"{{2^{5}}}" {gray,pos=0.08,scale=0.8,below left},"{{2^{5}}}" {gray,pos=0.92,scale=0.8,above right}] \\
H^{37} & \mathbb{Z}_{2} & (\mathbb{Z}_{2})^{78} & (\mathbb{Z}_{2})^{476} \ar[lu,"{{2}}" {gray,pos=0.08,scale=0.8,below left},"{{2}}" {gray,pos=0.92,scale=0.8,above right}] & (\mathbb{Z}_{2})^{789} \ar[lu,"{{2^{2}}}" {gray,pos=0.08,scale=0.8,below left},"{{2^{2}}}" {gray,pos=0.92,scale=0.8,above right}] & (\mathbb{Z}_{2})^{764} \ar[lu,"{{2^{3}}}" {gray,pos=0.08,scale=0.8,below left},"{{2^{3}}}" {gray,pos=0.92,scale=0.8,above right}] & (\mathbb{Z}_{2})^{599} \ar[lu,"{{2^{5}}}" {gray,pos=0.08,scale=0.8,below left},"{{2^{5}}}" {gray,pos=0.92,scale=0.8,above right}] & (\mathbb{Z}_{2})^{427} \ar[lu,"{{2^{6}}}" {gray,pos=0.08,scale=0.8,below left},"{{2^{6}}}" {gray,pos=0.92,scale=0.8,above right}] & (\mathbb{Z}_{2})^{295} \ar[lu,"{{2^{7}}}" {gray,pos=0.08,scale=0.8,below left},"{{2^{7}}}" {gray,pos=0.92,scale=0.8,above right}] \\
H^{38} & \mathbb{Z}_{2} & (\mathbb{Z}_{2})^{86} & (\mathbb{Z}_{2})^{543} & (\mathbb{Z}_{2})^{932} \ar[lu,"{{2}}" {gray,pos=0.08,scale=0.8,below left},"{{2}}" {gray,pos=0.92,scale=0.8,above right}] & (\mathbb{Z}_{2})^{919} \ar[lu,"{{2^{2}}}" {gray,pos=0.08,scale=0.8,below left},"{{2^{2}}}" {gray,pos=0.92,scale=0.8,above right}] & (\mathbb{Z}_{2})^{719} \ar[lu,"{{2^{3}}}" {gray,pos=0.08,scale=0.8,below left},"{{2^{3}}}" {gray,pos=0.92,scale=0.8,above right}] & (\mathbb{Z}_{2})^{518} \ar[lu,"{{2^{5}}}" {gray,pos=0.08,scale=0.8,below left},"{{2^{5}}}" {gray,pos=0.92,scale=0.8,above right}] & (\mathbb{Z}_{2})^{360} \ar[lu,"{{2^{6}}}" {gray,pos=0.08,scale=0.8,below left},"{{2^{6}}}" {gray,pos=0.92,scale=0.8,above right}] \\
H^{39} & \mathbb{Z}_{2} & (\mathbb{Z}_{2})^{92} & (\mathbb{Z}_{2})^{623} & (\mathbb{Z}_{2})^{1102} \ar[lu,"{{2}}" {gray,pos=0.08,scale=0.8,below left},"{{2}}" {gray,pos=0.92,scale=0.8,above right}] & (\mathbb{Z}_{2})^{1098} \ar[lu,"{{2^{3}}}" {gray,pos=0.08,scale=0.8,below left},"{{2^{3}}}" {gray,pos=0.92,scale=0.8,above right}] & (\mathbb{Z}_{2})^{869} \ar[lu,"{{2^{4}}}" {gray,pos=0.08,scale=0.8,below left},"{{2^{4}}}" {gray,pos=0.92,scale=0.8,above right}] & (\mathbb{Z}_{2})^{625} \ar[lu,"{{2^{6}}}" {gray,pos=0.08,scale=0.8,below left},"{{2^{6}}}" {gray,pos=0.92,scale=0.8,above right}] & (\mathbb{Z}_{2})^{434} \ar[lu,"{{2^{9}}}" {gray,pos=0.08,scale=0.8,below left},"{{2^{9}}}" {gray,pos=0.92,scale=0.8,above right}] \\
H^{40} & \mathbb{Z}_{2} & (\mathbb{Z}_{2})^{98} & (\mathbb{Z}_{2})^{712} & (\mathbb{Z}_{2})^{1299} & (\mathbb{Z}_{2})^{1313} \ar[lu,"{{2}}" {gray,pos=0.08,scale=0.8,below left},"{{2}}" {gray,pos=0.92,scale=0.8,above right}] & (\mathbb{Z}_{2})^{1047} \ar[lu,"{{2^{3}}}" {gray,pos=0.08,scale=0.8,below left},"{{2^{3}}}" {gray,pos=0.92,scale=0.8,above right}] & (\mathbb{Z}_{2})^{752} \ar[lu,"{{2^{4}}}" {gray,pos=0.08,scale=0.8,below left},"{{2^{4}}}" {gray,pos=0.92,scale=0.8,above right}] & (\mathbb{Z}_{2})^{523} \ar[lu,"{{2^{6}}}" {gray,pos=0.08,scale=0.8,below left},"{{2^{6}}}" {gray,pos=0.92,scale=0.8,above right}] \\
\end{tikzcd}

%% file: tables/omega-Z2-Z4.tex
\begin{tikzcd}[row sep=tiny,column sep=large]
& B\mathbb{Z}_{2} & B^{2}\mathbb{Z}_{2} & B^{3}\mathbb{Z}_{2} & B^{4}\mathbb{Z}_{2} & B^{5}\mathbb{Z}_{2} & B^{6}\mathbb{Z}_{2} \\
H^{0} & \mathbb{Z}_{4} & \mathbb{Z}_{4} & \mathbb{Z}_{4} & \mathbb{Z}_{4} & \mathbb{Z}_{4} & \mathbb{Z}_{4} \\
H^{1} & \mathbb{Z}_{2} & 0 & 0 & 0 & 0 & 0 \\
H^{2} & \mathbb{Z}_{2} & \mathbb{Z}_{2} \ar[lu,BrickRed,hook,two heads,"{{2}}" {gray,pos=0.08,scale=0.8,below left},"{{2}}" {gray,pos=0.92,scale=0.8,above right}] & 0 & 0 & 0 & 0 \\
H^{3} & \mathbb{Z}_{2} & \mathbb{Z}_{2} \ar[lu,hook,two heads,"{{2}}" {gray,pos=0.08,scale=0.8,below left},"{{2}}" {gray,pos=0.92,scale=0.8,above right}] & \mathbb{Z}_{2} \ar[lu,BrickRed,hook,two heads,"{{2}}" {gray,pos=0.08,scale=0.8,below left},"{{2}}" {gray,pos=0.92,scale=0.8,above right}] & 0 & 0 & 0 \\
H^{4} & \mathbb{Z}_{2} & \mathbb{Z}_{4} \ar[lu,two heads,"{{4}}" {gray,pos=0.08,scale=0.8,below left},"{{2}}" {gray,pos=0.92,scale=0.8,above right}] & \mathbb{Z}_{2} \ar[lu,BrickRed,hook,two heads,"{{2}}" {gray,pos=0.08,scale=0.8,below left},"{{2}}" {gray,pos=0.92,scale=0.8,above right}] & \mathbb{Z}_{2} \ar[lu,BrickRed,hook,two heads,"{{2}}" {gray,pos=0.08,scale=0.8,below left},"{{2}}" {gray,pos=0.92,scale=0.8,above right}] & 0 & 0 \\
H^{5} & \mathbb{Z}_{2} & \mathbb{Z}_{2} \oplus \mathbb{Z}_{4} \ar[lu,two heads,"{{4}}" {gray,pos=0.08,scale=0.8,below left},"{{2}}" {gray,pos=0.92,scale=0.8,above right}] & \mathbb{Z}_{2} \ar[lu,hook,"{{2}}" {gray,pos=0.08,scale=0.8,below left},"{{2}}" {gray,pos=0.92,scale=0.8,above right}] & \mathbb{Z}_{2} \ar[lu,BrickRed,hook,two heads,"{{2}}" {gray,pos=0.08,scale=0.8,below left},"{{2}}" {gray,pos=0.92,scale=0.8,above right}] & \mathbb{Z}_{2} \ar[lu,BrickRed,hook,two heads,"{{2}}" {gray,pos=0.08,scale=0.8,below left},"{{2}}" {gray,pos=0.92,scale=0.8,above right}] & 0 \\
H^{6} & \mathbb{Z}_{2} & (\mathbb{Z}_{2})^{2} & (\mathbb{Z}_{2})^{2} \ar[lu,hook,"{{2^{2}}}" {gray,pos=0.08,scale=0.8,below left},"{{2^{2}}}" {gray,pos=0.92,scale=0.8,above right}] & \mathbb{Z}_{2} \ar[lu,BrickRed,hook,two heads,"{{2}}" {gray,pos=0.08,scale=0.8,below left},"{{2}}" {gray,pos=0.92,scale=0.8,above right}] & \mathbb{Z}_{2} \ar[lu,BrickRed,hook,two heads,"{{2}}" {gray,pos=0.08,scale=0.8,below left},"{{2}}" {gray,pos=0.92,scale=0.8,above right}] & \mathbb{Z}_{2} \ar[lu,BrickRed,hook,two heads,"{{2}}" {gray,pos=0.08,scale=0.8,below left},"{{2}}" {gray,pos=0.92,scale=0.8,above right}] \\
H^{7} & \mathbb{Z}_{2} & (\mathbb{Z}_{2})^{2} & (\mathbb{Z}_{2})^{2} \ar[lu,"{{2}}" {gray,pos=0.08,scale=0.8,below left},"{{2}}" {gray,pos=0.92,scale=0.8,above right}] & (\mathbb{Z}_{2})^{2} \ar[lu,hook,two heads,"{{2^{2}}}" {gray,pos=0.08,scale=0.8,below left},"{{2^{2}}}" {gray,pos=0.92,scale=0.8,above right}] & \mathbb{Z}_{2} \ar[lu,BrickRed,hook,two heads,"{{2}}" {gray,pos=0.08,scale=0.8,below left},"{{2}}" {gray,pos=0.92,scale=0.8,above right}] & \mathbb{Z}_{2} \ar[lu,BrickRed,hook,two heads,"{{2}}" {gray,pos=0.08,scale=0.8,below left},"{{2}}" {gray,pos=0.92,scale=0.8,above right}] \\
H^{8} & \mathbb{Z}_{2} & (\mathbb{Z}_{2})^{2} \oplus \mathbb{Z}_{4} & (\mathbb{Z}_{2})^{2} & \mathbb{Z}_{2} \oplus \mathbb{Z}_{4} \ar[lu,two heads,"{{2,4}}" {gray,pos=0.08,scale=0.8,below left},"{{2^{2}}}" {gray,pos=0.92,scale=0.8,above right}] & (\mathbb{Z}_{2})^{2} \ar[lu,BrickRed,hook,two heads,"{{2^{2}}}" {gray,pos=0.08,scale=0.8,below left},"{{2^{2}}}" {gray,pos=0.92,scale=0.8,above right}] & \mathbb{Z}_{2} \ar[lu,BrickRed,hook,two heads,"{{2}}" {gray,pos=0.08,scale=0.8,below left},"{{2}}" {gray,pos=0.92,scale=0.8,above right}] \\
H^{9} & \mathbb{Z}_{2} & (\mathbb{Z}_{2})^{3} \oplus \mathbb{Z}_{4} & (\mathbb{Z}_{2})^{4} \ar[lu,"{{2}}" {gray,pos=0.08,scale=0.8,below left},"{{2}}" {gray,pos=0.92,scale=0.8,above right}] & \mathbb{Z}_{2} \oplus \mathbb{Z}_{4} \ar[lu,"{{4}}" {gray,pos=0.08,scale=0.8,below left},"{{2}}" {gray,pos=0.92,scale=0.8,above right}] & (\mathbb{Z}_{2})^{2} \ar[lu,hook,"{{2^{2}}}" {gray,pos=0.08,scale=0.8,below left},"{{2^{2}}}" {gray,pos=0.92,scale=0.8,above right}] & (\mathbb{Z}_{2})^{2} \ar[lu,BrickRed,hook,two heads,"{{2^{2}}}" {gray,pos=0.08,scale=0.8,below left},"{{2^{2}}}" {gray,pos=0.92,scale=0.8,above right}] \\
H^{10} & \mathbb{Z}_{2} & (\mathbb{Z}_{2})^{4} & (\mathbb{Z}_{2})^{5} \ar[lu,"{{2}}" {gray,pos=0.08,scale=0.8,below left},"{{2}}" {gray,pos=0.92,scale=0.8,above right}] & (\mathbb{Z}_{2})^{3} \ar[lu,"{{2}}" {gray,pos=0.08,scale=0.8,below left},"{{2}}" {gray,pos=0.92,scale=0.8,above right}] & (\mathbb{Z}_{2})^{2} \ar[lu,hook,"{{2^{2}}}" {gray,pos=0.08,scale=0.8,below left},"{{2^{2}}}" {gray,pos=0.92,scale=0.8,above right}] & (\mathbb{Z}_{2})^{2} \ar[lu,BrickRed,hook,two heads,"{{2^{2}}}" {gray,pos=0.08,scale=0.8,below left},"{{2^{2}}}" {gray,pos=0.92,scale=0.8,above right}] \\
H^{11} & \mathbb{Z}_{2} & (\mathbb{Z}_{2})^{5} & (\mathbb{Z}_{2})^{5} \ar[lu,"{{2}}" {gray,pos=0.08,scale=0.8,below left},"{{2}}" {gray,pos=0.92,scale=0.8,above right}] & (\mathbb{Z}_{2})^{5} \ar[lu,"{{2^{2}}}" {gray,pos=0.08,scale=0.8,below left},"{{2^{2}}}" {gray,pos=0.92,scale=0.8,above right}] & (\mathbb{Z}_{2})^{3} \ar[lu,"{{2^{2}}}" {gray,pos=0.08,scale=0.8,below left},"{{2^{2}}}" {gray,pos=0.92,scale=0.8,above right}] & (\mathbb{Z}_{2})^{2} \ar[lu,hook,two heads,"{{2^{2}}}" {gray,pos=0.08,scale=0.8,below left},"{{2^{2}}}" {gray,pos=0.92,scale=0.8,above right}] \\
H^{12} & \mathbb{Z}_{2} & (\mathbb{Z}_{2})^{5} \oplus \mathbb{Z}_{4} & (\mathbb{Z}_{2})^{6} \oplus \mathbb{Z}_{4} \ar[lu,"{{4}}" {gray,pos=0.08,scale=0.8,below left},"{{2}}" {gray,pos=0.92,scale=0.8,above right}] & (\mathbb{Z}_{2})^{5} \oplus \mathbb{Z}_{4} \ar[lu,"{{2,4}}" {gray,pos=0.08,scale=0.8,below left},"{{2^{2}}}" {gray,pos=0.92,scale=0.8,above right}] & (\mathbb{Z}_{2})^{4} \ar[lu,"{{2^{2}}}" {gray,pos=0.08,scale=0.8,below left},"{{2^{2}}}" {gray,pos=0.92,scale=0.8,above right}] & (\mathbb{Z}_{2})^{2} \oplus \mathbb{Z}_{4} \ar[lu,two heads,"{{2^{2},4}}" {gray,pos=0.08,scale=0.8,below left},"{{2^{3}}}" {gray,pos=0.92,scale=0.8,above right}] \\
H^{13} & \mathbb{Z}_{2} & (\mathbb{Z}_{2})^{5} \oplus \mathbb{Z}_{4} & (\mathbb{Z}_{2})^{8} \oplus \mathbb{Z}_{4} \ar[lu,"{{4}}" {gray,pos=0.08,scale=0.8,below left},"{{2}}" {gray,pos=0.92,scale=0.8,above right}] & (\mathbb{Z}_{2})^{6} \oplus \mathbb{Z}_{4} \ar[lu,"{{2,4}}" {gray,pos=0.08,scale=0.8,below left},"{{2^{2}}}" {gray,pos=0.92,scale=0.8,above right}] & (\mathbb{Z}_{2})^{5} \ar[lu,"{{2^{2}}}" {gray,pos=0.08,scale=0.8,below left},"{{2^{2}}}" {gray,pos=0.92,scale=0.8,above right}] & (\mathbb{Z}_{2})^{3} \oplus \mathbb{Z}_{4} \ar[lu,"{{2^{2},4}}" {gray,pos=0.08,scale=0.8,below left},"{{2^{3}}}" {gray,pos=0.92,scale=0.8,above right}] \\
H^{14} & \mathbb{Z}_{2} & (\mathbb{Z}_{2})^{8} & (\mathbb{Z}_{2})^{10} & (\mathbb{Z}_{2})^{9} \ar[lu,"{{2^{2}}}" {gray,pos=0.08,scale=0.8,below left},"{{2^{2}}}" {gray,pos=0.92,scale=0.8,above right}] & (\mathbb{Z}_{2})^{7} \ar[lu,"{{2^{3}}}" {gray,pos=0.08,scale=0.8,below left},"{{2^{3}}}" {gray,pos=0.92,scale=0.8,above right}] & (\mathbb{Z}_{2})^{4} \ar[lu,"{{2^{2}}}" {gray,pos=0.08,scale=0.8,below left},"{{2^{2}}}" {gray,pos=0.92,scale=0.8,above right}] \\
H^{15} & \mathbb{Z}_{2} & (\mathbb{Z}_{2})^{9} & (\mathbb{Z}_{2})^{13} & (\mathbb{Z}_{2})^{11} \ar[lu,"{{2}}" {gray,pos=0.08,scale=0.8,below left},"{{2}}" {gray,pos=0.92,scale=0.8,above right}] & (\mathbb{Z}_{2})^{9} \ar[lu,"{{2^{3}}}" {gray,pos=0.08,scale=0.8,below left},"{{2^{3}}}" {gray,pos=0.92,scale=0.8,above right}] & (\mathbb{Z}_{2})^{6} \ar[lu,"{{2^{3}}}" {gray,pos=0.08,scale=0.8,below left},"{{2^{3}}}" {gray,pos=0.92,scale=0.8,above right}] \\
H^{16} & \mathbb{Z}_{2} & (\mathbb{Z}_{2})^{8} \oplus \mathbb{Z}_{4} & (\mathbb{Z}_{2})^{16} & (\mathbb{Z}_{2})^{12} \oplus \mathbb{Z}_{4} & (\mathbb{Z}_{2})^{9} \oplus \mathbb{Z}_{4} \ar[lu,"{{2,4}}" {gray,pos=0.08,scale=0.8,below left},"{{2^{2}}}" {gray,pos=0.92,scale=0.8,above right}] & (\mathbb{Z}_{2})^{7} \oplus \mathbb{Z}_{4} \ar[lu,"{{2^{3},4}}" {gray,pos=0.08,scale=0.8,below left},"{{2^{4}}}" {gray,pos=0.92,scale=0.8,above right}] \\
H^{17} & \mathbb{Z}_{2} & (\mathbb{Z}_{2})^{11} \oplus \mathbb{Z}_{4} & (\mathbb{Z}_{2})^{19} \ar[lu,"{{2}}" {gray,pos=0.08,scale=0.8,below left},"{{2}}" {gray,pos=0.92,scale=0.8,above right}] & (\mathbb{Z}_{2})^{16} \oplus \mathbb{Z}_{4} & (\mathbb{Z}_{2})^{12} \oplus \mathbb{Z}_{4} \ar[lu,"{{2,4}}" {gray,pos=0.08,scale=0.8,below left},"{{2^{2}}}" {gray,pos=0.92,scale=0.8,above right}] & (\mathbb{Z}_{2})^{8} \oplus \mathbb{Z}_{4} \ar[lu,"{{2^{2},4}}" {gray,pos=0.08,scale=0.8,below left},"{{2^{3}}}" {gray,pos=0.92,scale=0.8,above right}] \\
H^{18} & \mathbb{Z}_{2} & (\mathbb{Z}_{2})^{13} & (\mathbb{Z}_{2})^{24} \ar[lu,"{{2}}" {gray,pos=0.08,scale=0.8,below left},"{{2}}" {gray,pos=0.92,scale=0.8,above right}] & (\mathbb{Z}_{2})^{23} \ar[lu,"{{2}}" {gray,pos=0.08,scale=0.8,below left},"{{2}}" {gray,pos=0.92,scale=0.8,above right}] & (\mathbb{Z}_{2})^{16} \ar[lu,"{{2}}" {gray,pos=0.08,scale=0.8,below left},"{{2}}" {gray,pos=0.92,scale=0.8,above right}] & (\mathbb{Z}_{2})^{12} \ar[lu,"{{2^{3}}}" {gray,pos=0.08,scale=0.8,below left},"{{2^{3}}}" {gray,pos=0.92,scale=0.8,above right}] \\
H^{19} & \mathbb{Z}_{2} & (\mathbb{Z}_{2})^{14} & (\mathbb{Z}_{2})^{28} \ar[lu,"{{2}}" {gray,pos=0.08,scale=0.8,below left},"{{2}}" {gray,pos=0.92,scale=0.8,above right}] & (\mathbb{Z}_{2})^{28} \ar[lu,"{{2^{2}}}" {gray,pos=0.08,scale=0.8,below left},"{{2^{2}}}" {gray,pos=0.92,scale=0.8,above right}] & (\mathbb{Z}_{2})^{20} \ar[lu,"{{2^{2}}}" {gray,pos=0.08,scale=0.8,below left},"{{2^{2}}}" {gray,pos=0.92,scale=0.8,above right}] & (\mathbb{Z}_{2})^{15} \ar[lu,"{{2^{3}}}" {gray,pos=0.08,scale=0.8,below left},"{{2^{3}}}" {gray,pos=0.92,scale=0.8,above right}] \\
H^{20} & \mathbb{Z}_{2} & (\mathbb{Z}_{2})^{16} \oplus \mathbb{Z}_{4} & (\mathbb{Z}_{2})^{32} \oplus \mathbb{Z}_{4} \ar[lu,"{{4}}" {gray,pos=0.08,scale=0.8,below left},"{{2}}" {gray,pos=0.92,scale=0.8,above right}] & (\mathbb{Z}_{2})^{31} \oplus (\mathbb{Z}_{4})^{2} \ar[lu,"{{2,4}}" {gray,pos=0.08,scale=0.8,below left},"{{2^{2}}}" {gray,pos=0.92,scale=0.8,above right}] & (\mathbb{Z}_{2})^{26} \oplus \mathbb{Z}_{4} \ar[lu,"{{2^{2},4}}" {gray,pos=0.08,scale=0.8,below left},"{{2^{3}}}" {gray,pos=0.92,scale=0.8,above right}] & (\mathbb{Z}_{2})^{17} \oplus \mathbb{Z}_{4} \ar[lu,"{{2^{2},4}}" {gray,pos=0.08,scale=0.8,below left},"{{2^{3}}}" {gray,pos=0.92,scale=0.8,above right}] \\
H^{21} & \mathbb{Z}_{2} & (\mathbb{Z}_{2})^{17} \oplus \mathbb{Z}_{4} & (\mathbb{Z}_{2})^{40} \oplus \mathbb{Z}_{4} \ar[lu,"{{4}}" {gray,pos=0.08,scale=0.8,below left},"{{2}}" {gray,pos=0.92,scale=0.8,above right}] & (\mathbb{Z}_{2})^{38} \oplus (\mathbb{Z}_{4})^{3} \ar[lu,"{{2,4}}" {gray,pos=0.08,scale=0.8,below left},"{{2^{2}}}" {gray,pos=0.92,scale=0.8,above right}] & (\mathbb{Z}_{2})^{32} \oplus \mathbb{Z}_{4} \ar[lu,"{{2^{2},4}}" {gray,pos=0.08,scale=0.8,below left},"{{2^{3}}}" {gray,pos=0.92,scale=0.8,above right}] & (\mathbb{Z}_{2})^{23} \oplus \mathbb{Z}_{4} \ar[lu,"{{2^{3},4}}" {gray,pos=0.08,scale=0.8,below left},"{{2^{4}}}" {gray,pos=0.92,scale=0.8,above right}] \\
H^{22} & \mathbb{Z}_{2} & (\mathbb{Z}_{2})^{20} & (\mathbb{Z}_{2})^{48} & (\mathbb{Z}_{2})^{49} \oplus \mathbb{Z}_{4} \ar[lu,"{{2^{2}}}" {gray,pos=0.08,scale=0.8,below left},"{{2^{2}}}" {gray,pos=0.92,scale=0.8,above right}] & (\mathbb{Z}_{2})^{40} \ar[lu,"{{2^{3}}}" {gray,pos=0.08,scale=0.8,below left},"{{2^{3}}}" {gray,pos=0.92,scale=0.8,above right}] & (\mathbb{Z}_{2})^{30} \ar[lu,"{{2^{4}}}" {gray,pos=0.08,scale=0.8,below left},"{{2^{4}}}" {gray,pos=0.92,scale=0.8,above right}] \\
H^{23} & \mathbb{Z}_{2} & (\mathbb{Z}_{2})^{23} & (\mathbb{Z}_{2})^{56} & (\mathbb{Z}_{2})^{60} \ar[lu,"{{2}}" {gray,pos=0.08,scale=0.8,below left},"{{2}}" {gray,pos=0.92,scale=0.8,above right}] & (\mathbb{Z}_{2})^{50} \ar[lu,"{{2^{3}}}" {gray,pos=0.08,scale=0.8,below left},"{{2^{3}}}" {gray,pos=0.92,scale=0.8,above right}] & (\mathbb{Z}_{2})^{36} \ar[lu,"{{2^{4}}}" {gray,pos=0.08,scale=0.8,below left},"{{2^{4}}}" {gray,pos=0.92,scale=0.8,above right}] \\
H^{24} & \mathbb{Z}_{2} & (\mathbb{Z}_{2})^{24} \oplus \mathbb{Z}_{4} & (\mathbb{Z}_{2})^{67} \oplus \mathbb{Z}_{4} & (\mathbb{Z}_{2})^{72} \oplus (\mathbb{Z}_{4})^{2} & (\mathbb{Z}_{2})^{59} \oplus \mathbb{Z}_{4} \ar[lu,"{{2,4}}" {gray,pos=0.08,scale=0.8,below left},"{{2^{2}}}" {gray,pos=0.92,scale=0.8,above right}] & (\mathbb{Z}_{2})^{43} \oplus (\mathbb{Z}_{4})^{2} \ar[lu,"{{2^{3},4}}" {gray,pos=0.08,scale=0.8,below left},"{{2^{4}}}" {gray,pos=0.92,scale=0.8,above right}] \\
H^{25} & \mathbb{Z}_{2} & (\mathbb{Z}_{2})^{26} \oplus \mathbb{Z}_{4} & (\mathbb{Z}_{2})^{79} \oplus \mathbb{Z}_{4} & (\mathbb{Z}_{2})^{91} \oplus (\mathbb{Z}_{4})^{2} \ar[lu,"{{2}}" {gray,pos=0.08,scale=0.8,below left},"{{2}}" {gray,pos=0.92,scale=0.8,above right}] & (\mathbb{Z}_{2})^{72} \oplus \mathbb{Z}_{4} \ar[lu,"{{2,4}}" {gray,pos=0.08,scale=0.8,below left},"{{2^{2}}}" {gray,pos=0.92,scale=0.8,above right}] & (\mathbb{Z}_{2})^{53} \oplus (\mathbb{Z}_{4})^{2} \ar[lu,"{{2^{2},4}}" {gray,pos=0.08,scale=0.8,below left},"{{2^{3}}}" {gray,pos=0.92,scale=0.8,above right}] \\
H^{26} & \mathbb{Z}_{2} & (\mathbb{Z}_{2})^{31} & (\mathbb{Z}_{2})^{92} & (\mathbb{Z}_{2})^{112} \ar[lu,"{{2}}" {gray,pos=0.08,scale=0.8,below left},"{{2}}" {gray,pos=0.92,scale=0.8,above right}] & (\mathbb{Z}_{2})^{92} \ar[lu,"{{2^{2}}}" {gray,pos=0.08,scale=0.8,below left},"{{2^{2}}}" {gray,pos=0.92,scale=0.8,above right}] & (\mathbb{Z}_{2})^{66} \ar[lu,"{{2^{3}}}" {gray,pos=0.08,scale=0.8,below left},"{{2^{3}}}" {gray,pos=0.92,scale=0.8,above right}] \\
H^{27} & \mathbb{Z}_{2} & (\mathbb{Z}_{2})^{34} & (\mathbb{Z}_{2})^{110} & (\mathbb{Z}_{2})^{132} \ar[lu,"{{2}}" {gray,pos=0.08,scale=0.8,below left},"{{2}}" {gray,pos=0.92,scale=0.8,above right}] & (\mathbb{Z}_{2})^{113} \ar[lu,"{{2^{3}}}" {gray,pos=0.08,scale=0.8,below left},"{{2^{3}}}" {gray,pos=0.92,scale=0.8,above right}] & (\mathbb{Z}_{2})^{83} \ar[lu,"{{2^{4}}}" {gray,pos=0.08,scale=0.8,below left},"{{2^{4}}}" {gray,pos=0.92,scale=0.8,above right}] \\
H^{28} & \mathbb{Z}_{2} & (\mathbb{Z}_{2})^{35} \oplus \mathbb{Z}_{4} & (\mathbb{Z}_{2})^{129} & (\mathbb{Z}_{2})^{156} \oplus (\mathbb{Z}_{4})^{3} \ar[lu,"{{4}}" {gray,pos=0.08,scale=0.8,below left},"{{2}}" {gray,pos=0.92,scale=0.8,above right}] & (\mathbb{Z}_{2})^{135} \oplus (\mathbb{Z}_{4})^{2} \ar[lu,"{{2,4^{2}}}" {gray,pos=0.08,scale=0.8,below left},"{{2^{3}}}" {gray,pos=0.92,scale=0.8,above right}] & (\mathbb{Z}_{2})^{100} \oplus (\mathbb{Z}_{4})^{2} \ar[lu,"{{2^{3},4}}" {gray,pos=0.08,scale=0.8,below left},"{{2^{4}}}" {gray,pos=0.92,scale=0.8,above right}] \\
H^{29} & \mathbb{Z}_{2} & (\mathbb{Z}_{2})^{40} \oplus \mathbb{Z}_{4} & (\mathbb{Z}_{2})^{148} & (\mathbb{Z}_{2})^{188} \oplus (\mathbb{Z}_{4})^{5} \ar[lu,"{{4}}" {gray,pos=0.08,scale=0.8,below left},"{{2}}" {gray,pos=0.92,scale=0.8,above right}] & (\mathbb{Z}_{2})^{167} \oplus (\mathbb{Z}_{4})^{2} \ar[lu,"{{2,4^{2}}}" {gray,pos=0.08,scale=0.8,below left},"{{2^{3}}}" {gray,pos=0.92,scale=0.8,above right}] & (\mathbb{Z}_{2})^{121} \oplus (\mathbb{Z}_{4})^{3} \ar[lu,"{{2^{3},4}}" {gray,pos=0.08,scale=0.8,below left},"{{2^{4}}}" {gray,pos=0.92,scale=0.8,above right}] \\
H^{30} & \mathbb{Z}_{2} & (\mathbb{Z}_{2})^{44} & (\mathbb{Z}_{2})^{175} & (\mathbb{Z}_{2})^{230} \oplus (\mathbb{Z}_{4})^{2} & (\mathbb{Z}_{2})^{204} \ar[lu,"{{2^{2}}}" {gray,pos=0.08,scale=0.8,below left},"{{2^{2}}}" {gray,pos=0.92,scale=0.8,above right}] & (\mathbb{Z}_{2})^{153} \oplus \mathbb{Z}_{4} \ar[lu,"{{2^{5}}}" {gray,pos=0.08,scale=0.8,below left},"{{2^{5}}}" {gray,pos=0.92,scale=0.8,above right}] \\
H^{31} & \mathbb{Z}_{2} & (\mathbb{Z}_{2})^{47} & (\mathbb{Z}_{2})^{203} & (\mathbb{Z}_{2})^{277} & (\mathbb{Z}_{2})^{245} \ar[lu,"{{2}}" {gray,pos=0.08,scale=0.8,below left},"{{2}}" {gray,pos=0.92,scale=0.8,above right}] & (\mathbb{Z}_{2})^{187} \ar[lu,"{{2^{4}}}" {gray,pos=0.08,scale=0.8,below left},"{{2^{4}}}" {gray,pos=0.92,scale=0.8,above right}] \\
H^{32} & \mathbb{Z}_{2} & (\mathbb{Z}_{2})^{52} \oplus \mathbb{Z}_{4} & (\mathbb{Z}_{2})^{231} \oplus \mathbb{Z}_{4} & (\mathbb{Z}_{2})^{331} \oplus (\mathbb{Z}_{4})^{3} & (\mathbb{Z}_{2})^{298} \oplus \mathbb{Z}_{4} & (\mathbb{Z}_{2})^{222} \oplus (\mathbb{Z}_{4})^{3} \ar[lu,"{{2,4}}" {gray,pos=0.08,scale=0.8,below left},"{{2^{2}}}" {gray,pos=0.92,scale=0.8,above right}] \\
H^{33} & \mathbb{Z}_{2} & (\mathbb{Z}_{2})^{56} \oplus \mathbb{Z}_{4} & (\mathbb{Z}_{2})^{269} \oplus (\mathbb{Z}_{4})^{2} \ar[lu,"{{2}}" {gray,pos=0.08,scale=0.8,below left},"{{2}}" {gray,pos=0.92,scale=0.8,above right}] & (\mathbb{Z}_{2})^{394} \oplus (\mathbb{Z}_{4})^{5} & (\mathbb{Z}_{2})^{361} \oplus \mathbb{Z}_{4} \ar[lu,"{{2}}" {gray,pos=0.08,scale=0.8,below left},"{{2}}" {gray,pos=0.92,scale=0.8,above right}] & (\mathbb{Z}_{2})^{272} \oplus (\mathbb{Z}_{4})^{4} \ar[lu,"{{2,4}}" {gray,pos=0.08,scale=0.8,below left},"{{2^{2}}}" {gray,pos=0.92,scale=0.8,above right}] \\
H^{34} & \mathbb{Z}_{2} & (\mathbb{Z}_{2})^{61} & (\mathbb{Z}_{2})^{313} \oplus \mathbb{Z}_{4} \ar[lu,"{{2}}" {gray,pos=0.08,scale=0.8,below left},"{{2}}" {gray,pos=0.92,scale=0.8,above right}] & (\mathbb{Z}_{2})^{469} \oplus (\mathbb{Z}_{4})^{2} \ar[lu,"{{2}}" {gray,pos=0.08,scale=0.8,below left},"{{2}}" {gray,pos=0.92,scale=0.8,above right}] & (\mathbb{Z}_{2})^{438} \ar[lu,"{{2}}" {gray,pos=0.08,scale=0.8,below left},"{{2}}" {gray,pos=0.92,scale=0.8,above right}] & (\mathbb{Z}_{2})^{335} \oplus \mathbb{Z}_{4} \ar[lu,"{{2^{2}}}" {gray,pos=0.08,scale=0.8,below left},"{{2^{2}}}" {gray,pos=0.92,scale=0.8,above right}] \\
H^{35} & \mathbb{Z}_{2} & (\mathbb{Z}_{2})^{68} & (\mathbb{Z}_{2})^{359} \ar[lu,"{{2}}" {gray,pos=0.08,scale=0.8,below left},"{{2}}" {gray,pos=0.92,scale=0.8,above right}] & (\mathbb{Z}_{2})^{559} \ar[lu,"{{2^{2}}}" {gray,pos=0.08,scale=0.8,below left},"{{2^{2}}}" {gray,pos=0.92,scale=0.8,above right}] & (\mathbb{Z}_{2})^{532} \ar[lu,"{{2^{2}}}" {gray,pos=0.08,scale=0.8,below left},"{{2^{2}}}" {gray,pos=0.92,scale=0.8,above right}] & (\mathbb{Z}_{2})^{405} \ar[lu,"{{2^{3}}}" {gray,pos=0.08,scale=0.8,below left},"{{2^{3}}}" {gray,pos=0.92,scale=0.8,above right}] \\
H^{36} & \mathbb{Z}_{2} & (\mathbb{Z}_{2})^{72} \oplus \mathbb{Z}_{4} & (\mathbb{Z}_{2})^{412} \oplus (\mathbb{Z}_{4})^{2} \ar[lu,"{{4}}" {gray,pos=0.08,scale=0.8,below left},"{{2}}" {gray,pos=0.92,scale=0.8,above right}] & (\mathbb{Z}_{2})^{660} \oplus (\mathbb{Z}_{4})^{5} \ar[lu,"{{2,4}}" {gray,pos=0.08,scale=0.8,below left},"{{2^{2}}}" {gray,pos=0.92,scale=0.8,above right}] & (\mathbb{Z}_{2})^{636} \oplus (\mathbb{Z}_{4})^{2} \ar[lu,"{{2^{2},4}}" {gray,pos=0.08,scale=0.8,below left},"{{2^{3}}}" {gray,pos=0.92,scale=0.8,above right}] & (\mathbb{Z}_{2})^{489} \oplus (\mathbb{Z}_{4})^{5} \ar[lu,"{{2^{2},4^{2}}}" {gray,pos=0.08,scale=0.8,below left},"{{2^{4}}}" {gray,pos=0.92,scale=0.8,above right}] \\
H^{37} & \mathbb{Z}_{2} & (\mathbb{Z}_{2})^{77} \oplus \mathbb{Z}_{4} & (\mathbb{Z}_{2})^{474} \oplus (\mathbb{Z}_{4})^{2} \ar[lu,"{{4}}" {gray,pos=0.08,scale=0.8,below left},"{{2}}" {gray,pos=0.92,scale=0.8,above right}] & (\mathbb{Z}_{2})^{781} \oplus (\mathbb{Z}_{4})^{8} \ar[lu,"{{2,4}}" {gray,pos=0.08,scale=0.8,below left},"{{2^{2}}}" {gray,pos=0.92,scale=0.8,above right}] & (\mathbb{Z}_{2})^{761} \oplus (\mathbb{Z}_{4})^{3} \ar[lu,"{{2^{2},4}}" {gray,pos=0.08,scale=0.8,below left},"{{2^{3}}}" {gray,pos=0.92,scale=0.8,above right}] & (\mathbb{Z}_{2})^{592} \oplus (\mathbb{Z}_{4})^{7} \ar[lu,"{{2^{3},4^{2}}}" {gray,pos=0.08,scale=0.8,below left},"{{2^{5}}}" {gray,pos=0.92,scale=0.8,above right}] \\
H^{38} & \mathbb{Z}_{2} & (\mathbb{Z}_{2})^{86} & (\mathbb{Z}_{2})^{543} & (\mathbb{Z}_{2})^{929} \oplus (\mathbb{Z}_{4})^{3} \ar[lu,"{{2^{2}}}" {gray,pos=0.08,scale=0.8,below left},"{{2^{2}}}" {gray,pos=0.92,scale=0.8,above right}] & (\mathbb{Z}_{2})^{918} \oplus \mathbb{Z}_{4} \ar[lu,"{{2^{3}}}" {gray,pos=0.08,scale=0.8,below left},"{{2^{3}}}" {gray,pos=0.92,scale=0.8,above right}] & (\mathbb{Z}_{2})^{717} \oplus (\mathbb{Z}_{4})^{2} \ar[lu,"{{2^{4}}}" {gray,pos=0.08,scale=0.8,below left},"{{2^{4}}}" {gray,pos=0.92,scale=0.8,above right}] \\
H^{39} & \mathbb{Z}_{2} & (\mathbb{Z}_{2})^{92} & (\mathbb{Z}_{2})^{623} & (\mathbb{Z}_{2})^{1102} \ar[lu,"{{2}}" {gray,pos=0.08,scale=0.8,below left},"{{2}}" {gray,pos=0.92,scale=0.8,above right}] & (\mathbb{Z}_{2})^{1098} \ar[lu,"{{2^{3}}}" {gray,pos=0.08,scale=0.8,below left},"{{2^{3}}}" {gray,pos=0.92,scale=0.8,above right}] & (\mathbb{Z}_{2})^{869} \ar[lu,"{{2^{4}}}" {gray,pos=0.08,scale=0.8,below left},"{{2^{4}}}" {gray,pos=0.92,scale=0.8,above right}] \\
H^{40} & \mathbb{Z}_{2} & (\mathbb{Z}_{2})^{97} \oplus \mathbb{Z}_{4} & (\mathbb{Z}_{2})^{711} \oplus \mathbb{Z}_{4} & (\mathbb{Z}_{2})^{1294} \oplus (\mathbb{Z}_{4})^{5} & (\mathbb{Z}_{2})^{1310} \oplus (\mathbb{Z}_{4})^{3} \ar[lu,"{{2,4}}" {gray,pos=0.08,scale=0.8,below left},"{{2^{2}}}" {gray,pos=0.92,scale=0.8,above right}] & (\mathbb{Z}_{2})^{1042} \oplus (\mathbb{Z}_{4})^{5} \ar[lu,"{{2^{3},4}}" {gray,pos=0.08,scale=0.8,below left},"{{2^{4}}}" {gray,pos=0.92,scale=0.8,above right}] \\
\end{tikzcd}

%% file: tables/omega-Z4-RZ.tex
\begin{tikzcd}[row sep=tiny,column sep=tiny]
& B\mathbb{Z}_{4} & B^{2}\mathbb{Z}_{4} & B^{3}\mathbb{Z}_{4} & B^{4}\mathbb{Z}_{4} & B^{5}\mathbb{Z}_{4} & B^{6}\mathbb{Z}_{4} & B^{7}\mathbb{Z}_{4} \\
H^{0} & \mathbb{R}/\mathbb{Z} & \mathbb{R}/\mathbb{Z} & \mathbb{R}/\mathbb{Z} & \mathbb{R}/\mathbb{Z} & \mathbb{R}/\mathbb{Z} & \mathbb{R}/\mathbb{Z} & \mathbb{R}/\mathbb{Z} \\
H^{1} & \mathbb{Z}_{4} & 0 & 0 & 0 & 0 & 0 & 0 \\
H^{2} & 0 & \mathbb{Z}_{4} \ar[lu,BrickRed,hook,two heads,"{{4}}" {gray,pos=0.08,scale=0.8,below left},"{{4}}" {gray,pos=0.92,scale=0.8,above right}] & 0 & 0 & 0 & 0 & 0 \\
H^{3} & \mathbb{Z}_{4} & 0 & \mathbb{Z}_{4} \ar[lu,BrickRed,hook,two heads,"{{4}}" {gray,pos=0.08,scale=0.8,below left},"{{4}}" {gray,pos=0.92,scale=0.8,above right}] & 0 & 0 & 0 & 0 \\
H^{4} & 0 & \mathbb{Z}_{8} \ar[lu,"{{8}}" {gray,pos=0.08,scale=0.8,below left},"{{2}}" {gray,pos=0.92,scale=0.8,above right}] & 0 & \mathbb{Z}_{4} \ar[lu,BrickRed,hook,two heads,"{{4}}" {gray,pos=0.08,scale=0.8,below left},"{{4}}" {gray,pos=0.92,scale=0.8,above right}] & 0 & 0 & 0 \\
H^{5} & \mathbb{Z}_{4} & \mathbb{Z}_{2} & \mathbb{Z}_{2} \ar[lu,hook,"{{2}}" {gray,pos=0.08,scale=0.8,below left},"{{2}}" {gray,pos=0.92,scale=0.8,above right}] & 0 & \mathbb{Z}_{4} \ar[lu,BrickRed,hook,two heads,"{{4}}" {gray,pos=0.08,scale=0.8,below left},"{{4}}" {gray,pos=0.92,scale=0.8,above right}] & 0 & 0 \\
H^{6} & 0 & \mathbb{Z}_{4} & \mathbb{Z}_{2} \ar[lu,hook,two heads,"{{2}}" {gray,pos=0.08,scale=0.8,below left},"{{2}}" {gray,pos=0.92,scale=0.8,above right}] & \mathbb{Z}_{2} \ar[lu,BrickRed,hook,two heads,"{{2}}" {gray,pos=0.08,scale=0.8,below left},"{{2}}" {gray,pos=0.92,scale=0.8,above right}] & 0 & \mathbb{Z}_{4} \ar[lu,BrickRed,hook,two heads,"{{4}}" {gray,pos=0.08,scale=0.8,below left},"{{4}}" {gray,pos=0.92,scale=0.8,above right}] & 0 \\
H^{7} & \mathbb{Z}_{4} & \mathbb{Z}_{2} & \mathbb{Z}_{4} & \mathbb{Z}_{2} \ar[lu,hook,two heads,"{{2}}" {gray,pos=0.08,scale=0.8,below left},"{{2}}" {gray,pos=0.92,scale=0.8,above right}] & \mathbb{Z}_{2} \ar[lu,BrickRed,hook,two heads,"{{2}}" {gray,pos=0.08,scale=0.8,below left},"{{2}}" {gray,pos=0.92,scale=0.8,above right}] & 0 & \mathbb{Z}_{4} \ar[lu,BrickRed,hook,two heads,"{{4}}" {gray,pos=0.08,scale=0.8,below left},"{{4}}" {gray,pos=0.92,scale=0.8,above right}] \\
H^{8} & 0 & \mathbb{Z}_{2} \oplus \mathbb{Z}_{16} & \mathbb{Z}_{2} & \mathbb{Z}_{8} \ar[lu,"{{8}}" {gray,pos=0.08,scale=0.8,below left},"{{2}}" {gray,pos=0.92,scale=0.8,above right}] & \mathbb{Z}_{2} \ar[lu,BrickRed,hook,two heads,"{{2}}" {gray,pos=0.08,scale=0.8,below left},"{{2}}" {gray,pos=0.92,scale=0.8,above right}] & \mathbb{Z}_{2} \ar[lu,BrickRed,hook,two heads,"{{2}}" {gray,pos=0.08,scale=0.8,below left},"{{2}}" {gray,pos=0.92,scale=0.8,above right}] & 0 \\
H^{9} & \mathbb{Z}_{4} & (\mathbb{Z}_{2})^{2} & (\mathbb{Z}_{2})^{3} \ar[lu,"{{2}}" {gray,pos=0.08,scale=0.8,below left},"{{2}}" {gray,pos=0.92,scale=0.8,above right}] & \mathbb{Z}_{2} & \mathbb{Z}_{2} \ar[lu,hook,"{{2}}" {gray,pos=0.08,scale=0.8,below left},"{{2}}" {gray,pos=0.92,scale=0.8,above right}] & \mathbb{Z}_{2} \ar[lu,BrickRed,hook,two heads,"{{2}}" {gray,pos=0.08,scale=0.8,below left},"{{2}}" {gray,pos=0.92,scale=0.8,above right}] & \mathbb{Z}_{2} \ar[lu,BrickRed,hook,two heads,"{{2}}" {gray,pos=0.08,scale=0.8,below left},"{{2}}" {gray,pos=0.92,scale=0.8,above right}] \\
H^{10} & 0 & \mathbb{Z}_{2} \oplus \mathbb{Z}_{4} & (\mathbb{Z}_{2})^{2} \ar[lu,"{{2}}" {gray,pos=0.08,scale=0.8,below left},"{{2}}" {gray,pos=0.92,scale=0.8,above right}] & (\mathbb{Z}_{2})^{2} \ar[lu,"{{2}}" {gray,pos=0.08,scale=0.8,below left},"{{2}}" {gray,pos=0.92,scale=0.8,above right}] & \mathbb{Z}_{2} \ar[lu,hook,two heads,"{{2}}" {gray,pos=0.08,scale=0.8,below left},"{{2}}" {gray,pos=0.92,scale=0.8,above right}] & \mathbb{Z}_{2} \ar[lu,BrickRed,hook,two heads,"{{2}}" {gray,pos=0.08,scale=0.8,below left},"{{2}}" {gray,pos=0.92,scale=0.8,above right}] & \mathbb{Z}_{2} \ar[lu,BrickRed,hook,two heads,"{{2}}" {gray,pos=0.08,scale=0.8,below left},"{{2}}" {gray,pos=0.92,scale=0.8,above right}] \\
H^{11} & \mathbb{Z}_{4} & (\mathbb{Z}_{2})^{3} & (\mathbb{Z}_{2})^{2} \oplus \mathbb{Z}_{4} & (\mathbb{Z}_{2})^{3} \ar[lu,"{{2}}" {gray,pos=0.08,scale=0.8,below left},"{{2}}" {gray,pos=0.92,scale=0.8,above right}] & \mathbb{Z}_{2} \oplus \mathbb{Z}_{4} \ar[lu,"{{2}}" {gray,pos=0.08,scale=0.8,below left},"{{2}}" {gray,pos=0.92,scale=0.8,above right}] & \mathbb{Z}_{2} \ar[lu,hook,two heads,"{{2}}" {gray,pos=0.08,scale=0.8,below left},"{{2}}" {gray,pos=0.92,scale=0.8,above right}] & \mathbb{Z}_{2} \ar[lu,BrickRed,hook,two heads,"{{2}}" {gray,pos=0.08,scale=0.8,below left},"{{2}}" {gray,pos=0.92,scale=0.8,above right}] \\
H^{12} & 0 & (\mathbb{Z}_{2})^{2} \oplus \mathbb{Z}_{8} & (\mathbb{Z}_{2})^{3} \oplus \mathbb{Z}_{4} \ar[lu,"{{4}}" {gray,pos=0.08,scale=0.8,below left},"{{2}}" {gray,pos=0.92,scale=0.8,above right}] & \mathbb{Z}_{2} \oplus (\mathbb{Z}_{4})^{2} \ar[lu,"{{4}}" {gray,pos=0.08,scale=0.8,below left},"{{2}}" {gray,pos=0.92,scale=0.8,above right}] & (\mathbb{Z}_{2})^{2} \ar[lu,"{{2}}" {gray,pos=0.08,scale=0.8,below left},"{{2}}" {gray,pos=0.92,scale=0.8,above right}] & \mathbb{Z}_{2} \oplus \mathbb{Z}_{8} \ar[lu,"{{2,8}}" {gray,pos=0.08,scale=0.8,below left},"{{2^{2}}}" {gray,pos=0.92,scale=0.8,above right}] & \mathbb{Z}_{2} \ar[lu,BrickRed,hook,two heads,"{{2}}" {gray,pos=0.08,scale=0.8,below left},"{{2}}" {gray,pos=0.92,scale=0.8,above right}] \\
H^{13} & \mathbb{Z}_{4} & (\mathbb{Z}_{2})^{3} & (\mathbb{Z}_{2})^{5} & (\mathbb{Z}_{2})^{4} \ar[lu,"{{2}}" {gray,pos=0.08,scale=0.8,below left},"{{2}}" {gray,pos=0.92,scale=0.8,above right}] & (\mathbb{Z}_{2})^{3} \ar[lu,"{{2}}" {gray,pos=0.08,scale=0.8,below left},"{{2}}" {gray,pos=0.92,scale=0.8,above right}] & (\mathbb{Z}_{2})^{2} \ar[lu,"{{2}}" {gray,pos=0.08,scale=0.8,below left},"{{2}}" {gray,pos=0.92,scale=0.8,above right}] & (\mathbb{Z}_{2})^{2} \ar[lu,hook,"{{2^{2}}}" {gray,pos=0.08,scale=0.8,below left},"{{2^{2}}}" {gray,pos=0.92,scale=0.8,above right}] \\
H^{14} & 0 & (\mathbb{Z}_{2})^{4} \oplus \mathbb{Z}_{4} & (\mathbb{Z}_{2})^{5} & (\mathbb{Z}_{2})^{5} \ar[lu,"{{2}}" {gray,pos=0.08,scale=0.8,below left},"{{2}}" {gray,pos=0.92,scale=0.8,above right}] & (\mathbb{Z}_{2})^{4} \ar[lu,"{{2^{2}}}" {gray,pos=0.08,scale=0.8,below left},"{{2^{2}}}" {gray,pos=0.92,scale=0.8,above right}] & (\mathbb{Z}_{2})^{2} \ar[lu,"{{2}}" {gray,pos=0.08,scale=0.8,below left},"{{2}}" {gray,pos=0.92,scale=0.8,above right}] & (\mathbb{Z}_{2})^{2} \ar[lu,hook,two heads,"{{2^{2}}}" {gray,pos=0.08,scale=0.8,below left},"{{2^{2}}}" {gray,pos=0.92,scale=0.8,above right}] \\
H^{15} & \mathbb{Z}_{4} & (\mathbb{Z}_{2})^{4} & (\mathbb{Z}_{2})^{6} \oplus (\mathbb{Z}_{4})^{2} & (\mathbb{Z}_{2})^{6} & (\mathbb{Z}_{2})^{5} \ar[lu,"{{2}}" {gray,pos=0.08,scale=0.8,below left},"{{2}}" {gray,pos=0.92,scale=0.8,above right}] & (\mathbb{Z}_{2})^{4} \ar[lu,"{{2^{2}}}" {gray,pos=0.08,scale=0.8,below left},"{{2^{2}}}" {gray,pos=0.92,scale=0.8,above right}] & \mathbb{Z}_{2} \oplus \mathbb{Z}_{4} \ar[lu,"{{2}}" {gray,pos=0.08,scale=0.8,below left},"{{2}}" {gray,pos=0.92,scale=0.8,above right}] \\
H^{16} & 0 & (\mathbb{Z}_{2})^{4} \oplus \mathbb{Z}_{32} & (\mathbb{Z}_{2})^{7} \oplus \mathbb{Z}_{4} & (\mathbb{Z}_{2})^{5} \oplus \mathbb{Z}_{4} \oplus \mathbb{Z}_{16} & (\mathbb{Z}_{2})^{4} \oplus \mathbb{Z}_{4} \ar[lu,"{{4}}" {gray,pos=0.08,scale=0.8,below left},"{{2}}" {gray,pos=0.92,scale=0.8,above right}] & (\mathbb{Z}_{2})^{3} \oplus \mathbb{Z}_{4} \ar[lu,"{{2,4}}" {gray,pos=0.08,scale=0.8,below left},"{{2^{2}}}" {gray,pos=0.92,scale=0.8,above right}] & (\mathbb{Z}_{2})^{3} \ar[lu,"{{2^{2}}}" {gray,pos=0.08,scale=0.8,below left},"{{2^{2}}}" {gray,pos=0.92,scale=0.8,above right}] \\
H^{17} & \mathbb{Z}_{4} & (\mathbb{Z}_{2})^{7} & (\mathbb{Z}_{2})^{11} \ar[lu,"{{2}}" {gray,pos=0.08,scale=0.8,below left},"{{2}}" {gray,pos=0.92,scale=0.8,above right}] & (\mathbb{Z}_{2})^{9} \oplus \mathbb{Z}_{4} & (\mathbb{Z}_{2})^{7} \oplus \mathbb{Z}_{4} \ar[lu,"{{2}}" {gray,pos=0.08,scale=0.8,below left},"{{2}}" {gray,pos=0.92,scale=0.8,above right}] & (\mathbb{Z}_{2})^{5} \ar[lu,"{{2}}" {gray,pos=0.08,scale=0.8,below left},"{{2}}" {gray,pos=0.92,scale=0.8,above right}] & (\mathbb{Z}_{2})^{4} \ar[lu,"{{2^{2}}}" {gray,pos=0.08,scale=0.8,below left},"{{2^{2}}}" {gray,pos=0.92,scale=0.8,above right}] \\
H^{18} & 0 & (\mathbb{Z}_{2})^{5} \oplus \mathbb{Z}_{4} & (\mathbb{Z}_{2})^{13} \ar[lu,"{{2}}" {gray,pos=0.08,scale=0.8,below left},"{{2}}" {gray,pos=0.92,scale=0.8,above right}] & (\mathbb{Z}_{2})^{13} \ar[lu,"{{2}}" {gray,pos=0.08,scale=0.8,below left},"{{2}}" {gray,pos=0.92,scale=0.8,above right}] & (\mathbb{Z}_{2})^{8} \ar[lu,"{{2}}" {gray,pos=0.08,scale=0.8,below left},"{{2}}" {gray,pos=0.92,scale=0.8,above right}] & (\mathbb{Z}_{2})^{6} \oplus \mathbb{Z}_{4} \ar[lu,"{{2^{2}}}" {gray,pos=0.08,scale=0.8,below left},"{{2^{2}}}" {gray,pos=0.92,scale=0.8,above right}] & (\mathbb{Z}_{2})^{4} \ar[lu,"{{2^{2}}}" {gray,pos=0.08,scale=0.8,below left},"{{2^{2}}}" {gray,pos=0.92,scale=0.8,above right}] \\
H^{19} & \mathbb{Z}_{4} & (\mathbb{Z}_{2})^{8} & (\mathbb{Z}_{2})^{13} \oplus (\mathbb{Z}_{4})^{2} & (\mathbb{Z}_{2})^{15} \ar[lu,"{{2}}" {gray,pos=0.08,scale=0.8,below left},"{{2}}" {gray,pos=0.92,scale=0.8,above right}] & (\mathbb{Z}_{2})^{12} \ar[lu,"{{2}}" {gray,pos=0.08,scale=0.8,below left},"{{2}}" {gray,pos=0.92,scale=0.8,above right}] & (\mathbb{Z}_{2})^{8} \ar[lu,"{{2}}" {gray,pos=0.08,scale=0.8,below left},"{{2}}" {gray,pos=0.92,scale=0.8,above right}] & (\mathbb{Z}_{2})^{6} \ar[lu,"{{2^{2}}}" {gray,pos=0.08,scale=0.8,below left},"{{2^{2}}}" {gray,pos=0.92,scale=0.8,above right}] \\
H^{20} & 0 & (\mathbb{Z}_{2})^{8} \oplus \mathbb{Z}_{8} & (\mathbb{Z}_{2})^{16} \oplus (\mathbb{Z}_{4})^{2} \ar[lu,"{{4}}" {gray,pos=0.08,scale=0.8,below left},"{{2}}" {gray,pos=0.92,scale=0.8,above right}] & (\mathbb{Z}_{2})^{15} \oplus (\mathbb{Z}_{4})^{3} \ar[lu,"{{4}}" {gray,pos=0.08,scale=0.8,below left},"{{2}}" {gray,pos=0.92,scale=0.8,above right}] & (\mathbb{Z}_{2})^{14} \oplus \mathbb{Z}_{4} \ar[lu,"{{2,4}}" {gray,pos=0.08,scale=0.8,below left},"{{2^{2}}}" {gray,pos=0.92,scale=0.8,above right}] & (\mathbb{Z}_{2})^{9} \oplus \mathbb{Z}_{4} \ar[lu,"{{2,4}}" {gray,pos=0.08,scale=0.8,below left},"{{2^{2}}}" {gray,pos=0.92,scale=0.8,above right}] & (\mathbb{Z}_{2})^{6} \oplus \mathbb{Z}_{4} \ar[lu,"{{2,4}}" {gray,pos=0.08,scale=0.8,below left},"{{2^{2}}}" {gray,pos=0.92,scale=0.8,above right}] \\
H^{21} & \mathbb{Z}_{4} & (\mathbb{Z}_{2})^{9} & (\mathbb{Z}_{2})^{23} & (\mathbb{Z}_{2})^{22} \oplus \mathbb{Z}_{4} \ar[lu,"{{2}}" {gray,pos=0.08,scale=0.8,below left},"{{2}}" {gray,pos=0.92,scale=0.8,above right}] & (\mathbb{Z}_{2})^{17} \oplus \mathbb{Z}_{4} \ar[lu,"{{2}}" {gray,pos=0.08,scale=0.8,below left},"{{2}}" {gray,pos=0.92,scale=0.8,above right}] & (\mathbb{Z}_{2})^{14} \ar[lu,"{{2^{2}}}" {gray,pos=0.08,scale=0.8,below left},"{{2^{2}}}" {gray,pos=0.92,scale=0.8,above right}] & (\mathbb{Z}_{2})^{10} \ar[lu,"{{2^{2}}}" {gray,pos=0.08,scale=0.8,below left},"{{2^{2}}}" {gray,pos=0.92,scale=0.8,above right}] \\
H^{22} & 0 & (\mathbb{Z}_{2})^{10} \oplus \mathbb{Z}_{4} & (\mathbb{Z}_{2})^{25} & (\mathbb{Z}_{2})^{27} \ar[lu,"{{2}}" {gray,pos=0.08,scale=0.8,below left},"{{2}}" {gray,pos=0.92,scale=0.8,above right}] & (\mathbb{Z}_{2})^{21} \oplus \mathbb{Z}_{4} \ar[lu,"{{2^{2}}}" {gray,pos=0.08,scale=0.8,below left},"{{2^{2}}}" {gray,pos=0.92,scale=0.8,above right}] & (\mathbb{Z}_{2})^{15} \oplus \mathbb{Z}_{4} \ar[lu,"{{2^{2}}}" {gray,pos=0.08,scale=0.8,below left},"{{2^{2}}}" {gray,pos=0.92,scale=0.8,above right}] & (\mathbb{Z}_{2})^{11} \ar[lu,"{{2^{3}}}" {gray,pos=0.08,scale=0.8,below left},"{{2^{3}}}" {gray,pos=0.92,scale=0.8,above right}] \\
H^{23} & \mathbb{Z}_{4} & (\mathbb{Z}_{2})^{12} & (\mathbb{Z}_{2})^{28} \oplus (\mathbb{Z}_{4})^{3} & (\mathbb{Z}_{2})^{33} & (\mathbb{Z}_{2})^{27} \oplus \mathbb{Z}_{4} \ar[lu,"{{2}}" {gray,pos=0.08,scale=0.8,below left},"{{2}}" {gray,pos=0.92,scale=0.8,above right}] & (\mathbb{Z}_{2})^{19} \oplus \mathbb{Z}_{4} \ar[lu,"{{2^{2}}}" {gray,pos=0.08,scale=0.8,below left},"{{2^{2}}}" {gray,pos=0.92,scale=0.8,above right}] & (\mathbb{Z}_{2})^{14} \oplus \mathbb{Z}_{4} \ar[lu,"{{2^{2}}}" {gray,pos=0.08,scale=0.8,below left},"{{2^{2}}}" {gray,pos=0.92,scale=0.8,above right}] \\
H^{24} & 0 & (\mathbb{Z}_{2})^{12} \oplus \mathbb{Z}_{16} & (\mathbb{Z}_{2})^{34} \oplus (\mathbb{Z}_{4})^{2} \oplus \mathbb{Z}_{8} & (\mathbb{Z}_{2})^{37} \oplus (\mathbb{Z}_{4})^{2} \oplus (\mathbb{Z}_{8})^{2} & (\mathbb{Z}_{2})^{31} \oplus \mathbb{Z}_{4} \ar[lu,"{{4}}" {gray,pos=0.08,scale=0.8,below left},"{{2}}" {gray,pos=0.92,scale=0.8,above right}] & (\mathbb{Z}_{2})^{23} \oplus \mathbb{Z}_{4} \oplus \mathbb{Z}_{16} \ar[lu,"{{2,4}}" {gray,pos=0.08,scale=0.8,below left},"{{2^{2}}}" {gray,pos=0.92,scale=0.8,above right}] & (\mathbb{Z}_{2})^{17} \oplus \mathbb{Z}_{4} \ar[lu,"{{2^{2},4}}" {gray,pos=0.08,scale=0.8,below left},"{{2^{3}}}" {gray,pos=0.92,scale=0.8,above right}] \\
H^{25} & \mathbb{Z}_{4} & (\mathbb{Z}_{2})^{14} & (\mathbb{Z}_{2})^{43} & (\mathbb{Z}_{2})^{50} \oplus (\mathbb{Z}_{4})^{2} \ar[lu,"{{2}}" {gray,pos=0.08,scale=0.8,below left},"{{2}}" {gray,pos=0.92,scale=0.8,above right}] & (\mathbb{Z}_{2})^{40} \oplus \mathbb{Z}_{4} \ar[lu,"{{2}}" {gray,pos=0.08,scale=0.8,below left},"{{2}}" {gray,pos=0.92,scale=0.8,above right}] & (\mathbb{Z}_{2})^{30} \ar[lu,"{{2}}" {gray,pos=0.08,scale=0.8,below left},"{{2}}" {gray,pos=0.92,scale=0.8,above right}] & (\mathbb{Z}_{2})^{22} \ar[lu,"{{2^{3}}}" {gray,pos=0.08,scale=0.8,below left},"{{2^{3}}}" {gray,pos=0.92,scale=0.8,above right}] \\
H^{26} & 0 & (\mathbb{Z}_{2})^{16} \oplus \mathbb{Z}_{4} & (\mathbb{Z}_{2})^{49} & (\mathbb{Z}_{2})^{60} \ar[lu,"{{2}}" {gray,pos=0.08,scale=0.8,below left},"{{2}}" {gray,pos=0.92,scale=0.8,above right}] & (\mathbb{Z}_{2})^{50} \oplus \mathbb{Z}_{4} \ar[lu,"{{2^{2}}}" {gray,pos=0.08,scale=0.8,below left},"{{2^{2}}}" {gray,pos=0.92,scale=0.8,above right}] & (\mathbb{Z}_{2})^{35} \oplus \mathbb{Z}_{4} \ar[lu,"{{2^{2}}}" {gray,pos=0.08,scale=0.8,below left},"{{2^{2}}}" {gray,pos=0.92,scale=0.8,above right}] & (\mathbb{Z}_{2})^{26} \ar[lu,"{{2^{3}}}" {gray,pos=0.08,scale=0.8,below left},"{{2^{3}}}" {gray,pos=0.92,scale=0.8,above right}] \\
H^{27} & \mathbb{Z}_{4} & (\mathbb{Z}_{2})^{17} & (\mathbb{Z}_{2})^{57} \oplus (\mathbb{Z}_{4})^{4} & (\mathbb{Z}_{2})^{72} & (\mathbb{Z}_{2})^{61} \oplus \mathbb{Z}_{4} \ar[lu,"{{2}}" {gray,pos=0.08,scale=0.8,below left},"{{2}}" {gray,pos=0.92,scale=0.8,above right}] & (\mathbb{Z}_{2})^{46} \oplus \mathbb{Z}_{4} \ar[lu,"{{2^{2}}}" {gray,pos=0.08,scale=0.8,below left},"{{2^{2}}}" {gray,pos=0.92,scale=0.8,above right}] & (\mathbb{Z}_{2})^{32} \oplus \mathbb{Z}_{4} \ar[lu,"{{2^{2}}}" {gray,pos=0.08,scale=0.8,below left},"{{2^{2}}}" {gray,pos=0.92,scale=0.8,above right}] \\
H^{28} & 0 & (\mathbb{Z}_{2})^{18} \oplus \mathbb{Z}_{8} & (\mathbb{Z}_{2})^{65} \oplus (\mathbb{Z}_{4})^{3} & (\mathbb{Z}_{2})^{82} \oplus (\mathbb{Z}_{4})^{5} \ar[lu,"{{4}}" {gray,pos=0.08,scale=0.8,below left},"{{2}}" {gray,pos=0.92,scale=0.8,above right}] & (\mathbb{Z}_{2})^{72} \oplus (\mathbb{Z}_{4})^{3} \ar[lu,"{{4^{2}}}" {gray,pos=0.08,scale=0.8,below left},"{{2^{2}}}" {gray,pos=0.92,scale=0.8,above right}] & (\mathbb{Z}_{2})^{53} \oplus (\mathbb{Z}_{4})^{2} \ar[lu,"{{2,4}}" {gray,pos=0.08,scale=0.8,below left},"{{2^{2}}}" {gray,pos=0.92,scale=0.8,above right}] & (\mathbb{Z}_{2})^{36} \oplus (\mathbb{Z}_{4})^{3} \ar[lu,"{{2^{2},4^{2}}}" {gray,pos=0.08,scale=0.8,below left},"{{2^{4}}}" {gray,pos=0.92,scale=0.8,above right}] \\
H^{29} & \mathbb{Z}_{4} & (\mathbb{Z}_{2})^{22} & (\mathbb{Z}_{2})^{80} & (\mathbb{Z}_{2})^{103} \oplus (\mathbb{Z}_{4})^{3} & (\mathbb{Z}_{2})^{92} \oplus (\mathbb{Z}_{4})^{2} \ar[lu,"{{2}}" {gray,pos=0.08,scale=0.8,below left},"{{2}}" {gray,pos=0.92,scale=0.8,above right}] & (\mathbb{Z}_{2})^{68} \oplus \mathbb{Z}_{4} \ar[lu,"{{2^{2}}}" {gray,pos=0.08,scale=0.8,below left},"{{2^{2}}}" {gray,pos=0.92,scale=0.8,above right}] & (\mathbb{Z}_{2})^{49} \ar[lu,"{{2^{2}}}" {gray,pos=0.08,scale=0.8,below left},"{{2^{2}}}" {gray,pos=0.92,scale=0.8,above right}] \\
H^{30} & 0 & (\mathbb{Z}_{2})^{21} \oplus \mathbb{Z}_{4} & (\mathbb{Z}_{2})^{95} & (\mathbb{Z}_{2})^{126} & (\mathbb{Z}_{2})^{109} \oplus \mathbb{Z}_{4} \ar[lu,"{{2}}" {gray,pos=0.08,scale=0.8,below left},"{{2}}" {gray,pos=0.92,scale=0.8,above right}] & (\mathbb{Z}_{2})^{83} \oplus (\mathbb{Z}_{4})^{2} \ar[lu,"{{2^{3}}}" {gray,pos=0.08,scale=0.8,below left},"{{2^{3}}}" {gray,pos=0.92,scale=0.8,above right}] & (\mathbb{Z}_{2})^{59} \ar[lu,"{{2^{3}}}" {gray,pos=0.08,scale=0.8,below left},"{{2^{3}}}" {gray,pos=0.92,scale=0.8,above right}] \\
H^{31} & \mathbb{Z}_{4} & (\mathbb{Z}_{2})^{25} & (\mathbb{Z}_{2})^{104} \oplus (\mathbb{Z}_{4})^{4} & (\mathbb{Z}_{2})^{151} & (\mathbb{Z}_{2})^{134} \oplus \mathbb{Z}_{4} & (\mathbb{Z}_{2})^{101} \oplus \mathbb{Z}_{4} \ar[lu,"{{2}}" {gray,pos=0.08,scale=0.8,below left},"{{2}}" {gray,pos=0.92,scale=0.8,above right}] & (\mathbb{Z}_{2})^{71} \oplus (\mathbb{Z}_{4})^{2} \ar[lu,"{{2^{3}}}" {gray,pos=0.08,scale=0.8,below left},"{{2^{3}}}" {gray,pos=0.92,scale=0.8,above right}] \\
H^{32} & 0 & (\mathbb{Z}_{2})^{27} \oplus \mathbb{Z}_{64} & (\mathbb{Z}_{2})^{120} \oplus (\mathbb{Z}_{4})^{4} & (\mathbb{Z}_{2})^{177} \oplus (\mathbb{Z}_{4})^{4} \oplus \mathbb{Z}_{8} \oplus \mathbb{Z}_{32} & (\mathbb{Z}_{2})^{162} \oplus \mathbb{Z}_{4} \oplus \mathbb{Z}_{8} & (\mathbb{Z}_{2})^{120} \oplus (\mathbb{Z}_{4})^{2} \oplus \mathbb{Z}_{8} \ar[lu,"{{4}}" {gray,pos=0.08,scale=0.8,below left},"{{2}}" {gray,pos=0.92,scale=0.8,above right}] & (\mathbb{Z}_{2})^{85} \oplus (\mathbb{Z}_{4})^{3} \ar[lu,"{{2,4^{2}}}" {gray,pos=0.08,scale=0.8,below left},"{{2^{3}}}" {gray,pos=0.92,scale=0.8,above right}] \\
H^{33} & \mathbb{Z}_{4} & (\mathbb{Z}_{2})^{29} & (\mathbb{Z}_{2})^{146} \oplus \mathbb{Z}_{4} \ar[lu,"{{2}}" {gray,pos=0.08,scale=0.8,below left},"{{2}}" {gray,pos=0.92,scale=0.8,above right}] & (\mathbb{Z}_{2})^{211} \oplus (\mathbb{Z}_{4})^{4} \oplus \mathbb{Z}_{8} & (\mathbb{Z}_{2})^{195} \oplus (\mathbb{Z}_{4})^{3} \ar[lu,"{{2}}" {gray,pos=0.08,scale=0.8,below left},"{{2}}" {gray,pos=0.92,scale=0.8,above right}] & (\mathbb{Z}_{2})^{152} \oplus \mathbb{Z}_{4} \ar[lu,"{{2}}" {gray,pos=0.08,scale=0.8,below left},"{{2}}" {gray,pos=0.92,scale=0.8,above right}] & (\mathbb{Z}_{2})^{108} \ar[lu,"{{2^{2}}}" {gray,pos=0.08,scale=0.8,below left},"{{2^{2}}}" {gray,pos=0.92,scale=0.8,above right}] \\
H^{34} & 0 & (\mathbb{Z}_{2})^{31} \oplus \mathbb{Z}_{4} & (\mathbb{Z}_{2})^{167} \ar[lu,"{{2}}" {gray,pos=0.08,scale=0.8,below left},"{{2}}" {gray,pos=0.92,scale=0.8,above right}] & (\mathbb{Z}_{2})^{255} \ar[lu,"{{2}}" {gray,pos=0.08,scale=0.8,below left},"{{2}}" {gray,pos=0.92,scale=0.8,above right}] & (\mathbb{Z}_{2})^{237} \oplus (\mathbb{Z}_{4})^{3} \ar[lu,"{{2}}" {gray,pos=0.08,scale=0.8,below left},"{{2}}" {gray,pos=0.92,scale=0.8,above right}] & (\mathbb{Z}_{2})^{181} \oplus (\mathbb{Z}_{4})^{2} \ar[lu,"{{2^{2}}}" {gray,pos=0.08,scale=0.8,below left},"{{2^{2}}}" {gray,pos=0.92,scale=0.8,above right}] & (\mathbb{Z}_{2})^{130} \ar[lu,"{{2^{3}}}" {gray,pos=0.08,scale=0.8,below left},"{{2^{3}}}" {gray,pos=0.92,scale=0.8,above right}] \\
H^{35} & \mathbb{Z}_{4} & (\mathbb{Z}_{2})^{36} & (\mathbb{Z}_{2})^{187} \oplus (\mathbb{Z}_{4})^{5} & (\mathbb{Z}_{2})^{304} \ar[lu,"{{2}}" {gray,pos=0.08,scale=0.8,below left},"{{2}}" {gray,pos=0.92,scale=0.8,above right}] & (\mathbb{Z}_{2})^{290} \oplus (\mathbb{Z}_{4})^{2} \ar[lu,"{{2}}" {gray,pos=0.08,scale=0.8,below left},"{{2}}" {gray,pos=0.92,scale=0.8,above right}] & (\mathbb{Z}_{2})^{220} \oplus (\mathbb{Z}_{4})^{2} \ar[lu,"{{2}}" {gray,pos=0.08,scale=0.8,below left},"{{2}}" {gray,pos=0.92,scale=0.8,above right}] & (\mathbb{Z}_{2})^{157} \oplus (\mathbb{Z}_{4})^{3} \ar[lu,"{{2^{2}}}" {gray,pos=0.08,scale=0.8,below left},"{{2^{2}}}" {gray,pos=0.92,scale=0.8,above right}] \\
H^{36} & 0 & (\mathbb{Z}_{2})^{36} \oplus \mathbb{Z}_{8} & (\mathbb{Z}_{2})^{215} \oplus (\mathbb{Z}_{4})^{7} \ar[lu,"{{4}}" {gray,pos=0.08,scale=0.8,below left},"{{2}}" {gray,pos=0.92,scale=0.8,above right}] & (\mathbb{Z}_{2})^{353} \oplus (\mathbb{Z}_{4})^{8} \ar[lu,"{{4}}" {gray,pos=0.08,scale=0.8,below left},"{{2}}" {gray,pos=0.92,scale=0.8,above right}] & (\mathbb{Z}_{2})^{343} \oplus (\mathbb{Z}_{4})^{3} \ar[lu,"{{2,4}}" {gray,pos=0.08,scale=0.8,below left},"{{2^{2}}}" {gray,pos=0.92,scale=0.8,above right}] & (\mathbb{Z}_{2})^{267} \oplus (\mathbb{Z}_{4})^{4} \oplus \mathbb{Z}_{8} \ar[lu,"{{2,4^{2}}}" {gray,pos=0.08,scale=0.8,below left},"{{2^{3}}}" {gray,pos=0.92,scale=0.8,above right}] & (\mathbb{Z}_{2})^{187} \oplus (\mathbb{Z}_{4})^{5} \ar[lu,"{{2,4^{2}}}" {gray,pos=0.08,scale=0.8,below left},"{{2^{3}}}" {gray,pos=0.92,scale=0.8,above right}] \\
H^{37} & \mathbb{Z}_{4} & (\mathbb{Z}_{2})^{41} & (\mathbb{Z}_{2})^{253} \oplus \mathbb{Z}_{4} & (\mathbb{Z}_{2})^{422} \oplus (\mathbb{Z}_{4})^{6} \ar[lu,"{{2}}" {gray,pos=0.08,scale=0.8,below left},"{{2}}" {gray,pos=0.92,scale=0.8,above right}] & (\mathbb{Z}_{2})^{415} \oplus (\mathbb{Z}_{4})^{3} \ar[lu,"{{2}}" {gray,pos=0.08,scale=0.8,below left},"{{2}}" {gray,pos=0.92,scale=0.8,above right}] & (\mathbb{Z}_{2})^{325} \oplus (\mathbb{Z}_{4})^{2} \ar[lu,"{{2^{2}}}" {gray,pos=0.08,scale=0.8,below left},"{{2^{2}}}" {gray,pos=0.92,scale=0.8,above right}] & (\mathbb{Z}_{2})^{235} \ar[lu,"{{2^{3}}}" {gray,pos=0.08,scale=0.8,below left},"{{2^{3}}}" {gray,pos=0.92,scale=0.8,above right}] \\
H^{38} & 0 & (\mathbb{Z}_{2})^{44} \oplus \mathbb{Z}_{4} & (\mathbb{Z}_{2})^{289} & (\mathbb{Z}_{2})^{503} \oplus \mathbb{Z}_{4} \ar[lu,"{{2}}" {gray,pos=0.08,scale=0.8,below left},"{{2}}" {gray,pos=0.92,scale=0.8,above right}] & (\mathbb{Z}_{2})^{499} \oplus (\mathbb{Z}_{4})^{2} \ar[lu,"{{2^{2}}}" {gray,pos=0.08,scale=0.8,below left},"{{2^{2}}}" {gray,pos=0.92,scale=0.8,above right}] & (\mathbb{Z}_{2})^{389} \oplus (\mathbb{Z}_{4})^{3} \ar[lu,"{{2^{2}}}" {gray,pos=0.08,scale=0.8,below left},"{{2^{2}}}" {gray,pos=0.92,scale=0.8,above right}] & (\mathbb{Z}_{2})^{283} \ar[lu,"{{2^{4}}}" {gray,pos=0.08,scale=0.8,below left},"{{2^{4}}}" {gray,pos=0.92,scale=0.8,above right}] \\
H^{39} & \mathbb{Z}_{4} & (\mathbb{Z}_{2})^{47} & (\mathbb{Z}_{2})^{327} \oplus (\mathbb{Z}_{4})^{7} & (\mathbb{Z}_{2})^{598} & (\mathbb{Z}_{2})^{594} \oplus (\mathbb{Z}_{4})^{3} \ar[lu,"{{2}}" {gray,pos=0.08,scale=0.8,below left},"{{2}}" {gray,pos=0.92,scale=0.8,above right}] & (\mathbb{Z}_{2})^{474} \oplus (\mathbb{Z}_{4})^{3} \ar[lu,"{{2^{2}}}" {gray,pos=0.08,scale=0.8,below left},"{{2^{2}}}" {gray,pos=0.92,scale=0.8,above right}] & (\mathbb{Z}_{2})^{338} \oplus (\mathbb{Z}_{4})^{4} \ar[lu,"{{2^{2}}}" {gray,pos=0.08,scale=0.8,below left},"{{2^{2}}}" {gray,pos=0.92,scale=0.8,above right}] \\
H^{40} & 0 & (\mathbb{Z}_{2})^{50} \oplus \mathbb{Z}_{16} & (\mathbb{Z}_{2})^{370} \oplus (\mathbb{Z}_{4})^{7} \oplus \mathbb{Z}_{8} & (\mathbb{Z}_{2})^{691} \oplus (\mathbb{Z}_{4})^{7} \oplus (\mathbb{Z}_{8})^{3} & (\mathbb{Z}_{2})^{710} \oplus (\mathbb{Z}_{4})^{5} \oplus \mathbb{Z}_{8} \ar[lu,"{{4}}" {gray,pos=0.08,scale=0.8,below left},"{{2}}" {gray,pos=0.92,scale=0.8,above right}] & (\mathbb{Z}_{2})^{565} \oplus (\mathbb{Z}_{4})^{4} \oplus \mathbb{Z}_{8} \ar[lu,"{{2,4}}" {gray,pos=0.08,scale=0.8,below left},"{{2^{2}}}" {gray,pos=0.92,scale=0.8,above right}] & (\mathbb{Z}_{2})^{405} \oplus (\mathbb{Z}_{4})^{4} \oplus \mathbb{Z}_{8} \ar[lu,"{{2^{2},4}}" {gray,pos=0.08,scale=0.8,below left},"{{2^{3}}}" {gray,pos=0.92,scale=0.8,above right}] \\
\end{tikzcd}

%% file: tables/omega-Z4-Z4.tex
\begin{tikzcd}[row sep=tiny,column sep=large]
& B\mathbb{Z}_{4} & B^{2}\mathbb{Z}_{4} & B^{3}\mathbb{Z}_{4} & B^{4}\mathbb{Z}_{4} & B^{5}\mathbb{Z}_{4} & B^{6}\mathbb{Z}_{4} \\
H^{0} & \mathbb{Z}_{4} & \mathbb{Z}_{4} & \mathbb{Z}_{4} & \mathbb{Z}_{4} & \mathbb{Z}_{4} & \mathbb{Z}_{4} \\
H^{1} & \mathbb{Z}_{4} & 0 & 0 & 0 & 0 & 0 \\
H^{2} & \mathbb{Z}_{4} & \mathbb{Z}_{4} \ar[lu,BrickRed,hook,two heads,"{{4}}" {gray,pos=0.08,scale=0.8,below left},"{{4}}" {gray,pos=0.92,scale=0.8,above right}] & 0 & 0 & 0 & 0 \\
H^{3} & \mathbb{Z}_{4} & \mathbb{Z}_{4} \ar[lu,hook,two heads,"{{4}}" {gray,pos=0.08,scale=0.8,below left},"{{4}}" {gray,pos=0.92,scale=0.8,above right}] & \mathbb{Z}_{4} \ar[lu,BrickRed,hook,two heads,"{{4}}" {gray,pos=0.08,scale=0.8,below left},"{{4}}" {gray,pos=0.92,scale=0.8,above right}] & 0 & 0 & 0 \\
H^{4} & \mathbb{Z}_{4} & \mathbb{Z}_{4} & \mathbb{Z}_{4} \ar[lu,BrickRed,hook,two heads,"{{4}}" {gray,pos=0.08,scale=0.8,below left},"{{4}}" {gray,pos=0.92,scale=0.8,above right}] & \mathbb{Z}_{4} \ar[lu,BrickRed,hook,two heads,"{{4}}" {gray,pos=0.08,scale=0.8,below left},"{{4}}" {gray,pos=0.92,scale=0.8,above right}] & 0 & 0 \\
H^{5} & \mathbb{Z}_{4} & \mathbb{Z}_{2} \oplus \mathbb{Z}_{4} \ar[lu,"{{2}}" {gray,pos=0.08,scale=0.8,below left},"{{2}}" {gray,pos=0.92,scale=0.8,above right}] & \mathbb{Z}_{2} \ar[lu,hook,"{{2}}" {gray,pos=0.08,scale=0.8,below left},"{{2}}" {gray,pos=0.92,scale=0.8,above right}] & \mathbb{Z}_{4} \ar[lu,BrickRed,hook,two heads,"{{4}}" {gray,pos=0.08,scale=0.8,below left},"{{4}}" {gray,pos=0.92,scale=0.8,above right}] & \mathbb{Z}_{4} \ar[lu,BrickRed,hook,two heads,"{{4}}" {gray,pos=0.08,scale=0.8,below left},"{{4}}" {gray,pos=0.92,scale=0.8,above right}] & 0 \\
H^{6} & \mathbb{Z}_{4} & \mathbb{Z}_{2} \oplus \mathbb{Z}_{4} & (\mathbb{Z}_{2})^{2} \ar[lu,"{{2}}" {gray,pos=0.08,scale=0.8,below left},"{{2}}" {gray,pos=0.92,scale=0.8,above right}] & \mathbb{Z}_{2} \ar[lu,BrickRed,hook,two heads,"{{2}}" {gray,pos=0.08,scale=0.8,below left},"{{2}}" {gray,pos=0.92,scale=0.8,above right}] & \mathbb{Z}_{4} \ar[lu,BrickRed,hook,two heads,"{{4}}" {gray,pos=0.08,scale=0.8,below left},"{{4}}" {gray,pos=0.92,scale=0.8,above right}] & \mathbb{Z}_{4} \ar[lu,BrickRed,hook,two heads,"{{4}}" {gray,pos=0.08,scale=0.8,below left},"{{4}}" {gray,pos=0.92,scale=0.8,above right}] \\
H^{7} & \mathbb{Z}_{4} & \mathbb{Z}_{2} \oplus \mathbb{Z}_{4} & \mathbb{Z}_{2} \oplus \mathbb{Z}_{4} \ar[lu,"{{2}}" {gray,pos=0.08,scale=0.8,below left},"{{2}}" {gray,pos=0.92,scale=0.8,above right}] & (\mathbb{Z}_{2})^{2} \ar[lu,hook,two heads,"{{2^{2}}}" {gray,pos=0.08,scale=0.8,below left},"{{2^{2}}}" {gray,pos=0.92,scale=0.8,above right}] & \mathbb{Z}_{2} \ar[lu,BrickRed,hook,two heads,"{{2}}" {gray,pos=0.08,scale=0.8,below left},"{{2}}" {gray,pos=0.92,scale=0.8,above right}] & \mathbb{Z}_{4} \ar[lu,BrickRed,hook,two heads,"{{4}}" {gray,pos=0.08,scale=0.8,below left},"{{4}}" {gray,pos=0.92,scale=0.8,above right}] \\
H^{8} & \mathbb{Z}_{4} & (\mathbb{Z}_{2})^{2} \oplus \mathbb{Z}_{4} & \mathbb{Z}_{2} \oplus \mathbb{Z}_{4} & \mathbb{Z}_{2} \oplus \mathbb{Z}_{4} \ar[lu,"{{2}}" {gray,pos=0.08,scale=0.8,below left},"{{2}}" {gray,pos=0.92,scale=0.8,above right}] & (\mathbb{Z}_{2})^{2} \ar[lu,BrickRed,hook,two heads,"{{2^{2}}}" {gray,pos=0.08,scale=0.8,below left},"{{2^{2}}}" {gray,pos=0.92,scale=0.8,above right}] & \mathbb{Z}_{2} \ar[lu,BrickRed,hook,two heads,"{{2}}" {gray,pos=0.08,scale=0.8,below left},"{{2}}" {gray,pos=0.92,scale=0.8,above right}] \\
H^{9} & \mathbb{Z}_{4} & (\mathbb{Z}_{2})^{3} \oplus \mathbb{Z}_{4} \ar[lu,"{{2}}" {gray,pos=0.08,scale=0.8,below left},"{{2}}" {gray,pos=0.92,scale=0.8,above right}] & (\mathbb{Z}_{2})^{4} \ar[lu,"{{2}}" {gray,pos=0.08,scale=0.8,below left},"{{2}}" {gray,pos=0.92,scale=0.8,above right}] & \mathbb{Z}_{2} \oplus \mathbb{Z}_{4} \ar[lu,"{{2}}" {gray,pos=0.08,scale=0.8,below left},"{{2}}" {gray,pos=0.92,scale=0.8,above right}] & (\mathbb{Z}_{2})^{2} \ar[lu,hook,"{{2^{2}}}" {gray,pos=0.08,scale=0.8,below left},"{{2^{2}}}" {gray,pos=0.92,scale=0.8,above right}] & (\mathbb{Z}_{2})^{2} \ar[lu,BrickRed,hook,two heads,"{{2^{2}}}" {gray,pos=0.08,scale=0.8,below left},"{{2^{2}}}" {gray,pos=0.92,scale=0.8,above right}] \\
H^{10} & \mathbb{Z}_{4} & (\mathbb{Z}_{2})^{3} \oplus \mathbb{Z}_{4} & (\mathbb{Z}_{2})^{5} \ar[lu,"{{2}}" {gray,pos=0.08,scale=0.8,below left},"{{2}}" {gray,pos=0.92,scale=0.8,above right}] & (\mathbb{Z}_{2})^{3} \ar[lu,"{{2}}" {gray,pos=0.08,scale=0.8,below left},"{{2}}" {gray,pos=0.92,scale=0.8,above right}] & (\mathbb{Z}_{2})^{2} \ar[lu,"{{2}}" {gray,pos=0.08,scale=0.8,below left},"{{2}}" {gray,pos=0.92,scale=0.8,above right}] & (\mathbb{Z}_{2})^{2} \ar[lu,BrickRed,hook,two heads,"{{2^{2}}}" {gray,pos=0.08,scale=0.8,below left},"{{2^{2}}}" {gray,pos=0.92,scale=0.8,above right}] \\
H^{11} & \mathbb{Z}_{4} & (\mathbb{Z}_{2})^{4} \oplus \mathbb{Z}_{4} & (\mathbb{Z}_{2})^{4} \oplus \mathbb{Z}_{4} \ar[lu,"{{2}}" {gray,pos=0.08,scale=0.8,below left},"{{2}}" {gray,pos=0.92,scale=0.8,above right}] & (\mathbb{Z}_{2})^{5} \ar[lu,"{{2^{2}}}" {gray,pos=0.08,scale=0.8,below left},"{{2^{2}}}" {gray,pos=0.92,scale=0.8,above right}] & (\mathbb{Z}_{2})^{2} \oplus \mathbb{Z}_{4} \ar[lu,"{{2^{2}}}" {gray,pos=0.08,scale=0.8,below left},"{{2^{2}}}" {gray,pos=0.92,scale=0.8,above right}] & (\mathbb{Z}_{2})^{2} \ar[lu,hook,two heads,"{{2^{2}}}" {gray,pos=0.08,scale=0.8,below left},"{{2^{2}}}" {gray,pos=0.92,scale=0.8,above right}] \\
H^{12} & \mathbb{Z}_{4} & (\mathbb{Z}_{2})^{5} \oplus \mathbb{Z}_{4} & (\mathbb{Z}_{2})^{5} \oplus (\mathbb{Z}_{4})^{2} \ar[lu,"{{4}}" {gray,pos=0.08,scale=0.8,below left},"{{2}}" {gray,pos=0.92,scale=0.8,above right}] & (\mathbb{Z}_{2})^{4} \oplus (\mathbb{Z}_{4})^{2} \ar[lu,"{{2,4}}" {gray,pos=0.08,scale=0.8,below left},"{{2^{2}}}" {gray,pos=0.92,scale=0.8,above right}] & (\mathbb{Z}_{2})^{3} \oplus \mathbb{Z}_{4} \ar[lu,"{{2^{2}}}" {gray,pos=0.08,scale=0.8,below left},"{{2^{2}}}" {gray,pos=0.92,scale=0.8,above right}] & (\mathbb{Z}_{2})^{2} \oplus \mathbb{Z}_{4} \ar[lu,"{{2^{2}}}" {gray,pos=0.08,scale=0.8,below left},"{{2^{2}}}" {gray,pos=0.92,scale=0.8,above right}] \\
H^{13} & \mathbb{Z}_{4} & (\mathbb{Z}_{2})^{5} \oplus \mathbb{Z}_{4} & (\mathbb{Z}_{2})^{8} \oplus \mathbb{Z}_{4} \ar[lu,"{{4}}" {gray,pos=0.08,scale=0.8,below left},"{{2}}" {gray,pos=0.92,scale=0.8,above right}] & (\mathbb{Z}_{2})^{5} \oplus (\mathbb{Z}_{4})^{2} \ar[lu,"{{2,4}}" {gray,pos=0.08,scale=0.8,below left},"{{2^{2}}}" {gray,pos=0.92,scale=0.8,above right}] & (\mathbb{Z}_{2})^{5} \ar[lu,"{{2^{2}}}" {gray,pos=0.08,scale=0.8,below left},"{{2^{2}}}" {gray,pos=0.92,scale=0.8,above right}] & (\mathbb{Z}_{2})^{3} \oplus \mathbb{Z}_{4} \ar[lu,"{{2^{3}}}" {gray,pos=0.08,scale=0.8,below left},"{{2^{3}}}" {gray,pos=0.92,scale=0.8,above right}] \\
H^{14} & \mathbb{Z}_{4} & (\mathbb{Z}_{2})^{7} \oplus \mathbb{Z}_{4} & (\mathbb{Z}_{2})^{10} & (\mathbb{Z}_{2})^{9} \ar[lu,"{{2^{2}}}" {gray,pos=0.08,scale=0.8,below left},"{{2^{2}}}" {gray,pos=0.92,scale=0.8,above right}] & (\mathbb{Z}_{2})^{7} \ar[lu,"{{2^{3}}}" {gray,pos=0.08,scale=0.8,below left},"{{2^{3}}}" {gray,pos=0.92,scale=0.8,above right}] & (\mathbb{Z}_{2})^{4} \ar[lu,"{{2^{2}}}" {gray,pos=0.08,scale=0.8,below left},"{{2^{2}}}" {gray,pos=0.92,scale=0.8,above right}] \\
H^{15} & \mathbb{Z}_{4} & (\mathbb{Z}_{2})^{8} \oplus \mathbb{Z}_{4} & (\mathbb{Z}_{2})^{11} \oplus (\mathbb{Z}_{4})^{2} & (\mathbb{Z}_{2})^{11} \ar[lu,"{{2}}" {gray,pos=0.08,scale=0.8,below left},"{{2}}" {gray,pos=0.92,scale=0.8,above right}] & (\mathbb{Z}_{2})^{9} \ar[lu,"{{2^{3}}}" {gray,pos=0.08,scale=0.8,below left},"{{2^{3}}}" {gray,pos=0.92,scale=0.8,above right}] & (\mathbb{Z}_{2})^{6} \ar[lu,"{{2^{3}}}" {gray,pos=0.08,scale=0.8,below left},"{{2^{3}}}" {gray,pos=0.92,scale=0.8,above right}] \\
H^{16} & \mathbb{Z}_{4} & (\mathbb{Z}_{2})^{8} \oplus \mathbb{Z}_{4} & (\mathbb{Z}_{2})^{13} \oplus (\mathbb{Z}_{4})^{3} & (\mathbb{Z}_{2})^{11} \oplus (\mathbb{Z}_{4})^{2} & (\mathbb{Z}_{2})^{9} \oplus \mathbb{Z}_{4} \ar[lu,"{{2,4}}" {gray,pos=0.08,scale=0.8,below left},"{{2^{2}}}" {gray,pos=0.92,scale=0.8,above right}] & (\mathbb{Z}_{2})^{7} \oplus \mathbb{Z}_{4} \ar[lu,"{{2^{3},4}}" {gray,pos=0.08,scale=0.8,below left},"{{2^{4}}}" {gray,pos=0.92,scale=0.8,above right}] \\
H^{17} & \mathbb{Z}_{4} & (\mathbb{Z}_{2})^{11} \oplus \mathbb{Z}_{4} \ar[lu,"{{2}}" {gray,pos=0.08,scale=0.8,below left},"{{2}}" {gray,pos=0.92,scale=0.8,above right}] & (\mathbb{Z}_{2})^{18} \oplus \mathbb{Z}_{4} \ar[lu,"{{2}}" {gray,pos=0.08,scale=0.8,below left},"{{2}}" {gray,pos=0.92,scale=0.8,above right}] & (\mathbb{Z}_{2})^{14} \oplus (\mathbb{Z}_{4})^{3} \ar[lu,"{{2}}" {gray,pos=0.08,scale=0.8,below left},"{{2}}" {gray,pos=0.92,scale=0.8,above right}] & (\mathbb{Z}_{2})^{11} \oplus (\mathbb{Z}_{4})^{2} \ar[lu,"{{2,4}}" {gray,pos=0.08,scale=0.8,below left},"{{2^{2}}}" {gray,pos=0.92,scale=0.8,above right}] & (\mathbb{Z}_{2})^{8} \oplus \mathbb{Z}_{4} \ar[lu,"{{2^{2},4}}" {gray,pos=0.08,scale=0.8,below left},"{{2^{3}}}" {gray,pos=0.92,scale=0.8,above right}] \\
H^{18} & \mathbb{Z}_{4} & (\mathbb{Z}_{2})^{12} \oplus \mathbb{Z}_{4} & (\mathbb{Z}_{2})^{24} \ar[lu,"{{2}}" {gray,pos=0.08,scale=0.8,below left},"{{2}}" {gray,pos=0.92,scale=0.8,above right}] & (\mathbb{Z}_{2})^{22} \oplus \mathbb{Z}_{4} \ar[lu,"{{2}}" {gray,pos=0.08,scale=0.8,below left},"{{2}}" {gray,pos=0.92,scale=0.8,above right}] & (\mathbb{Z}_{2})^{15} \oplus \mathbb{Z}_{4} \ar[lu,"{{2}}" {gray,pos=0.08,scale=0.8,below left},"{{2}}" {gray,pos=0.92,scale=0.8,above right}] & (\mathbb{Z}_{2})^{11} \oplus \mathbb{Z}_{4} \ar[lu,"{{2^{3}}}" {gray,pos=0.08,scale=0.8,below left},"{{2^{3}}}" {gray,pos=0.92,scale=0.8,above right}] \\
H^{19} & \mathbb{Z}_{4} & (\mathbb{Z}_{2})^{13} \oplus \mathbb{Z}_{4} & (\mathbb{Z}_{2})^{26} \oplus (\mathbb{Z}_{4})^{2} \ar[lu,"{{2}}" {gray,pos=0.08,scale=0.8,below left},"{{2}}" {gray,pos=0.92,scale=0.8,above right}] & (\mathbb{Z}_{2})^{28} \ar[lu,"{{2^{2}}}" {gray,pos=0.08,scale=0.8,below left},"{{2^{2}}}" {gray,pos=0.92,scale=0.8,above right}] & (\mathbb{Z}_{2})^{20} \ar[lu,"{{2^{2}}}" {gray,pos=0.08,scale=0.8,below left},"{{2^{2}}}" {gray,pos=0.92,scale=0.8,above right}] & (\mathbb{Z}_{2})^{14} \oplus \mathbb{Z}_{4} \ar[lu,"{{2^{3}}}" {gray,pos=0.08,scale=0.8,below left},"{{2^{3}}}" {gray,pos=0.92,scale=0.8,above right}] \\
H^{20} & \mathbb{Z}_{4} & (\mathbb{Z}_{2})^{16} \oplus \mathbb{Z}_{4} & (\mathbb{Z}_{2})^{29} \oplus (\mathbb{Z}_{4})^{4} \ar[lu,"{{4}}" {gray,pos=0.08,scale=0.8,below left},"{{2}}" {gray,pos=0.92,scale=0.8,above right}] & (\mathbb{Z}_{2})^{30} \oplus (\mathbb{Z}_{4})^{3} \ar[lu,"{{2,4}}" {gray,pos=0.08,scale=0.8,below left},"{{2^{2}}}" {gray,pos=0.92,scale=0.8,above right}] & (\mathbb{Z}_{2})^{26} \oplus \mathbb{Z}_{4} \ar[lu,"{{2^{2},4}}" {gray,pos=0.08,scale=0.8,below left},"{{2^{3}}}" {gray,pos=0.92,scale=0.8,above right}] & (\mathbb{Z}_{2})^{17} \oplus \mathbb{Z}_{4} \ar[lu,"{{2^{2},4}}" {gray,pos=0.08,scale=0.8,below left},"{{2^{3}}}" {gray,pos=0.92,scale=0.8,above right}] \\
H^{21} & \mathbb{Z}_{4} & (\mathbb{Z}_{2})^{17} \oplus \mathbb{Z}_{4} & (\mathbb{Z}_{2})^{39} \oplus (\mathbb{Z}_{4})^{2} \ar[lu,"{{4}}" {gray,pos=0.08,scale=0.8,below left},"{{2}}" {gray,pos=0.92,scale=0.8,above right}] & (\mathbb{Z}_{2})^{37} \oplus (\mathbb{Z}_{4})^{4} \ar[lu,"{{2,4}}" {gray,pos=0.08,scale=0.8,below left},"{{2^{2}}}" {gray,pos=0.92,scale=0.8,above right}] & (\mathbb{Z}_{2})^{31} \oplus (\mathbb{Z}_{4})^{2} \ar[lu,"{{2^{2},4}}" {gray,pos=0.08,scale=0.8,below left},"{{2^{3}}}" {gray,pos=0.92,scale=0.8,above right}] & (\mathbb{Z}_{2})^{23} \oplus \mathbb{Z}_{4} \ar[lu,"{{2^{3},4}}" {gray,pos=0.08,scale=0.8,below left},"{{2^{4}}}" {gray,pos=0.92,scale=0.8,above right}] \\
H^{22} & \mathbb{Z}_{4} & (\mathbb{Z}_{2})^{19} \oplus \mathbb{Z}_{4} & (\mathbb{Z}_{2})^{48} & (\mathbb{Z}_{2})^{49} \oplus \mathbb{Z}_{4} \ar[lu,"{{2^{2}}}" {gray,pos=0.08,scale=0.8,below left},"{{2^{2}}}" {gray,pos=0.92,scale=0.8,above right}] & (\mathbb{Z}_{2})^{38} \oplus (\mathbb{Z}_{4})^{2} \ar[lu,"{{2^{3}}}" {gray,pos=0.08,scale=0.8,below left},"{{2^{3}}}" {gray,pos=0.92,scale=0.8,above right}] & (\mathbb{Z}_{2})^{29} \oplus \mathbb{Z}_{4} \ar[lu,"{{2^{4}}}" {gray,pos=0.08,scale=0.8,below left},"{{2^{4}}}" {gray,pos=0.92,scale=0.8,above right}] \\
H^{23} & \mathbb{Z}_{4} & (\mathbb{Z}_{2})^{22} \oplus \mathbb{Z}_{4} & (\mathbb{Z}_{2})^{53} \oplus (\mathbb{Z}_{4})^{3} & (\mathbb{Z}_{2})^{60} \ar[lu,"{{2}}" {gray,pos=0.08,scale=0.8,below left},"{{2}}" {gray,pos=0.92,scale=0.8,above right}] & (\mathbb{Z}_{2})^{48} \oplus (\mathbb{Z}_{4})^{2} \ar[lu,"{{2^{3}}}" {gray,pos=0.08,scale=0.8,below left},"{{2^{3}}}" {gray,pos=0.92,scale=0.8,above right}] & (\mathbb{Z}_{2})^{34} \oplus (\mathbb{Z}_{4})^{2} \ar[lu,"{{2^{4}}}" {gray,pos=0.08,scale=0.8,below left},"{{2^{4}}}" {gray,pos=0.92,scale=0.8,above right}] \\
H^{24} & \mathbb{Z}_{4} & (\mathbb{Z}_{2})^{24} \oplus \mathbb{Z}_{4} & (\mathbb{Z}_{2})^{62} \oplus (\mathbb{Z}_{4})^{6} & (\mathbb{Z}_{2})^{70} \oplus (\mathbb{Z}_{4})^{4} & (\mathbb{Z}_{2})^{58} \oplus (\mathbb{Z}_{4})^{2} \ar[lu,"{{2,4}}" {gray,pos=0.08,scale=0.8,below left},"{{2^{2}}}" {gray,pos=0.92,scale=0.8,above right}] & (\mathbb{Z}_{2})^{42} \oplus (\mathbb{Z}_{4})^{3} \ar[lu,"{{2^{3},4}}" {gray,pos=0.08,scale=0.8,below left},"{{2^{4}}}" {gray,pos=0.92,scale=0.8,above right}] \\
H^{25} & \mathbb{Z}_{4} & (\mathbb{Z}_{2})^{26} \oplus \mathbb{Z}_{4} & (\mathbb{Z}_{2})^{77} \oplus (\mathbb{Z}_{4})^{3} & (\mathbb{Z}_{2})^{87} \oplus (\mathbb{Z}_{4})^{6} \ar[lu,"{{2}}" {gray,pos=0.08,scale=0.8,below left},"{{2}}" {gray,pos=0.92,scale=0.8,above right}] & (\mathbb{Z}_{2})^{71} \oplus (\mathbb{Z}_{4})^{2} \ar[lu,"{{2,4}}" {gray,pos=0.08,scale=0.8,below left},"{{2^{2}}}" {gray,pos=0.92,scale=0.8,above right}] & (\mathbb{Z}_{2})^{53} \oplus (\mathbb{Z}_{4})^{2} \ar[lu,"{{2^{3},4}}" {gray,pos=0.08,scale=0.8,below left},"{{2^{4}}}" {gray,pos=0.92,scale=0.8,above right}] \\
H^{26} & \mathbb{Z}_{4} & (\mathbb{Z}_{2})^{30} \oplus \mathbb{Z}_{4} & (\mathbb{Z}_{2})^{92} & (\mathbb{Z}_{2})^{110} \oplus (\mathbb{Z}_{4})^{2} \ar[lu,"{{2}}" {gray,pos=0.08,scale=0.8,below left},"{{2}}" {gray,pos=0.92,scale=0.8,above right}] & (\mathbb{Z}_{2})^{90} \oplus (\mathbb{Z}_{4})^{2} \ar[lu,"{{2^{2}}}" {gray,pos=0.08,scale=0.8,below left},"{{2^{2}}}" {gray,pos=0.92,scale=0.8,above right}] & (\mathbb{Z}_{2})^{65} \oplus \mathbb{Z}_{4} \ar[lu,"{{2^{3}}}" {gray,pos=0.08,scale=0.8,below left},"{{2^{3}}}" {gray,pos=0.92,scale=0.8,above right}] \\
H^{27} & \mathbb{Z}_{4} & (\mathbb{Z}_{2})^{33} \oplus \mathbb{Z}_{4} & (\mathbb{Z}_{2})^{106} \oplus (\mathbb{Z}_{4})^{4} & (\mathbb{Z}_{2})^{132} \ar[lu,"{{2}}" {gray,pos=0.08,scale=0.8,below left},"{{2}}" {gray,pos=0.92,scale=0.8,above right}] & (\mathbb{Z}_{2})^{111} \oplus (\mathbb{Z}_{4})^{2} \ar[lu,"{{2^{3}}}" {gray,pos=0.08,scale=0.8,below left},"{{2^{3}}}" {gray,pos=0.92,scale=0.8,above right}] & (\mathbb{Z}_{2})^{81} \oplus (\mathbb{Z}_{4})^{2} \ar[lu,"{{2^{4}}}" {gray,pos=0.08,scale=0.8,below left},"{{2^{4}}}" {gray,pos=0.92,scale=0.8,above right}] \\
H^{28} & \mathbb{Z}_{4} & (\mathbb{Z}_{2})^{35} \oplus \mathbb{Z}_{4} & (\mathbb{Z}_{2})^{122} \oplus (\mathbb{Z}_{4})^{7} & (\mathbb{Z}_{2})^{154} \oplus (\mathbb{Z}_{4})^{5} \ar[lu,"{{4}}" {gray,pos=0.08,scale=0.8,below left},"{{2}}" {gray,pos=0.92,scale=0.8,above right}] & (\mathbb{Z}_{2})^{133} \oplus (\mathbb{Z}_{4})^{4} \ar[lu,"{{2,4^{2}}}" {gray,pos=0.08,scale=0.8,below left},"{{2^{3}}}" {gray,pos=0.92,scale=0.8,above right}] & (\mathbb{Z}_{2})^{99} \oplus (\mathbb{Z}_{4})^{3} \ar[lu,"{{2^{3},4}}" {gray,pos=0.08,scale=0.8,below left},"{{2^{4}}}" {gray,pos=0.92,scale=0.8,above right}] \\
H^{29} & \mathbb{Z}_{4} & (\mathbb{Z}_{2})^{40} \oplus \mathbb{Z}_{4} & (\mathbb{Z}_{2})^{145} \oplus (\mathbb{Z}_{4})^{3} & (\mathbb{Z}_{2})^{185} \oplus (\mathbb{Z}_{4})^{8} \ar[lu,"{{4}}" {gray,pos=0.08,scale=0.8,below left},"{{2}}" {gray,pos=0.92,scale=0.8,above right}] & (\mathbb{Z}_{2})^{164} \oplus (\mathbb{Z}_{4})^{5} \ar[lu,"{{2,4^{2}}}" {gray,pos=0.08,scale=0.8,below left},"{{2^{3}}}" {gray,pos=0.92,scale=0.8,above right}] & (\mathbb{Z}_{2})^{121} \oplus (\mathbb{Z}_{4})^{3} \ar[lu,"{{2^{3},4}}" {gray,pos=0.08,scale=0.8,below left},"{{2^{4}}}" {gray,pos=0.92,scale=0.8,above right}] \\
H^{30} & \mathbb{Z}_{4} & (\mathbb{Z}_{2})^{43} \oplus \mathbb{Z}_{4} & (\mathbb{Z}_{2})^{175} & (\mathbb{Z}_{2})^{229} \oplus (\mathbb{Z}_{4})^{3} & (\mathbb{Z}_{2})^{201} \oplus (\mathbb{Z}_{4})^{3} \ar[lu,"{{2^{2}}}" {gray,pos=0.08,scale=0.8,below left},"{{2^{2}}}" {gray,pos=0.92,scale=0.8,above right}] & (\mathbb{Z}_{2})^{151} \oplus (\mathbb{Z}_{4})^{3} \ar[lu,"{{2^{5}}}" {gray,pos=0.08,scale=0.8,below left},"{{2^{5}}}" {gray,pos=0.92,scale=0.8,above right}] \\
H^{31} & \mathbb{Z}_{4} & (\mathbb{Z}_{2})^{46} \oplus \mathbb{Z}_{4} & (\mathbb{Z}_{2})^{199} \oplus (\mathbb{Z}_{4})^{4} & (\mathbb{Z}_{2})^{277} & (\mathbb{Z}_{2})^{243} \oplus (\mathbb{Z}_{4})^{2} \ar[lu,"{{2}}" {gray,pos=0.08,scale=0.8,below left},"{{2}}" {gray,pos=0.92,scale=0.8,above right}] & (\mathbb{Z}_{2})^{184} \oplus (\mathbb{Z}_{4})^{3} \ar[lu,"{{2^{4}}}" {gray,pos=0.08,scale=0.8,below left},"{{2^{4}}}" {gray,pos=0.92,scale=0.8,above right}] \\
H^{32} & \mathbb{Z}_{4} & (\mathbb{Z}_{2})^{52} \oplus \mathbb{Z}_{4} & (\mathbb{Z}_{2})^{224} \oplus (\mathbb{Z}_{4})^{8} & (\mathbb{Z}_{2})^{328} \oplus (\mathbb{Z}_{4})^{6} & (\mathbb{Z}_{2})^{296} \oplus (\mathbb{Z}_{4})^{3} & (\mathbb{Z}_{2})^{221} \oplus (\mathbb{Z}_{4})^{4} \ar[lu,"{{2,4}}" {gray,pos=0.08,scale=0.8,below left},"{{2^{2}}}" {gray,pos=0.92,scale=0.8,above right}] \\
H^{33} & \mathbb{Z}_{4} & (\mathbb{Z}_{2})^{56} \oplus \mathbb{Z}_{4} \ar[lu,"{{2}}" {gray,pos=0.08,scale=0.8,below left},"{{2}}" {gray,pos=0.92,scale=0.8,above right}] & (\mathbb{Z}_{2})^{266} \oplus (\mathbb{Z}_{4})^{5} \ar[lu,"{{2}}" {gray,pos=0.08,scale=0.8,below left},"{{2}}" {gray,pos=0.92,scale=0.8,above right}] & (\mathbb{Z}_{2})^{388} \oplus (\mathbb{Z}_{4})^{11} \ar[lu,"{{2}}" {gray,pos=0.08,scale=0.8,below left},"{{2}}" {gray,pos=0.92,scale=0.8,above right}] & (\mathbb{Z}_{2})^{357} \oplus (\mathbb{Z}_{4})^{5} \ar[lu,"{{2}}" {gray,pos=0.08,scale=0.8,below left},"{{2}}" {gray,pos=0.92,scale=0.8,above right}] & (\mathbb{Z}_{2})^{272} \oplus (\mathbb{Z}_{4})^{4} \ar[lu,"{{2,4}}" {gray,pos=0.08,scale=0.8,below left},"{{2^{2}}}" {gray,pos=0.92,scale=0.8,above right}] \\
H^{34} & \mathbb{Z}_{4} & (\mathbb{Z}_{2})^{60} \oplus \mathbb{Z}_{4} & (\mathbb{Z}_{2})^{313} \oplus \mathbb{Z}_{4} \ar[lu,"{{2}}" {gray,pos=0.08,scale=0.8,below left},"{{2}}" {gray,pos=0.92,scale=0.8,above right}] & (\mathbb{Z}_{2})^{466} \oplus (\mathbb{Z}_{4})^{5} \ar[lu,"{{2}}" {gray,pos=0.08,scale=0.8,below left},"{{2}}" {gray,pos=0.92,scale=0.8,above right}] & (\mathbb{Z}_{2})^{432} \oplus (\mathbb{Z}_{4})^{6} \ar[lu,"{{2}}" {gray,pos=0.08,scale=0.8,below left},"{{2}}" {gray,pos=0.92,scale=0.8,above right}] & (\mathbb{Z}_{2})^{333} \oplus (\mathbb{Z}_{4})^{3} \ar[lu,"{{2^{2}}}" {gray,pos=0.08,scale=0.8,below left},"{{2^{2}}}" {gray,pos=0.92,scale=0.8,above right}] \\
H^{35} & \mathbb{Z}_{4} & (\mathbb{Z}_{2})^{67} \oplus \mathbb{Z}_{4} & (\mathbb{Z}_{2})^{354} \oplus (\mathbb{Z}_{4})^{5} \ar[lu,"{{2}}" {gray,pos=0.08,scale=0.8,below left},"{{2}}" {gray,pos=0.92,scale=0.8,above right}] & (\mathbb{Z}_{2})^{559} \ar[lu,"{{2^{2}}}" {gray,pos=0.08,scale=0.8,below left},"{{2^{2}}}" {gray,pos=0.92,scale=0.8,above right}] & (\mathbb{Z}_{2})^{527} \oplus (\mathbb{Z}_{4})^{5} \ar[lu,"{{2^{2}}}" {gray,pos=0.08,scale=0.8,below left},"{{2^{2}}}" {gray,pos=0.92,scale=0.8,above right}] & (\mathbb{Z}_{2})^{401} \oplus (\mathbb{Z}_{4})^{4} \ar[lu,"{{2^{3}}}" {gray,pos=0.08,scale=0.8,below left},"{{2^{3}}}" {gray,pos=0.92,scale=0.8,above right}] \\
H^{36} & \mathbb{Z}_{4} & (\mathbb{Z}_{2})^{72} \oplus \mathbb{Z}_{4} & (\mathbb{Z}_{2})^{402} \oplus (\mathbb{Z}_{4})^{12} \ar[lu,"{{4}}" {gray,pos=0.08,scale=0.8,below left},"{{2}}" {gray,pos=0.92,scale=0.8,above right}] & (\mathbb{Z}_{2})^{657} \oplus (\mathbb{Z}_{4})^{8} \ar[lu,"{{2,4}}" {gray,pos=0.08,scale=0.8,below left},"{{2^{2}}}" {gray,pos=0.92,scale=0.8,above right}] & (\mathbb{Z}_{2})^{633} \oplus (\mathbb{Z}_{4})^{5} \ar[lu,"{{2^{2},4}}" {gray,pos=0.08,scale=0.8,below left},"{{2^{3}}}" {gray,pos=0.92,scale=0.8,above right}] & (\mathbb{Z}_{2})^{487} \oplus (\mathbb{Z}_{4})^{7} \ar[lu,"{{2^{2},4^{2}}}" {gray,pos=0.08,scale=0.8,below left},"{{2^{4}}}" {gray,pos=0.92,scale=0.8,above right}] \\
H^{37} & \mathbb{Z}_{4} & (\mathbb{Z}_{2})^{77} \oplus \mathbb{Z}_{4} & (\mathbb{Z}_{2})^{468} \oplus (\mathbb{Z}_{4})^{8} \ar[lu,"{{4}}" {gray,pos=0.08,scale=0.8,below left},"{{2}}" {gray,pos=0.92,scale=0.8,above right}] & (\mathbb{Z}_{2})^{775} \oplus (\mathbb{Z}_{4})^{14} \ar[lu,"{{2,4}}" {gray,pos=0.08,scale=0.8,below left},"{{2^{2}}}" {gray,pos=0.92,scale=0.8,above right}] & (\mathbb{Z}_{2})^{758} \oplus (\mathbb{Z}_{4})^{6} \ar[lu,"{{2^{2},4}}" {gray,pos=0.08,scale=0.8,below left},"{{2^{3}}}" {gray,pos=0.92,scale=0.8,above right}] & (\mathbb{Z}_{2})^{592} \oplus (\mathbb{Z}_{4})^{7} \ar[lu,"{{2^{3},4^{2}}}" {gray,pos=0.08,scale=0.8,below left},"{{2^{5}}}" {gray,pos=0.92,scale=0.8,above right}] \\
H^{38} & \mathbb{Z}_{4} & (\mathbb{Z}_{2})^{85} \oplus \mathbb{Z}_{4} & (\mathbb{Z}_{2})^{542} \oplus \mathbb{Z}_{4} & (\mathbb{Z}_{2})^{925} \oplus (\mathbb{Z}_{4})^{7} \ar[lu,"{{2^{2}}}" {gray,pos=0.08,scale=0.8,below left},"{{2^{2}}}" {gray,pos=0.92,scale=0.8,above right}] & (\mathbb{Z}_{2})^{914} \oplus (\mathbb{Z}_{4})^{5} \ar[lu,"{{2^{3}}}" {gray,pos=0.08,scale=0.8,below left},"{{2^{3}}}" {gray,pos=0.92,scale=0.8,above right}] & (\mathbb{Z}_{2})^{714} \oplus (\mathbb{Z}_{4})^{5} \ar[lu,"{{2^{4}}}" {gray,pos=0.08,scale=0.8,below left},"{{2^{4}}}" {gray,pos=0.92,scale=0.8,above right}] \\
H^{39} & \mathbb{Z}_{4} & (\mathbb{Z}_{2})^{91} \oplus \mathbb{Z}_{4} & (\mathbb{Z}_{2})^{616} \oplus (\mathbb{Z}_{4})^{7} & (\mathbb{Z}_{2})^{1101} \oplus \mathbb{Z}_{4} \ar[lu,"{{2}}" {gray,pos=0.08,scale=0.8,below left},"{{2}}" {gray,pos=0.92,scale=0.8,above right}] & (\mathbb{Z}_{2})^{1093} \oplus (\mathbb{Z}_{4})^{5} \ar[lu,"{{2^{3}}}" {gray,pos=0.08,scale=0.8,below left},"{{2^{3}}}" {gray,pos=0.92,scale=0.8,above right}] & (\mathbb{Z}_{2})^{863} \oplus (\mathbb{Z}_{4})^{6} \ar[lu,"{{2^{4}}}" {gray,pos=0.08,scale=0.8,below left},"{{2^{4}}}" {gray,pos=0.92,scale=0.8,above right}] \\
H^{40} & \mathbb{Z}_{4} & (\mathbb{Z}_{2})^{97} \oplus \mathbb{Z}_{4} & (\mathbb{Z}_{2})^{697} \oplus (\mathbb{Z}_{4})^{15} & (\mathbb{Z}_{2})^{1289} \oplus (\mathbb{Z}_{4})^{10} & (\mathbb{Z}_{2})^{1304} \oplus (\mathbb{Z}_{4})^{9} \ar[lu,"{{2,4}}" {gray,pos=0.08,scale=0.8,below left},"{{2^{2}}}" {gray,pos=0.92,scale=0.8,above right}] & (\mathbb{Z}_{2})^{1039} \oplus (\mathbb{Z}_{4})^{8} \ar[lu,"{{2^{3},4}}" {gray,pos=0.08,scale=0.8,below left},"{{2^{4}}}" {gray,pos=0.92,scale=0.8,above right}] \\
\end{tikzcd}

%% file: tables/omega-Z3-RZ.tex
\begin{tikzcd}[row sep=tiny,column sep=tiny]
& B\mathbb{Z}_{3} & B^{2}\mathbb{Z}_{3} & B^{3}\mathbb{Z}_{3} & B^{4}\mathbb{Z}_{3} & B^{5}\mathbb{Z}_{3} & B^{6}\mathbb{Z}_{3} & B^{7}\mathbb{Z}_{3} & B^{8}\mathbb{Z}_{3} & B^{9}\mathbb{Z}_{3} \\
H^{0} & \mathbb{R}/\mathbb{Z} & \mathbb{R}/\mathbb{Z} & \mathbb{R}/\mathbb{Z} & \mathbb{R}/\mathbb{Z} & \mathbb{R}/\mathbb{Z} & \mathbb{R}/\mathbb{Z} & \mathbb{R}/\mathbb{Z} & \mathbb{R}/\mathbb{Z} & \mathbb{R}/\mathbb{Z} \\
H^{1} & \mathbb{Z}_{3} & 0 & 0 & 0 & 0 & 0 & 0 & 0 & 0 \\
H^{2} & 0 & \mathbb{Z}_{3} \ar[lu,BrickRed,hook,two heads,"{{3}}" {gray,pos=0.08,scale=0.8,below left},"{{3}}" {gray,pos=0.92,scale=0.8,above right}] & 0 & 0 & 0 & 0 & 0 & 0 & 0 \\
H^{3} & \mathbb{Z}_{3} & 0 & \mathbb{Z}_{3} \ar[lu,BrickRed,hook,two heads,"{{3}}" {gray,pos=0.08,scale=0.8,below left},"{{3}}" {gray,pos=0.92,scale=0.8,above right}] & 0 & 0 & 0 & 0 & 0 & 0 \\
H^{4} & 0 & \mathbb{Z}_{3} & 0 & \mathbb{Z}_{3} \ar[lu,BrickRed,hook,two heads,"{{3}}" {gray,pos=0.08,scale=0.8,below left},"{{3}}" {gray,pos=0.92,scale=0.8,above right}] & 0 & 0 & 0 & 0 & 0 \\
H^{5} & \mathbb{Z}_{3} & 0 & 0 & 0 & \mathbb{Z}_{3} \ar[lu,BrickRed,hook,two heads,"{{3}}" {gray,pos=0.08,scale=0.8,below left},"{{3}}" {gray,pos=0.92,scale=0.8,above right}] & 0 & 0 & 0 & 0 \\
H^{6} & 0 & \mathbb{Z}_{9} & 0 & 0 & 0 & \mathbb{Z}_{3} \ar[lu,BrickRed,hook,two heads,"{{3}}" {gray,pos=0.08,scale=0.8,below left},"{{3}}" {gray,pos=0.92,scale=0.8,above right}] & 0 & 0 & 0 \\
H^{7} & \mathbb{Z}_{3} & \mathbb{Z}_{3} & (\mathbb{Z}_{3})^{2} \ar[lu,"{{3}}" {gray,pos=0.08,scale=0.8,below left},"{{3}}" {gray,pos=0.92,scale=0.8,above right}] & 0 & 0 & 0 & \mathbb{Z}_{3} \ar[lu,BrickRed,hook,two heads,"{{3}}" {gray,pos=0.08,scale=0.8,below left},"{{3}}" {gray,pos=0.92,scale=0.8,above right}] & 0 & 0 \\
H^{8} & 0 & \mathbb{Z}_{3} & \mathbb{Z}_{3} \ar[lu,hook,two heads,"{{3}}" {gray,pos=0.08,scale=0.8,below left},"{{3}}" {gray,pos=0.92,scale=0.8,above right}] & (\mathbb{Z}_{3})^{2} \ar[lu,"{{3}}" {gray,pos=0.08,scale=0.8,below left},"{{3}}" {gray,pos=0.92,scale=0.8,above right}] & 0 & 0 & 0 & \mathbb{Z}_{3} \ar[lu,BrickRed,hook,two heads,"{{3}}" {gray,pos=0.08,scale=0.8,below left},"{{3}}" {gray,pos=0.92,scale=0.8,above right}] & 0 \\
H^{9} & \mathbb{Z}_{3} & \mathbb{Z}_{3} & 0 & \mathbb{Z}_{3} \ar[lu,hook,two heads,"{{3}}" {gray,pos=0.08,scale=0.8,below left},"{{3}}" {gray,pos=0.92,scale=0.8,above right}] & \mathbb{Z}_{3} \ar[lu,hook,"{{3}}" {gray,pos=0.08,scale=0.8,below left},"{{3}}" {gray,pos=0.92,scale=0.8,above right}] & 0 & 0 & 0 & \mathbb{Z}_{3} \ar[lu,BrickRed,hook,two heads,"{{3}}" {gray,pos=0.08,scale=0.8,below left},"{{3}}" {gray,pos=0.92,scale=0.8,above right}] \\
H^{10} & 0 & (\mathbb{Z}_{3})^{2} & \mathbb{Z}_{3} & 0 & \mathbb{Z}_{3} \ar[lu,hook,two heads,"{{3}}" {gray,pos=0.08,scale=0.8,below left},"{{3}}" {gray,pos=0.92,scale=0.8,above right}] & \mathbb{Z}_{3} \ar[lu,BrickRed,hook,two heads,"{{3}}" {gray,pos=0.08,scale=0.8,below left},"{{3}}" {gray,pos=0.92,scale=0.8,above right}] & 0 & 0 & 0 \\
H^{11} & \mathbb{Z}_{3} & \mathbb{Z}_{3} & (\mathbb{Z}_{3})^{3} & 0 & \mathbb{Z}_{3} & \mathbb{Z}_{3} \ar[lu,hook,two heads,"{{3}}" {gray,pos=0.08,scale=0.8,below left},"{{3}}" {gray,pos=0.92,scale=0.8,above right}] & \mathbb{Z}_{3} \ar[lu,BrickRed,hook,two heads,"{{3}}" {gray,pos=0.08,scale=0.8,below left},"{{3}}" {gray,pos=0.92,scale=0.8,above right}] & 0 & 0 \\
H^{12} & 0 & \mathbb{Z}_{3} \oplus \mathbb{Z}_{9} & \mathbb{Z}_{3} & \mathbb{Z}_{3} \oplus \mathbb{Z}_{9} & 0 & \mathbb{Z}_{3} & \mathbb{Z}_{3} \ar[lu,BrickRed,hook,two heads,"{{3}}" {gray,pos=0.08,scale=0.8,below left},"{{3}}" {gray,pos=0.92,scale=0.8,above right}] & \mathbb{Z}_{3} \ar[lu,BrickRed,hook,two heads,"{{3}}" {gray,pos=0.08,scale=0.8,below left},"{{3}}" {gray,pos=0.92,scale=0.8,above right}] & 0 \\
H^{13} & \mathbb{Z}_{3} & \mathbb{Z}_{3} & 0 & (\mathbb{Z}_{3})^{3} & \mathbb{Z}_{3} \ar[lu,hook,"{{3}}" {gray,pos=0.08,scale=0.8,below left},"{{3}}" {gray,pos=0.92,scale=0.8,above right}] & 0 & 0 & \mathbb{Z}_{3} \ar[lu,BrickRed,hook,two heads,"{{3}}" {gray,pos=0.08,scale=0.8,below left},"{{3}}" {gray,pos=0.92,scale=0.8,above right}] & \mathbb{Z}_{3} \ar[lu,BrickRed,hook,two heads,"{{3}}" {gray,pos=0.08,scale=0.8,below left},"{{3}}" {gray,pos=0.92,scale=0.8,above right}] \\
H^{14} & 0 & (\mathbb{Z}_{3})^{2} & \mathbb{Z}_{3} & \mathbb{Z}_{3} & (\mathbb{Z}_{3})^{2} \ar[lu,"{{3}}" {gray,pos=0.08,scale=0.8,below left},"{{3}}" {gray,pos=0.92,scale=0.8,above right}] & \mathbb{Z}_{3} \ar[lu,hook,two heads,"{{3}}" {gray,pos=0.08,scale=0.8,below left},"{{3}}" {gray,pos=0.92,scale=0.8,above right}] & 0 & 0 & \mathbb{Z}_{3} \ar[lu,BrickRed,hook,two heads,"{{3}}" {gray,pos=0.08,scale=0.8,below left},"{{3}}" {gray,pos=0.92,scale=0.8,above right}] \\
H^{15} & \mathbb{Z}_{3} & (\mathbb{Z}_{3})^{2} & (\mathbb{Z}_{3})^{5} & 0 & (\mathbb{Z}_{3})^{2} & \mathbb{Z}_{3} \ar[lu,hook,"{{3}}" {gray,pos=0.08,scale=0.8,below left},"{{3}}" {gray,pos=0.92,scale=0.8,above right}] & (\mathbb{Z}_{3})^{2} \ar[lu,two heads,"{{3}}" {gray,pos=0.08,scale=0.8,below left},"{{3}}" {gray,pos=0.92,scale=0.8,above right}] & 0 & 0 \\
H^{16} & 0 & (\mathbb{Z}_{3})^{2} & (\mathbb{Z}_{3})^{3} & (\mathbb{Z}_{3})^{3} & \mathbb{Z}_{3} & \mathbb{Z}_{3} & \mathbb{Z}_{3} \ar[lu,hook,two heads,"{{3}}" {gray,pos=0.08,scale=0.8,below left},"{{3}}" {gray,pos=0.92,scale=0.8,above right}] & (\mathbb{Z}_{3})^{2} \ar[lu,"{{3}}" {gray,pos=0.08,scale=0.8,below left},"{{3}}" {gray,pos=0.92,scale=0.8,above right}] & 0 \\
H^{17} & \mathbb{Z}_{3} & (\mathbb{Z}_{3})^{2} & 0 & (\mathbb{Z}_{3})^{4} & \mathbb{Z}_{3} & (\mathbb{Z}_{3})^{2} & 0 & \mathbb{Z}_{3} \ar[lu,hook,two heads,"{{3}}" {gray,pos=0.08,scale=0.8,below left},"{{3}}" {gray,pos=0.92,scale=0.8,above right}] & \mathbb{Z}_{3} \ar[lu,hook,"{{3}}" {gray,pos=0.08,scale=0.8,below left},"{{3}}" {gray,pos=0.92,scale=0.8,above right}] \\
H^{18} & 0 & (\mathbb{Z}_{3})^{2} \oplus \mathbb{Z}_{27} & (\mathbb{Z}_{3})^{3} & (\mathbb{Z}_{3})^{3} & \mathbb{Z}_{3} & \mathbb{Z}_{3} \oplus \mathbb{Z}_{9} & \mathbb{Z}_{3} & 0 & \mathbb{Z}_{3} \ar[lu,hook,two heads,"{{3}}" {gray,pos=0.08,scale=0.8,below left},"{{3}}" {gray,pos=0.92,scale=0.8,above right}] \\
H^{19} & \mathbb{Z}_{3} & (\mathbb{Z}_{3})^{3} & (\mathbb{Z}_{3})^{8} \ar[lu,"{{3}}" {gray,pos=0.08,scale=0.8,below left},"{{3}}" {gray,pos=0.92,scale=0.8,above right}] & \mathbb{Z}_{3} & (\mathbb{Z}_{3})^{4} & \mathbb{Z}_{3} & (\mathbb{Z}_{3})^{3} \ar[lu,"{{3}}" {gray,pos=0.08,scale=0.8,below left},"{{3}}" {gray,pos=0.92,scale=0.8,above right}] & 0 & \mathbb{Z}_{3} \\
H^{20} & 0 & (\mathbb{Z}_{3})^{3} & (\mathbb{Z}_{3})^{4} \ar[lu,"{{3}}" {gray,pos=0.08,scale=0.8,below left},"{{3}}" {gray,pos=0.92,scale=0.8,above right}] & (\mathbb{Z}_{3})^{4} \ar[lu,"{{3}}" {gray,pos=0.08,scale=0.8,below left},"{{3}}" {gray,pos=0.92,scale=0.8,above right}] & (\mathbb{Z}_{3})^{4} & (\mathbb{Z}_{3})^{2} & (\mathbb{Z}_{3})^{2} \ar[lu,two heads,"{{3}}" {gray,pos=0.08,scale=0.8,below left},"{{3}}" {gray,pos=0.92,scale=0.8,above right}] & (\mathbb{Z}_{3})^{2} \ar[lu,"{{3}}" {gray,pos=0.08,scale=0.8,below left},"{{3}}" {gray,pos=0.92,scale=0.8,above right}] & 0 \\
H^{21} & \mathbb{Z}_{3} & (\mathbb{Z}_{3})^{3} & 0 & (\mathbb{Z}_{3})^{7} \ar[lu,"{{3}}" {gray,pos=0.08,scale=0.8,below left},"{{3}}" {gray,pos=0.92,scale=0.8,above right}] & (\mathbb{Z}_{3})^{3} \ar[lu,"{{3}}" {gray,pos=0.08,scale=0.8,below left},"{{3}}" {gray,pos=0.92,scale=0.8,above right}] & (\mathbb{Z}_{3})^{3} & 0 & (\mathbb{Z}_{3})^{3} \ar[lu,"{{3}}" {gray,pos=0.08,scale=0.8,below left},"{{3}}" {gray,pos=0.92,scale=0.8,above right}] & \mathbb{Z}_{3} \ar[lu,hook,"{{3}}" {gray,pos=0.08,scale=0.8,below left},"{{3}}" {gray,pos=0.92,scale=0.8,above right}] \\
H^{22} & 0 & (\mathbb{Z}_{3})^{4} & (\mathbb{Z}_{3})^{4} & (\mathbb{Z}_{3})^{6} & (\mathbb{Z}_{3})^{3} \ar[lu,"{{3}}" {gray,pos=0.08,scale=0.8,below left},"{{3}}" {gray,pos=0.92,scale=0.8,above right}] & (\mathbb{Z}_{3})^{4} \ar[lu,"{{3}}" {gray,pos=0.08,scale=0.8,below left},"{{3}}" {gray,pos=0.92,scale=0.8,above right}] & \mathbb{Z}_{3} & \mathbb{Z}_{3} & (\mathbb{Z}_{3})^{2} \ar[lu,"{{3}}" {gray,pos=0.08,scale=0.8,below left},"{{3}}" {gray,pos=0.92,scale=0.8,above right}] \\
H^{23} & \mathbb{Z}_{3} & (\mathbb{Z}_{3})^{4} & (\mathbb{Z}_{3})^{12} & (\mathbb{Z}_{3})^{3} & (\mathbb{Z}_{3})^{4} & (\mathbb{Z}_{3})^{4} \ar[lu,"{{3}}" {gray,pos=0.08,scale=0.8,below left},"{{3}}" {gray,pos=0.92,scale=0.8,above right}] & (\mathbb{Z}_{3})^{6} \ar[lu,"{{3}}" {gray,pos=0.08,scale=0.8,below left},"{{3}}" {gray,pos=0.92,scale=0.8,above right}] & 0 & (\mathbb{Z}_{3})^{2} \\
H^{24} & 0 & (\mathbb{Z}_{3})^{3} \oplus \mathbb{Z}_{9} & (\mathbb{Z}_{3})^{6} \oplus \mathbb{Z}_{9} & (\mathbb{Z}_{3})^{4} \oplus (\mathbb{Z}_{9})^{2} & (\mathbb{Z}_{3})^{6} & (\mathbb{Z}_{3})^{3} & (\mathbb{Z}_{3})^{4} \ar[lu,"{{3}}" {gray,pos=0.08,scale=0.8,below left},"{{3}}" {gray,pos=0.92,scale=0.8,above right}] & (\mathbb{Z}_{3})^{3} \oplus \mathbb{Z}_{9} \ar[lu,"{{3}}" {gray,pos=0.08,scale=0.8,below left},"{{3}}" {gray,pos=0.92,scale=0.8,above right}] & \mathbb{Z}_{3} \\
H^{25} & \mathbb{Z}_{3} & (\mathbb{Z}_{3})^{4} & \mathbb{Z}_{3} & (\mathbb{Z}_{3})^{11} \ar[lu,"{{3}}" {gray,pos=0.08,scale=0.8,below left},"{{3}}" {gray,pos=0.92,scale=0.8,above right}] & (\mathbb{Z}_{3})^{8} \ar[lu,"{{3}}" {gray,pos=0.08,scale=0.8,below left},"{{3}}" {gray,pos=0.92,scale=0.8,above right}] & (\mathbb{Z}_{3})^{4} & 0 & (\mathbb{Z}_{3})^{5} \ar[lu,"{{3}}" {gray,pos=0.08,scale=0.8,below left},"{{3}}" {gray,pos=0.92,scale=0.8,above right}] & (\mathbb{Z}_{3})^{2} \ar[lu,hook,"{{3^{2}}}" {gray,pos=0.08,scale=0.8,below left},"{{3^{2}}}" {gray,pos=0.92,scale=0.8,above right}] \\
H^{26} & 0 & (\mathbb{Z}_{3})^{6} & (\mathbb{Z}_{3})^{8} & (\mathbb{Z}_{3})^{10} \ar[lu,two heads,"{{3}}" {gray,pos=0.08,scale=0.8,below left},"{{3}}" {gray,pos=0.92,scale=0.8,above right}] & (\mathbb{Z}_{3})^{7} \ar[lu,"{{3^{2}}}" {gray,pos=0.08,scale=0.8,below left},"{{3^{2}}}" {gray,pos=0.92,scale=0.8,above right}] & (\mathbb{Z}_{3})^{7} \ar[lu,"{{3}}" {gray,pos=0.08,scale=0.8,below left},"{{3}}" {gray,pos=0.92,scale=0.8,above right}] & (\mathbb{Z}_{3})^{3} & (\mathbb{Z}_{3})^{2} & (\mathbb{Z}_{3})^{3} \ar[lu,"{{3^{2}}}" {gray,pos=0.08,scale=0.8,below left},"{{3^{2}}}" {gray,pos=0.92,scale=0.8,above right}] \\
H^{27} & \mathbb{Z}_{3} & (\mathbb{Z}_{3})^{5} & (\mathbb{Z}_{3})^{18} & (\mathbb{Z}_{3})^{7} & (\mathbb{Z}_{3})^{8} \ar[lu,"{{3}}" {gray,pos=0.08,scale=0.8,below left},"{{3}}" {gray,pos=0.92,scale=0.8,above right}] & (\mathbb{Z}_{3})^{7} \ar[lu,"{{3^{2}}}" {gray,pos=0.08,scale=0.8,below left},"{{3^{2}}}" {gray,pos=0.92,scale=0.8,above right}] & (\mathbb{Z}_{3})^{8} \ar[lu,"{{3}}" {gray,pos=0.08,scale=0.8,below left},"{{3}}" {gray,pos=0.92,scale=0.8,above right}] & \mathbb{Z}_{3} & (\mathbb{Z}_{3})^{4} \\
H^{28} & 0 & (\mathbb{Z}_{3})^{6} & (\mathbb{Z}_{3})^{11} & (\mathbb{Z}_{3})^{9} & (\mathbb{Z}_{3})^{8} & (\mathbb{Z}_{3})^{7} \ar[lu,"{{3}}" {gray,pos=0.08,scale=0.8,below left},"{{3}}" {gray,pos=0.92,scale=0.8,above right}] & (\mathbb{Z}_{3})^{7} \ar[lu,"{{3^{2}}}" {gray,pos=0.08,scale=0.8,below left},"{{3^{2}}}" {gray,pos=0.92,scale=0.8,above right}] & (\mathbb{Z}_{3})^{4} \ar[lu,"{{3}}" {gray,pos=0.08,scale=0.8,below left},"{{3}}" {gray,pos=0.92,scale=0.8,above right}] & (\mathbb{Z}_{3})^{3} \\
H^{29} & \mathbb{Z}_{3} & (\mathbb{Z}_{3})^{6} & (\mathbb{Z}_{3})^{3} & (\mathbb{Z}_{3})^{17} & (\mathbb{Z}_{3})^{10} & (\mathbb{Z}_{3})^{8} & (\mathbb{Z}_{3})^{2} \ar[lu,"{{3}}" {gray,pos=0.08,scale=0.8,below left},"{{3}}" {gray,pos=0.92,scale=0.8,above right}] & (\mathbb{Z}_{3})^{9} \ar[lu,"{{3^{2}}}" {gray,pos=0.08,scale=0.8,below left},"{{3^{2}}}" {gray,pos=0.92,scale=0.8,above right}] & (\mathbb{Z}_{3})^{2} \ar[lu,"{{3}}" {gray,pos=0.08,scale=0.8,below left},"{{3}}" {gray,pos=0.92,scale=0.8,above right}] \\
H^{30} & 0 & (\mathbb{Z}_{3})^{6} \oplus \mathbb{Z}_{9} & (\mathbb{Z}_{3})^{11} & (\mathbb{Z}_{3})^{18} & (\mathbb{Z}_{3})^{11} \oplus \mathbb{Z}_{9} & (\mathbb{Z}_{3})^{8} \oplus \mathbb{Z}_{9} & (\mathbb{Z}_{3})^{5} & (\mathbb{Z}_{3})^{6} \ar[lu,"{{3}}" {gray,pos=0.08,scale=0.8,below left},"{{3}}" {gray,pos=0.92,scale=0.8,above right}] & (\mathbb{Z}_{3})^{4} \ar[lu,"{{3^{2}}}" {gray,pos=0.08,scale=0.8,below left},"{{3^{2}}}" {gray,pos=0.92,scale=0.8,above right}] \\
H^{31} & \mathbb{Z}_{3} & (\mathbb{Z}_{3})^{7} & (\mathbb{Z}_{3})^{25} & (\mathbb{Z}_{3})^{12} & (\mathbb{Z}_{3})^{14} & (\mathbb{Z}_{3})^{10} \ar[lu,"{{3}}" {gray,pos=0.08,scale=0.8,below left},"{{3}}" {gray,pos=0.92,scale=0.8,above right}] & (\mathbb{Z}_{3})^{15} \ar[lu,"{{3}}" {gray,pos=0.08,scale=0.8,below left},"{{3}}" {gray,pos=0.92,scale=0.8,above right}] & \mathbb{Z}_{3} & (\mathbb{Z}_{3})^{6} \ar[lu,"{{3}}" {gray,pos=0.08,scale=0.8,below left},"{{3}}" {gray,pos=0.92,scale=0.8,above right}] \\
H^{32} & 0 & (\mathbb{Z}_{3})^{7} & (\mathbb{Z}_{3})^{17} & (\mathbb{Z}_{3})^{14} & (\mathbb{Z}_{3})^{15} & (\mathbb{Z}_{3})^{13} \ar[lu,"{{3}}" {gray,pos=0.08,scale=0.8,below left},"{{3}}" {gray,pos=0.92,scale=0.8,above right}] & (\mathbb{Z}_{3})^{11} \ar[lu,"{{3^{2}}}" {gray,pos=0.08,scale=0.8,below left},"{{3^{2}}}" {gray,pos=0.92,scale=0.8,above right}] & (\mathbb{Z}_{3})^{7} \ar[lu,"{{3}}" {gray,pos=0.08,scale=0.8,below left},"{{3}}" {gray,pos=0.92,scale=0.8,above right}] & (\mathbb{Z}_{3})^{5} \\
H^{33} & \mathbb{Z}_{3} & (\mathbb{Z}_{3})^{7} & (\mathbb{Z}_{3})^{5} & (\mathbb{Z}_{3})^{24} & (\mathbb{Z}_{3})^{17} & (\mathbb{Z}_{3})^{16} & (\mathbb{Z}_{3})^{3} \ar[lu,"{{3}}" {gray,pos=0.08,scale=0.8,below left},"{{3}}" {gray,pos=0.92,scale=0.8,above right}] & (\mathbb{Z}_{3})^{13} \ar[lu,"{{3^{2}}}" {gray,pos=0.08,scale=0.8,below left},"{{3^{2}}}" {gray,pos=0.92,scale=0.8,above right}] & (\mathbb{Z}_{3})^{4} \ar[lu,"{{3}}" {gray,pos=0.08,scale=0.8,below left},"{{3}}" {gray,pos=0.92,scale=0.8,above right}] \\
H^{34} & 0 & (\mathbb{Z}_{3})^{9} & (\mathbb{Z}_{3})^{18} & (\mathbb{Z}_{3})^{31} & (\mathbb{Z}_{3})^{19} & (\mathbb{Z}_{3})^{17} & (\mathbb{Z}_{3})^{9} & (\mathbb{Z}_{3})^{10} \ar[lu,"{{3}}" {gray,pos=0.08,scale=0.8,below left},"{{3}}" {gray,pos=0.92,scale=0.8,above right}] & (\mathbb{Z}_{3})^{6} \ar[lu,"{{3^{2}}}" {gray,pos=0.08,scale=0.8,below left},"{{3^{2}}}" {gray,pos=0.92,scale=0.8,above right}] \\
H^{35} & \mathbb{Z}_{3} & (\mathbb{Z}_{3})^{8} & (\mathbb{Z}_{3})^{36} & (\mathbb{Z}_{3})^{24} & (\mathbb{Z}_{3})^{23} & (\mathbb{Z}_{3})^{16} & (\mathbb{Z}_{3})^{22} & (\mathbb{Z}_{3})^{4} & (\mathbb{Z}_{3})^{10} \ar[lu,"{{3}}" {gray,pos=0.08,scale=0.8,below left},"{{3}}" {gray,pos=0.92,scale=0.8,above right}] \\
H^{36} & 0 & (\mathbb{Z}_{3})^{8} \oplus \mathbb{Z}_{27} & (\mathbb{Z}_{3})^{25} & (\mathbb{Z}_{3})^{20} \oplus \mathbb{Z}_{9} \oplus \mathbb{Z}_{27} & (\mathbb{Z}_{3})^{27} & (\mathbb{Z}_{3})^{16} \oplus \mathbb{Z}_{9} & (\mathbb{Z}_{3})^{17} \oplus \mathbb{Z}_{9} & (\mathbb{Z}_{3})^{7} \oplus \mathbb{Z}_{9} & (\mathbb{Z}_{3})^{10} \\
H^{37} & \mathbb{Z}_{3} & (\mathbb{Z}_{3})^{9} & (\mathbb{Z}_{3})^{10} & (\mathbb{Z}_{3})^{33} \oplus \mathbb{Z}_{9} & (\mathbb{Z}_{3})^{31} \ar[lu,"{{3}}" {gray,pos=0.08,scale=0.8,below left},"{{3}}" {gray,pos=0.92,scale=0.8,above right}] & (\mathbb{Z}_{3})^{25} & (\mathbb{Z}_{3})^{6} & (\mathbb{Z}_{3})^{21} \ar[lu,"{{3}}" {gray,pos=0.08,scale=0.8,below left},"{{3}}" {gray,pos=0.92,scale=0.8,above right}] & (\mathbb{Z}_{3})^{8} \ar[lu,"{{3}}" {gray,pos=0.08,scale=0.8,below left},"{{3}}" {gray,pos=0.92,scale=0.8,above right}] \\
H^{38} & 0 & (\mathbb{Z}_{3})^{10} & (\mathbb{Z}_{3})^{24} & (\mathbb{Z}_{3})^{45} & (\mathbb{Z}_{3})^{31} \ar[lu,"{{3}}" {gray,pos=0.08,scale=0.8,below left},"{{3}}" {gray,pos=0.92,scale=0.8,above right}] & (\mathbb{Z}_{3})^{32} \ar[lu,"{{3}}" {gray,pos=0.08,scale=0.8,below left},"{{3}}" {gray,pos=0.92,scale=0.8,above right}] & (\mathbb{Z}_{3})^{15} & (\mathbb{Z}_{3})^{21} \ar[lu,"{{3}}" {gray,pos=0.08,scale=0.8,below left},"{{3}}" {gray,pos=0.92,scale=0.8,above right}] & (\mathbb{Z}_{3})^{9} \ar[lu,"{{3^{2}}}" {gray,pos=0.08,scale=0.8,below left},"{{3^{2}}}" {gray,pos=0.92,scale=0.8,above right}] \\
H^{39} & \mathbb{Z}_{3} & (\mathbb{Z}_{3})^{11} & (\mathbb{Z}_{3})^{49} & (\mathbb{Z}_{3})^{42} & (\mathbb{Z}_{3})^{35} & (\mathbb{Z}_{3})^{31} \ar[lu,"{{3}}" {gray,pos=0.08,scale=0.8,below left},"{{3}}" {gray,pos=0.92,scale=0.8,above right}] & (\mathbb{Z}_{3})^{38} \ar[lu,"{{3}}" {gray,pos=0.08,scale=0.8,below left},"{{3}}" {gray,pos=0.92,scale=0.8,above right}] & (\mathbb{Z}_{3})^{9} & (\mathbb{Z}_{3})^{15} \ar[lu,"{{3}}" {gray,pos=0.08,scale=0.8,below left},"{{3}}" {gray,pos=0.92,scale=0.8,above right}] \\
H^{40} & 0 & (\mathbb{Z}_{3})^{10} & (\mathbb{Z}_{3})^{36} & (\mathbb{Z}_{3})^{37} & (\mathbb{Z}_{3})^{41} & (\mathbb{Z}_{3})^{31} & (\mathbb{Z}_{3})^{31} \ar[lu,"{{3}}" {gray,pos=0.08,scale=0.8,below left},"{{3}}" {gray,pos=0.92,scale=0.8,above right}] & (\mathbb{Z}_{3})^{14} \ar[lu,"{{3}}" {gray,pos=0.08,scale=0.8,below left},"{{3}}" {gray,pos=0.92,scale=0.8,above right}] & (\mathbb{Z}_{3})^{17} \\
\end{tikzcd}

%% file: tables/omega-Z3-Z3.tex
\begin{tikzcd}[row sep=tiny,column sep=large]
& B\mathbb{Z}_{3} & B^{2}\mathbb{Z}_{3} & B^{3}\mathbb{Z}_{3} & B^{4}\mathbb{Z}_{3} & B^{5}\mathbb{Z}_{3} & B^{6}\mathbb{Z}_{3} & B^{7}\mathbb{Z}_{3} & B^{8}\mathbb{Z}_{3} & B^{9}\mathbb{Z}_{3} \\
H^{0} & \mathbb{Z}_{3} & \mathbb{Z}_{3} & \mathbb{Z}_{3} & \mathbb{Z}_{3} & \mathbb{Z}_{3} & \mathbb{Z}_{3} & \mathbb{Z}_{3} & \mathbb{Z}_{3} & \mathbb{Z}_{3} \\
H^{1} & \mathbb{Z}_{3} & 0 & 0 & 0 & 0 & 0 & 0 & 0 & 0 \\
H^{2} & \mathbb{Z}_{3} & \mathbb{Z}_{3} \ar[lu,BrickRed,hook,two heads,"{{3}}" {gray,pos=0.08,scale=0.8,below left},"{{3}}" {gray,pos=0.92,scale=0.8,above right}] & 0 & 0 & 0 & 0 & 0 & 0 & 0 \\
H^{3} & \mathbb{Z}_{3} & \mathbb{Z}_{3} \ar[lu,hook,two heads,"{{3}}" {gray,pos=0.08,scale=0.8,below left},"{{3}}" {gray,pos=0.92,scale=0.8,above right}] & \mathbb{Z}_{3} \ar[lu,BrickRed,hook,two heads,"{{3}}" {gray,pos=0.08,scale=0.8,below left},"{{3}}" {gray,pos=0.92,scale=0.8,above right}] & 0 & 0 & 0 & 0 & 0 & 0 \\
H^{4} & \mathbb{Z}_{3} & \mathbb{Z}_{3} & \mathbb{Z}_{3} \ar[lu,BrickRed,hook,two heads,"{{3}}" {gray,pos=0.08,scale=0.8,below left},"{{3}}" {gray,pos=0.92,scale=0.8,above right}] & \mathbb{Z}_{3} \ar[lu,BrickRed,hook,two heads,"{{3}}" {gray,pos=0.08,scale=0.8,below left},"{{3}}" {gray,pos=0.92,scale=0.8,above right}] & 0 & 0 & 0 & 0 & 0 \\
H^{5} & \mathbb{Z}_{3} & \mathbb{Z}_{3} & 0 & \mathbb{Z}_{3} \ar[lu,BrickRed,hook,two heads,"{{3}}" {gray,pos=0.08,scale=0.8,below left},"{{3}}" {gray,pos=0.92,scale=0.8,above right}] & \mathbb{Z}_{3} \ar[lu,BrickRed,hook,two heads,"{{3}}" {gray,pos=0.08,scale=0.8,below left},"{{3}}" {gray,pos=0.92,scale=0.8,above right}] & 0 & 0 & 0 & 0 \\
H^{6} & \mathbb{Z}_{3} & \mathbb{Z}_{3} & 0 & 0 & \mathbb{Z}_{3} \ar[lu,BrickRed,hook,two heads,"{{3}}" {gray,pos=0.08,scale=0.8,below left},"{{3}}" {gray,pos=0.92,scale=0.8,above right}] & \mathbb{Z}_{3} \ar[lu,BrickRed,hook,two heads,"{{3}}" {gray,pos=0.08,scale=0.8,below left},"{{3}}" {gray,pos=0.92,scale=0.8,above right}] & 0 & 0 & 0 \\
H^{7} & \mathbb{Z}_{3} & (\mathbb{Z}_{3})^{2} \ar[lu,two heads,"{{3}}" {gray,pos=0.08,scale=0.8,below left},"{{3}}" {gray,pos=0.92,scale=0.8,above right}] & (\mathbb{Z}_{3})^{2} \ar[lu,two heads,"{{3}}" {gray,pos=0.08,scale=0.8,below left},"{{3}}" {gray,pos=0.92,scale=0.8,above right}] & 0 & 0 & \mathbb{Z}_{3} \ar[lu,BrickRed,hook,two heads,"{{3}}" {gray,pos=0.08,scale=0.8,below left},"{{3}}" {gray,pos=0.92,scale=0.8,above right}] & \mathbb{Z}_{3} \ar[lu,BrickRed,hook,two heads,"{{3}}" {gray,pos=0.08,scale=0.8,below left},"{{3}}" {gray,pos=0.92,scale=0.8,above right}] & 0 & 0 \\
H^{8} & \mathbb{Z}_{3} & (\mathbb{Z}_{3})^{2} & (\mathbb{Z}_{3})^{3} \ar[lu,"{{3}}" {gray,pos=0.08,scale=0.8,below left},"{{3}}" {gray,pos=0.92,scale=0.8,above right}] & (\mathbb{Z}_{3})^{2} \ar[lu,"{{3}}" {gray,pos=0.08,scale=0.8,below left},"{{3}}" {gray,pos=0.92,scale=0.8,above right}] & 0 & 0 & \mathbb{Z}_{3} \ar[lu,BrickRed,hook,two heads,"{{3}}" {gray,pos=0.08,scale=0.8,below left},"{{3}}" {gray,pos=0.92,scale=0.8,above right}] & \mathbb{Z}_{3} \ar[lu,BrickRed,hook,two heads,"{{3}}" {gray,pos=0.08,scale=0.8,below left},"{{3}}" {gray,pos=0.92,scale=0.8,above right}] & 0 \\
H^{9} & \mathbb{Z}_{3} & (\mathbb{Z}_{3})^{2} & \mathbb{Z}_{3} \ar[lu,hook,"{{3}}" {gray,pos=0.08,scale=0.8,below left},"{{3}}" {gray,pos=0.92,scale=0.8,above right}] & (\mathbb{Z}_{3})^{3} \ar[lu,"{{3^{2}}}" {gray,pos=0.08,scale=0.8,below left},"{{3^{2}}}" {gray,pos=0.92,scale=0.8,above right}] & \mathbb{Z}_{3} \ar[lu,hook,"{{3}}" {gray,pos=0.08,scale=0.8,below left},"{{3}}" {gray,pos=0.92,scale=0.8,above right}] & 0 & 0 & \mathbb{Z}_{3} \ar[lu,BrickRed,hook,two heads,"{{3}}" {gray,pos=0.08,scale=0.8,below left},"{{3}}" {gray,pos=0.92,scale=0.8,above right}] & \mathbb{Z}_{3} \ar[lu,BrickRed,hook,two heads,"{{3}}" {gray,pos=0.08,scale=0.8,below left},"{{3}}" {gray,pos=0.92,scale=0.8,above right}] \\
H^{10} & \mathbb{Z}_{3} & (\mathbb{Z}_{3})^{3} & \mathbb{Z}_{3} & \mathbb{Z}_{3} \ar[lu,hook,two heads,"{{3}}" {gray,pos=0.08,scale=0.8,below left},"{{3}}" {gray,pos=0.92,scale=0.8,above right}] & (\mathbb{Z}_{3})^{2} \ar[lu,hook,"{{3^{2}}}" {gray,pos=0.08,scale=0.8,below left},"{{3^{2}}}" {gray,pos=0.92,scale=0.8,above right}] & \mathbb{Z}_{3} \ar[lu,BrickRed,hook,two heads,"{{3}}" {gray,pos=0.08,scale=0.8,below left},"{{3}}" {gray,pos=0.92,scale=0.8,above right}] & 0 & 0 & \mathbb{Z}_{3} \ar[lu,BrickRed,hook,two heads,"{{3}}" {gray,pos=0.08,scale=0.8,below left},"{{3}}" {gray,pos=0.92,scale=0.8,above right}] \\
H^{11} & \mathbb{Z}_{3} & (\mathbb{Z}_{3})^{3} & (\mathbb{Z}_{3})^{4} & 0 & (\mathbb{Z}_{3})^{2} \ar[lu,two heads,"{{3}}" {gray,pos=0.08,scale=0.8,below left},"{{3}}" {gray,pos=0.92,scale=0.8,above right}] & (\mathbb{Z}_{3})^{2} \ar[lu,hook,two heads,"{{3^{2}}}" {gray,pos=0.08,scale=0.8,below left},"{{3^{2}}}" {gray,pos=0.92,scale=0.8,above right}] & \mathbb{Z}_{3} \ar[lu,BrickRed,hook,two heads,"{{3}}" {gray,pos=0.08,scale=0.8,below left},"{{3}}" {gray,pos=0.92,scale=0.8,above right}] & 0 & 0 \\
H^{12} & \mathbb{Z}_{3} & (\mathbb{Z}_{3})^{3} & (\mathbb{Z}_{3})^{4} & (\mathbb{Z}_{3})^{2} & \mathbb{Z}_{3} & (\mathbb{Z}_{3})^{2} \ar[lu,"{{3}}" {gray,pos=0.08,scale=0.8,below left},"{{3}}" {gray,pos=0.92,scale=0.8,above right}] & (\mathbb{Z}_{3})^{2} \ar[lu,BrickRed,hook,two heads,"{{3^{2}}}" {gray,pos=0.08,scale=0.8,below left},"{{3^{2}}}" {gray,pos=0.92,scale=0.8,above right}] & \mathbb{Z}_{3} \ar[lu,BrickRed,hook,two heads,"{{3}}" {gray,pos=0.08,scale=0.8,below left},"{{3}}" {gray,pos=0.92,scale=0.8,above right}] & 0 \\
H^{13} & \mathbb{Z}_{3} & (\mathbb{Z}_{3})^{3} & \mathbb{Z}_{3} & (\mathbb{Z}_{3})^{5} \ar[lu,"{{3}}" {gray,pos=0.08,scale=0.8,below left},"{{3}}" {gray,pos=0.92,scale=0.8,above right}] & \mathbb{Z}_{3} \ar[lu,hook,"{{3}}" {gray,pos=0.08,scale=0.8,below left},"{{3}}" {gray,pos=0.92,scale=0.8,above right}] & \mathbb{Z}_{3} & \mathbb{Z}_{3} \ar[lu,hook,"{{3}}" {gray,pos=0.08,scale=0.8,below left},"{{3}}" {gray,pos=0.92,scale=0.8,above right}] & (\mathbb{Z}_{3})^{2} \ar[lu,BrickRed,hook,two heads,"{{3^{2}}}" {gray,pos=0.08,scale=0.8,below left},"{{3^{2}}}" {gray,pos=0.92,scale=0.8,above right}] & \mathbb{Z}_{3} \ar[lu,BrickRed,hook,two heads,"{{3}}" {gray,pos=0.08,scale=0.8,below left},"{{3}}" {gray,pos=0.92,scale=0.8,above right}] \\
H^{14} & \mathbb{Z}_{3} & (\mathbb{Z}_{3})^{3} & \mathbb{Z}_{3} & (\mathbb{Z}_{3})^{4} & (\mathbb{Z}_{3})^{3} \ar[lu,"{{3}}" {gray,pos=0.08,scale=0.8,below left},"{{3}}" {gray,pos=0.92,scale=0.8,above right}] & \mathbb{Z}_{3} \ar[lu,hook,two heads,"{{3}}" {gray,pos=0.08,scale=0.8,below left},"{{3}}" {gray,pos=0.92,scale=0.8,above right}] & 0 & \mathbb{Z}_{3} \ar[lu,BrickRed,hook,two heads,"{{3}}" {gray,pos=0.08,scale=0.8,below left},"{{3}}" {gray,pos=0.92,scale=0.8,above right}] & (\mathbb{Z}_{3})^{2} \ar[lu,BrickRed,hook,two heads,"{{3^{2}}}" {gray,pos=0.08,scale=0.8,below left},"{{3^{2}}}" {gray,pos=0.92,scale=0.8,above right}] \\
H^{15} & \mathbb{Z}_{3} & (\mathbb{Z}_{3})^{4} & (\mathbb{Z}_{3})^{6} & \mathbb{Z}_{3} & (\mathbb{Z}_{3})^{4} \ar[lu,"{{3}}" {gray,pos=0.08,scale=0.8,below left},"{{3}}" {gray,pos=0.92,scale=0.8,above right}] & (\mathbb{Z}_{3})^{2} \ar[lu,hook,"{{3^{2}}}" {gray,pos=0.08,scale=0.8,below left},"{{3^{2}}}" {gray,pos=0.92,scale=0.8,above right}] & (\mathbb{Z}_{3})^{2} \ar[lu,two heads,"{{3}}" {gray,pos=0.08,scale=0.8,below left},"{{3}}" {gray,pos=0.92,scale=0.8,above right}] & 0 & \mathbb{Z}_{3} \ar[lu,BrickRed,hook,two heads,"{{3}}" {gray,pos=0.08,scale=0.8,below left},"{{3}}" {gray,pos=0.92,scale=0.8,above right}] \\
H^{16} & \mathbb{Z}_{3} & (\mathbb{Z}_{3})^{4} & (\mathbb{Z}_{3})^{8} & (\mathbb{Z}_{3})^{3} & (\mathbb{Z}_{3})^{3} & (\mathbb{Z}_{3})^{2} \ar[lu,"{{3}}" {gray,pos=0.08,scale=0.8,below left},"{{3}}" {gray,pos=0.92,scale=0.8,above right}] & (\mathbb{Z}_{3})^{3} \ar[lu,two heads,"{{3^{2}}}" {gray,pos=0.08,scale=0.8,below left},"{{3^{2}}}" {gray,pos=0.92,scale=0.8,above right}] & (\mathbb{Z}_{3})^{2} \ar[lu,"{{3}}" {gray,pos=0.08,scale=0.8,below left},"{{3}}" {gray,pos=0.92,scale=0.8,above right}] & 0 \\
H^{17} & \mathbb{Z}_{3} & (\mathbb{Z}_{3})^{4} & (\mathbb{Z}_{3})^{3} & (\mathbb{Z}_{3})^{7} & (\mathbb{Z}_{3})^{2} & (\mathbb{Z}_{3})^{3} & \mathbb{Z}_{3} \ar[lu,hook,"{{3}}" {gray,pos=0.08,scale=0.8,below left},"{{3}}" {gray,pos=0.92,scale=0.8,above right}] & (\mathbb{Z}_{3})^{3} \ar[lu,"{{3^{2}}}" {gray,pos=0.08,scale=0.8,below left},"{{3^{2}}}" {gray,pos=0.92,scale=0.8,above right}] & \mathbb{Z}_{3} \ar[lu,hook,"{{3}}" {gray,pos=0.08,scale=0.8,below left},"{{3}}" {gray,pos=0.92,scale=0.8,above right}] \\
H^{18} & \mathbb{Z}_{3} & (\mathbb{Z}_{3})^{5} & (\mathbb{Z}_{3})^{3} & (\mathbb{Z}_{3})^{7} & (\mathbb{Z}_{3})^{2} & (\mathbb{Z}_{3})^{4} & \mathbb{Z}_{3} & \mathbb{Z}_{3} \ar[lu,hook,two heads,"{{3}}" {gray,pos=0.08,scale=0.8,below left},"{{3}}" {gray,pos=0.92,scale=0.8,above right}] & (\mathbb{Z}_{3})^{2} \ar[lu,hook,"{{3^{2}}}" {gray,pos=0.08,scale=0.8,below left},"{{3^{2}}}" {gray,pos=0.92,scale=0.8,above right}] \\
H^{19} & \mathbb{Z}_{3} & (\mathbb{Z}_{3})^{6} \ar[lu,two heads,"{{3}}" {gray,pos=0.08,scale=0.8,below left},"{{3}}" {gray,pos=0.92,scale=0.8,above right}] & (\mathbb{Z}_{3})^{11} \ar[lu,"{{3}}" {gray,pos=0.08,scale=0.8,below left},"{{3}}" {gray,pos=0.92,scale=0.8,above right}] & (\mathbb{Z}_{3})^{4} & (\mathbb{Z}_{3})^{5} & (\mathbb{Z}_{3})^{3} \ar[lu,"{{3}}" {gray,pos=0.08,scale=0.8,below left},"{{3}}" {gray,pos=0.92,scale=0.8,above right}] & (\mathbb{Z}_{3})^{4} \ar[lu,"{{3}}" {gray,pos=0.08,scale=0.8,below left},"{{3}}" {gray,pos=0.92,scale=0.8,above right}] & 0 & (\mathbb{Z}_{3})^{2} \ar[lu,two heads,"{{3}}" {gray,pos=0.08,scale=0.8,below left},"{{3}}" {gray,pos=0.92,scale=0.8,above right}] \\
H^{20} & \mathbb{Z}_{3} & (\mathbb{Z}_{3})^{6} & (\mathbb{Z}_{3})^{12} \ar[lu,"{{3}}" {gray,pos=0.08,scale=0.8,below left},"{{3}}" {gray,pos=0.92,scale=0.8,above right}] & (\mathbb{Z}_{3})^{5} \ar[lu,"{{3}}" {gray,pos=0.08,scale=0.8,below left},"{{3}}" {gray,pos=0.92,scale=0.8,above right}] & (\mathbb{Z}_{3})^{8} & (\mathbb{Z}_{3})^{3} & (\mathbb{Z}_{3})^{5} \ar[lu,"{{3}}" {gray,pos=0.08,scale=0.8,below left},"{{3}}" {gray,pos=0.92,scale=0.8,above right}] & (\mathbb{Z}_{3})^{2} \ar[lu,"{{3}}" {gray,pos=0.08,scale=0.8,below left},"{{3}}" {gray,pos=0.92,scale=0.8,above right}] & \mathbb{Z}_{3} \\
H^{21} & \mathbb{Z}_{3} & (\mathbb{Z}_{3})^{6} & (\mathbb{Z}_{3})^{4} \ar[lu,"{{3}}" {gray,pos=0.08,scale=0.8,below left},"{{3}}" {gray,pos=0.92,scale=0.8,above right}] & (\mathbb{Z}_{3})^{11} \ar[lu,"{{3^{2}}}" {gray,pos=0.08,scale=0.8,below left},"{{3^{2}}}" {gray,pos=0.92,scale=0.8,above right}] & (\mathbb{Z}_{3})^{7} \ar[lu,"{{3}}" {gray,pos=0.08,scale=0.8,below left},"{{3}}" {gray,pos=0.92,scale=0.8,above right}] & (\mathbb{Z}_{3})^{5} & (\mathbb{Z}_{3})^{2} \ar[lu,"{{3}}" {gray,pos=0.08,scale=0.8,below left},"{{3}}" {gray,pos=0.92,scale=0.8,above right}] & (\mathbb{Z}_{3})^{5} \ar[lu,"{{3^{2}}}" {gray,pos=0.08,scale=0.8,below left},"{{3^{2}}}" {gray,pos=0.92,scale=0.8,above right}] & \mathbb{Z}_{3} \ar[lu,hook,"{{3}}" {gray,pos=0.08,scale=0.8,below left},"{{3}}" {gray,pos=0.92,scale=0.8,above right}] \\
H^{22} & \mathbb{Z}_{3} & (\mathbb{Z}_{3})^{7} & (\mathbb{Z}_{3})^{4} & (\mathbb{Z}_{3})^{13} \ar[lu,"{{3}}" {gray,pos=0.08,scale=0.8,below left},"{{3}}" {gray,pos=0.92,scale=0.8,above right}] & (\mathbb{Z}_{3})^{6} \ar[lu,"{{3^{2}}}" {gray,pos=0.08,scale=0.8,below left},"{{3^{2}}}" {gray,pos=0.92,scale=0.8,above right}] & (\mathbb{Z}_{3})^{7} \ar[lu,"{{3}}" {gray,pos=0.08,scale=0.8,below left},"{{3}}" {gray,pos=0.92,scale=0.8,above right}] & \mathbb{Z}_{3} & (\mathbb{Z}_{3})^{4} \ar[lu,"{{3}}" {gray,pos=0.08,scale=0.8,below left},"{{3}}" {gray,pos=0.92,scale=0.8,above right}] & (\mathbb{Z}_{3})^{3} \ar[lu,"{{3^{2}}}" {gray,pos=0.08,scale=0.8,below left},"{{3^{2}}}" {gray,pos=0.92,scale=0.8,above right}] \\
H^{23} & \mathbb{Z}_{3} & (\mathbb{Z}_{3})^{8} & (\mathbb{Z}_{3})^{16} & (\mathbb{Z}_{3})^{9} & (\mathbb{Z}_{3})^{7} \ar[lu,"{{3}}" {gray,pos=0.08,scale=0.8,below left},"{{3}}" {gray,pos=0.92,scale=0.8,above right}] & (\mathbb{Z}_{3})^{8} \ar[lu,"{{3^{2}}}" {gray,pos=0.08,scale=0.8,below left},"{{3^{2}}}" {gray,pos=0.92,scale=0.8,above right}] & (\mathbb{Z}_{3})^{7} \ar[lu,"{{3}}" {gray,pos=0.08,scale=0.8,below left},"{{3}}" {gray,pos=0.92,scale=0.8,above right}] & \mathbb{Z}_{3} & (\mathbb{Z}_{3})^{4} \ar[lu,"{{3}}" {gray,pos=0.08,scale=0.8,below left},"{{3}}" {gray,pos=0.92,scale=0.8,above right}] \\
H^{24} & \mathbb{Z}_{3} & (\mathbb{Z}_{3})^{8} & (\mathbb{Z}_{3})^{19} & (\mathbb{Z}_{3})^{9} & (\mathbb{Z}_{3})^{10} & (\mathbb{Z}_{3})^{7} \ar[lu,"{{3}}" {gray,pos=0.08,scale=0.8,below left},"{{3}}" {gray,pos=0.92,scale=0.8,above right}] & (\mathbb{Z}_{3})^{10} \ar[lu,"{{3^{2}}}" {gray,pos=0.08,scale=0.8,below left},"{{3^{2}}}" {gray,pos=0.92,scale=0.8,above right}] & (\mathbb{Z}_{3})^{4} \ar[lu,"{{3}}" {gray,pos=0.08,scale=0.8,below left},"{{3}}" {gray,pos=0.92,scale=0.8,above right}] & (\mathbb{Z}_{3})^{3} \\
H^{25} & \mathbb{Z}_{3} & (\mathbb{Z}_{3})^{8} & (\mathbb{Z}_{3})^{8} \ar[lu,"{{3}}" {gray,pos=0.08,scale=0.8,below left},"{{3}}" {gray,pos=0.92,scale=0.8,above right}] & (\mathbb{Z}_{3})^{17} \ar[lu,"{{3^{2}}}" {gray,pos=0.08,scale=0.8,below left},"{{3^{2}}}" {gray,pos=0.92,scale=0.8,above right}] & (\mathbb{Z}_{3})^{14} \ar[lu,"{{3}}" {gray,pos=0.08,scale=0.8,below left},"{{3}}" {gray,pos=0.92,scale=0.8,above right}] & (\mathbb{Z}_{3})^{7} & (\mathbb{Z}_{3})^{4} \ar[lu,"{{3}}" {gray,pos=0.08,scale=0.8,below left},"{{3}}" {gray,pos=0.92,scale=0.8,above right}] & (\mathbb{Z}_{3})^{9} \ar[lu,"{{3^{3}}}" {gray,pos=0.08,scale=0.8,below left},"{{3^{3}}}" {gray,pos=0.92,scale=0.8,above right}] & (\mathbb{Z}_{3})^{3} \ar[lu,"{{3^{2}}}" {gray,pos=0.08,scale=0.8,below left},"{{3^{2}}}" {gray,pos=0.92,scale=0.8,above right}] \\
H^{26} & \mathbb{Z}_{3} & (\mathbb{Z}_{3})^{10} & (\mathbb{Z}_{3})^{9} & (\mathbb{Z}_{3})^{21} \ar[lu,"{{3}}" {gray,pos=0.08,scale=0.8,below left},"{{3}}" {gray,pos=0.92,scale=0.8,above right}] & (\mathbb{Z}_{3})^{15} \ar[lu,"{{3^{2}}}" {gray,pos=0.08,scale=0.8,below left},"{{3^{2}}}" {gray,pos=0.92,scale=0.8,above right}] & (\mathbb{Z}_{3})^{11} \ar[lu,"{{3}}" {gray,pos=0.08,scale=0.8,below left},"{{3}}" {gray,pos=0.92,scale=0.8,above right}] & (\mathbb{Z}_{3})^{3} & (\mathbb{Z}_{3})^{7} \ar[lu,"{{3}}" {gray,pos=0.08,scale=0.8,below left},"{{3}}" {gray,pos=0.92,scale=0.8,above right}] & (\mathbb{Z}_{3})^{5} \ar[lu,"{{3^{3}}}" {gray,pos=0.08,scale=0.8,below left},"{{3^{3}}}" {gray,pos=0.92,scale=0.8,above right}] \\
H^{27} & \mathbb{Z}_{3} & (\mathbb{Z}_{3})^{11} & (\mathbb{Z}_{3})^{26} & (\mathbb{Z}_{3})^{17} \ar[lu,"{{3}}" {gray,pos=0.08,scale=0.8,below left},"{{3}}" {gray,pos=0.92,scale=0.8,above right}] & (\mathbb{Z}_{3})^{15} \ar[lu,"{{3^{3}}}" {gray,pos=0.08,scale=0.8,below left},"{{3^{3}}}" {gray,pos=0.92,scale=0.8,above right}] & (\mathbb{Z}_{3})^{14} \ar[lu,"{{3^{3}}}" {gray,pos=0.08,scale=0.8,below left},"{{3^{3}}}" {gray,pos=0.92,scale=0.8,above right}] & (\mathbb{Z}_{3})^{11} \ar[lu,"{{3}}" {gray,pos=0.08,scale=0.8,below left},"{{3}}" {gray,pos=0.92,scale=0.8,above right}] & (\mathbb{Z}_{3})^{3} & (\mathbb{Z}_{3})^{7} \ar[lu,"{{3^{2}}}" {gray,pos=0.08,scale=0.8,below left},"{{3^{2}}}" {gray,pos=0.92,scale=0.8,above right}] \\
H^{28} & \mathbb{Z}_{3} & (\mathbb{Z}_{3})^{11} & (\mathbb{Z}_{3})^{29} & (\mathbb{Z}_{3})^{16} & (\mathbb{Z}_{3})^{16} \ar[lu,"{{3}}" {gray,pos=0.08,scale=0.8,below left},"{{3}}" {gray,pos=0.92,scale=0.8,above right}] & (\mathbb{Z}_{3})^{14} \ar[lu,"{{3^{3}}}" {gray,pos=0.08,scale=0.8,below left},"{{3^{3}}}" {gray,pos=0.92,scale=0.8,above right}] & (\mathbb{Z}_{3})^{15} \ar[lu,"{{3^{3}}}" {gray,pos=0.08,scale=0.8,below left},"{{3^{3}}}" {gray,pos=0.92,scale=0.8,above right}] & (\mathbb{Z}_{3})^{5} \ar[lu,"{{3}}" {gray,pos=0.08,scale=0.8,below left},"{{3}}" {gray,pos=0.92,scale=0.8,above right}] & (\mathbb{Z}_{3})^{7} \\
H^{29} & \mathbb{Z}_{3} & (\mathbb{Z}_{3})^{12} & (\mathbb{Z}_{3})^{14} & (\mathbb{Z}_{3})^{26} & (\mathbb{Z}_{3})^{18} & (\mathbb{Z}_{3})^{15} \ar[lu,"{{3}}" {gray,pos=0.08,scale=0.8,below left},"{{3}}" {gray,pos=0.92,scale=0.8,above right}] & (\mathbb{Z}_{3})^{9} \ar[lu,"{{3^{3}}}" {gray,pos=0.08,scale=0.8,below left},"{{3^{3}}}" {gray,pos=0.92,scale=0.8,above right}] & (\mathbb{Z}_{3})^{13} \ar[lu,"{{3^{3}}}" {gray,pos=0.08,scale=0.8,below left},"{{3^{3}}}" {gray,pos=0.92,scale=0.8,above right}] & (\mathbb{Z}_{3})^{5} \ar[lu,"{{3}}" {gray,pos=0.08,scale=0.8,below left},"{{3}}" {gray,pos=0.92,scale=0.8,above right}] \\
H^{30} & \mathbb{Z}_{3} & (\mathbb{Z}_{3})^{13} & (\mathbb{Z}_{3})^{14} & (\mathbb{Z}_{3})^{35} & (\mathbb{Z}_{3})^{22} & (\mathbb{Z}_{3})^{17} & (\mathbb{Z}_{3})^{7} \ar[lu,"{{3}}" {gray,pos=0.08,scale=0.8,below left},"{{3}}" {gray,pos=0.92,scale=0.8,above right}] & (\mathbb{Z}_{3})^{15} \ar[lu,"{{3^{3}}}" {gray,pos=0.08,scale=0.8,below left},"{{3^{3}}}" {gray,pos=0.92,scale=0.8,above right}] & (\mathbb{Z}_{3})^{6} \ar[lu,"{{3^{3}}}" {gray,pos=0.08,scale=0.8,below left},"{{3^{3}}}" {gray,pos=0.92,scale=0.8,above right}] \\
H^{31} & \mathbb{Z}_{3} & (\mathbb{Z}_{3})^{14} & (\mathbb{Z}_{3})^{36} & (\mathbb{Z}_{3})^{30} & (\mathbb{Z}_{3})^{26} \ar[lu,"{{3}}" {gray,pos=0.08,scale=0.8,below left},"{{3}}" {gray,pos=0.92,scale=0.8,above right}] & (\mathbb{Z}_{3})^{19} \ar[lu,"{{3^{2}}}" {gray,pos=0.08,scale=0.8,below left},"{{3^{2}}}" {gray,pos=0.92,scale=0.8,above right}] & (\mathbb{Z}_{3})^{20} \ar[lu,"{{3}}" {gray,pos=0.08,scale=0.8,below left},"{{3}}" {gray,pos=0.92,scale=0.8,above right}] & (\mathbb{Z}_{3})^{7} \ar[lu,"{{3}}" {gray,pos=0.08,scale=0.8,below left},"{{3}}" {gray,pos=0.92,scale=0.8,above right}] & (\mathbb{Z}_{3})^{10} \ar[lu,"{{3^{3}}}" {gray,pos=0.08,scale=0.8,below left},"{{3^{3}}}" {gray,pos=0.92,scale=0.8,above right}] \\
H^{32} & \mathbb{Z}_{3} & (\mathbb{Z}_{3})^{14} & (\mathbb{Z}_{3})^{42} & (\mathbb{Z}_{3})^{26} & (\mathbb{Z}_{3})^{29} & (\mathbb{Z}_{3})^{23} \ar[lu,"{{3}}" {gray,pos=0.08,scale=0.8,below left},"{{3}}" {gray,pos=0.92,scale=0.8,above right}] & (\mathbb{Z}_{3})^{26} \ar[lu,"{{3^{2}}}" {gray,pos=0.08,scale=0.8,below left},"{{3^{2}}}" {gray,pos=0.92,scale=0.8,above right}] & (\mathbb{Z}_{3})^{8} \ar[lu,"{{3}}" {gray,pos=0.08,scale=0.8,below left},"{{3}}" {gray,pos=0.92,scale=0.8,above right}] & (\mathbb{Z}_{3})^{11} \ar[lu,"{{3}}" {gray,pos=0.08,scale=0.8,below left},"{{3}}" {gray,pos=0.92,scale=0.8,above right}] \\
H^{33} & \mathbb{Z}_{3} & (\mathbb{Z}_{3})^{14} & (\mathbb{Z}_{3})^{22} & (\mathbb{Z}_{3})^{38} & (\mathbb{Z}_{3})^{32} & (\mathbb{Z}_{3})^{29} \ar[lu,"{{3}}" {gray,pos=0.08,scale=0.8,below left},"{{3}}" {gray,pos=0.92,scale=0.8,above right}] & (\mathbb{Z}_{3})^{14} \ar[lu,"{{3^{3}}}" {gray,pos=0.08,scale=0.8,below left},"{{3^{3}}}" {gray,pos=0.92,scale=0.8,above right}] & (\mathbb{Z}_{3})^{20} \ar[lu,"{{3^{3}}}" {gray,pos=0.08,scale=0.8,below left},"{{3^{3}}}" {gray,pos=0.92,scale=0.8,above right}] & (\mathbb{Z}_{3})^{9} \ar[lu,"{{3}}" {gray,pos=0.08,scale=0.8,below left},"{{3}}" {gray,pos=0.92,scale=0.8,above right}] \\
H^{34} & \mathbb{Z}_{3} & (\mathbb{Z}_{3})^{16} & (\mathbb{Z}_{3})^{23} & (\mathbb{Z}_{3})^{55} & (\mathbb{Z}_{3})^{36} & (\mathbb{Z}_{3})^{33} & (\mathbb{Z}_{3})^{12} \ar[lu,"{{3}}" {gray,pos=0.08,scale=0.8,below left},"{{3}}" {gray,pos=0.92,scale=0.8,above right}] & (\mathbb{Z}_{3})^{23} \ar[lu,"{{3^{3}}}" {gray,pos=0.08,scale=0.8,below left},"{{3^{3}}}" {gray,pos=0.92,scale=0.8,above right}] & (\mathbb{Z}_{3})^{10} \ar[lu,"{{3^{3}}}" {gray,pos=0.08,scale=0.8,below left},"{{3^{3}}}" {gray,pos=0.92,scale=0.8,above right}] \\
H^{35} & \mathbb{Z}_{3} & (\mathbb{Z}_{3})^{17} & (\mathbb{Z}_{3})^{54} & (\mathbb{Z}_{3})^{55} & (\mathbb{Z}_{3})^{42} & (\mathbb{Z}_{3})^{33} & (\mathbb{Z}_{3})^{31} & (\mathbb{Z}_{3})^{14} \ar[lu,"{{3}}" {gray,pos=0.08,scale=0.8,below left},"{{3}}" {gray,pos=0.92,scale=0.8,above right}] & (\mathbb{Z}_{3})^{16} \ar[lu,"{{3^{3}}}" {gray,pos=0.08,scale=0.8,below left},"{{3^{3}}}" {gray,pos=0.92,scale=0.8,above right}] \\
H^{36} & \mathbb{Z}_{3} & (\mathbb{Z}_{3})^{17} & (\mathbb{Z}_{3})^{61} & (\mathbb{Z}_{3})^{46} & (\mathbb{Z}_{3})^{50} & (\mathbb{Z}_{3})^{33} & (\mathbb{Z}_{3})^{40} & (\mathbb{Z}_{3})^{12} & (\mathbb{Z}_{3})^{20} \ar[lu,"{{3}}" {gray,pos=0.08,scale=0.8,below left},"{{3}}" {gray,pos=0.92,scale=0.8,above right}] \\
H^{37} & \mathbb{Z}_{3} & (\mathbb{Z}_{3})^{18} & (\mathbb{Z}_{3})^{35} & (\mathbb{Z}_{3})^{56} \ar[lu,"{{3}}" {gray,pos=0.08,scale=0.8,below left},"{{3}}" {gray,pos=0.92,scale=0.8,above right}] & (\mathbb{Z}_{3})^{58} \ar[lu,"{{3}}" {gray,pos=0.08,scale=0.8,below left},"{{3}}" {gray,pos=0.92,scale=0.8,above right}] & (\mathbb{Z}_{3})^{42} & (\mathbb{Z}_{3})^{24} \ar[lu,"{{3}}" {gray,pos=0.08,scale=0.8,below left},"{{3}}" {gray,pos=0.92,scale=0.8,above right}] & (\mathbb{Z}_{3})^{29} \ar[lu,"{{3^{2}}}" {gray,pos=0.08,scale=0.8,below left},"{{3^{2}}}" {gray,pos=0.92,scale=0.8,above right}] & (\mathbb{Z}_{3})^{18} \ar[lu,"{{3}}" {gray,pos=0.08,scale=0.8,below left},"{{3}}" {gray,pos=0.92,scale=0.8,above right}] \\
H^{38} & \mathbb{Z}_{3} & (\mathbb{Z}_{3})^{19} & (\mathbb{Z}_{3})^{34} & (\mathbb{Z}_{3})^{79} & (\mathbb{Z}_{3})^{62} \ar[lu,"{{3}}" {gray,pos=0.08,scale=0.8,below left},"{{3}}" {gray,pos=0.92,scale=0.8,above right}] & (\mathbb{Z}_{3})^{57} \ar[lu,"{{3}}" {gray,pos=0.08,scale=0.8,below left},"{{3}}" {gray,pos=0.92,scale=0.8,above right}] & (\mathbb{Z}_{3})^{21} & (\mathbb{Z}_{3})^{42} \ar[lu,"{{3}}" {gray,pos=0.08,scale=0.8,below left},"{{3}}" {gray,pos=0.92,scale=0.8,above right}] & (\mathbb{Z}_{3})^{17} \ar[lu,"{{3^{2}}}" {gray,pos=0.08,scale=0.8,below left},"{{3^{2}}}" {gray,pos=0.92,scale=0.8,above right}] \\
H^{39} & \mathbb{Z}_{3} & (\mathbb{Z}_{3})^{21} & (\mathbb{Z}_{3})^{73} & (\mathbb{Z}_{3})^{87} & (\mathbb{Z}_{3})^{66} \ar[lu,"{{3}}" {gray,pos=0.08,scale=0.8,below left},"{{3}}" {gray,pos=0.92,scale=0.8,above right}] & (\mathbb{Z}_{3})^{63} \ar[lu,"{{3^{2}}}" {gray,pos=0.08,scale=0.8,below left},"{{3^{2}}}" {gray,pos=0.92,scale=0.8,above right}] & (\mathbb{Z}_{3})^{53} \ar[lu,"{{3}}" {gray,pos=0.08,scale=0.8,below left},"{{3}}" {gray,pos=0.92,scale=0.8,above right}] & (\mathbb{Z}_{3})^{30} \ar[lu,"{{3}}" {gray,pos=0.08,scale=0.8,below left},"{{3}}" {gray,pos=0.92,scale=0.8,above right}] & (\mathbb{Z}_{3})^{24} \ar[lu,"{{3^{3}}}" {gray,pos=0.08,scale=0.8,below left},"{{3^{3}}}" {gray,pos=0.92,scale=0.8,above right}] \\
H^{40} & \mathbb{Z}_{3} & (\mathbb{Z}_{3})^{21} & (\mathbb{Z}_{3})^{85} & (\mathbb{Z}_{3})^{79} & (\mathbb{Z}_{3})^{76} & (\mathbb{Z}_{3})^{62} \ar[lu,"{{3}}" {gray,pos=0.08,scale=0.8,below left},"{{3}}" {gray,pos=0.92,scale=0.8,above right}] & (\mathbb{Z}_{3})^{69} \ar[lu,"{{3^{2}}}" {gray,pos=0.08,scale=0.8,below left},"{{3^{2}}}" {gray,pos=0.92,scale=0.8,above right}] & (\mathbb{Z}_{3})^{23} \ar[lu,"{{3}}" {gray,pos=0.08,scale=0.8,below left},"{{3}}" {gray,pos=0.92,scale=0.8,above right}] & (\mathbb{Z}_{3})^{32} \ar[lu,"{{3}}" {gray,pos=0.08,scale=0.8,below left},"{{3}}" {gray,pos=0.92,scale=0.8,above right}] \\
\end{tikzcd}

%% file: tables/omega-Z3-Z9.tex
\begin{tikzcd}[row sep=tiny,column sep=large]
& B\mathbb{Z}_{3} & B^{2}\mathbb{Z}_{3} & B^{3}\mathbb{Z}_{3} & B^{4}\mathbb{Z}_{3} & B^{5}\mathbb{Z}_{3} & B^{6}\mathbb{Z}_{3} & B^{7}\mathbb{Z}_{3} \\
H^{0} & \mathbb{Z}_{9} & \mathbb{Z}_{9} & \mathbb{Z}_{9} & \mathbb{Z}_{9} & \mathbb{Z}_{9} & \mathbb{Z}_{9} & \mathbb{Z}_{9} \\
H^{1} & \mathbb{Z}_{3} & 0 & 0 & 0 & 0 & 0 & 0 \\
H^{2} & \mathbb{Z}_{3} & \mathbb{Z}_{3} \ar[lu,BrickRed,hook,two heads,"{{3}}" {gray,pos=0.08,scale=0.8,below left},"{{3}}" {gray,pos=0.92,scale=0.8,above right}] & 0 & 0 & 0 & 0 & 0 \\
H^{3} & \mathbb{Z}_{3} & \mathbb{Z}_{3} \ar[lu,hook,two heads,"{{3}}" {gray,pos=0.08,scale=0.8,below left},"{{3}}" {gray,pos=0.92,scale=0.8,above right}] & \mathbb{Z}_{3} \ar[lu,BrickRed,hook,two heads,"{{3}}" {gray,pos=0.08,scale=0.8,below left},"{{3}}" {gray,pos=0.92,scale=0.8,above right}] & 0 & 0 & 0 & 0 \\
H^{4} & \mathbb{Z}_{3} & \mathbb{Z}_{3} & \mathbb{Z}_{3} \ar[lu,BrickRed,hook,two heads,"{{3}}" {gray,pos=0.08,scale=0.8,below left},"{{3}}" {gray,pos=0.92,scale=0.8,above right}] & \mathbb{Z}_{3} \ar[lu,BrickRed,hook,two heads,"{{3}}" {gray,pos=0.08,scale=0.8,below left},"{{3}}" {gray,pos=0.92,scale=0.8,above right}] & 0 & 0 & 0 \\
H^{5} & \mathbb{Z}_{3} & \mathbb{Z}_{3} & 0 & \mathbb{Z}_{3} \ar[lu,BrickRed,hook,two heads,"{{3}}" {gray,pos=0.08,scale=0.8,below left},"{{3}}" {gray,pos=0.92,scale=0.8,above right}] & \mathbb{Z}_{3} \ar[lu,BrickRed,hook,two heads,"{{3}}" {gray,pos=0.08,scale=0.8,below left},"{{3}}" {gray,pos=0.92,scale=0.8,above right}] & 0 & 0 \\
H^{6} & \mathbb{Z}_{3} & \mathbb{Z}_{9} & 0 & 0 & \mathbb{Z}_{3} \ar[lu,BrickRed,hook,two heads,"{{3}}" {gray,pos=0.08,scale=0.8,below left},"{{3}}" {gray,pos=0.92,scale=0.8,above right}] & \mathbb{Z}_{3} \ar[lu,BrickRed,hook,two heads,"{{3}}" {gray,pos=0.08,scale=0.8,below left},"{{3}}" {gray,pos=0.92,scale=0.8,above right}] & 0 \\
H^{7} & \mathbb{Z}_{3} & \mathbb{Z}_{3} \oplus \mathbb{Z}_{9} & (\mathbb{Z}_{3})^{2} \ar[lu,"{{3}}" {gray,pos=0.08,scale=0.8,below left},"{{3}}" {gray,pos=0.92,scale=0.8,above right}] & 0 & 0 & \mathbb{Z}_{3} \ar[lu,BrickRed,hook,two heads,"{{3}}" {gray,pos=0.08,scale=0.8,below left},"{{3}}" {gray,pos=0.92,scale=0.8,above right}] & \mathbb{Z}_{3} \ar[lu,BrickRed,hook,two heads,"{{3}}" {gray,pos=0.08,scale=0.8,below left},"{{3}}" {gray,pos=0.92,scale=0.8,above right}] \\
H^{8} & \mathbb{Z}_{3} & (\mathbb{Z}_{3})^{2} & (\mathbb{Z}_{3})^{3} \ar[lu,"{{3^{2}}}" {gray,pos=0.08,scale=0.8,below left},"{{3^{2}}}" {gray,pos=0.92,scale=0.8,above right}] & (\mathbb{Z}_{3})^{2} \ar[lu,"{{3}}" {gray,pos=0.08,scale=0.8,below left},"{{3}}" {gray,pos=0.92,scale=0.8,above right}] & 0 & 0 & \mathbb{Z}_{3} \ar[lu,BrickRed,hook,two heads,"{{3}}" {gray,pos=0.08,scale=0.8,below left},"{{3}}" {gray,pos=0.92,scale=0.8,above right}] \\
H^{9} & \mathbb{Z}_{3} & (\mathbb{Z}_{3})^{2} & \mathbb{Z}_{3} \ar[lu,hook,"{{3}}" {gray,pos=0.08,scale=0.8,below left},"{{3}}" {gray,pos=0.92,scale=0.8,above right}] & (\mathbb{Z}_{3})^{3} \ar[lu,"{{3^{2}}}" {gray,pos=0.08,scale=0.8,below left},"{{3^{2}}}" {gray,pos=0.92,scale=0.8,above right}] & \mathbb{Z}_{3} \ar[lu,hook,"{{3}}" {gray,pos=0.08,scale=0.8,below left},"{{3}}" {gray,pos=0.92,scale=0.8,above right}] & 0 & 0 \\
H^{10} & \mathbb{Z}_{3} & (\mathbb{Z}_{3})^{3} & \mathbb{Z}_{3} & \mathbb{Z}_{3} \ar[lu,hook,two heads,"{{3}}" {gray,pos=0.08,scale=0.8,below left},"{{3}}" {gray,pos=0.92,scale=0.8,above right}] & (\mathbb{Z}_{3})^{2} \ar[lu,hook,"{{3^{2}}}" {gray,pos=0.08,scale=0.8,below left},"{{3^{2}}}" {gray,pos=0.92,scale=0.8,above right}] & \mathbb{Z}_{3} \ar[lu,BrickRed,hook,two heads,"{{3}}" {gray,pos=0.08,scale=0.8,below left},"{{3}}" {gray,pos=0.92,scale=0.8,above right}] & 0 \\
H^{11} & \mathbb{Z}_{3} & (\mathbb{Z}_{3})^{3} & (\mathbb{Z}_{3})^{4} & 0 & (\mathbb{Z}_{3})^{2} \ar[lu,two heads,"{{3}}" {gray,pos=0.08,scale=0.8,below left},"{{3}}" {gray,pos=0.92,scale=0.8,above right}] & (\mathbb{Z}_{3})^{2} \ar[lu,hook,two heads,"{{3^{2}}}" {gray,pos=0.08,scale=0.8,below left},"{{3^{2}}}" {gray,pos=0.92,scale=0.8,above right}] & \mathbb{Z}_{3} \ar[lu,BrickRed,hook,two heads,"{{3}}" {gray,pos=0.08,scale=0.8,below left},"{{3}}" {gray,pos=0.92,scale=0.8,above right}] \\
H^{12} & \mathbb{Z}_{3} & (\mathbb{Z}_{3})^{2} \oplus \mathbb{Z}_{9} & (\mathbb{Z}_{3})^{4} & \mathbb{Z}_{3} \oplus \mathbb{Z}_{9} & \mathbb{Z}_{3} & (\mathbb{Z}_{3})^{2} \ar[lu,"{{3}}" {gray,pos=0.08,scale=0.8,below left},"{{3}}" {gray,pos=0.92,scale=0.8,above right}] & (\mathbb{Z}_{3})^{2} \ar[lu,BrickRed,hook,two heads,"{{3^{2}}}" {gray,pos=0.08,scale=0.8,below left},"{{3^{2}}}" {gray,pos=0.92,scale=0.8,above right}] \\
H^{13} & \mathbb{Z}_{3} & (\mathbb{Z}_{3})^{2} \oplus \mathbb{Z}_{9} & \mathbb{Z}_{3} & (\mathbb{Z}_{3})^{4} \oplus \mathbb{Z}_{9} & \mathbb{Z}_{3} \ar[lu,hook,"{{3}}" {gray,pos=0.08,scale=0.8,below left},"{{3}}" {gray,pos=0.92,scale=0.8,above right}] & \mathbb{Z}_{3} & \mathbb{Z}_{3} \ar[lu,hook,"{{3}}" {gray,pos=0.08,scale=0.8,below left},"{{3}}" {gray,pos=0.92,scale=0.8,above right}] \\
H^{14} & \mathbb{Z}_{3} & (\mathbb{Z}_{3})^{3} & \mathbb{Z}_{3} & (\mathbb{Z}_{3})^{4} & (\mathbb{Z}_{3})^{3} \ar[lu,"{{3^{2}}}" {gray,pos=0.08,scale=0.8,below left},"{{3^{2}}}" {gray,pos=0.92,scale=0.8,above right}] & \mathbb{Z}_{3} \ar[lu,hook,two heads,"{{3}}" {gray,pos=0.08,scale=0.8,below left},"{{3}}" {gray,pos=0.92,scale=0.8,above right}] & 0 \\
H^{15} & \mathbb{Z}_{3} & (\mathbb{Z}_{3})^{4} & (\mathbb{Z}_{3})^{6} & \mathbb{Z}_{3} & (\mathbb{Z}_{3})^{4} \ar[lu,"{{3}}" {gray,pos=0.08,scale=0.8,below left},"{{3}}" {gray,pos=0.92,scale=0.8,above right}] & (\mathbb{Z}_{3})^{2} \ar[lu,hook,"{{3^{2}}}" {gray,pos=0.08,scale=0.8,below left},"{{3^{2}}}" {gray,pos=0.92,scale=0.8,above right}] & (\mathbb{Z}_{3})^{2} \ar[lu,two heads,"{{3}}" {gray,pos=0.08,scale=0.8,below left},"{{3}}" {gray,pos=0.92,scale=0.8,above right}] \\
H^{16} & \mathbb{Z}_{3} & (\mathbb{Z}_{3})^{4} & (\mathbb{Z}_{3})^{8} & (\mathbb{Z}_{3})^{3} & (\mathbb{Z}_{3})^{3} & (\mathbb{Z}_{3})^{2} \ar[lu,"{{3}}" {gray,pos=0.08,scale=0.8,below left},"{{3}}" {gray,pos=0.92,scale=0.8,above right}] & (\mathbb{Z}_{3})^{3} \ar[lu,two heads,"{{3^{2}}}" {gray,pos=0.08,scale=0.8,below left},"{{3^{2}}}" {gray,pos=0.92,scale=0.8,above right}] \\
H^{17} & \mathbb{Z}_{3} & (\mathbb{Z}_{3})^{4} & (\mathbb{Z}_{3})^{3} & (\mathbb{Z}_{3})^{7} & (\mathbb{Z}_{3})^{2} & (\mathbb{Z}_{3})^{3} & \mathbb{Z}_{3} \ar[lu,hook,"{{3}}" {gray,pos=0.08,scale=0.8,below left},"{{3}}" {gray,pos=0.92,scale=0.8,above right}] \\
H^{18} & \mathbb{Z}_{3} & (\mathbb{Z}_{3})^{4} \oplus \mathbb{Z}_{9} & (\mathbb{Z}_{3})^{3} & (\mathbb{Z}_{3})^{7} & (\mathbb{Z}_{3})^{2} & (\mathbb{Z}_{3})^{3} \oplus \mathbb{Z}_{9} & \mathbb{Z}_{3} \\
H^{19} & \mathbb{Z}_{3} & (\mathbb{Z}_{3})^{5} \oplus \mathbb{Z}_{9} & (\mathbb{Z}_{3})^{11} \ar[lu,"{{3}}" {gray,pos=0.08,scale=0.8,below left},"{{3}}" {gray,pos=0.92,scale=0.8,above right}] & (\mathbb{Z}_{3})^{4} & (\mathbb{Z}_{3})^{5} & (\mathbb{Z}_{3})^{2} \oplus \mathbb{Z}_{9} & (\mathbb{Z}_{3})^{4} \ar[lu,"{{3}}" {gray,pos=0.08,scale=0.8,below left},"{{3}}" {gray,pos=0.92,scale=0.8,above right}] \\
H^{20} & \mathbb{Z}_{3} & (\mathbb{Z}_{3})^{6} & (\mathbb{Z}_{3})^{12} \ar[lu,"{{3}}" {gray,pos=0.08,scale=0.8,below left},"{{3}}" {gray,pos=0.92,scale=0.8,above right}] & (\mathbb{Z}_{3})^{5} \ar[lu,"{{3}}" {gray,pos=0.08,scale=0.8,below left},"{{3}}" {gray,pos=0.92,scale=0.8,above right}] & (\mathbb{Z}_{3})^{8} & (\mathbb{Z}_{3})^{3} & (\mathbb{Z}_{3})^{5} \ar[lu,"{{3^{2}}}" {gray,pos=0.08,scale=0.8,below left},"{{3^{2}}}" {gray,pos=0.92,scale=0.8,above right}] \\
H^{21} & \mathbb{Z}_{3} & (\mathbb{Z}_{3})^{6} & (\mathbb{Z}_{3})^{4} \ar[lu,"{{3}}" {gray,pos=0.08,scale=0.8,below left},"{{3}}" {gray,pos=0.92,scale=0.8,above right}] & (\mathbb{Z}_{3})^{11} \ar[lu,"{{3^{2}}}" {gray,pos=0.08,scale=0.8,below left},"{{3^{2}}}" {gray,pos=0.92,scale=0.8,above right}] & (\mathbb{Z}_{3})^{7} \ar[lu,"{{3}}" {gray,pos=0.08,scale=0.8,below left},"{{3}}" {gray,pos=0.92,scale=0.8,above right}] & (\mathbb{Z}_{3})^{5} & (\mathbb{Z}_{3})^{2} \ar[lu,"{{3}}" {gray,pos=0.08,scale=0.8,below left},"{{3}}" {gray,pos=0.92,scale=0.8,above right}] \\
H^{22} & \mathbb{Z}_{3} & (\mathbb{Z}_{3})^{7} & (\mathbb{Z}_{3})^{4} & (\mathbb{Z}_{3})^{13} \ar[lu,"{{3}}" {gray,pos=0.08,scale=0.8,below left},"{{3}}" {gray,pos=0.92,scale=0.8,above right}] & (\mathbb{Z}_{3})^{6} \ar[lu,"{{3^{2}}}" {gray,pos=0.08,scale=0.8,below left},"{{3^{2}}}" {gray,pos=0.92,scale=0.8,above right}] & (\mathbb{Z}_{3})^{7} \ar[lu,"{{3}}" {gray,pos=0.08,scale=0.8,below left},"{{3}}" {gray,pos=0.92,scale=0.8,above right}] & \mathbb{Z}_{3} \\
H^{23} & \mathbb{Z}_{3} & (\mathbb{Z}_{3})^{8} & (\mathbb{Z}_{3})^{16} & (\mathbb{Z}_{3})^{9} & (\mathbb{Z}_{3})^{7} \ar[lu,"{{3}}" {gray,pos=0.08,scale=0.8,below left},"{{3}}" {gray,pos=0.92,scale=0.8,above right}] & (\mathbb{Z}_{3})^{8} \ar[lu,"{{3^{2}}}" {gray,pos=0.08,scale=0.8,below left},"{{3^{2}}}" {gray,pos=0.92,scale=0.8,above right}] & (\mathbb{Z}_{3})^{7} \ar[lu,"{{3}}" {gray,pos=0.08,scale=0.8,below left},"{{3}}" {gray,pos=0.92,scale=0.8,above right}] \\
H^{24} & \mathbb{Z}_{3} & (\mathbb{Z}_{3})^{7} \oplus \mathbb{Z}_{9} & (\mathbb{Z}_{3})^{18} \oplus \mathbb{Z}_{9} & (\mathbb{Z}_{3})^{7} \oplus (\mathbb{Z}_{9})^{2} & (\mathbb{Z}_{3})^{10} & (\mathbb{Z}_{3})^{7} \ar[lu,"{{3}}" {gray,pos=0.08,scale=0.8,below left},"{{3}}" {gray,pos=0.92,scale=0.8,above right}] & (\mathbb{Z}_{3})^{10} \ar[lu,"{{3^{2}}}" {gray,pos=0.08,scale=0.8,below left},"{{3^{2}}}" {gray,pos=0.92,scale=0.8,above right}] \\
H^{25} & \mathbb{Z}_{3} & (\mathbb{Z}_{3})^{7} \oplus \mathbb{Z}_{9} & (\mathbb{Z}_{3})^{7} \oplus \mathbb{Z}_{9} & (\mathbb{Z}_{3})^{15} \oplus (\mathbb{Z}_{9})^{2} \ar[lu,"{{3}}" {gray,pos=0.08,scale=0.8,below left},"{{3}}" {gray,pos=0.92,scale=0.8,above right}] & (\mathbb{Z}_{3})^{14} \ar[lu,"{{3}}" {gray,pos=0.08,scale=0.8,below left},"{{3}}" {gray,pos=0.92,scale=0.8,above right}] & (\mathbb{Z}_{3})^{7} & (\mathbb{Z}_{3})^{4} \ar[lu,"{{3}}" {gray,pos=0.08,scale=0.8,below left},"{{3}}" {gray,pos=0.92,scale=0.8,above right}] \\
H^{26} & \mathbb{Z}_{3} & (\mathbb{Z}_{3})^{10} & (\mathbb{Z}_{3})^{9} & (\mathbb{Z}_{3})^{21} \ar[lu,"{{3^{2}}}" {gray,pos=0.08,scale=0.8,below left},"{{3^{2}}}" {gray,pos=0.92,scale=0.8,above right}] & (\mathbb{Z}_{3})^{15} \ar[lu,"{{3^{3}}}" {gray,pos=0.08,scale=0.8,below left},"{{3^{3}}}" {gray,pos=0.92,scale=0.8,above right}] & (\mathbb{Z}_{3})^{11} \ar[lu,"{{3}}" {gray,pos=0.08,scale=0.8,below left},"{{3}}" {gray,pos=0.92,scale=0.8,above right}] & (\mathbb{Z}_{3})^{3} \\
H^{27} & \mathbb{Z}_{3} & (\mathbb{Z}_{3})^{11} & (\mathbb{Z}_{3})^{26} & (\mathbb{Z}_{3})^{17} \ar[lu,"{{3}}" {gray,pos=0.08,scale=0.8,below left},"{{3}}" {gray,pos=0.92,scale=0.8,above right}] & (\mathbb{Z}_{3})^{15} \ar[lu,"{{3^{3}}}" {gray,pos=0.08,scale=0.8,below left},"{{3^{3}}}" {gray,pos=0.92,scale=0.8,above right}] & (\mathbb{Z}_{3})^{14} \ar[lu,"{{3^{3}}}" {gray,pos=0.08,scale=0.8,below left},"{{3^{3}}}" {gray,pos=0.92,scale=0.8,above right}] & (\mathbb{Z}_{3})^{11} \ar[lu,"{{3}}" {gray,pos=0.08,scale=0.8,below left},"{{3}}" {gray,pos=0.92,scale=0.8,above right}] \\
H^{28} & \mathbb{Z}_{3} & (\mathbb{Z}_{3})^{11} & (\mathbb{Z}_{3})^{29} & (\mathbb{Z}_{3})^{16} & (\mathbb{Z}_{3})^{16} \ar[lu,"{{3}}" {gray,pos=0.08,scale=0.8,below left},"{{3}}" {gray,pos=0.92,scale=0.8,above right}] & (\mathbb{Z}_{3})^{14} \ar[lu,"{{3^{3}}}" {gray,pos=0.08,scale=0.8,below left},"{{3^{3}}}" {gray,pos=0.92,scale=0.8,above right}] & (\mathbb{Z}_{3})^{15} \ar[lu,"{{3^{3}}}" {gray,pos=0.08,scale=0.8,below left},"{{3^{3}}}" {gray,pos=0.92,scale=0.8,above right}] \\
H^{29} & \mathbb{Z}_{3} & (\mathbb{Z}_{3})^{12} & (\mathbb{Z}_{3})^{14} & (\mathbb{Z}_{3})^{26} & (\mathbb{Z}_{3})^{18} & (\mathbb{Z}_{3})^{15} \ar[lu,"{{3}}" {gray,pos=0.08,scale=0.8,below left},"{{3}}" {gray,pos=0.92,scale=0.8,above right}] & (\mathbb{Z}_{3})^{9} \ar[lu,"{{3^{3}}}" {gray,pos=0.08,scale=0.8,below left},"{{3^{3}}}" {gray,pos=0.92,scale=0.8,above right}] \\
H^{30} & \mathbb{Z}_{3} & (\mathbb{Z}_{3})^{12} \oplus \mathbb{Z}_{9} & (\mathbb{Z}_{3})^{14} & (\mathbb{Z}_{3})^{35} & (\mathbb{Z}_{3})^{21} \oplus \mathbb{Z}_{9} & (\mathbb{Z}_{3})^{16} \oplus \mathbb{Z}_{9} & (\mathbb{Z}_{3})^{7} \ar[lu,"{{3}}" {gray,pos=0.08,scale=0.8,below left},"{{3}}" {gray,pos=0.92,scale=0.8,above right}] \\
H^{31} & \mathbb{Z}_{3} & (\mathbb{Z}_{3})^{13} \oplus \mathbb{Z}_{9} & (\mathbb{Z}_{3})^{36} & (\mathbb{Z}_{3})^{30} & (\mathbb{Z}_{3})^{25} \oplus \mathbb{Z}_{9} & (\mathbb{Z}_{3})^{18} \oplus \mathbb{Z}_{9} \ar[lu,"{{3}}" {gray,pos=0.08,scale=0.8,below left},"{{3}}" {gray,pos=0.92,scale=0.8,above right}] & (\mathbb{Z}_{3})^{20} \ar[lu,"{{3}}" {gray,pos=0.08,scale=0.8,below left},"{{3}}" {gray,pos=0.92,scale=0.8,above right}] \\
H^{32} & \mathbb{Z}_{3} & (\mathbb{Z}_{3})^{14} & (\mathbb{Z}_{3})^{42} & (\mathbb{Z}_{3})^{26} & (\mathbb{Z}_{3})^{29} & (\mathbb{Z}_{3})^{23} \ar[lu,"{{3^{2}}}" {gray,pos=0.08,scale=0.8,below left},"{{3^{2}}}" {gray,pos=0.92,scale=0.8,above right}] & (\mathbb{Z}_{3})^{26} \ar[lu,"{{3^{3}}}" {gray,pos=0.08,scale=0.8,below left},"{{3^{3}}}" {gray,pos=0.92,scale=0.8,above right}] \\
H^{33} & \mathbb{Z}_{3} & (\mathbb{Z}_{3})^{14} & (\mathbb{Z}_{3})^{22} & (\mathbb{Z}_{3})^{38} & (\mathbb{Z}_{3})^{32} & (\mathbb{Z}_{3})^{29} \ar[lu,"{{3}}" {gray,pos=0.08,scale=0.8,below left},"{{3}}" {gray,pos=0.92,scale=0.8,above right}] & (\mathbb{Z}_{3})^{14} \ar[lu,"{{3^{3}}}" {gray,pos=0.08,scale=0.8,below left},"{{3^{3}}}" {gray,pos=0.92,scale=0.8,above right}] \\
H^{34} & \mathbb{Z}_{3} & (\mathbb{Z}_{3})^{16} & (\mathbb{Z}_{3})^{23} & (\mathbb{Z}_{3})^{55} & (\mathbb{Z}_{3})^{36} & (\mathbb{Z}_{3})^{33} & (\mathbb{Z}_{3})^{12} \ar[lu,"{{3}}" {gray,pos=0.08,scale=0.8,below left},"{{3}}" {gray,pos=0.92,scale=0.8,above right}] \\
H^{35} & \mathbb{Z}_{3} & (\mathbb{Z}_{3})^{17} & (\mathbb{Z}_{3})^{54} & (\mathbb{Z}_{3})^{55} & (\mathbb{Z}_{3})^{42} & (\mathbb{Z}_{3})^{33} & (\mathbb{Z}_{3})^{31} \\
H^{36} & \mathbb{Z}_{3} & (\mathbb{Z}_{3})^{16} \oplus \mathbb{Z}_{9} & (\mathbb{Z}_{3})^{61} & (\mathbb{Z}_{3})^{44} \oplus (\mathbb{Z}_{9})^{2} & (\mathbb{Z}_{3})^{50} & (\mathbb{Z}_{3})^{32} \oplus \mathbb{Z}_{9} & (\mathbb{Z}_{3})^{39} \oplus \mathbb{Z}_{9} \\
H^{37} & \mathbb{Z}_{3} & (\mathbb{Z}_{3})^{17} \oplus \mathbb{Z}_{9} & (\mathbb{Z}_{3})^{35} & (\mathbb{Z}_{3})^{53} \oplus (\mathbb{Z}_{9})^{3} & (\mathbb{Z}_{3})^{58} \ar[lu,"{{3}}" {gray,pos=0.08,scale=0.8,below left},"{{3}}" {gray,pos=0.92,scale=0.8,above right}] & (\mathbb{Z}_{3})^{41} \oplus \mathbb{Z}_{9} & (\mathbb{Z}_{3})^{23} \oplus \mathbb{Z}_{9} \\
H^{38} & \mathbb{Z}_{3} & (\mathbb{Z}_{3})^{19} & (\mathbb{Z}_{3})^{34} & (\mathbb{Z}_{3})^{78} \oplus \mathbb{Z}_{9} & (\mathbb{Z}_{3})^{62} \ar[lu,"{{3}}" {gray,pos=0.08,scale=0.8,below left},"{{3}}" {gray,pos=0.92,scale=0.8,above right}] & (\mathbb{Z}_{3})^{57} \ar[lu,"{{3}}" {gray,pos=0.08,scale=0.8,below left},"{{3}}" {gray,pos=0.92,scale=0.8,above right}] & (\mathbb{Z}_{3})^{21} \\
H^{39} & \mathbb{Z}_{3} & (\mathbb{Z}_{3})^{21} & (\mathbb{Z}_{3})^{73} & (\mathbb{Z}_{3})^{87} & (\mathbb{Z}_{3})^{66} \ar[lu,"{{3}}" {gray,pos=0.08,scale=0.8,below left},"{{3}}" {gray,pos=0.92,scale=0.8,above right}] & (\mathbb{Z}_{3})^{63} \ar[lu,"{{3^{2}}}" {gray,pos=0.08,scale=0.8,below left},"{{3^{2}}}" {gray,pos=0.92,scale=0.8,above right}] & (\mathbb{Z}_{3})^{53} \ar[lu,"{{3}}" {gray,pos=0.08,scale=0.8,below left},"{{3}}" {gray,pos=0.92,scale=0.8,above right}] \\
H^{40} & \mathbb{Z}_{3} & (\mathbb{Z}_{3})^{21} & (\mathbb{Z}_{3})^{85} & (\mathbb{Z}_{3})^{79} & (\mathbb{Z}_{3})^{76} & (\mathbb{Z}_{3})^{62} \ar[lu,"{{3}}" {gray,pos=0.08,scale=0.8,below left},"{{3}}" {gray,pos=0.92,scale=0.8,above right}] & (\mathbb{Z}_{3})^{69} \ar[lu,"{{3^{2}}}" {gray,pos=0.08,scale=0.8,below left},"{{3^{2}}}" {gray,pos=0.92,scale=0.8,above right}] \\
\end{tikzcd}

%% file: tables/omega-Z5-RZ.tex
\begin{tikzcd}[row sep=tiny,column sep=tiny]
& B\mathbb{Z}_{5} & B^{2}\mathbb{Z}_{5} & B^{3}\mathbb{Z}_{5} & B^{4}\mathbb{Z}_{5} & B^{5}\mathbb{Z}_{5} & B^{6}\mathbb{Z}_{5} & B^{7}\mathbb{Z}_{5} & B^{8}\mathbb{Z}_{5} & B^{9}\mathbb{Z}_{5} & B^{10}\mathbb{Z}_{5} & B^{11}\mathbb{Z}_{5} \\
H^{0} & \mathbb{R}/\mathbb{Z} & \mathbb{R}/\mathbb{Z} & \mathbb{R}/\mathbb{Z} & \mathbb{R}/\mathbb{Z} & \mathbb{R}/\mathbb{Z} & \mathbb{R}/\mathbb{Z} & \mathbb{R}/\mathbb{Z} & \mathbb{R}/\mathbb{Z} & \mathbb{R}/\mathbb{Z} & \mathbb{R}/\mathbb{Z} & \mathbb{R}/\mathbb{Z} \\
H^{1} & \mathbb{Z}_{5} & 0 & 0 & 0 & 0 & 0 & 0 & 0 & 0 & 0 & 0 \\
H^{2} & 0 & \mathbb{Z}_{5} \ar[lu,BrickRed,hook,two heads,"{{5}}" {gray,pos=0.08,scale=0.8,below left},"{{5}}" {gray,pos=0.92,scale=0.8,above right}] & 0 & 0 & 0 & 0 & 0 & 0 & 0 & 0 & 0 \\
H^{3} & \mathbb{Z}_{5} & 0 & \mathbb{Z}_{5} \ar[lu,BrickRed,hook,two heads,"{{5}}" {gray,pos=0.08,scale=0.8,below left},"{{5}}" {gray,pos=0.92,scale=0.8,above right}] & 0 & 0 & 0 & 0 & 0 & 0 & 0 & 0 \\
H^{4} & 0 & \mathbb{Z}_{5} & 0 & \mathbb{Z}_{5} \ar[lu,BrickRed,hook,two heads,"{{5}}" {gray,pos=0.08,scale=0.8,below left},"{{5}}" {gray,pos=0.92,scale=0.8,above right}] & 0 & 0 & 0 & 0 & 0 & 0 & 0 \\
H^{5} & \mathbb{Z}_{5} & 0 & 0 & 0 & \mathbb{Z}_{5} \ar[lu,BrickRed,hook,two heads,"{{5}}" {gray,pos=0.08,scale=0.8,below left},"{{5}}" {gray,pos=0.92,scale=0.8,above right}] & 0 & 0 & 0 & 0 & 0 & 0 \\
H^{6} & 0 & \mathbb{Z}_{5} & 0 & 0 & 0 & \mathbb{Z}_{5} \ar[lu,BrickRed,hook,two heads,"{{5}}" {gray,pos=0.08,scale=0.8,below left},"{{5}}" {gray,pos=0.92,scale=0.8,above right}] & 0 & 0 & 0 & 0 & 0 \\
H^{7} & \mathbb{Z}_{5} & 0 & \mathbb{Z}_{5} & 0 & 0 & 0 & \mathbb{Z}_{5} \ar[lu,BrickRed,hook,two heads,"{{5}}" {gray,pos=0.08,scale=0.8,below left},"{{5}}" {gray,pos=0.92,scale=0.8,above right}] & 0 & 0 & 0 & 0 \\
H^{8} & 0 & \mathbb{Z}_{5} & 0 & \mathbb{Z}_{5} & 0 & 0 & 0 & \mathbb{Z}_{5} \ar[lu,BrickRed,hook,two heads,"{{5}}" {gray,pos=0.08,scale=0.8,below left},"{{5}}" {gray,pos=0.92,scale=0.8,above right}] & 0 & 0 & 0 \\
H^{9} & \mathbb{Z}_{5} & 0 & 0 & 0 & 0 & 0 & 0 & 0 & \mathbb{Z}_{5} \ar[lu,BrickRed,hook,two heads,"{{5}}" {gray,pos=0.08,scale=0.8,below left},"{{5}}" {gray,pos=0.92,scale=0.8,above right}] & 0 & 0 \\
H^{10} & 0 & \mathbb{Z}_{25} & 0 & 0 & 0 & 0 & 0 & 0 & 0 & \mathbb{Z}_{5} \ar[lu,BrickRed,hook,two heads,"{{5}}" {gray,pos=0.08,scale=0.8,below left},"{{5}}" {gray,pos=0.92,scale=0.8,above right}] & 0 \\
H^{11} & \mathbb{Z}_{5} & \mathbb{Z}_{5} & (\mathbb{Z}_{5})^{2} \ar[lu,"{{5}}" {gray,pos=0.08,scale=0.8,below left},"{{5}}" {gray,pos=0.92,scale=0.8,above right}] & 0 & \mathbb{Z}_{5} & 0 & 0 & 0 & 0 & 0 & \mathbb{Z}_{5} \ar[lu,BrickRed,hook,two heads,"{{5}}" {gray,pos=0.08,scale=0.8,below left},"{{5}}" {gray,pos=0.92,scale=0.8,above right}] \\
H^{12} & 0 & \mathbb{Z}_{5} & \mathbb{Z}_{5} \ar[lu,hook,two heads,"{{5}}" {gray,pos=0.08,scale=0.8,below left},"{{5}}" {gray,pos=0.92,scale=0.8,above right}] & (\mathbb{Z}_{5})^{2} \ar[lu,"{{5}}" {gray,pos=0.08,scale=0.8,below left},"{{5}}" {gray,pos=0.92,scale=0.8,above right}] & 0 & \mathbb{Z}_{5} & 0 & 0 & 0 & 0 & 0 \\
H^{13} & \mathbb{Z}_{5} & \mathbb{Z}_{5} & 0 & \mathbb{Z}_{5} \ar[lu,hook,two heads,"{{5}}" {gray,pos=0.08,scale=0.8,below left},"{{5}}" {gray,pos=0.92,scale=0.8,above right}] & \mathbb{Z}_{5} \ar[lu,hook,"{{5}}" {gray,pos=0.08,scale=0.8,below left},"{{5}}" {gray,pos=0.92,scale=0.8,above right}] & 0 & 0 & 0 & 0 & 0 & 0 \\
H^{14} & 0 & (\mathbb{Z}_{5})^{2} & \mathbb{Z}_{5} & 0 & \mathbb{Z}_{5} \ar[lu,hook,two heads,"{{5}}" {gray,pos=0.08,scale=0.8,below left},"{{5}}" {gray,pos=0.92,scale=0.8,above right}] & \mathbb{Z}_{5} \ar[lu,hook,two heads,"{{5}}" {gray,pos=0.08,scale=0.8,below left},"{{5}}" {gray,pos=0.92,scale=0.8,above right}] & 0 & 0 & 0 & 0 & 0 \\
H^{15} & \mathbb{Z}_{5} & \mathbb{Z}_{5} & (\mathbb{Z}_{5})^{3} & 0 & 0 & \mathbb{Z}_{5} \ar[lu,hook,two heads,"{{5}}" {gray,pos=0.08,scale=0.8,below left},"{{5}}" {gray,pos=0.92,scale=0.8,above right}] & (\mathbb{Z}_{5})^{2} \ar[lu,two heads,"{{5}}" {gray,pos=0.08,scale=0.8,below left},"{{5}}" {gray,pos=0.92,scale=0.8,above right}] & 0 & 0 & 0 & 0 \\
H^{16} & 0 & (\mathbb{Z}_{5})^{2} & \mathbb{Z}_{5} & (\mathbb{Z}_{5})^{2} & 0 & 0 & \mathbb{Z}_{5} \ar[lu,hook,two heads,"{{5}}" {gray,pos=0.08,scale=0.8,below left},"{{5}}" {gray,pos=0.92,scale=0.8,above right}] & (\mathbb{Z}_{5})^{2} \ar[lu,"{{5}}" {gray,pos=0.08,scale=0.8,below left},"{{5}}" {gray,pos=0.92,scale=0.8,above right}] & 0 & 0 & 0 \\
H^{17} & \mathbb{Z}_{5} & \mathbb{Z}_{5} & 0 & (\mathbb{Z}_{5})^{2} & \mathbb{Z}_{5} & 0 & 0 & \mathbb{Z}_{5} \ar[lu,hook,two heads,"{{5}}" {gray,pos=0.08,scale=0.8,below left},"{{5}}" {gray,pos=0.92,scale=0.8,above right}] & \mathbb{Z}_{5} \ar[lu,hook,"{{5}}" {gray,pos=0.08,scale=0.8,below left},"{{5}}" {gray,pos=0.92,scale=0.8,above right}] & 0 & 0 \\
H^{18} & 0 & (\mathbb{Z}_{5})^{2} & \mathbb{Z}_{5} & \mathbb{Z}_{5} & \mathbb{Z}_{5} & \mathbb{Z}_{5} & 0 & 0 & \mathbb{Z}_{5} \ar[lu,hook,two heads,"{{5}}" {gray,pos=0.08,scale=0.8,below left},"{{5}}" {gray,pos=0.92,scale=0.8,above right}] & \mathbb{Z}_{5} \ar[lu,BrickRed,hook,two heads,"{{5}}" {gray,pos=0.08,scale=0.8,below left},"{{5}}" {gray,pos=0.92,scale=0.8,above right}] & 0 \\
H^{19} & \mathbb{Z}_{5} & \mathbb{Z}_{5} & (\mathbb{Z}_{5})^{3} & 0 & (\mathbb{Z}_{5})^{2} & 0 & 0 & 0 & \mathbb{Z}_{5} & \mathbb{Z}_{5} \ar[lu,hook,two heads,"{{5}}" {gray,pos=0.08,scale=0.8,below left},"{{5}}" {gray,pos=0.92,scale=0.8,above right}] & \mathbb{Z}_{5} \ar[lu,BrickRed,hook,two heads,"{{5}}" {gray,pos=0.08,scale=0.8,below left},"{{5}}" {gray,pos=0.92,scale=0.8,above right}] \\
H^{20} & 0 & \mathbb{Z}_{5} \oplus \mathbb{Z}_{25} & \mathbb{Z}_{5} & \mathbb{Z}_{5} \oplus \mathbb{Z}_{25} & \mathbb{Z}_{5} & \mathbb{Z}_{5} & 0 & 0 & 0 & \mathbb{Z}_{5} & \mathbb{Z}_{5} \ar[lu,BrickRed,hook,two heads,"{{5}}" {gray,pos=0.08,scale=0.8,below left},"{{5}}" {gray,pos=0.92,scale=0.8,above right}] \\
H^{21} & \mathbb{Z}_{5} & \mathbb{Z}_{5} & 0 & (\mathbb{Z}_{5})^{3} & \mathbb{Z}_{5} \ar[lu,hook,"{{5}}" {gray,pos=0.08,scale=0.8,below left},"{{5}}" {gray,pos=0.92,scale=0.8,above right}] & (\mathbb{Z}_{5})^{2} & 0 & 0 & 0 & 0 & 0 \\
H^{22} & 0 & (\mathbb{Z}_{5})^{2} & \mathbb{Z}_{5} & \mathbb{Z}_{5} & \mathbb{Z}_{5} \ar[lu,hook,"{{5}}" {gray,pos=0.08,scale=0.8,below left},"{{5}}" {gray,pos=0.92,scale=0.8,above right}] & (\mathbb{Z}_{5})^{2} \ar[lu,two heads,"{{5}}" {gray,pos=0.08,scale=0.8,below left},"{{5}}" {gray,pos=0.92,scale=0.8,above right}] & \mathbb{Z}_{5} & 0 & 0 & 0 & 0 \\
H^{23} & \mathbb{Z}_{5} & (\mathbb{Z}_{5})^{2} & (\mathbb{Z}_{5})^{5} & 0 & \mathbb{Z}_{5} & \mathbb{Z}_{5} \ar[lu,hook,two heads,"{{5}}" {gray,pos=0.08,scale=0.8,below left},"{{5}}" {gray,pos=0.92,scale=0.8,above right}] & (\mathbb{Z}_{5})^{4} \ar[lu,"{{5}}" {gray,pos=0.08,scale=0.8,below left},"{{5}}" {gray,pos=0.92,scale=0.8,above right}] & 0 & 0 & 0 & \mathbb{Z}_{5} \\
H^{24} & 0 & (\mathbb{Z}_{5})^{2} & (\mathbb{Z}_{5})^{3} & (\mathbb{Z}_{5})^{3} & \mathbb{Z}_{5} & \mathbb{Z}_{5} & (\mathbb{Z}_{5})^{2} \ar[lu,two heads,"{{5}}" {gray,pos=0.08,scale=0.8,below left},"{{5}}" {gray,pos=0.92,scale=0.8,above right}] & (\mathbb{Z}_{5})^{3} \ar[lu,"{{5}}" {gray,pos=0.08,scale=0.8,below left},"{{5}}" {gray,pos=0.92,scale=0.8,above right}] & 0 & 0 & 0 \\
H^{25} & \mathbb{Z}_{5} & (\mathbb{Z}_{5})^{2} & 0 & (\mathbb{Z}_{5})^{4} & (\mathbb{Z}_{5})^{2} & 0 & 0 & (\mathbb{Z}_{5})^{3} \ar[lu,"{{5}}" {gray,pos=0.08,scale=0.8,below left},"{{5}}" {gray,pos=0.92,scale=0.8,above right}] & \mathbb{Z}_{5} \ar[lu,hook,"{{5}}" {gray,pos=0.08,scale=0.8,below left},"{{5}}" {gray,pos=0.92,scale=0.8,above right}] & 0 & 0 \\
H^{26} & 0 & (\mathbb{Z}_{5})^{3} & (\mathbb{Z}_{5})^{3} & (\mathbb{Z}_{5})^{3} & (\mathbb{Z}_{5})^{2} & \mathbb{Z}_{5} & 0 & \mathbb{Z}_{5} & (\mathbb{Z}_{5})^{2} \ar[lu,"{{5}}" {gray,pos=0.08,scale=0.8,below left},"{{5}}" {gray,pos=0.92,scale=0.8,above right}] & \mathbb{Z}_{5} \ar[lu,hook,two heads,"{{5}}" {gray,pos=0.08,scale=0.8,below left},"{{5}}" {gray,pos=0.92,scale=0.8,above right}] & 0 \\
H^{27} & \mathbb{Z}_{5} & (\mathbb{Z}_{5})^{2} & (\mathbb{Z}_{5})^{7} & \mathbb{Z}_{5} & (\mathbb{Z}_{5})^{4} & (\mathbb{Z}_{5})^{2} & 0 & 0 & (\mathbb{Z}_{5})^{2} & \mathbb{Z}_{5} \ar[lu,hook,"{{5}}" {gray,pos=0.08,scale=0.8,below left},"{{5}}" {gray,pos=0.92,scale=0.8,above right}] & \mathbb{Z}_{5} \ar[lu,hook,two heads,"{{5}}" {gray,pos=0.08,scale=0.8,below left},"{{5}}" {gray,pos=0.92,scale=0.8,above right}] \\
H^{28} & 0 & (\mathbb{Z}_{5})^{3} & (\mathbb{Z}_{5})^{3} & (\mathbb{Z}_{5})^{3} & (\mathbb{Z}_{5})^{3} & (\mathbb{Z}_{5})^{3} & 0 & 0 & \mathbb{Z}_{5} & \mathbb{Z}_{5} & \mathbb{Z}_{5} \ar[lu,hook,two heads,"{{5}}" {gray,pos=0.08,scale=0.8,below left},"{{5}}" {gray,pos=0.92,scale=0.8,above right}] \\
H^{29} & \mathbb{Z}_{5} & (\mathbb{Z}_{5})^{2} & 0 & (\mathbb{Z}_{5})^{5} & \mathbb{Z}_{5} & (\mathbb{Z}_{5})^{3} & 0 & 0 & \mathbb{Z}_{5} & (\mathbb{Z}_{5})^{2} & 0 \\
H^{30} & 0 & (\mathbb{Z}_{5})^{2} \oplus \mathbb{Z}_{25} & (\mathbb{Z}_{5})^{3} & (\mathbb{Z}_{5})^{4} & \mathbb{Z}_{5} & (\mathbb{Z}_{5})^{2} \oplus \mathbb{Z}_{25} & (\mathbb{Z}_{5})^{2} & 0 & 0 & (\mathbb{Z}_{5})^{2} & \mathbb{Z}_{5} \\
H^{31} & \mathbb{Z}_{5} & (\mathbb{Z}_{5})^{2} & (\mathbb{Z}_{5})^{7} & (\mathbb{Z}_{5})^{2} & (\mathbb{Z}_{5})^{2} & (\mathbb{Z}_{5})^{2} & (\mathbb{Z}_{5})^{8} \ar[lu,"{{5}}" {gray,pos=0.08,scale=0.8,below left},"{{5}}" {gray,pos=0.92,scale=0.8,above right}] & 0 & 0 & 0 & (\mathbb{Z}_{5})^{2} \\
H^{32} & 0 & (\mathbb{Z}_{5})^{3} & (\mathbb{Z}_{5})^{3} & (\mathbb{Z}_{5})^{4} & (\mathbb{Z}_{5})^{4} & \mathbb{Z}_{5} & (\mathbb{Z}_{5})^{5} \ar[lu,"{{5}}" {gray,pos=0.08,scale=0.8,below left},"{{5}}" {gray,pos=0.92,scale=0.8,above right}] & (\mathbb{Z}_{5})^{5} \ar[lu,"{{5}}" {gray,pos=0.08,scale=0.8,below left},"{{5}}" {gray,pos=0.92,scale=0.8,above right}] & 0 & 0 & \mathbb{Z}_{5} \\
H^{33} & \mathbb{Z}_{5} & (\mathbb{Z}_{5})^{2} & 0 & (\mathbb{Z}_{5})^{6} & (\mathbb{Z}_{5})^{6} & (\mathbb{Z}_{5})^{2} & 0 & (\mathbb{Z}_{5})^{6} \ar[lu,"{{5}}" {gray,pos=0.08,scale=0.8,below left},"{{5}}" {gray,pos=0.92,scale=0.8,above right}] & \mathbb{Z}_{5} \ar[lu,hook,"{{5}}" {gray,pos=0.08,scale=0.8,below left},"{{5}}" {gray,pos=0.92,scale=0.8,above right}] & 0 & 0 \\
H^{34} & 0 & (\mathbb{Z}_{5})^{3} & (\mathbb{Z}_{5})^{3} & (\mathbb{Z}_{5})^{6} & (\mathbb{Z}_{5})^{4} & (\mathbb{Z}_{5})^{3} & 0 & (\mathbb{Z}_{5})^{3} & (\mathbb{Z}_{5})^{2} \ar[lu,"{{5}}" {gray,pos=0.08,scale=0.8,below left},"{{5}}" {gray,pos=0.92,scale=0.8,above right}] & \mathbb{Z}_{5} \ar[lu,hook,two heads,"{{5}}" {gray,pos=0.08,scale=0.8,below left},"{{5}}" {gray,pos=0.92,scale=0.8,above right}] & 0 \\
H^{35} & \mathbb{Z}_{5} & (\mathbb{Z}_{5})^{3} & (\mathbb{Z}_{5})^{10} & (\mathbb{Z}_{5})^{3} & (\mathbb{Z}_{5})^{4} & (\mathbb{Z}_{5})^{4} & 0 & \mathbb{Z}_{5} & (\mathbb{Z}_{5})^{4} & \mathbb{Z}_{5} \ar[lu,hook,"{{5}}" {gray,pos=0.08,scale=0.8,below left},"{{5}}" {gray,pos=0.92,scale=0.8,above right}] & (\mathbb{Z}_{5})^{2} \ar[lu,two heads,"{{5}}" {gray,pos=0.08,scale=0.8,below left},"{{5}}" {gray,pos=0.92,scale=0.8,above right}] \\
H^{36} & 0 & (\mathbb{Z}_{5})^{3} & (\mathbb{Z}_{5})^{6} & (\mathbb{Z}_{5})^{5} & (\mathbb{Z}_{5})^{4} & (\mathbb{Z}_{5})^{5} & 0 & 0 & (\mathbb{Z}_{5})^{4} & (\mathbb{Z}_{5})^{2} & \mathbb{Z}_{5} \ar[lu,hook,two heads,"{{5}}" {gray,pos=0.08,scale=0.8,below left},"{{5}}" {gray,pos=0.92,scale=0.8,above right}] \\
H^{37} & \mathbb{Z}_{5} & (\mathbb{Z}_{5})^{3} & 0 & (\mathbb{Z}_{5})^{7} & (\mathbb{Z}_{5})^{3} & (\mathbb{Z}_{5})^{6} & 0 & 0 & (\mathbb{Z}_{5})^{2} & (\mathbb{Z}_{5})^{3} & 0 \\
H^{38} & 0 & (\mathbb{Z}_{5})^{4} & (\mathbb{Z}_{5})^{6} & (\mathbb{Z}_{5})^{8} & (\mathbb{Z}_{5})^{4} & (\mathbb{Z}_{5})^{6} & (\mathbb{Z}_{5})^{6} & 0 & \mathbb{Z}_{5} & (\mathbb{Z}_{5})^{3} & \mathbb{Z}_{5} \\
H^{39} & \mathbb{Z}_{5} & (\mathbb{Z}_{5})^{3} & (\mathbb{Z}_{5})^{13} & (\mathbb{Z}_{5})^{6} & (\mathbb{Z}_{5})^{7} & (\mathbb{Z}_{5})^{3} & (\mathbb{Z}_{5})^{14} & 0 & \mathbb{Z}_{5} & (\mathbb{Z}_{5})^{3} & (\mathbb{Z}_{5})^{4} \\
H^{40} & 0 & (\mathbb{Z}_{5})^{3} \oplus \mathbb{Z}_{25} & (\mathbb{Z}_{5})^{6} & (\mathbb{Z}_{5})^{6} \oplus \mathbb{Z}_{25} & (\mathbb{Z}_{5})^{7} & (\mathbb{Z}_{5})^{3} & (\mathbb{Z}_{5})^{8} & (\mathbb{Z}_{5})^{5} \oplus \mathbb{Z}_{25} & 0 & (\mathbb{Z}_{5})^{2} & (\mathbb{Z}_{5})^{3} \\
\end{tikzcd}

%% file: tables/omega-Z5-Z5.tex
\begin{tikzcd}[row sep=tiny,column sep=large]
& B\mathbb{Z}_{5} & B^{2}\mathbb{Z}_{5} & B^{3}\mathbb{Z}_{5} & B^{4}\mathbb{Z}_{5} & B^{5}\mathbb{Z}_{5} & B^{6}\mathbb{Z}_{5} & B^{7}\mathbb{Z}_{5} & B^{8}\mathbb{Z}_{5} & B^{9}\mathbb{Z}_{5} \\
H^{0} & \mathbb{Z}_{5} & \mathbb{Z}_{5} & \mathbb{Z}_{5} & \mathbb{Z}_{5} & \mathbb{Z}_{5} & \mathbb{Z}_{5} & \mathbb{Z}_{5} & \mathbb{Z}_{5} & \mathbb{Z}_{5} \\
H^{1} & \mathbb{Z}_{5} & 0 & 0 & 0 & 0 & 0 & 0 & 0 & 0 \\
H^{2} & \mathbb{Z}_{5} & \mathbb{Z}_{5} \ar[lu,BrickRed,hook,two heads,"{{5}}" {gray,pos=0.08,scale=0.8,below left},"{{5}}" {gray,pos=0.92,scale=0.8,above right}] & 0 & 0 & 0 & 0 & 0 & 0 & 0 \\
H^{3} & \mathbb{Z}_{5} & \mathbb{Z}_{5} \ar[lu,hook,two heads,"{{5}}" {gray,pos=0.08,scale=0.8,below left},"{{5}}" {gray,pos=0.92,scale=0.8,above right}] & \mathbb{Z}_{5} \ar[lu,BrickRed,hook,two heads,"{{5}}" {gray,pos=0.08,scale=0.8,below left},"{{5}}" {gray,pos=0.92,scale=0.8,above right}] & 0 & 0 & 0 & 0 & 0 & 0 \\
H^{4} & \mathbb{Z}_{5} & \mathbb{Z}_{5} & \mathbb{Z}_{5} \ar[lu,BrickRed,hook,two heads,"{{5}}" {gray,pos=0.08,scale=0.8,below left},"{{5}}" {gray,pos=0.92,scale=0.8,above right}] & \mathbb{Z}_{5} \ar[lu,BrickRed,hook,two heads,"{{5}}" {gray,pos=0.08,scale=0.8,below left},"{{5}}" {gray,pos=0.92,scale=0.8,above right}] & 0 & 0 & 0 & 0 & 0 \\
H^{5} & \mathbb{Z}_{5} & \mathbb{Z}_{5} & 0 & \mathbb{Z}_{5} \ar[lu,BrickRed,hook,two heads,"{{5}}" {gray,pos=0.08,scale=0.8,below left},"{{5}}" {gray,pos=0.92,scale=0.8,above right}] & \mathbb{Z}_{5} \ar[lu,BrickRed,hook,two heads,"{{5}}" {gray,pos=0.08,scale=0.8,below left},"{{5}}" {gray,pos=0.92,scale=0.8,above right}] & 0 & 0 & 0 & 0 \\
H^{6} & \mathbb{Z}_{5} & \mathbb{Z}_{5} & 0 & 0 & \mathbb{Z}_{5} \ar[lu,BrickRed,hook,two heads,"{{5}}" {gray,pos=0.08,scale=0.8,below left},"{{5}}" {gray,pos=0.92,scale=0.8,above right}] & \mathbb{Z}_{5} \ar[lu,BrickRed,hook,two heads,"{{5}}" {gray,pos=0.08,scale=0.8,below left},"{{5}}" {gray,pos=0.92,scale=0.8,above right}] & 0 & 0 & 0 \\
H^{7} & \mathbb{Z}_{5} & \mathbb{Z}_{5} & \mathbb{Z}_{5} & 0 & 0 & \mathbb{Z}_{5} \ar[lu,BrickRed,hook,two heads,"{{5}}" {gray,pos=0.08,scale=0.8,below left},"{{5}}" {gray,pos=0.92,scale=0.8,above right}] & \mathbb{Z}_{5} \ar[lu,BrickRed,hook,two heads,"{{5}}" {gray,pos=0.08,scale=0.8,below left},"{{5}}" {gray,pos=0.92,scale=0.8,above right}] & 0 & 0 \\
H^{8} & \mathbb{Z}_{5} & \mathbb{Z}_{5} & \mathbb{Z}_{5} & \mathbb{Z}_{5} & 0 & 0 & \mathbb{Z}_{5} \ar[lu,BrickRed,hook,two heads,"{{5}}" {gray,pos=0.08,scale=0.8,below left},"{{5}}" {gray,pos=0.92,scale=0.8,above right}] & \mathbb{Z}_{5} \ar[lu,BrickRed,hook,two heads,"{{5}}" {gray,pos=0.08,scale=0.8,below left},"{{5}}" {gray,pos=0.92,scale=0.8,above right}] & 0 \\
H^{9} & \mathbb{Z}_{5} & \mathbb{Z}_{5} & 0 & \mathbb{Z}_{5} & 0 & 0 & 0 & \mathbb{Z}_{5} \ar[lu,BrickRed,hook,two heads,"{{5}}" {gray,pos=0.08,scale=0.8,below left},"{{5}}" {gray,pos=0.92,scale=0.8,above right}] & \mathbb{Z}_{5} \ar[lu,BrickRed,hook,two heads,"{{5}}" {gray,pos=0.08,scale=0.8,below left},"{{5}}" {gray,pos=0.92,scale=0.8,above right}] \\
H^{10} & \mathbb{Z}_{5} & \mathbb{Z}_{5} & 0 & 0 & 0 & 0 & 0 & 0 & \mathbb{Z}_{5} \ar[lu,BrickRed,hook,two heads,"{{5}}" {gray,pos=0.08,scale=0.8,below left},"{{5}}" {gray,pos=0.92,scale=0.8,above right}] \\
H^{11} & \mathbb{Z}_{5} & (\mathbb{Z}_{5})^{2} \ar[lu,two heads,"{{5}}" {gray,pos=0.08,scale=0.8,below left},"{{5}}" {gray,pos=0.92,scale=0.8,above right}] & (\mathbb{Z}_{5})^{2} \ar[lu,two heads,"{{5}}" {gray,pos=0.08,scale=0.8,below left},"{{5}}" {gray,pos=0.92,scale=0.8,above right}] & 0 & \mathbb{Z}_{5} & 0 & 0 & 0 & 0 \\
H^{12} & \mathbb{Z}_{5} & (\mathbb{Z}_{5})^{2} & (\mathbb{Z}_{5})^{3} \ar[lu,"{{5}}" {gray,pos=0.08,scale=0.8,below left},"{{5}}" {gray,pos=0.92,scale=0.8,above right}] & (\mathbb{Z}_{5})^{2} \ar[lu,"{{5}}" {gray,pos=0.08,scale=0.8,below left},"{{5}}" {gray,pos=0.92,scale=0.8,above right}] & \mathbb{Z}_{5} & \mathbb{Z}_{5} & 0 & 0 & 0 \\
H^{13} & \mathbb{Z}_{5} & (\mathbb{Z}_{5})^{2} & \mathbb{Z}_{5} \ar[lu,hook,"{{5}}" {gray,pos=0.08,scale=0.8,below left},"{{5}}" {gray,pos=0.92,scale=0.8,above right}] & (\mathbb{Z}_{5})^{3} \ar[lu,"{{5^{2}}}" {gray,pos=0.08,scale=0.8,below left},"{{5^{2}}}" {gray,pos=0.92,scale=0.8,above right}] & \mathbb{Z}_{5} \ar[lu,hook,"{{5}}" {gray,pos=0.08,scale=0.8,below left},"{{5}}" {gray,pos=0.92,scale=0.8,above right}] & \mathbb{Z}_{5} & 0 & 0 & 0 \\
H^{14} & \mathbb{Z}_{5} & (\mathbb{Z}_{5})^{3} & \mathbb{Z}_{5} & \mathbb{Z}_{5} \ar[lu,hook,two heads,"{{5}}" {gray,pos=0.08,scale=0.8,below left},"{{5}}" {gray,pos=0.92,scale=0.8,above right}] & (\mathbb{Z}_{5})^{2} \ar[lu,hook,"{{5^{2}}}" {gray,pos=0.08,scale=0.8,below left},"{{5^{2}}}" {gray,pos=0.92,scale=0.8,above right}] & \mathbb{Z}_{5} \ar[lu,hook,two heads,"{{5}}" {gray,pos=0.08,scale=0.8,below left},"{{5}}" {gray,pos=0.92,scale=0.8,above right}] & 0 & 0 & 0 \\
H^{15} & \mathbb{Z}_{5} & (\mathbb{Z}_{5})^{3} & (\mathbb{Z}_{5})^{4} & 0 & \mathbb{Z}_{5} \ar[lu,hook,two heads,"{{5}}" {gray,pos=0.08,scale=0.8,below left},"{{5}}" {gray,pos=0.92,scale=0.8,above right}] & (\mathbb{Z}_{5})^{2} \ar[lu,hook,two heads,"{{5^{2}}}" {gray,pos=0.08,scale=0.8,below left},"{{5^{2}}}" {gray,pos=0.92,scale=0.8,above right}] & (\mathbb{Z}_{5})^{2} \ar[lu,two heads,"{{5}}" {gray,pos=0.08,scale=0.8,below left},"{{5}}" {gray,pos=0.92,scale=0.8,above right}] & 0 & 0 \\
H^{16} & \mathbb{Z}_{5} & (\mathbb{Z}_{5})^{3} & (\mathbb{Z}_{5})^{4} & (\mathbb{Z}_{5})^{2} & 0 & \mathbb{Z}_{5} \ar[lu,hook,two heads,"{{5}}" {gray,pos=0.08,scale=0.8,below left},"{{5}}" {gray,pos=0.92,scale=0.8,above right}] & (\mathbb{Z}_{5})^{3} \ar[lu,two heads,"{{5^{2}}}" {gray,pos=0.08,scale=0.8,below left},"{{5^{2}}}" {gray,pos=0.92,scale=0.8,above right}] & (\mathbb{Z}_{5})^{2} \ar[lu,"{{5}}" {gray,pos=0.08,scale=0.8,below left},"{{5}}" {gray,pos=0.92,scale=0.8,above right}] & 0 \\
H^{17} & \mathbb{Z}_{5} & (\mathbb{Z}_{5})^{3} & \mathbb{Z}_{5} & (\mathbb{Z}_{5})^{4} & \mathbb{Z}_{5} & 0 & \mathbb{Z}_{5} \ar[lu,hook,two heads,"{{5}}" {gray,pos=0.08,scale=0.8,below left},"{{5}}" {gray,pos=0.92,scale=0.8,above right}] & (\mathbb{Z}_{5})^{3} \ar[lu,"{{5^{2}}}" {gray,pos=0.08,scale=0.8,below left},"{{5^{2}}}" {gray,pos=0.92,scale=0.8,above right}] & \mathbb{Z}_{5} \ar[lu,hook,"{{5}}" {gray,pos=0.08,scale=0.8,below left},"{{5}}" {gray,pos=0.92,scale=0.8,above right}] \\
H^{18} & \mathbb{Z}_{5} & (\mathbb{Z}_{5})^{3} & \mathbb{Z}_{5} & (\mathbb{Z}_{5})^{3} & (\mathbb{Z}_{5})^{2} & \mathbb{Z}_{5} & 0 & \mathbb{Z}_{5} \ar[lu,hook,two heads,"{{5}}" {gray,pos=0.08,scale=0.8,below left},"{{5}}" {gray,pos=0.92,scale=0.8,above right}] & (\mathbb{Z}_{5})^{2} \ar[lu,hook,"{{5^{2}}}" {gray,pos=0.08,scale=0.8,below left},"{{5^{2}}}" {gray,pos=0.92,scale=0.8,above right}] \\
H^{19} & \mathbb{Z}_{5} & (\mathbb{Z}_{5})^{3} & (\mathbb{Z}_{5})^{4} & \mathbb{Z}_{5} & (\mathbb{Z}_{5})^{3} & \mathbb{Z}_{5} & 0 & 0 & (\mathbb{Z}_{5})^{2} \ar[lu,two heads,"{{5}}" {gray,pos=0.08,scale=0.8,below left},"{{5}}" {gray,pos=0.92,scale=0.8,above right}] \\
H^{20} & \mathbb{Z}_{5} & (\mathbb{Z}_{5})^{3} & (\mathbb{Z}_{5})^{4} & (\mathbb{Z}_{5})^{2} & (\mathbb{Z}_{5})^{3} & \mathbb{Z}_{5} & 0 & 0 & \mathbb{Z}_{5} \\
H^{21} & \mathbb{Z}_{5} & (\mathbb{Z}_{5})^{3} & \mathbb{Z}_{5} & (\mathbb{Z}_{5})^{5} \ar[lu,"{{5}}" {gray,pos=0.08,scale=0.8,below left},"{{5}}" {gray,pos=0.92,scale=0.8,above right}] & (\mathbb{Z}_{5})^{2} \ar[lu,"{{5}}" {gray,pos=0.08,scale=0.8,below left},"{{5}}" {gray,pos=0.92,scale=0.8,above right}] & (\mathbb{Z}_{5})^{3} & 0 & 0 & 0 \\
H^{22} & \mathbb{Z}_{5} & (\mathbb{Z}_{5})^{3} & \mathbb{Z}_{5} & (\mathbb{Z}_{5})^{4} & (\mathbb{Z}_{5})^{2} \ar[lu,"{{5}}" {gray,pos=0.08,scale=0.8,below left},"{{5}}" {gray,pos=0.92,scale=0.8,above right}] & (\mathbb{Z}_{5})^{4} \ar[lu,"{{5}}" {gray,pos=0.08,scale=0.8,below left},"{{5}}" {gray,pos=0.92,scale=0.8,above right}] & \mathbb{Z}_{5} & 0 & 0 \\
H^{23} & \mathbb{Z}_{5} & (\mathbb{Z}_{5})^{4} & (\mathbb{Z}_{5})^{6} & \mathbb{Z}_{5} & (\mathbb{Z}_{5})^{2} \ar[lu,"{{5}}" {gray,pos=0.08,scale=0.8,below left},"{{5}}" {gray,pos=0.92,scale=0.8,above right}] & (\mathbb{Z}_{5})^{3} \ar[lu,two heads,"{{5^{2}}}" {gray,pos=0.08,scale=0.8,below left},"{{5^{2}}}" {gray,pos=0.92,scale=0.8,above right}] & (\mathbb{Z}_{5})^{5} \ar[lu,"{{5}}" {gray,pos=0.08,scale=0.8,below left},"{{5}}" {gray,pos=0.92,scale=0.8,above right}] & 0 & 0 \\
H^{24} & \mathbb{Z}_{5} & (\mathbb{Z}_{5})^{4} & (\mathbb{Z}_{5})^{8} & (\mathbb{Z}_{5})^{3} & (\mathbb{Z}_{5})^{2} & (\mathbb{Z}_{5})^{2} \ar[lu,"{{5}}" {gray,pos=0.08,scale=0.8,below left},"{{5}}" {gray,pos=0.92,scale=0.8,above right}] & (\mathbb{Z}_{5})^{6} \ar[lu,"{{5^{2}}}" {gray,pos=0.08,scale=0.8,below left},"{{5^{2}}}" {gray,pos=0.92,scale=0.8,above right}] & (\mathbb{Z}_{5})^{3} \ar[lu,"{{5}}" {gray,pos=0.08,scale=0.8,below left},"{{5}}" {gray,pos=0.92,scale=0.8,above right}] & 0 \\
H^{25} & \mathbb{Z}_{5} & (\mathbb{Z}_{5})^{4} & (\mathbb{Z}_{5})^{3} & (\mathbb{Z}_{5})^{7} & (\mathbb{Z}_{5})^{3} & \mathbb{Z}_{5} & (\mathbb{Z}_{5})^{2} \ar[lu,"{{5}}" {gray,pos=0.08,scale=0.8,below left},"{{5}}" {gray,pos=0.92,scale=0.8,above right}] & (\mathbb{Z}_{5})^{6} \ar[lu,"{{5^{2}}}" {gray,pos=0.08,scale=0.8,below left},"{{5^{2}}}" {gray,pos=0.92,scale=0.8,above right}] & \mathbb{Z}_{5} \ar[lu,hook,"{{5}}" {gray,pos=0.08,scale=0.8,below left},"{{5}}" {gray,pos=0.92,scale=0.8,above right}] \\
H^{26} & \mathbb{Z}_{5} & (\mathbb{Z}_{5})^{5} & (\mathbb{Z}_{5})^{3} & (\mathbb{Z}_{5})^{7} & (\mathbb{Z}_{5})^{4} & \mathbb{Z}_{5} & 0 & (\mathbb{Z}_{5})^{4} \ar[lu,"{{5}}" {gray,pos=0.08,scale=0.8,below left},"{{5}}" {gray,pos=0.92,scale=0.8,above right}] & (\mathbb{Z}_{5})^{3} \ar[lu,"{{5^{2}}}" {gray,pos=0.08,scale=0.8,below left},"{{5^{2}}}" {gray,pos=0.92,scale=0.8,above right}] \\
H^{27} & \mathbb{Z}_{5} & (\mathbb{Z}_{5})^{5} & (\mathbb{Z}_{5})^{10} & (\mathbb{Z}_{5})^{4} & (\mathbb{Z}_{5})^{6} & (\mathbb{Z}_{5})^{3} & 0 & \mathbb{Z}_{5} & (\mathbb{Z}_{5})^{4} \ar[lu,"{{5}}" {gray,pos=0.08,scale=0.8,below left},"{{5}}" {gray,pos=0.92,scale=0.8,above right}] \\
H^{28} & \mathbb{Z}_{5} & (\mathbb{Z}_{5})^{5} & (\mathbb{Z}_{5})^{10} & (\mathbb{Z}_{5})^{4} & (\mathbb{Z}_{5})^{7} & (\mathbb{Z}_{5})^{5} & 0 & 0 & (\mathbb{Z}_{5})^{3} \\
H^{29} & \mathbb{Z}_{5} & (\mathbb{Z}_{5})^{5} & (\mathbb{Z}_{5})^{3} & (\mathbb{Z}_{5})^{8} & (\mathbb{Z}_{5})^{4} & (\mathbb{Z}_{5})^{6} & 0 & 0 & (\mathbb{Z}_{5})^{2} \\
H^{30} & \mathbb{Z}_{5} & (\mathbb{Z}_{5})^{5} & (\mathbb{Z}_{5})^{3} & (\mathbb{Z}_{5})^{9} & (\mathbb{Z}_{5})^{2} & (\mathbb{Z}_{5})^{6} & (\mathbb{Z}_{5})^{2} & 0 & \mathbb{Z}_{5} \\
H^{31} & \mathbb{Z}_{5} & (\mathbb{Z}_{5})^{5} & (\mathbb{Z}_{5})^{10} & (\mathbb{Z}_{5})^{6} & (\mathbb{Z}_{5})^{3} & (\mathbb{Z}_{5})^{5} \ar[lu,"{{5}}" {gray,pos=0.08,scale=0.8,below left},"{{5}}" {gray,pos=0.92,scale=0.8,above right}] & (\mathbb{Z}_{5})^{10} \ar[lu,"{{5}}" {gray,pos=0.08,scale=0.8,below left},"{{5}}" {gray,pos=0.92,scale=0.8,above right}] & 0 & 0 \\
H^{32} & \mathbb{Z}_{5} & (\mathbb{Z}_{5})^{5} & (\mathbb{Z}_{5})^{10} & (\mathbb{Z}_{5})^{6} & (\mathbb{Z}_{5})^{6} & (\mathbb{Z}_{5})^{3} & (\mathbb{Z}_{5})^{13} \ar[lu,"{{5}}" {gray,pos=0.08,scale=0.8,below left},"{{5}}" {gray,pos=0.92,scale=0.8,above right}] & (\mathbb{Z}_{5})^{5} \ar[lu,"{{5}}" {gray,pos=0.08,scale=0.8,below left},"{{5}}" {gray,pos=0.92,scale=0.8,above right}] & 0 \\
H^{33} & \mathbb{Z}_{5} & (\mathbb{Z}_{5})^{5} & (\mathbb{Z}_{5})^{3} & (\mathbb{Z}_{5})^{10} & (\mathbb{Z}_{5})^{10} & (\mathbb{Z}_{5})^{3} & (\mathbb{Z}_{5})^{5} \ar[lu,"{{5}}" {gray,pos=0.08,scale=0.8,below left},"{{5}}" {gray,pos=0.92,scale=0.8,above right}] & (\mathbb{Z}_{5})^{11} \ar[lu,"{{5^{2}}}" {gray,pos=0.08,scale=0.8,below left},"{{5^{2}}}" {gray,pos=0.92,scale=0.8,above right}] & \mathbb{Z}_{5} \ar[lu,hook,"{{5}}" {gray,pos=0.08,scale=0.8,below left},"{{5}}" {gray,pos=0.92,scale=0.8,above right}] \\
H^{34} & \mathbb{Z}_{5} & (\mathbb{Z}_{5})^{5} & (\mathbb{Z}_{5})^{3} & (\mathbb{Z}_{5})^{12} & (\mathbb{Z}_{5})^{10} & (\mathbb{Z}_{5})^{5} & 0 & (\mathbb{Z}_{5})^{9} \ar[lu,"{{5}}" {gray,pos=0.08,scale=0.8,below left},"{{5}}" {gray,pos=0.92,scale=0.8,above right}] & (\mathbb{Z}_{5})^{3} \ar[lu,"{{5^{2}}}" {gray,pos=0.08,scale=0.8,below left},"{{5^{2}}}" {gray,pos=0.92,scale=0.8,above right}] \\
H^{35} & \mathbb{Z}_{5} & (\mathbb{Z}_{5})^{6} & (\mathbb{Z}_{5})^{13} & (\mathbb{Z}_{5})^{9} & (\mathbb{Z}_{5})^{8} & (\mathbb{Z}_{5})^{7} & 0 & (\mathbb{Z}_{5})^{4} & (\mathbb{Z}_{5})^{6} \ar[lu,"{{5}}" {gray,pos=0.08,scale=0.8,below left},"{{5}}" {gray,pos=0.92,scale=0.8,above right}] \\
H^{36} & \mathbb{Z}_{5} & (\mathbb{Z}_{5})^{6} & (\mathbb{Z}_{5})^{16} & (\mathbb{Z}_{5})^{8} & (\mathbb{Z}_{5})^{8} & (\mathbb{Z}_{5})^{9} & 0 & \mathbb{Z}_{5} & (\mathbb{Z}_{5})^{8} \\
H^{37} & \mathbb{Z}_{5} & (\mathbb{Z}_{5})^{6} & (\mathbb{Z}_{5})^{6} & (\mathbb{Z}_{5})^{12} & (\mathbb{Z}_{5})^{7} & (\mathbb{Z}_{5})^{11} & 0 & 0 & (\mathbb{Z}_{5})^{6} \\
H^{38} & \mathbb{Z}_{5} & (\mathbb{Z}_{5})^{7} & (\mathbb{Z}_{5})^{6} & (\mathbb{Z}_{5})^{15} & (\mathbb{Z}_{5})^{7} & (\mathbb{Z}_{5})^{12} & (\mathbb{Z}_{5})^{6} & 0 & (\mathbb{Z}_{5})^{3} \\
H^{39} & \mathbb{Z}_{5} & (\mathbb{Z}_{5})^{7} & (\mathbb{Z}_{5})^{19} & (\mathbb{Z}_{5})^{14} & (\mathbb{Z}_{5})^{11} & (\mathbb{Z}_{5})^{9} & (\mathbb{Z}_{5})^{20} & 0 & (\mathbb{Z}_{5})^{2} \\
H^{40} & \mathbb{Z}_{5} & (\mathbb{Z}_{5})^{7} & (\mathbb{Z}_{5})^{19} & (\mathbb{Z}_{5})^{13} & (\mathbb{Z}_{5})^{14} & (\mathbb{Z}_{5})^{6} & (\mathbb{Z}_{5})^{22} & (\mathbb{Z}_{5})^{6} & \mathbb{Z}_{5} \\
\end{tikzcd}

%% file: tables/omega-Z5-Z25.tex
\begin{tikzcd}[row sep=tiny,column sep=large]
& B\mathbb{Z}_{5} & B^{2}\mathbb{Z}_{5} & B^{3}\mathbb{Z}_{5} & B^{4}\mathbb{Z}_{5} & B^{5}\mathbb{Z}_{5} & B^{6}\mathbb{Z}_{5} & B^{7}\mathbb{Z}_{5} \\
H^{0} & \mathbb{Z}_{25} & \mathbb{Z}_{25} & \mathbb{Z}_{25} & \mathbb{Z}_{25} & \mathbb{Z}_{25} & \mathbb{Z}_{25} & \mathbb{Z}_{25} \\
H^{1} & \mathbb{Z}_{5} & 0 & 0 & 0 & 0 & 0 & 0 \\
H^{2} & \mathbb{Z}_{5} & \mathbb{Z}_{5} \ar[lu,BrickRed,hook,two heads,"{{5}}" {gray,pos=0.08,scale=0.8,below left},"{{5}}" {gray,pos=0.92,scale=0.8,above right}] & 0 & 0 & 0 & 0 & 0 \\
H^{3} & \mathbb{Z}_{5} & \mathbb{Z}_{5} \ar[lu,hook,two heads,"{{5}}" {gray,pos=0.08,scale=0.8,below left},"{{5}}" {gray,pos=0.92,scale=0.8,above right}] & \mathbb{Z}_{5} \ar[lu,BrickRed,hook,two heads,"{{5}}" {gray,pos=0.08,scale=0.8,below left},"{{5}}" {gray,pos=0.92,scale=0.8,above right}] & 0 & 0 & 0 & 0 \\
H^{4} & \mathbb{Z}_{5} & \mathbb{Z}_{5} & \mathbb{Z}_{5} \ar[lu,BrickRed,hook,two heads,"{{5}}" {gray,pos=0.08,scale=0.8,below left},"{{5}}" {gray,pos=0.92,scale=0.8,above right}] & \mathbb{Z}_{5} \ar[lu,BrickRed,hook,two heads,"{{5}}" {gray,pos=0.08,scale=0.8,below left},"{{5}}" {gray,pos=0.92,scale=0.8,above right}] & 0 & 0 & 0 \\
H^{5} & \mathbb{Z}_{5} & \mathbb{Z}_{5} & 0 & \mathbb{Z}_{5} \ar[lu,BrickRed,hook,two heads,"{{5}}" {gray,pos=0.08,scale=0.8,below left},"{{5}}" {gray,pos=0.92,scale=0.8,above right}] & \mathbb{Z}_{5} \ar[lu,BrickRed,hook,two heads,"{{5}}" {gray,pos=0.08,scale=0.8,below left},"{{5}}" {gray,pos=0.92,scale=0.8,above right}] & 0 & 0 \\
H^{6} & \mathbb{Z}_{5} & \mathbb{Z}_{5} & 0 & 0 & \mathbb{Z}_{5} \ar[lu,BrickRed,hook,two heads,"{{5}}" {gray,pos=0.08,scale=0.8,below left},"{{5}}" {gray,pos=0.92,scale=0.8,above right}] & \mathbb{Z}_{5} \ar[lu,BrickRed,hook,two heads,"{{5}}" {gray,pos=0.08,scale=0.8,below left},"{{5}}" {gray,pos=0.92,scale=0.8,above right}] & 0 \\
H^{7} & \mathbb{Z}_{5} & \mathbb{Z}_{5} & \mathbb{Z}_{5} & 0 & 0 & \mathbb{Z}_{5} \ar[lu,BrickRed,hook,two heads,"{{5}}" {gray,pos=0.08,scale=0.8,below left},"{{5}}" {gray,pos=0.92,scale=0.8,above right}] & \mathbb{Z}_{5} \ar[lu,BrickRed,hook,two heads,"{{5}}" {gray,pos=0.08,scale=0.8,below left},"{{5}}" {gray,pos=0.92,scale=0.8,above right}] \\
H^{8} & \mathbb{Z}_{5} & \mathbb{Z}_{5} & \mathbb{Z}_{5} & \mathbb{Z}_{5} & 0 & 0 & \mathbb{Z}_{5} \ar[lu,BrickRed,hook,two heads,"{{5}}" {gray,pos=0.08,scale=0.8,below left},"{{5}}" {gray,pos=0.92,scale=0.8,above right}] \\
H^{9} & \mathbb{Z}_{5} & \mathbb{Z}_{5} & 0 & \mathbb{Z}_{5} & 0 & 0 & 0 \\
H^{10} & \mathbb{Z}_{5} & \mathbb{Z}_{25} & 0 & 0 & 0 & 0 & 0 \\
H^{11} & \mathbb{Z}_{5} & \mathbb{Z}_{5} \oplus \mathbb{Z}_{25} & (\mathbb{Z}_{5})^{2} \ar[lu,"{{5}}" {gray,pos=0.08,scale=0.8,below left},"{{5}}" {gray,pos=0.92,scale=0.8,above right}] & 0 & \mathbb{Z}_{5} & 0 & 0 \\
H^{12} & \mathbb{Z}_{5} & (\mathbb{Z}_{5})^{2} & (\mathbb{Z}_{5})^{3} \ar[lu,"{{5^{2}}}" {gray,pos=0.08,scale=0.8,below left},"{{5^{2}}}" {gray,pos=0.92,scale=0.8,above right}] & (\mathbb{Z}_{5})^{2} \ar[lu,"{{5}}" {gray,pos=0.08,scale=0.8,below left},"{{5}}" {gray,pos=0.92,scale=0.8,above right}] & \mathbb{Z}_{5} & \mathbb{Z}_{5} & 0 \\
H^{13} & \mathbb{Z}_{5} & (\mathbb{Z}_{5})^{2} & \mathbb{Z}_{5} \ar[lu,hook,"{{5}}" {gray,pos=0.08,scale=0.8,below left},"{{5}}" {gray,pos=0.92,scale=0.8,above right}] & (\mathbb{Z}_{5})^{3} \ar[lu,"{{5^{2}}}" {gray,pos=0.08,scale=0.8,below left},"{{5^{2}}}" {gray,pos=0.92,scale=0.8,above right}] & \mathbb{Z}_{5} \ar[lu,hook,"{{5}}" {gray,pos=0.08,scale=0.8,below left},"{{5}}" {gray,pos=0.92,scale=0.8,above right}] & \mathbb{Z}_{5} & 0 \\
H^{14} & \mathbb{Z}_{5} & (\mathbb{Z}_{5})^{3} & \mathbb{Z}_{5} & \mathbb{Z}_{5} \ar[lu,hook,two heads,"{{5}}" {gray,pos=0.08,scale=0.8,below left},"{{5}}" {gray,pos=0.92,scale=0.8,above right}] & (\mathbb{Z}_{5})^{2} \ar[lu,hook,"{{5^{2}}}" {gray,pos=0.08,scale=0.8,below left},"{{5^{2}}}" {gray,pos=0.92,scale=0.8,above right}] & \mathbb{Z}_{5} \ar[lu,hook,two heads,"{{5}}" {gray,pos=0.08,scale=0.8,below left},"{{5}}" {gray,pos=0.92,scale=0.8,above right}] & 0 \\
H^{15} & \mathbb{Z}_{5} & (\mathbb{Z}_{5})^{3} & (\mathbb{Z}_{5})^{4} & 0 & \mathbb{Z}_{5} \ar[lu,hook,two heads,"{{5}}" {gray,pos=0.08,scale=0.8,below left},"{{5}}" {gray,pos=0.92,scale=0.8,above right}] & (\mathbb{Z}_{5})^{2} \ar[lu,hook,two heads,"{{5^{2}}}" {gray,pos=0.08,scale=0.8,below left},"{{5^{2}}}" {gray,pos=0.92,scale=0.8,above right}] & (\mathbb{Z}_{5})^{2} \ar[lu,two heads,"{{5}}" {gray,pos=0.08,scale=0.8,below left},"{{5}}" {gray,pos=0.92,scale=0.8,above right}] \\
H^{16} & \mathbb{Z}_{5} & (\mathbb{Z}_{5})^{3} & (\mathbb{Z}_{5})^{4} & (\mathbb{Z}_{5})^{2} & 0 & \mathbb{Z}_{5} \ar[lu,hook,two heads,"{{5}}" {gray,pos=0.08,scale=0.8,below left},"{{5}}" {gray,pos=0.92,scale=0.8,above right}] & (\mathbb{Z}_{5})^{3} \ar[lu,two heads,"{{5^{2}}}" {gray,pos=0.08,scale=0.8,below left},"{{5^{2}}}" {gray,pos=0.92,scale=0.8,above right}] \\
H^{17} & \mathbb{Z}_{5} & (\mathbb{Z}_{5})^{3} & \mathbb{Z}_{5} & (\mathbb{Z}_{5})^{4} & \mathbb{Z}_{5} & 0 & \mathbb{Z}_{5} \ar[lu,hook,two heads,"{{5}}" {gray,pos=0.08,scale=0.8,below left},"{{5}}" {gray,pos=0.92,scale=0.8,above right}] \\
H^{18} & \mathbb{Z}_{5} & (\mathbb{Z}_{5})^{3} & \mathbb{Z}_{5} & (\mathbb{Z}_{5})^{3} & (\mathbb{Z}_{5})^{2} & \mathbb{Z}_{5} & 0 \\
H^{19} & \mathbb{Z}_{5} & (\mathbb{Z}_{5})^{3} & (\mathbb{Z}_{5})^{4} & \mathbb{Z}_{5} & (\mathbb{Z}_{5})^{3} & \mathbb{Z}_{5} & 0 \\
H^{20} & \mathbb{Z}_{5} & (\mathbb{Z}_{5})^{2} \oplus \mathbb{Z}_{25} & (\mathbb{Z}_{5})^{4} & \mathbb{Z}_{5} \oplus \mathbb{Z}_{25} & (\mathbb{Z}_{5})^{3} & \mathbb{Z}_{5} & 0 \\
H^{21} & \mathbb{Z}_{5} & (\mathbb{Z}_{5})^{2} \oplus \mathbb{Z}_{25} & \mathbb{Z}_{5} & (\mathbb{Z}_{5})^{4} \oplus \mathbb{Z}_{25} & (\mathbb{Z}_{5})^{2} \ar[lu,"{{5}}" {gray,pos=0.08,scale=0.8,below left},"{{5}}" {gray,pos=0.92,scale=0.8,above right}] & (\mathbb{Z}_{5})^{3} & 0 \\
H^{22} & \mathbb{Z}_{5} & (\mathbb{Z}_{5})^{3} & \mathbb{Z}_{5} & (\mathbb{Z}_{5})^{4} & (\mathbb{Z}_{5})^{2} \ar[lu,hook,"{{5^{2}}}" {gray,pos=0.08,scale=0.8,below left},"{{5^{2}}}" {gray,pos=0.92,scale=0.8,above right}] & (\mathbb{Z}_{5})^{4} \ar[lu,"{{5}}" {gray,pos=0.08,scale=0.8,below left},"{{5}}" {gray,pos=0.92,scale=0.8,above right}] & \mathbb{Z}_{5} \\
H^{23} & \mathbb{Z}_{5} & (\mathbb{Z}_{5})^{4} & (\mathbb{Z}_{5})^{6} & \mathbb{Z}_{5} & (\mathbb{Z}_{5})^{2} \ar[lu,"{{5}}" {gray,pos=0.08,scale=0.8,below left},"{{5}}" {gray,pos=0.92,scale=0.8,above right}] & (\mathbb{Z}_{5})^{3} \ar[lu,two heads,"{{5^{2}}}" {gray,pos=0.08,scale=0.8,below left},"{{5^{2}}}" {gray,pos=0.92,scale=0.8,above right}] & (\mathbb{Z}_{5})^{5} \ar[lu,"{{5}}" {gray,pos=0.08,scale=0.8,below left},"{{5}}" {gray,pos=0.92,scale=0.8,above right}] \\
H^{24} & \mathbb{Z}_{5} & (\mathbb{Z}_{5})^{4} & (\mathbb{Z}_{5})^{8} & (\mathbb{Z}_{5})^{3} & (\mathbb{Z}_{5})^{2} & (\mathbb{Z}_{5})^{2} \ar[lu,"{{5}}" {gray,pos=0.08,scale=0.8,below left},"{{5}}" {gray,pos=0.92,scale=0.8,above right}] & (\mathbb{Z}_{5})^{6} \ar[lu,"{{5^{2}}}" {gray,pos=0.08,scale=0.8,below left},"{{5^{2}}}" {gray,pos=0.92,scale=0.8,above right}] \\
H^{25} & \mathbb{Z}_{5} & (\mathbb{Z}_{5})^{4} & (\mathbb{Z}_{5})^{3} & (\mathbb{Z}_{5})^{7} & (\mathbb{Z}_{5})^{3} & \mathbb{Z}_{5} & (\mathbb{Z}_{5})^{2} \ar[lu,"{{5}}" {gray,pos=0.08,scale=0.8,below left},"{{5}}" {gray,pos=0.92,scale=0.8,above right}] \\
H^{26} & \mathbb{Z}_{5} & (\mathbb{Z}_{5})^{5} & (\mathbb{Z}_{5})^{3} & (\mathbb{Z}_{5})^{7} & (\mathbb{Z}_{5})^{4} & \mathbb{Z}_{5} & 0 \\
H^{27} & \mathbb{Z}_{5} & (\mathbb{Z}_{5})^{5} & (\mathbb{Z}_{5})^{10} & (\mathbb{Z}_{5})^{4} & (\mathbb{Z}_{5})^{6} & (\mathbb{Z}_{5})^{3} & 0 \\
H^{28} & \mathbb{Z}_{5} & (\mathbb{Z}_{5})^{5} & (\mathbb{Z}_{5})^{10} & (\mathbb{Z}_{5})^{4} & (\mathbb{Z}_{5})^{7} & (\mathbb{Z}_{5})^{5} & 0 \\
H^{29} & \mathbb{Z}_{5} & (\mathbb{Z}_{5})^{5} & (\mathbb{Z}_{5})^{3} & (\mathbb{Z}_{5})^{8} & (\mathbb{Z}_{5})^{4} & (\mathbb{Z}_{5})^{6} & 0 \\
H^{30} & \mathbb{Z}_{5} & (\mathbb{Z}_{5})^{4} \oplus \mathbb{Z}_{25} & (\mathbb{Z}_{5})^{3} & (\mathbb{Z}_{5})^{9} & (\mathbb{Z}_{5})^{2} & (\mathbb{Z}_{5})^{5} \oplus \mathbb{Z}_{25} & (\mathbb{Z}_{5})^{2} \\
H^{31} & \mathbb{Z}_{5} & (\mathbb{Z}_{5})^{4} \oplus \mathbb{Z}_{25} & (\mathbb{Z}_{5})^{10} & (\mathbb{Z}_{5})^{6} & (\mathbb{Z}_{5})^{3} & (\mathbb{Z}_{5})^{4} \oplus \mathbb{Z}_{25} & (\mathbb{Z}_{5})^{10} \ar[lu,"{{5}}" {gray,pos=0.08,scale=0.8,below left},"{{5}}" {gray,pos=0.92,scale=0.8,above right}] \\
H^{32} & \mathbb{Z}_{5} & (\mathbb{Z}_{5})^{5} & (\mathbb{Z}_{5})^{10} & (\mathbb{Z}_{5})^{6} & (\mathbb{Z}_{5})^{6} & (\mathbb{Z}_{5})^{3} & (\mathbb{Z}_{5})^{13} \ar[lu,"{{5^{2}}}" {gray,pos=0.08,scale=0.8,below left},"{{5^{2}}}" {gray,pos=0.92,scale=0.8,above right}] \\
H^{33} & \mathbb{Z}_{5} & (\mathbb{Z}_{5})^{5} & (\mathbb{Z}_{5})^{3} & (\mathbb{Z}_{5})^{10} & (\mathbb{Z}_{5})^{10} & (\mathbb{Z}_{5})^{3} & (\mathbb{Z}_{5})^{5} \ar[lu,"{{5}}" {gray,pos=0.08,scale=0.8,below left},"{{5}}" {gray,pos=0.92,scale=0.8,above right}] \\
H^{34} & \mathbb{Z}_{5} & (\mathbb{Z}_{5})^{5} & (\mathbb{Z}_{5})^{3} & (\mathbb{Z}_{5})^{12} & (\mathbb{Z}_{5})^{10} & (\mathbb{Z}_{5})^{5} & 0 \\
H^{35} & \mathbb{Z}_{5} & (\mathbb{Z}_{5})^{6} & (\mathbb{Z}_{5})^{13} & (\mathbb{Z}_{5})^{9} & (\mathbb{Z}_{5})^{8} & (\mathbb{Z}_{5})^{7} & 0 \\
H^{36} & \mathbb{Z}_{5} & (\mathbb{Z}_{5})^{6} & (\mathbb{Z}_{5})^{16} & (\mathbb{Z}_{5})^{8} & (\mathbb{Z}_{5})^{8} & (\mathbb{Z}_{5})^{9} & 0 \\
H^{37} & \mathbb{Z}_{5} & (\mathbb{Z}_{5})^{6} & (\mathbb{Z}_{5})^{6} & (\mathbb{Z}_{5})^{12} & (\mathbb{Z}_{5})^{7} & (\mathbb{Z}_{5})^{11} & 0 \\
H^{38} & \mathbb{Z}_{5} & (\mathbb{Z}_{5})^{7} & (\mathbb{Z}_{5})^{6} & (\mathbb{Z}_{5})^{15} & (\mathbb{Z}_{5})^{7} & (\mathbb{Z}_{5})^{12} & (\mathbb{Z}_{5})^{6} \\
H^{39} & \mathbb{Z}_{5} & (\mathbb{Z}_{5})^{7} & (\mathbb{Z}_{5})^{19} & (\mathbb{Z}_{5})^{14} & (\mathbb{Z}_{5})^{11} & (\mathbb{Z}_{5})^{9} & (\mathbb{Z}_{5})^{20} \\
H^{40} & \mathbb{Z}_{5} & (\mathbb{Z}_{5})^{6} \oplus \mathbb{Z}_{25} & (\mathbb{Z}_{5})^{19} & (\mathbb{Z}_{5})^{12} \oplus \mathbb{Z}_{25} & (\mathbb{Z}_{5})^{14} & (\mathbb{Z}_{5})^{6} & (\mathbb{Z}_{5})^{22} \\
\end{tikzcd}